\documentclass{article}
\usepackage{amssymb}
\usepackage{amsmath}
\usepackage{graphicx}
\usepackage{amsfonts}

\newtheorem{theorem}{Theorem}

\newtheorem{conjecture}[theorem]{Conjecture}
\newtheorem{corollary}[theorem]{Corollary}

\newtheorem{definition}[theorem]{Definition}
\newtheorem{example}[theorem]{Example}

\newtheorem{lemma}[theorem]{Lemma}

\newtheorem{problem}[theorem]{Problem}
\newtheorem{proposition}[theorem]{Proposition}
\newtheorem{remark}[theorem]{Remark}

\newtheorem{summary}[theorem]{Summary}
\newenvironment{proof}[1][Proof]{\noindent\textbf{#1.} }{\ \rule{0.5em}{0.5em}}
\input{tcilatex}

\begin{document}

\title{Kinematic geometry of triangles and the study of the three-body \
problem }
\author{Wu-Yi Hsiang \\
Department of mathematics\\
University of California, Berkeley \and Eldar Straume \\
Department of mathematical sciences\\
Norwegian University of Science and Technology\\
Trondheim, Norway\\
e-mail : eldars$@$math.ntnu.no}
\maketitle
\tableofcontents

\section{ Introduction}

The classical three-body problem studies the motion of a system with three
point masses under the action of the Newtonian gravitational potential. Let $%
\mathbf{a}_{i}=\overrightarrow{OP_{i}}$, $i=1,2,3,$ be the position vectors
of the points $P_{i}$ with masses $m_{i}>0,$ with respect to a chosen
inertial coordinate system for the Euclidean 3-space $\mathbb{R}^{3}$. Then
a motion of the three point masses will be described as a curve in the
(unrestricted) \emph{configuration space}, namely the Euclidean space $%
\mathbb{R}^{9}$ consisting of all triples $(\mathbf{a}_{1},\mathbf{a}_{2},%
\mathbf{a}_{3})$.

Newton's equation of motion is the following second order system of three
vector differential equations 
\begin{equation}
m_{i}\mathbf{\ddot{a}}_{i}=\frac{\partial U}{\partial\mathbf{a}_{i}}=\ \frac{%
m_{i}m_{j}\ }{\ r_{ij}^{3}\ }(\mathbf{a}_{j}-\mathbf{a}_{i})+\frac{m_{i}m_{k}%
}{r_{ik}^{3}}(\mathbf{a}_{k}-\mathbf{a}_{i})\text{ \ }\   \label{Newton1}
\end{equation}
where $\left\{ i,j,k\right\} =\left\{ 1,2,3\right\} ,$ and 
\begin{equation}
U=\frac{m_{1}m_{2}}{r_{12}}+\frac{m_{2}m_{3}}{r_{23}}+\frac{m_{1}m_{3}}{%
r_{13}},\text{ \ \ }r_{ij}=\left\| \mathbf{a}_{i}-\mathbf{a}_{j}\right\| , 
\label{U1}
\end{equation}
is the Newtonian potential function. In brief, the central problem is to
understand both the geometry and the analysis of the solutions of the above
system (\ref{Newton1}), where each solution curve (or trajectory) is
uniquely determined by the initial positions and velocities of the point
masses.

\subsection{The classical conservation laws}

The equations (\ref{Newton1}) amount to solving a dynamical system in phase
space of dimension 18. It is easily seen, however, that the classical 3-body
problem (in fact, the n-body problem for all $n$) is invariant under
Galilean transformations, namely the 10-dimensional Galilean group of
4-dimensional space-time $\mathbb{R}^{3}\times\mathbb{R}$. Accordingly,
there are 10 first integrals or \emph{conservation laws, }and they actually
reduce the integration problem to one of order $18-10=8$. These are the
conservation of linear momentum, angular momentum and total energy, and they
are easily deduced as follows. First, by adding the three vector equations (%
\ref{Newton1}) we have 
\begin{equation*}
m_{1}\mathbf{\ddot{a}}_{1}+m_{2}\mathbf{\ddot{a}}_{2}+m_{3}\mathbf{\ddot{a}}%
_{3}=0 
\end{equation*}
and hence the center of mass has the uniform motion 
\begin{equation*}
\mathbf{a}_{\text{\textsc{CM}}}=\frac{1}{\bar{m}}(m_{1}\mathbf{a}_{1}+m_{2}%
\mathbf{a}_{2}+m_{3}\mathbf{a}_{3})=\mathbf{p}_{0}+t\mathbf{v}_{0},\text{ \ }%
\bar{m}=\tsum m_{i}
\end{equation*}
where $\mathbf{p}_{0}$ and $\mathbf{v}_{0}$ are two constant vectors (and
hence six scalar conservation laws) determined by the initial data. To make
effective use of these we recall the Galilean principle of relativity of
Newtonian mechanics, according to which one may choose an equivalent
inertial frame of reference with origin at $\mathbf{a}_{\text{\textsc{CM}}}$
and axes in the same directions as before (or rotated). This justifies using
a center of mass inertial reference frame, which effectively reduces the
actual configuration space to the following $6$-dimensional Euclidean space 
\begin{equation}
M_{0}=\mathbb{R}^{6}\subset\mathbb{R}^{9}:\sum m_{i}\mathbf{a}_{i}=0 
\label{M0}
\end{equation}

Next, the vector 
\begin{equation}
\mathbf{\Omega=}\ m_{1}\mathbf{a}_{1}\times\mathbf{\dot{a}}_{1}+m_{2}\mathbf{%
a}_{2}\times\mathbf{\dot{a}}_{2}+m_{3}\mathbf{a}_{3}\times \mathbf{\dot{a}}%
_{3}   \label{angmom}
\end{equation}
is the total \emph{angular momentum }of the system. It varies covariantly
with rotations of 3-space, and it is also constant along a trajectory since
by (\ref{Newton1}) 
\begin{equation*}
\mathbf{\dot{\Omega}=}\tsum \limits_{i}m_{i}\mathbf{a}_{i}\times\mathbf{%
\ddot{a}}_{i}=\tsum \limits_{i}m_{i}\mathbf{a}_{i}\times\mathbf{(}\tsum
\limits_{j\neq i}\frac{m_{j}\ }{\ r_{ij}^{3}\ }(\mathbf{a}_{j}-\mathbf{a}%
_{i}))=0\text{ \ }
\end{equation*}
This is the law of conservation of angular momentum, whose geometric
significance lies much deeper than that of conservation of linear momentum.

Finally, the total energy is defined to be 
\begin{equation}
h=T-U   \label{h}
\end{equation}
where 
\begin{equation}
T=\frac{1}{2}(m_{1}\left\vert \mathbf{\dot{a}}_{1}\right\vert
^{2}+m_{2}\left\vert \mathbf{\dot{a}}_{2}\right\vert ^{2}+m_{3}\left\vert 
\mathbf{\dot{a}}_{3}\right\vert ^{2})   \label{T}
\end{equation}
is the \emph{kinetic energy }and $-U$ is the \emph{potential energy.}
Straightforward differentiation using Newton's equation (\ref{Newton1})
gives 
\begin{equation*}
\dot{h}=\dot{T}-\dot{U}=\tsum m_{i}\mathbf{\dot{a}}_{i}\cdot\mathbf{\ddot{a}}%
_{i}-\tsum \frac{\partial U}{\partial\mathbf{a}_{i}}\cdot\mathbf{\dot{a}}%
_{i}=\tsum \mathbf{\dot{a}}_{i}\cdot(m_{i}\mathbf{\ddot{a}}_{i}-\frac{%
\partial U}{\partial\mathbf{a}_{i}})=0 
\end{equation*}
Hence, $h$ is constant along a solution curve of (\ref{Newton1}) and this is
the law of conservation of energy.

\subsection{Least action principles}

Newton's equation provides a characterization of the motion from the
''differential '' viewpoint, but it is also useful to characterize the
motion as a boundary value problem, namely we ask

\begin{description}
\item \qquad\emph{for a given pair }$\left\{ p,q\right\} $ \emph{in the
configuration space} \emph{and time interval }$(t_{1},t_{2})$, \emph{what
are those trajectories }$\Gamma(t),t_{1}\leq t\leq t_{2}$, \emph{with }$%
\Gamma(t_{1})=p,\Gamma(t_{2})=q?$
\end{description}

The idea of seeking solutions of the above problem as the extremals of a 
\emph{variational principle }applied to virtual motions dates back to the
17th century, inspired by the success of Fermat's \emph{principle of least
time }in geometric optics. Thus, a type of least action principle for
classical mechanics was proposed already by Leibniz, Euler and Maupertuis.
It was, however, Lagrange who finally provided a precise mathematical
formulation :

\begin{description}
\item \textbf{Lagrange's least action principle. }The solutions of the above
boundary value problem are characterized by the variational principle of
extremizing the action 
\begin{equation}
J_{1}\left[ \Gamma\right] =\int_{\Gamma}Tdt   \label{J10}
\end{equation}

among all virtual motions between a given pair of points and with the same
constant total energy $h.$
\end{description}

It should be noted that time is allowed to vary in the above integral, that
is, the limit of integration is not fixed. This awkwardness led Jacobi to
reformulate the least action principle to the problem of determining the
geodesics on a suitably defined Riemannian manifold (in modern terminology).
In his famous lectures on mechanics $\cite{Jacobi}$, Jacobi essentially
introduced the notion of \emph{kinematic metric} on the configuration space $%
M_{0}$, namely the kinetic energy expression (\ref{T}) defines a mass
dependent Euclidean metric 
\begin{equation}
ds^{2}=2Tdt^{2}=\tsum m_{i}(dx_{i}^{2}+dy_{i}^{2}+dz_{i}^{2}) 
\label{metric1}
\end{equation}
where $\mathbf{a}_{i}=(x_{i},y_{i},z_{i})$ is the position vector of the
point $P_{i}$ with mass $m_{i}.$

On the other hand, $T=U+h$ is also a function on $M_{0}$ for a fixed energy
level $h$, and this explains the following step:

\begin{description}
\item \textbf{Jacobi's reformulation of Lagrange's least action principle}.
The action integral \textbf{\ } 
\begin{equation}
\sqrt{2}J_{1}\left[ \Gamma\right] =\sqrt{2}\int_{\Gamma}Tdt=\int_{\Gamma }%
\sqrt{T}ds=\int_{\Gamma}\sqrt{U+h}ds=\int_{\Gamma}ds_{h}   \label{J11}
\end{equation}
is the arc-length of the virtual motion $\Gamma$ in the metric space $%
(M_{h},ds_{h}^{2})$, namely in the subspace 
\begin{equation}
M_{h}=\left\{ p\in M_{0};U(p)+h>0\right\} \subset M_{0}   \label{Mh}
\end{equation}
with the squared arc-length element 
\begin{equation}
ds_{h}^{2}=(U+h)ds^{2}   \label{metric4}
\end{equation}
\end{description}

Thus, according to Jacobi's ''geometrization trick'', trajectories of
Newton's equation are precisely the geodesics in the above Riemannian space (%
$M_{h},ds_{h}^{2})$. This also demonstrates why Jacobi in his study of
mechanics, in fact, anticipated the general notion of a Riemannian metric.
For more information on these issues we refer to L\"{u}tzen\cite{Lutzen}.

On the other hand, in 1840 Hamilton formulated another least action
principle, also inspired by the results of geometric optics.

\begin{description}
\item \textbf{Hamilton's principle of least action. }The solutions of the
above boundary value problem are characterized by the variational principle
of extremizing the action integral 
\begin{equation}
J_{2}\left[ \Gamma\right] =\int_{t_{1}}^{t_{2}}Ldt\text{ , \ \ }L=T+U 
\label{J2}
\end{equation}
among all virtual motions $\Gamma(t)$ between a given pair of points, for a
fixed time interval $\left[ t_{1},t_{2}\right] .$
\end{description}

In general, the validity of the ''integral''\ viewpoint represented by any
chosen variational principle is verified by calculating its infinitesimal
limit, which must coincide with (or be equivalent to) Newton's equation. In
the case of (\ref{J11}) and (\ref{J2}) respectively, this amounts to the
calculation of the geodesic equations of the metric $ds_{h}$ and the
associated \emph{Euler-Lagrange} equations of the above Lagrangian function $%
L$, respectively. In both cases it is easily checked that these are
equivalent to Newton's equation.

\subsection{An alternative geometric approach}

Traditionally, the three-body problem is usually studied in the framework of
Hamiltonian mechanics, canonical transformations and symplectic geometry,
based on the least action principle of Hamilton and the Hamilton-Jacobi
theory. Moreover, the specific dynamics due to the Newtonian forces is
usually assumed from the very beginning. Our present approach is, however,
different from this, roughly for two major reasons :

\begin{itemize}
\item Firstly, we focus attention on the purely kinematic properties of
virtual three-body motions in a Riemannian geometric setting and in the
framework of equivariant differential geometry, and

\item secondly, the Newtonian dynamics is introduced as the final step,
involving geometric reduction and conformal modification of the kinematic
Riemannian structure, based on the least action principle of Lagrange and
Jacobi (cf. Jacobi\cite{Jacobi}, Lecture 6).
\end{itemize}

Guided by the above program, the first author initiated studies in 1993 and
was joined by the second author in 1994. The basic material, covering their
work up to the winter of 1995, was presented in the two preprints \cite{1994}%
, \cite{1995}, and further studies of the n-body problem continued in the
following years. However, the two basic preprints were never published and,
unfortunately, they have had a rather limited circulation in the
mathematical community.

On the other hand, in the recent years new and beautiful results on the
three-body problem, and the more general n-body problem as well, have
appeared in the literature, some of which are deeply related to the above
geometric approach. This clearly suggests that new and unsolved problems
along these lines are now becoming more feasible. To further stimulate this
trend we propose hereby a review of the works \cite{1994}, \cite{1995} from
1994-95, and Chapter 1 -7 of the present monograph is, indeed, merely a
faithful presentation of their actual contents.

The exposition has been updated by some changes in notation and terminology,
together with a restructuring of ideas and proofs in order to unify and
enhance the readiblity of the presentation. For the convenience of the
reader, central results are now formulated as main theorems, such as Theorem
A, B, ...,F,\ G presented in Section 2.2.

In the final Chapter 8 we have included some additional and selected
unpublished material from 1995, mostly concerning the moduli curves of
triple collision motions in the special case of energy $h=0$ (and, for
technical reasons, the case of uniform mass distribution). The more general
case is stated as an open problem in the last section. In this chapter we
have done some explicit calculations which also serve as an illustration of
how to apply the setting and the results up to Chapter 7.

The three-body problem has a vast literature\ with many excellent papers. In
particular, the theoretical analysis of the problem and its influence during
the last century is overwhelming. Our interest in the problem started with a
study of Siegel's monumental analysis of triple collisions involving clever
applications of canonical transformations in the Hamiltonian setting.
However, we were also astounded by the lack of basic geometric reasoning
more directly linked to the kinematic geometry of three-body configurations,
and apparently, this seemed to be typical in the more recent literature
(such as Marchal's book), of which we had only superficial knowledge.
Perhaps it was, after all, worthwhile having a closer look at the underlying
geometric structure of mass triangles and their motions in 3-space ?

With this ambition we started, hopefully with no prejudice due to the
existing literature, and this lead us to the purely kinematic study
described in the first five chapters of this monograph, together with some
preliminary investigations of the ensuing dynamics due to gravitational
forces, most of which is presented in Chapter 7 and 8.

Thus, in our ''blindfolded'' study during 1994-95 we deliberately avoided
and did not consult any paper on the three-body problem (except Siegel's
work).This explains why the list of references were almost empty, and we
must apologize for that. The present list purposely reflects this former
state of affairs, but now there are at least some relevant titles which were
available prior to 1995 and which may be useful for the reader. Some of
these are also standard references of historical interest. In retrospect,
various topics and results discussed in this monograph are certainly more or
less treated by the many authors who have contributed to the rich literature
in classical or celestial mechanics. Therefore, we also apologize to those
authors who may feel that we have failed to make the appropriate reference
to their work appearing before 1995. On the other hand, in the present
monograph references newer than 1995 have not been considered.

\section{The basic setting and a presentation of the Main Theorems\ \ \ }

\subsection{Basic notions and terminology}

Let $\mathbb{R}^{3}$ denote Euclidean 3-space with the standard basis $%
\left\{ \mathbf{i,j,k}\right\} $. A three-body system consists of three
labelled point masses ($P_{i},m_{i})$ in 3-space, and its geometric model is
the triangle with vertices $P_{i}$ and the mass $m_{i}$ $>0$ attached to $%
P_{i}$, which we shall refer to as an \emph{m-triangle (}or simply a \emph{%
triangle)} or \emph{configuration. }An m-triangle will also be identified
with its triple $(\mathbf{a}_{1},\mathbf{a}_{2},\mathbf{a}_{3})$ of position
vectors $\mathbf{a}_{i}=\overrightarrow{OP}_{i}$. These triples are usually
denoted by boldface letters such as $\mathbf{X,Y}$, but occasionally we also
use the notation $\delta,\delta_{1}$ etc. It is tacitly assumed that a fixed
mass distribution $(m_{1},m_{2},m_{3})$ is given, and it is normalized so
that $m_{1}+m_{2}+m_{3}=1$.

Thus, the abundance of individual motions of three point masses jointly
combine to a rich variety of \emph{virtual motions }of m-triangles, through
which the triangle changes its kinematic invariants such as size, shape and
orientation (position) and velocity. This is our starting point for a
systematic investigation of the kinematics of m-triangles, as a basis for a
geometric approach to the three-body problem.

In addition to previously defined quantities, set 
\begin{align}
\alpha_{j} & =\text{ the central angle (at origin) opposite to the vertex }%
P_{j}  \notag \\
\omega_{j} & =\text{ the (scalar) angular velocity of }\mathbf{a}_{j}\text{
for planary motion }  \notag \\
I_{j} & =m_{j}\left| \mathbf{a}_{j}\right| ^{2}\text{,\ \ \ \ }%
I=I_{1}+I_{2}+I_{3}=\rho^{2}  \label{notation2} \\
C_{1} & =-m_{1}I_{1}+m_{2}I_{2}+m_{3}I_{3}\text{ \ etc. (cyclic permutation
of indices)}  \notag \\
T_{j} & =\frac{1}{2}m_{j}\left| \mathbf{\dot{a}}_{j}\right| ^{2}\text{,\qquad%
}T=T_{1}+T_{2}+T_{3},\text{ \ cf. (\ref{T})}  \notag \\
\mathbf{\Omega}_{j} & =m_{j}\mathbf{a}_{j}\mathbf{\times\dot{a}}_{j}\text{,}%
\qquad\mathbf{\Omega=\Omega}_{1}+\mathbf{\Omega}_{2}+\mathbf{\Omega }_{3},\ 
\text{cf. (\ref{angmom})}  \notag \\
\Delta & =\text{the area of }\Delta(P_{1},P_{2},P_{3})\text{,\ \ }\Delta _{1}%
\text{ = the area of }\Delta(O,P_{2},P_{3}\text{)\ etc.}  \notag
\end{align}
where $I$ (resp. $I_{j}$) is the total (resp. individual) polar moment of
inertia, and similarly $T,\mathbf{\Omega}$ (resp. $T_{j},\mathbf{\Omega}%
_{j}) $ denote kinetic energy and angular momentum as in Section 1.1. See
Figure 1.

Certain functions of the mass distribution appear frequently, so we
introduce the notation 
\begin{align}
m_{j}^{\ast} & =\frac{1}{2}(1-m_{j})\text{: the dual mass distribution, with 
}\sum m_{j}^{\ast}=1  \label{mass2} \\
\hat{m}_{1} & =m_{2}m_{3}\text{ , }\hat{m}_{2}=m_{3}m_{1}\text{\ , }\hat {m}%
_{3}=m_{1}m_{2}\text{ \ }  \notag
\end{align}
and the basic elementary symmetric functions of the symbols $m_{i}$ are 
\begin{equation}
\sum m_{i}=1\text{, \ \ }\hat{m}=\tsum \text{\ }\hat{m}_{i}\text{, \ }\bar{m}%
=m_{1}m_{2}m_{3}   \label{symm}
\end{equation}

\subsubsection{Vector algebra and kinematics in the Euclidean space $M_{0}$}

We will assume a center of mass reference frame as in Section 1.1, and hence
an m-triangle $\mathbf{X}$ is a vector of the 6-dimensional Euclidean
configuration space $M_{0}$, cf. (\ref{M0}). The zero vector $\mathbf{X}=0$
represents the one-point triangle (or the \textquotedblright triple
collision\textquotedblright\ configuration), and we say $\mathbf{X}$ is 
\emph{collinear} or is an \emph{eclipse }configuration (resp. is
non-degenerate) if the subspace $\Pi(\mathbf{X})$ $\subset\mathbb{R}^{3}$
spanned by the position vectors $\mathbf{a}_{i}$ has dimension $1$ (resp. $%
2) $. A \emph{virtual motion }$\mathbf{X(}t)$ (or $\delta(t))$ is a time
parametrized curve in $M_{0}$, assumed to be (piecewise) differentiable so
that its kinetic energy (\ref{T}) is defined. The \emph{size} of an
m-triangle $\mathbf{X}$ is naturally measured by the Euclidean length 
\begin{equation}
\rho=\sqrt{I}=\left\vert \mathbf{X}\right\vert   \label{size}
\end{equation}
in $M_{0}$ with the \emph{kinematic metric }(\ref{metric1}), equivalently
given by the following inner product of Jacobi type 
\begin{equation}
\mathbf{X\cdot Y=}\ m_{1}\mathbf{a}_{1}\cdot\mathbf{b}_{1}+m_{2}\mathbf{a}%
_{2}\cdot\mathbf{b}_{2}+m_{3}\mathbf{a}_{3}\cdot\mathbf{b}_{3}\text{, \ \ } 
\label{metric2}
\end{equation}
where $\mathbf{X=}\ (\mathbf{a}_{1},\mathbf{a}_{2},\mathbf{a}_{3}),$ $%
\mathbf{Y=}\ (\mathbf{b}_{1},\mathbf{b}_{2},\mathbf{b}_{3})$. For
convenience, we define 
\begin{align}
\mathbf{\omega\times X} & =\mathbf{(\omega\times a}_{1},\mathbf{\omega\times
a}_{2},\mathbf{\omega\times a}_{3})\text{, \ \ }\mathbf{\omega\in}\mathbb{R}%
^{3}  \label{notation1} \\
\mathbf{X\times Y} & =\sum m_{i}(\mathbf{a}_{i}\times\mathbf{b}_{i})  \notag
\end{align}
and observe the general triple product identity 
\begin{equation}
\mathbf{\omega\times X\cdot Y=\omega\cdot X\times Y}\text{, \ \ }\mathbf{%
\omega\in}\mathbb{R}^{3}   \label{triple}
\end{equation}

The infinitesimal generators of the $SO(3)$-action on $M_{0}$ are the
rotational (or Killing) vector fields 
\begin{equation*}
\mathbf{X\rightarrow\omega\times X}\text{, \ }\mathbf{\omega\in}\mathbb{R}%
^{3}\simeq so(3) 
\end{equation*}
of fixed angular velocity $\mathbf{\omega}$. These vectors are tangential to
the $SO(3)$-orbits. Thus, at each $\mathbf{X}$ the tangent space $T_{\mathbf{%
X}}M_{0}\simeq M_{0}$ has an orthogonal decomposition into \emph{vertical }%
and \emph{horizontal }vectors, where the vertical ones are the above Killing
vectors $\mathbf{\omega\times X}$ and the horizontal vectors $\mathbf{Y}$
are characterized by $\mathbf{X\times Y}=0$, due to (\ref{triple}).

For any virtual motion $\mathbf{X}(t)$ in $M_{0}$, the velocity vector at
each time $t$ has the above type of splitting, namely 
\begin{equation}
\mathbf{\dot{X}}=\frac{d}{dt}\mathbf{X}=\mathbf{\dot{X}}^{\omega}+\mathbf{%
\dot{X}}^{h}=(\mathbf{\omega}\times\mathbf{X)+\dot{X}}^{h}   \label{Xdot}
\end{equation}
where $\mathbf{\omega=\omega}(t)$ is commonly referred to as the
(instantaneous)\emph{\ angular velocity }of the motion. Correspondingly,
kinetic energy splits as the sum 
\begin{equation}
T=\frac{1}{2}\left\vert \mathbf{\omega}\times\mathbf{X}\right\vert ^{2}+%
\frac{1}{2}\left\vert \mathbf{\dot{X}}^{h}\right\vert ^{2}=T^{\omega}+T^{h} 
\label{Tsplit}
\end{equation}
of purely rotational and horizontal kinetic energy, respectively. The motion
is called \emph{horizontal} if the velocity is always horizontal.

Using (\ref{triple}) we also deduce the following relationship between the
angular momentum and angular velocity of a virtual motion, namely 
\begin{equation}
\mathbf{\Omega=X\times\dot{X}=X\times\dot{X}}^{\omega}=\mathbf{%
X\times(\omega }\times\mathbf{X)}   \label{omega1}
\end{equation}
Indeed, to each m-triangle $\mathbf{X}$ is associated the \emph{inertia
operator} 
\begin{equation}
\mathbb{I}_{\mathbf{X}}:\mathbb{R}^{3}\rightarrow\mathbb{R}^{3}\text{, \ \ }%
\mathbf{\omega\rightarrow X\times(\omega}\times\mathbf{X)}   \label{inert-op}
\end{equation}
relating the two vectors $\mathbf{\omega}$ and $\mathbf{\Omega}$. This
operator is invertible when $\mathbf{X}$ is nondegenerate, whereas $\mathbf{%
\Omega}$ determines $\mathbf{\omega}$ modulo a summand along the line $\Pi(%
\mathbf{X)}$ when $\mathbf{X}$ is an eclipse configuration. In any case, the
rotational velocity component $\mathbf{\omega}\times\mathbf{X}$ in (\ref%
{Xdot}) is uniquely determined by $\mathbf{\Omega}$ and $\mathbf{X}$.
Consequently, the motion is horizontal if and only if $\mathbf{\Omega}$
vanishes; in particular, such a motion must be planary (see Remark \ref%
{Weier} ).

The above inertia operator corresponds uniquely to the associated \emph{%
inertia tensor } 
\begin{equation}
B_{\mathbf{X}}(\mathbf{u,v)=(u\times X)\cdot(v\times X)=}\sum m_{j}(\mathbf{%
u\times a}_{j})\cdot(\mathbf{v\times a}_{j})\text{, \ \ }\mathbf{u,v\in}%
\mathbb{R}^{3}   \label{B}
\end{equation}
which is a bilinear symmetric form on Euclidean 3-space. They are related by
the identity 
\begin{equation*}
B_{\mathbf{X}}(\mathbf{u,v)}=\mathbb{I}_{\mathbf{X}}(\mathbf{u})\cdot 
\mathbf{v}
\end{equation*}
For example, they provide an orthonormal eigenframe for each m-triangle, and
hence a moving eigenframe for a motion of m-triangles, see Theorem D and
Section 3.4.

\subsubsection{Oriented m-triangles and their configuration space $M$}

Since triangles in 3-space can be oriented we propose the following
''refinement'' of the notion of an m-triangle. Define an \emph{oriented
m-triangle} to be a pair $\mathbf{(X,n),}$ where $\mathbf{n\in}S^{2}\subset%
\mathbb{R}^{3}$ is a unit vector perpendicular to $\Pi(\mathbf{X})$. In
particular, a nondegenerate triangle can be oriented in two ways, namely we
say the orientation is \emph{positive} (resp. \emph{negative}) if $(\mathbf{a%
}_{1},\mathbf{a}_{2},\mathbf{n})$ is a right-handed (resp. left-handed)
frame.

Clearly, for an m-triangle motion the orientation may be chosen so that the
normals $\mathbf{n}(t)$ vary continuously along the motion, and then the
orientation changes to the opposite one as the motion passes (transversely)
through an \emph{eclipse} configuration. In the study of \emph{planary }%
motions we will (tacitly) assume the plane to be the xy-plane, with unit
normals $\mathbf{n=\pm k}$.

Set $M_{+}$ (resp. $M_{-})$ to be the set of positively (resp. negatively)
oriented m-triangles, together with the set $E$ of oriented eclipse
configurations, where the latter includes the 2-sphere of orientations of
the one-point triangle $\mathbf{X}=0.$ Then the \emph{configuration space of
oriented m-triangles }is\emph{\ }the union 
\begin{equation}
M=M_{+}\cup M_{-}\subset M_{0}\times S^{2},\text{ \ \ \ }E=M_{+}\cap M_{-} 
\label{M}
\end{equation}
and a virtual motion of oriented m-triangles is a parametrized curve on $M$ 
\begin{equation}
\Gamma(t)=\mathbf{(X}(t)\mathbf{,n(}t)\mathbf{)}   \label{virtual}
\end{equation}

As a submanifold of $M_{0}\times\mathbb{R}^{3}\simeq\mathbb{R}^{9}$, $M$
inherits a Riemannian structure and a natural isometric action of $SO(3)$,
also referred to as the \emph{congruence group}. Let us have a closer look
at this manifold and the two projection\ maps 
\begin{equation*}
M_{0}\overset{\pi_{1}}{\longleftarrow}M\overset{\pi_{2}}{\rightarrow}S^{2}, 
\end{equation*}
where $\pi_{1}$ is a 2-fold covering over the non-degenerate m-triangles. On
the other side, each $\mathbf{(X,n)}$ is $SO(3)$-equivalent to some $(%
\mathbf{Y},\mathbf{k})$, where $\mathbf{Y}$ lies in the xy-plane $\mathbb{R}%
^{2}$ and hence belongs to 
\begin{equation}
\mathbb{R}^{4}\subset M_{0}:\sum m_{i}\mathbf{a}_{i}=0\text{ , }\mathbf{a}%
_{i}\in\mathbb{R}^{2}   \label{fiber}
\end{equation}
Then it is a useful observation (not mentioned in the 1994-95 preprints)
that $\pi_{2}$ is the $SO(3)$-equivariant projection of a homogeneous
4-plane bundle 
\begin{equation}
\mathbb{R}^{4}\rightarrow M\simeq SO(3)\times_{SO(2)}\mathbb{R}%
^{4}\rightarrow SO(3)/SO(2)=S^{2}   \label{4-bundle}
\end{equation}
where the 4-space in (\ref{fiber}) is the fiber over the unit normal $%
\mathbf{k}$ of the xy-plane, and $SO(2)$ is the rotation group fixing the
z-axis.

In (\ref{4-bundle}) $M$ is expressed as the $SO(2)$-orbit space of $%
SO(3)\times\mathbb{R}^{4},$ where $h\in SO(2)$ acts by $(g,\mathbf{Y}%
)\rightarrow(gh^{-1},h\mathbf{Y),}$ and the $SO(2)$-orbit of $(g,\mathbf{Y)} 
$ is identified with the oriented m-triangle $(g\mathbf{Y},g\mathbf{k).}$
This is an example of a well known \textquotedblright twisted
product\textquotedblright\ construction.

\subsubsection{Congruence moduli space and shape space}

The orbit space 
\begin{equation}
\bar{M}=M/SO(3)=\bar{M}_{+}\cup\bar{M}_{-}\text{ , \ \ }\bar{E}=\bar{M}%
_{+}\cap\bar{M}_{-}   \label{Mbar}
\end{equation}
is the (congruence)\emph{\ moduli space}, where the union corresponds to the 
$SO(3)$-invariant splitting (\ref{M}) of $M$, and the \emph{shape space} is
the subspace 
\begin{equation}
M^{\ast}=M_{+}^{\ast}\cup M_{-}^{\ast}\text{ , \ \ \ }E^{\ast}=M_{+}^{\ast
}\cap M_{-}^{\ast}   \label{Mstar}
\end{equation}
corresponding to m-triangles of fixed size $\rho=1.$ Since an m-triangle
(and its congruence classes) is scaled by the size function $\rho\geq0$ (\ref%
{size}), $\bar{M}=C(M^{\ast})$ has a natural structure of a cone over $%
M^{\ast}$. \ So, we may regard $\bar{M}$ as the union of two identical cones
or ''half-spaces'' $\bar{M}_{\pm}$ glued together along their common
boundary $\bar{E}$. This will be made more precise below.

Let us first investigate the topology of the above spaces from a
trigonometric viewpoint, using the quadratic form (\ref{Qform}) representing
the squared area of m-triangles. The triple $(I_{1},I_{2},I_{3})$ is,
indeed, a complete congruence invariant for (unoriented) m-triangles. There
is only one ''half-space'' in the unoriented case and we may express it as
the following cone in $\left\{ I_{j}\right\} $-coordinate 3-space 
\begin{equation}
\bar{M}_{\pm}\simeq\left\{ (I_{1},I_{2},I_{3})\mid
I_{j}\geq0,Q(I_{1},I_{2},I_{3})\geq0\right\}   \label{halfspace1}
\end{equation}
where the corresponding shape space $M_{\pm}^{\ast}$ is cut out by the plane 
$I_{1}+I_{2}+I_{3}=1$. The eclipse variety $\bar{E}$, defined by the
condition $Q=0$, is the cone over $E^{\ast}$.

Now it is not difficult to see that $M_{\pm}^{\ast}$ is topologically a
closed 2-disk with $E^{\ast}$ as boundary circle, and consequently the full
shape space (\ref{Mstar}) is a 2-sphere $M^{\ast}\approx S^{2}$ with a
distinguished \emph{equator}\ circle $E^{\ast}$ separating the two
hemispheres $M_{\pm }^{\ast}.$ Note that the triple of central angles 
\begin{equation*}
(\alpha_{1,}\alpha_{2},\alpha_{3}),\tsum \alpha_{j}=2\pi 
\end{equation*}
is a complete system of invariants for the shape of unoriented m-triangles
and hence these angles also yield coordinates for each of the hemispheres.

It follows that the full moduli space $\bar{M}$ is the cone over a 2-sphere
and hence is homeomorphic to 3-space, 
\begin{equation}
\bar{M}=C(M^{\ast})\simeq C(S^{2})=\mathbb{R}^{3},   \label{Mbar2}
\end{equation}
in such a way that $\bar{E}$ $\simeq\mathbb{R}^{2}$ is the coordinate plane $%
z=0$, $\bar{M}_{+}$ is the upper half space $z\geq0$ and $\bar{M}_{-}$ is
the lower half-space $z\leq0$.

Finally, from the viewpoint of equivariant geometry, we observe that the
pair $\bar{M}\supset M^{\ast}$ is the $SO(3)$-orbit space of the vector
bundle (\ref{4-bundle}) and its sphere bundle, namely 
\begin{align}
M^{\ast} & =(SO(3)\times_{SO(2)}S^{3})/SO(3)=S^{3}/SO(2)\simeq S^{2}
\label{Mbar3} \\
\bar{M} & =(SO(3)\times_{SO(2)}\mathbb{R}^{4})/SO(3)=\mathbb{R}%
^{4}/SO(2)=C(S^{3}/SO(2))\approx\mathbb{R}^{3}  \notag
\end{align}
For comparison reasons, if we only consider unoriented m-triangles, then the
corresponding calculation of the moduli space as an orbit space will yield
the closed half-space 
\begin{equation}
\bar{M}_{\pm}=M_{0}/SO(3)=M_{0}/O(3)=\mathbb{R}^{4}/O(2)\approx\mathbb{R}%
_{\pm}^{3}   \label{half}
\end{equation}

\subsection{Statement of the Main Theorems}

In this summary we focus attention on six main topics, each of which is
centered around one or two main theorems, labelled by $A,B,C1,C2...$

\subsubsection{Kinematic geometry of m-triangles and universal sphericality}

For a virtual 3-body motion with vanishing angular momentum, that is, a
horizontal motion $\Gamma(t)$ in $M$, the kinetic energy $T$ depends solely
on the moduli curve $\bar{\Gamma}(t)$ in $\bar{M}$, namely in terms of the
local coordinates $(I_{1},I_{2},I_{3})$ it is the following ''differential''
expression 
\begin{equation}
\bar{T}=\frac{1}{8}\frac{\dot{I}}{I}^{2}+\frac{1}{2IQ}\left( \sum_{\text{i
mod 3}}m_{i}I_{i+1}I_{i+2}\dot{I}_{i}^{2}\ -C_{i}I_{i}\dot{I}_{i+1}\dot {I}%
_{i+2}\right)   \label{Tbar}
\end{equation}
where $Q=16m_{1}^{2}m_{2}^{2}m_{3}^{2}\Delta^{2}$. This is, indeed, a
positive definite quadratic form on the tangent bundle of the moduli space $%
\bar{M}$ and thus naturally defines a \emph{kinematic Riemannian metric} 
\begin{equation}
d\bar{s}^{2}=2\bar{T}dt^{2}   \label{dsbar}
\end{equation}

For a general virtual motion the same expression (\ref{Tbar}) is, in fact,
obtained from $T$ by removing the rotational kinetic energy. Therefore, by (%
\ref{Tsplit}), the horizontal kinetic energy 
\begin{equation}
\bar{T}=T^{h}=\frac{1}{2}\left| \frac{d}{dt}\bar{\Gamma}(t)\right|
^{2}=T-T^{\omega}   \label{Tbar1}
\end{equation}
may well be referred to as the kinetic energy in the moduli space.

Both the definition and the formula for the above metric (\ref{dsbar}) are
dependent on the given mass distribution in a rather intricate manner.
Therefore, it is a pleasant surprise that such a kinematically defined
Riemannian structure $(\bar{M},d\bar{s}^{2})$ turns out to be not only
independent of the mass distribution, but it is, in fact, isometric to the
Riemannian cone of the Euclidean sphere of radius 1/2, namely

\ \ \ \ 

\textbf{Theorem A \ }\emph{Let }$I=\rho^{2}\emph{\ }$\emph{be the moment of
inertia and }$M^{\ast}$\emph{\ be the subspace of }$\bar{M}$ \emph{with }$%
I=1 $\emph{, and set }$d\sigma^{2}=d\bar{s}^{2}\left| _{M^{\ast}}\right. $%
\emph{to be the restriction of the kinematic metric. Then } 
\begin{align}
d\bar{s}^{2} & =d\rho^{2}+\rho^{2}d\sigma^{2}  \label{dsbar1} \\
(M^{\ast},d\sigma^{2}) & \simeq S^{3}(1)/U(1)=\mathbb{C}P^{1}\simeq
S^{2}(1/2)   \label{dsbar1a}
\end{align}
\emph{where }$S^{3}(1)$\emph{\ is the 3-sphere of radius }$1$\emph{\ and }$%
S^{3}(1)\rightarrow\mathbb{C}P^{1}$\emph{\ is the classical Hopf fibration. }

\ \ \ \ \ \ \ \ 

The surprising emergence of spherical symmetry in the kinematic Riemannian
space ($\bar{M},d\bar{s}^{2})$ for arbitrary mass distribution naturally
brings in the classical spherical geometry as a useful tool in the study of
the three-body problem. We propose to call this fundamental fact the \emph{%
universal sphericality} of the kinematic geometry of m-triangles.

The orientation reversing map $\mathbf{(X,n})\rightarrow(\mathbf{X,-n})$ of
oriented m-triangles induces an isometric involution of $(M^{\ast},d\sigma
^{2})$ with $E^{\ast}$ as its fixed point set, namely the distinguished
equator which divides $M^{\ast}$ into two hemispheres $M_{\pm}^{\ast}$. On
this circle lie the three points $\mathfrak{p}_{ij}$ representing the shape
of the three types of binary collisions, cf. (\ref{binary}). Indeed, their
relative positions on the circle determine the mass distribution uniquely,
see Section 4.4 and (\ref{mass}).

\subsubsection{Unique lifting property}

\textbf{Theorem B}{\LARGE \ } \emph{To a given curve }$\bar{\Gamma}(t)$\emph{%
\ in the moduli space }$\bar{M}\smallsetminus\left\{ 0\right\} $, \emph{%
together with a given constant vector }$\Omega$\emph{\ and initial
configuration }$\Gamma(t_{0})$\emph{, there exists a unique curve }$\Gamma
(t)$\emph{\ in }$M$\emph{\ with }$\bar{\Gamma}(t)$\emph{\ as its moduli
curve and with }$\Omega$\emph{\ as its conserved angular momentum. Moreover,
the curve }$\Gamma(t)$\emph{\ can be computed in terms of the }$C^{1}$\emph{%
-data of \ }$\bar{\Gamma}(t)$\emph{. }

\ \ \ \ \ \ 

Consider the orbit map $\pi:M\rightarrow\bar{M}$, and observe that $SO(3)$
acts freely outside the sphere $\pi^{-1}(0)\simeq S^{2}$ and defines a
principal bundle 
\begin{equation}
\pi:M\smallsetminus\pi^{-1}(0)\rightarrow\bar{M}\smallsetminus\left\{
0\right\}   \label{princip}
\end{equation}
In particular, above the half-spaces $\bar{M}_{\pm}\simeq\mathbb{R}%
_{\pm}^{3} $ there are locally trivializing diffeomorphisms 
\begin{equation}
SO(3)\times\bar{M}_{\pm}\rightarrow M_{\pm}   \label{princ}
\end{equation}
\qquad Geometrically speaking, a motion of m-triangles can be represented by
a time parametrized curve $\Gamma(t)$ in $M$ (or $M_{\pm})$ which (locally)
consists of two components, namely a \emph{moduli} \emph{curve} $\bar{\Gamma 
}(t)$ that records the change of size and shape of the oriented m-triangles,
and a \emph{position curve} $\gamma(t)$ in $SO(3)$ that records the change
of position. The latter curve is, of course, constrained by the fixed
angular momentum.

In the special case of $\mathbf{\Omega}=0$, namely the horizontal lifting of
moduli curves, the proof of Theorem B follows directly from the standard
theory of principal $G$-bundles with a connection (cf. e.g. \cite{K-N}),
applied to the above principal $SO(3)$-bundle. Therefore, we shall rather
focus on the general case with $\mathbf{\Omega}\neq0$ and present two
different proofs. The first proof involves the inertia operator (\ref%
{inert-op}), and the second proof is an application of Theorem D stated
below. We refer to Section 5.2.2

\subsubsection{Angular velocities and kinematic Gauss-Bonnet formula}

\textbf{Theorem C1 }\emph{For a planary motion\ of oriented m-triangles with
normal vector }$\mathbf{k}$ \emph{and} \emph{angular momentum }$\mathbf{%
\Omega }=\Omega\mathbf{k}$\emph{, let }$\omega_{i}=\dot{\phi}_{i}$\emph{\ } 
\emph{be the individual (scalar) angular velocity of the position vector }$%
\mathbf{a}_{i}=\overrightarrow{OP}_{i}$\emph{. Then } 
\begin{equation}
\omega_{i}=\omega_{i}^{0}+\frac{\Omega}{I},\text{ \ }i=1,2,3 
\label{angvel2}
\end{equation}
\emph{where }$\omega_{i}^{0}$ \emph{is a ''differential'' expression purely
at the moduli space level, namely\ \ \ } 
\begin{align}
\omega_{1}^{0} & =\frac{1}{I}(I_{3}\dot{\alpha}_{2}-I_{2}\dot{\alpha}_{3})%
\text{\emph{\ etc. (cyclic permutation of indices)}}  \label{kin1} \\
& =\frac{1}{8m_{1}m_{2}m_{3}\Delta I}[(C_{3}I_{3}-C_{2}I_{2})\frac{\dot {I}%
_{1}}{I_{1}}-(C_{1}+2m_{2}I_{3})\dot{I}_{2}  \label{kin2} \\
& +(C_{1}+2m_{3}I_{2})\dot{I}_{3}]\text{ \ \emph{etc.}}  \notag
\end{align}

\begin{remark}
\label{RemC1}The proof of Theorem C1 holds for non-planary motions as well,
that is, the plane $\Pi(t)$ of the m-triangle is time dependent. Then $%
\omega_{i}$ stands for the (scalar) angular velocity of the velocity
component of $\mathbf{a}_{i}$ in $\Pi(t)$, and in formula (\ref{angvel2}) $%
\Omega$ must be replaced by the normal component $\mathbf{\Omega}$ $\cdot%
\mathbf{n}(t)$ (cf. Theorem D). We refer to Section 3.2.1.
\end{remark}

We introduce the following three \emph{kinematic 1-forms }on the moduli
space $\bar{M}-\left\{ 0\right\} :$ \emph{\ } 
\begin{align}
\Theta_{1} & =\frac{1}{I}(I_{3}d\alpha_{2}-I_{2}d\alpha_{3})\text{ \ \ \emph{%
etc. (cyclic permutation of indices)}}  \notag \\
& =\frac{1}{8m_{1}m_{2}m_{3}\Delta I}[(C_{3}I_{3}-C_{2}I_{2})\frac{dI_{1}}{%
I_{1}}-(C_{1}+2m_{2}I_{3})dI_{2}  \label{1forms} \\
& +(C_{1}+2m_{3}I_{2})dI_{3}]\text{ \ \emph{etc.}}  \notag
\end{align}
In fact, they are invariant under scaling and may therefore be regarded as
1-forms on the shape space $M^{\ast}$ via the canonical retraction $\bar
{M}%
\smallsetminus\left\{ 0\right\} \rightarrow M^{\ast}.$ They share the basic
property 
\begin{equation}
d\Theta_{i}=2dA,\text{ \ }i=1,2,3   \label{2form}
\end{equation}
where $dA$ is the area form of the 2-sphere $M^{\ast}$ $\simeq S^{2}(1/2)$.
Evidently, the 1-forms have singularities on the eclipse circle $E^{\ast}$.

By suitably combining the kinematic 1-forms on appropriate regions on $%
M^{\ast}$ and applying Green's theorem, the \emph{kinematic} \emph{%
Gauss-Bonnet }version\emph{\ }as described by the next theorem follows
immediately from Theorem C1 and (\ref{2form}).

\ \ 

\textbf{Theorem C2 }\emph{Let the shape curve of a piecewise differentiable
motion of oriented m-triangles with }$\Omega=0$\emph{\ constitute the
oriented boundary of a region }$D$\emph{\ in }$M^{\ast}\simeq S^{2}(1/2)$%
\emph{. Then the total change of position of the triangle is a rotation of
angle equal to twice the oriented area of }$D$\emph{, namely} 
\begin{equation}
\Delta\phi_{i}=\tint \limits_{t_{0}}^{t_{1}}\omega_{i}^{0}dt=\tint
\limits_{\partial D}\Theta_{i}=\tiint \limits_{D}2dA   \label{GB}
\end{equation}

\ \ \ \ 

The above type of integral (\ref{GB}) is an example of the \emph{geometric
phase }in the literature. Its value depends only on the shape curve and is
independent of its parametrization. In the case of a planary motion with
nonzero angular momentum, however, the total change of position in the above
case (\ref{GB}) has an additional term called the \emph{dynamical phase},
namely as a consequence of (\ref{angvel2}) 
\begin{equation}
\Delta\phi_{i}=\tiint \limits_{D}2dA+\tint \limits_{t_{0}}^{t_{1}}\frac{%
\Omega}{I}dt   \label{GB2}
\end{equation}

\subsubsection{Moving eigenframe and Euler equations for m-triangles}

Let $(\mathbf{X,n)}$ be a nondegenerate oriented m-triangle. Then we can
choose eigenvectors of the inertia tensor (\ref{B}) which constitute a
positive orthonormal frame 
\begin{equation*}
(\mathbf{u}_{1},\mathbf{u}_{2},\mathbf{n)\in}SO(3), 
\end{equation*}
where $\mathbf{u}_{1},\mathbf{u}_{2}$ lie in the plane $\Pi(\mathbf{X)}$ and 
$\mathbf{u}_{1}\times\mathbf{u}_{2}=\mathbf{n}$. By definition, 
\begin{equation}
B_{\mathbf{X}}(\mathbf{u}_{1},\mathbf{u}_{1})=\lambda_{1}\text{, \ }B_{%
\mathbf{X}}(\mathbf{u}_{2},\mathbf{u}_{2})=\lambda_{2}\text{, \ \ }B_{%
\mathbf{X}}(\mathbf{u}_{1},\mathbf{u}_{2})=0\text{\ }   \label{eigen1}
\end{equation}
where the two eigenvalues $\lambda_{i}$ may be expressed (cf. Section 3.4)
as 
\begin{equation}
\left\{ \lambda_{1},\lambda_{2}\right\} =\frac{1}{2}(I\pm\sqrt{%
I^{2}-16m_{1}m_{2}m_{3}\Delta^{2}})=\frac{I}{2}(1\pm\sin\varphi), 
\label{eigen2}
\end{equation}
using spherical polar coordinates $(\varphi,\theta)$ on the 2-sphere $%
M^{\ast }$ centered at the north pole $\mathcal{N}$, where $\varphi$ is the
colatitude with $\varphi=0$ at the pole. The eigenvalue in the normal
direction $\pm\mathbf{n}$ is the largest eigenvalue 
\begin{equation*}
\lambda_{3}=\lambda_{1}+\lambda_{2}=I. 
\end{equation*}
To a continuous motion of oriented m-triangles we may choose such an
eigenframe 
\begin{equation}
\mathfrak{F}(t)=\left\{ \mathbf{u}_{1}(t),\mathbf{u}_{2}(t),\mathbf{n(}%
t)\right\}   \label{mov}
\end{equation}
varying continuously with the motion. In particular, $t\rightarrow$ $%
\mathfrak{F}(t)$ is also a parametrized curve in $SO(3).$

\textbf{Theorem D}\emph{\ Let }$\mathfrak{F}(t)$ in (\ref{mov}) \emph{be a
moving eigenframe attached to a differentiable motion }$\Gamma(t)$ \emph{of
m-triangles, with }$\mathbf{\Omega}$\emph{\ as the conserved angular
momentum. Then the triple of inner products} 
\begin{equation}
g_{1}=\mathbf{\Omega\cdot u}_{1}\text{, \ \ }g_{2}=\mathbf{\Omega\cdot u}_{2}%
\text{, \ \ \ }g_{3}=\mathbf{\Omega\cdot n}   \label{g123}
\end{equation}
\emph{satisfy the following system of ODE, namely} 
\begin{align}
\dot{g}_{1} & =g_{2}\left[ (\frac{1}{\lambda_{3}}-\frac{1}{\lambda_{2}}%
)g_{3}+\frac{1}{2}\dot{\theta}\cos\varphi\right]  \notag \\
\dot{g}_{2} & =g_{1}\left[ (\frac{1}{\lambda_{1}}-\frac{1}{\lambda_{3}}%
)g_{3}-\frac{1}{2}\dot{\theta}\cos\varphi\right]  \label{Euler} \\
\dot{g}_{3} & =g_{1}g_{2}(\frac{1}{\lambda_{2}}-\frac{1}{\lambda_{1}}) 
\notag
\end{align}
\emph{where the numbers }$\lambda_{i}(t)$ \emph{are the eigenvalues of the
inertia tensor of }$\Gamma(t)$\emph{\ and depend solely on the moduli curve }%
$\bar{\Gamma}(t)=(\rho(t),\varphi(t),\theta(t)).$

\begin{corollary}
\label{planary}It follows from (\ref{Euler}) that $%
g_{1}(t_{0})=g_{2}(t_{0})=0$ at just one time $t_{0}$ implies that $%
g_{1}(t)=g_{2}(t)=0$ for all time. Thus, such a motion is planary if and
only if the angular momentum vector is perpendicular to the m-triangle at
just one time $t_{0}$.
\end{corollary}

\begin{remark}
The system (\ref{Euler}) is the exact generalization of the classical Euler
equations for a rigid body, see e.g. Arnold\cite{Arnold}, p.143, where $%
M_{i} $ and the fixed numbers $I_{i}$ correspond to our $g_{i}$ and $%
\lambda_{i}$, respectively. In (\ref{Euler}) the additional terms are\ due
to the change of shape, and the system is singular where the motion passes
through an eclipse configuration, say, with $\lambda_{1}=0$ and hence also $%
g_{1}=0$. In particular, the eclipse takes place along a line perpendicular
to $\mathbf{\Omega}$.
\end{remark}

The triple $(g_{1,}g_{2},g_{3})$ is the coordinate vector, with respect to
the moving eigenframe, of the constant vector $\mathbf{\Omega}$ $=\Omega 
\mathbf{k}$. It determines the position of the m-triangle, in particular its
normal vector $\mathbf{n}$, up to a rotation around the $\mathbf{k}$-axis by
a specific \emph{precession\ angle} $\chi(t)$. This angle is calculated by
quadrature from the formula \ 
\begin{equation}
\dot{\chi}=\frac{\mathbf{\dot{n}\cdot k\times n}}{\left| \mathbf{k\times n}%
\right| ^{2}}=\frac{\Omega}{g_{1}^{2}+g_{2}^{2}}(\frac{g_{1}^{2}}{\lambda_{1}%
}+\frac{g_{2}^{2}}{\lambda_{2}})\text{ ,\ \ \ \ \ cf. Section 5.2.1\ } 
\label{prec}
\end{equation}
It follows that the m-triangle motion \emph{\ }$\Gamma(t)$ is largely
described by two curves on the 2-sphere, namely the shape curve $%
\Gamma^{\ast }(t)=(\varphi(t),\theta(t))$ and the \emph{precession curve},
that is, the curve traced out by the normal vector $\mathbf{n}(t)$. Thus,
for the study of non-planar motions it is a basic problem to investigate the
relationship between these two curves.

\subsubsection{The reduced Newton's equations}

Now, let us focus attention on the dynamics of three-body motions, namely
the Newtonian equation of motion (\ref{Newton1}). Such motions are, of
course, a very special subclass of all the virtual three-body motions
considered before.

The $\Omega$\emph{-reduced Newton's equation} is a second order system of
ODE for the moduli curves of three-body motions with a fixed angular
momentum vector $\mathbf{\Omega}$, namely the three equations (with index $i 
$ mod $3)$ 
\begin{align}
\ddot{I}_{i} & =4T_{i}-\left( \frac{m_{i}+2m_{i+1}}{r_{i,i+1}^{3}}+\frac{%
m_{i}+2m_{i+2}}{r_{i,i+2}^{3}}\right) I_{i}  \label{redu1} \\
& -\left( \frac{1}{r_{i,i+1}^{3}}-\frac{1}{r_{i,i+2}^{3}}\right) \left(
m_{i+1}I_{i+1}-m_{i+2}I_{i+2}\right)  \notag
\end{align}
which are easily derived from the system (\ref{Newton1}). However, the
individual kinetic energy terms $T_{i}$ depend on $\mathbf{\Omega}$, of
course, but are otherwise expressed solely at the level of the moduli space $%
\bar{M}$.

The two cases of planary and non-planary motions differ substantially in
complexity, so we will consider them separately. In the following two
theorems we assume the initial position $\Gamma(t_{0})$ is not a collinear
configuration (since otherwise the given initial data will be incomplete).

\textbf{Theorem E1 }\emph{A planary three-body motion }$\Gamma(t)$\emph{\ is
completely determined by its moduli curve }$\bar{\Gamma}(t)$, \emph{initial
position }$\Gamma(t_{0})$\emph{\ and angular momentum vector. The curve }$%
\bar{\Gamma}(t)$ \emph{is a solution of the }$\Omega$\emph{-reduced Newton's
equation (\ref{redu1}), with kinetic energy terms} 
\begin{equation}
T_{i}=\frac{\dot{I}_{i}^{2}}{8I_{i}}+\frac{1}{2}\omega_{i}^{2}I_{i} 
\label{Ti}
\end{equation}
\emph{where }$\omega_{i}$\emph{\ is the i-th individual angular velocity (%
\ref{angvel2}). }

\emph{Conversely, each solution curve of this }$\Omega$\emph{-reduced
Newton's equation can be realized as the moduli curve of a three-body motion
in the xy-plane, with a given initial position and normal vector }$\Omega%
\mathbf{k}$ \emph{as the conserved angular momentum. }

\begin{remark}
The above $\Omega$-reduced Newton's equation may, of course, be expressed
purely in terms of the coordinates $\left\{ I_{j}\right\} $ or the mutual
distances $\left\{ r_{ij}\right\} $, see (\ref{r/I}). In fact, such an $%
\Omega$-reduced Newton's equation in terms of coordinates $\left\{
r_{ij}\right\} $ was derived by Lagrange$\cite{Lagrange}$. We refer to
Section 4.3.1 for another version in terms of spherical coordinates of $\bar{%
M}\simeq\mathbb{R}^{3}$ as \ a cone over the 2-sphere.
\end{remark}

For the statement of the general (e.g. non-planary) version of the above
theorem, let $\left\{ \mathbf{u}_{1}\mathbf{,u}_{2}\mathbf{,n}\right\} $
denote a continuous eigenframe associated with the motion $\Gamma(t)$, and
let 
\begin{equation}
(g_{1},g_{2},g_{3}),\text{ \ \ where }g_{1}^{2}+g_{2}^{2}+g_{3}^{2}=\Omega
^{2},   \label{g}
\end{equation}
be the coordinate vector of $\mathbf{\Omega}$ relative to this frame, as in
Theorem D. The individual kinetic energies depend on the components $g_{k}$,
more precisely, they split into a tangential and normal component 
\begin{equation}
T_{i}=T_{i}^{\tau}+T_{i}^{\eta}   \label{Ti-split}
\end{equation}
where the tangential term $T_{i}^{\tau}$ depends on the normal component $%
g_{3}$ and $T_{i}^{\eta}$ depends on $g_{1}$ and $g_{2}$. We refer to
Section 5.1 and 5.2.3 for a precise description of these quantities.

\textbf{Theorem E2}\emph{\ A general three-body motion }$\Gamma(t)$\emph{\
is completely determined by its moduli curve }$\bar{\Gamma}(t)$, \emph{%
initial position }$\Gamma(t_{0})$\emph{\ and angular momentum vector }$%
\mathbf{\Omega }$.\emph{\ The curve }$\bar{\Gamma}(t)$ \emph{is
characterized by the }$\Omega$\emph{-reduced Newton's equations (\ref{redu1}%
) with kinetic energy terms }$T_{i}$ \emph{(\ref{Ti-split}) depending on the
moving frame coordinates (\ref{g}) of }$\mathbf{\Omega}$ \emph{and are thus
coupled with the Euler equations, namely the first order ODE (\ref{Euler}). }

\subsubsection{Reduction of the least action principles}

\ \ Here we will only consider planary three-body motions and state the
associated $\Omega$-reduced least action principles, whose extremals are
precisely the moduli curves of those planary three-body motions with a fixed
angular momentum $\mathbf{\Omega}=\Omega\mathbf{k}$. In this case the total
kinetic energy 
\begin{equation}
T=\bar{T}+T^{\omega}=\bar{T}+\frac{\Omega^{2}}{2I}   \label{T5}
\end{equation}
and the Lagrange function $L=T+U$ are, in fact, defined at the level of the
moduli space $\bar{M}$, and therefore the two action integrals 
\begin{equation}
\bar{J}_{1,\Omega}=\tint \limits_{\bar{\Gamma}}Tdt\text{ , \ \ \ }\bar{J}%
_{2,\Omega}=\tint \limits_{t_{1}}^{t_{2}}Ldt   \label{red.action}
\end{equation}
apply to moduli curves $\bar{\Gamma}(t).$

\textbf{Theorem F }\emph{The solution curves of the planary }$\Omega $\emph{%
-reduced Newton's equation can be characterized as the extremal curves of }$%
\bar{J}_{1,\Omega}$\emph{\ (resp. }$\bar{J}_{2,\Omega})$\emph{\ applied to
curves }$\bar{\Gamma}$\emph{\ in }$\bar{M}$\emph{\ with fixed end points
together with fixed energy }$h$\emph{\ (resp. fixed time interval }$\left[
t_{1},t_{2}\right] $\emph{\ )}$.$

\subsubsection{Shape curves of triple collision trajectories}

In Chapter 8 we initiate a general study of the geometry of moduli curves in
the vicinity of a triple collision. According to a classical result of
Sundman and Siegel, towards the collision these curves $\bar{\Gamma}(t)$
approach a ray solution, which in the generic case has the shape $\pm\mathbf{%
\hat{p}}_{0}$ of an (oriented) equilateral triangle. In the simplest case of
equal masses, $\pm\mathbf{\hat{p}}_{0}$ are the poles of the 2-sphere $%
M^{\ast }\simeq S^{2}$, and hence the correponding shape curves approach one
of the poles. Thus, it is natural to focus attention on the family of shape
curves representing a triple collision with the limit shape of $\mathbf{\hat{%
p}}_{0}$. Here we state a theorem which is a simplified version of Theorem G$%
_{1}$ stated in Section 8.1.

\ \ \ \ \ \ 

\textbf{Theorem G} \emph{Assume uniform mass distribution and zero total
energy, and consider the family }$\mathfrak{S}$\emph{\ of arc-length
parametrized shape curves }$\Gamma^{\ast}(s),s\geq0$\emph{, representing
3-body motions with a triple collision at }$s=0$\emph{, say }$\Gamma^{\ast
}(0)$\emph{\ is the north pole of }$S^{2}$\emph{. The family has the
following properties :}

\emph{(i) There is a unique curve }$\Gamma_{\theta_{0}}^{\ast}$\emph{\ for
each initial longitude direction }$\theta_{0}$\emph{\ at the pole, and }$%
\Gamma_{\theta_{0}}^{\ast}$\emph{\ and }$\Gamma_{\theta_{0}+\pi/3}^{\ast}$%
\emph{\ are congruent modulo a rotation of the sphere.}

\emph{(ii) The six meridians representing isosceles triangles belong to the
family }$\mathfrak{S}.$\emph{\ They divide the sphere into six congruent
sectors of angular width }$\pi/3$\emph{, and each curve }$%
\Gamma_{\theta_{0}}^{\ast}$\emph{\ stays within a sector, at least until the
first eclipse (at the equator).}

\emph{(iii) The curves }$\Gamma_{\theta_{0}}^{\ast}(s)$\emph{\ are analytic
in }$s$\emph{, with no singularity before the first eclipse, and }$\Gamma
_{\theta_{0}+\pi}^{\ast}(s)=$\emph{\ }$\Gamma_{\theta_{0}}^{\ast}(-s)$\emph{.%
}

\emph{(iv) The sign of the curvature of the above shape curves is the same
inside a sector, and the sign is the opposite in neigboring sectors. }

\section{Basic geometric and kinematic invariants of m-triangles}

\subsection{Ceva-type trigonometry}

In classical Greek geometry individual triangles - not their motions and
kinematic relations - are the geometric objects of basic importance. A\
triangle $\Delta(P_{1},P_{2},P_{3})$ is specified by its three vertices $%
P_{i}$, and its congruence properties involve the fundamental geometric
concepts \textquotedblright side\textquotedblright, \textquotedblright
angle\textquotedblright\ and \textquotedblright area\textquotedblright,
whose relationships are expressed by trigonometric identities and congruence
theorems. In our study, however, we are rather concerned with m-triangles,
that is, a positive mass $m_{i}$ is attached to $P_{i}$. Thus it is natural
and useful to reformulate the usual trigonometry into a kind of $\emph{Ceva}$%
-\emph{trigonometry}, depending on the given mass distribution.

Let us first establish the following three identities (cf. (\ref{notation2}%
)) : 
\begin{equation}
\text{\emph{Ceva-area law : }}\Delta_{j}=m_{j}\Delta   \label{area}
\end{equation}
\begin{equation}
\text{\emph{Ceva-sine law : }\ }\frac{\sin\alpha_{i}}{m_{i}\left\Vert 
\mathbf{a}_{i}\right\Vert }=\frac{2\Delta}{\left\Vert \mathbf{a}%
_{1}\right\Vert \left\Vert \mathbf{a}_{2}\right\Vert \left\Vert \mathbf{a}%
_{3}\right\Vert }\text{ , }i=1,2,3   \label{Ceva-sine}
\end{equation}
\begin{equation}
\text{\emph{Ceva-cosine\ law}}\emph{\ :\ \ }2\sqrt{m_{i}m_{j}}\sqrt{%
I_{i}I_{j}}\cos\alpha_{k}=-C_{k}\ \ \   \label{Ceva-cos}
\end{equation}
By calculating cross products such as 
\begin{align*}
0 & =\mathbf{a}_{1}\times\sum m_{j}\mathbf{a}_{j}=m_{2}\mathbf{a}_{1}\times%
\mathbf{a}_{2}+m_{3}\mathbf{a}_{1}\times\mathbf{a}_{3} \\
& =(2m_{2}\Delta_{3}-2m_{3}\Delta_{2})\mathbf{n}
\end{align*}
the first law (\ref{area}) follows directly, and then the sine law (\ref%
{Ceva-sine}) follows : 
\begin{equation*}
2m_{3}\Delta=2\Delta_{3}=\left\vert \mathbf{a}_{1}\right\vert \left\vert 
\mathbf{a}_{2}\right\vert \sin\alpha_{3}\text{ \ }\Longrightarrow\frac {%
\sin\alpha_{3}}{m_{3}\left\vert \mathbf{a}_{3}\right\vert }\text{ =}\frac{%
2\Delta}{\left\Vert \mathbf{a}_{1}\right\Vert \left\Vert \mathbf{a}%
_{2}\right\Vert \left\Vert \mathbf{a}_{3}\right\Vert }
\end{equation*}
Furthermore, consider the \textquotedblright small\ triangle" with side
vectors $\left\{ m_{i}\mathbf{a}_{i}\right\} $, say, with one vertex at the
center of mass $O$ and an adjacent side along $OP_{j}$ for some $j$. The
triangle has outer angles $\alpha_{i}$, and by applying the usual cosine law
to it we deduce the cosine law (\ref{Ceva-cos}).

Next, by combining the usual cosine law and its Ceva version, the
relationship between the mutual distances $s_{i}$ and the moments of inertia 
$I_{i}$ is 
\begin{equation}
s_{i}^{2}=r_{jk}^{2}=\frac{(1-m_{i})I-I_{i}}{m_{j}m_{k}}\text{, \ \ }\left\{
i,j,k\right\} =\left\{ 1,2,3\right\} ,   \label{r/I}
\end{equation}
from which we also deduce 
\begin{equation}
I=\ \sum_{i<j}m_{i}m_{j}r_{ij}^{2}\ =\frac{m_{1}^{\ast}}{m_{2}m_{3}}C_{1}+%
\frac{m_{2}^{\ast}}{m_{3}m_{1}}C_{2}+\frac{m_{3}^{\ast}}{m_{1}m_{2}}C_{3} 
\label{Isum}
\end{equation}
where the first identity is known as Lagrange's formula for the total moment
of inertia with respect to the center of mass, and $m_{i}^{\ast}$ are the
dual masses (\ref{mass2}).

Finally, consider the \textquotedblright Heron\textquotedblright\ quadratic
form\ 
\begin{equation}
H(a,b,c)=2(ab+bc+ca)-(a^{2}+b^{2}+c^{2})   \label{H}
\end{equation}
and recall the classical Heron's formula for the area $\Delta$%
\begin{equation}
H(s_{1}^{2},s_{2}^{2},s_{3}^{2})=16\Delta^{2}   \label{Heron}
\end{equation}
Set 
\begin{align}
Q(I_{1},I_{2},I_{3}) &
=H(m_{1}I_{1},m_{2}I_{2},m_{3}I_{3})=2\sum_{i<j}m_{i}m_{j}I_{i}I_{j}-%
\sum_{j}m_{j}^{2}I_{j}^{2}  \label{Qform} \\
& =\sum_{i<j}C_{i}C_{j}=4m_{1}m_{2}I_{1}I_{2}-C_{3}^{2}\text{ \ etc.}  \notag
\end{align}
and consider again the \textquotedblright small\ triangle" with side vectors 
$m_{i}\mathbf{a}_{i}$. On the one hand, its area $\hat{\Delta}$ is related
to $\Delta$ by 
\begin{equation*}
4\hat{\Delta}^{2}=m_{1}^{2}m_{2}^{2}\left\vert \mathbf{a}_{1}\times \mathbf{a%
}_{2}\right\vert
^{2}=4m_{1}^{2}m_{2}^{2}\Delta_{3}^{2}=4m_{1}^{2}m_{2}^{2}m_{3}^{2}\Delta^{2}
\end{equation*}
and on the other hand, $16\hat{\Delta}^{2}=$ $Q(I_{1},I_{2},I_{3})$ by (\ref%
{Heron}) and (\ref{Qform}). Consequently, we obtain the 
\begin{equation}
\text{\emph{Ceva-Heron formula : }}%
Q(I_{1},I_{2},I_{3})=16m_{1}^{2}m_{2}^{2}m_{3}^{2}\Delta^{2}   \label{C-H}
\end{equation}

\subsubsection{A simple torque formula}

As a simple application of the Ceva-area law (\ref{area}) we prove the
following result concerning the individual torques due to gravitational
forces acting at the vertices $P_{i}$ of a nondegenerate m-triangle $\Delta
(P_{1},P_{2},P_{3}).$

\begin{lemma}
Let $\mathbf{t}_{i}$ be the torque of the Newtonian gravitational forces at $%
P_{i},i=1,2,3,$with respect to the center of mass $O$. \ Then 
\begin{equation*}
\mathbf{t}_{i}=\mathbf{\dot{\Omega}}_{i}=2m_{1}m_{2}m_{3}\Delta(\frac {1}{%
r_{i,i+1}^{3}}-\frac{1}{r_{i,i+2}^{3}})\mathbf{n}\text{ , \ i mod 3}
\end{equation*}
where $\mathbf{n}$ is the unit normal vector so that $(\mathbf{a}_{1},%
\mathbf{a}_{2},\mathbf{n})$ is a right-handed frame.
\end{lemma}

\begin{proof}
Let $\mathbf{F}_{12}$ and $\mathbf{F}_{13}$ be the gravitational forces due
to the mass points $P_{2}$ and $P_{3}$ acting on $P_{1}$, namely 
\begin{equation*}
\mathbf{F}_{12}=\frac{m_{1}m_{2}}{r_{12}^{3}}(\mathbf{a}_{2}-\mathbf{a}_{1})%
\text{, \ }\mathbf{F}_{13}=\frac{m_{1}m_{3}}{r_{13}^{3}}(\mathbf{a}_{3}-%
\mathbf{a}_{1}) 
\end{equation*}
Then, by definition of torque and the area law (\ref{area}) 
\begin{align*}
\mathbf{t}_{1} & =\mathbf{a}_{1}\times(\mathbf{F}_{12}+\mathbf{F}_{13})=%
\frac{m_{1}m_{2}}{r_{12}^{3}}(\mathbf{a}_{1}\times\mathbf{a}_{2})+\frac{%
m_{1}m_{3}}{r_{13}^{3}}(\mathbf{a}_{1}\times\mathbf{a}_{3}) \\
& =(\frac{m_{1}m_{2}}{r_{12}^{3}}2m_{3}\Delta)\mathbf{n}-(\frac{m_{1}m_{3}}{%
r_{13}^{3}}2m_{2}\Delta)\mathbf{n=}\ 2m_{1}m_{2}m_{3}\Delta(\frac{1}{%
r_{12}^{3}}-\frac{1}{r_{13}^{3}})\mathbf{n}
\end{align*}
and similarly at the other two vertices.
\end{proof}

\begin{corollary}
Corollary $\mathbf{t}_{1}=0$ if and only if $r_{12}=r_{13}$, and all $%
\mathbf{t}_{i}=0$ if and only if the triangle is regular (i.e. equilateral).
\end{corollary}

\subsection{Analysis of angular velocities and kinetic energies}

In the orthogonal splitting (\ref{Xdot}) of the velocity of a virtual motion 
$\mathbf{X}(t)=(\mathbf{a}_{1}(t),\mathbf{a}_{2}(t),\mathbf{a}_{3}(t))$, the
horizontal component further splits into two summands 
\begin{equation}
\mathbf{\dot{X}}^{h}=\mathbf{\dot{X}}^{\rho}+\mathbf{\dot{X}}^{\sigma}=\frac{%
\dot{\rho}}{\rho}\mathbf{X+\dot{X}}^{\sigma}   \label{Xdot1}
\end{equation}
representing the change of size and shape, respectively, and correspondingly
the total kinetic energy splits as 
\begin{equation}
T=T^{\omega}+T^{h}=T^{\omega}+(T^{\rho}+T^{\sigma})=\frac{1}{2}\left| 
\mathbf{\omega\times X}\right| ^{2}+(\frac{1}{2}\dot{\rho}^{2}+T^{\sigma }) 
\label{Tspace}
\end{equation}
In this chapter we will show that $T^{h}$ actually equals the expression in (%
\ref{Tbar}), and in particular it depends only on the velocity of the image
curve in $\bar{M}$. This will justify our definition of $T^{h}$ as the
kinetic energy $\bar{T}$ of the moduli curve, hence also our definition of
the kinematic Riemannian metric on $\bar{M}$ 
\begin{equation}
d\bar{s}^{2}=2\bar{T}dt^{2}=d\rho^{2}+2T^{\sigma}dt^{2}   \label{metric6}
\end{equation}
Our first proof of Theorem A is by showing that the metric (\ref{metric6})
actually transforms to the metric (\ref{dsbar3}).

The differential expression (\ref{Tbar}), as a function on the tangent
bundle of $\bar{M}$, is calculated by eliminating from $T$ its dependence on
the angular momentum, namely the rotational energy. In fact, it suffices to
consider a class of virtual motions whose term $T^{\omega}$ is easy to
calculate and hence eliminate. For this single purpose we could as well
assume $T^{\omega}=0$ from the outset and simply express $T$ at the moduli
space level. However, it is also illuminating to analyze the class of
planary motions with a broader perspective.

\subsubsection{Kinematics of planary motions and proof of Theorem C1 and C2}

We assume the motion takes place in the xy-plane and write 
\begin{equation*}
\mathbf{\Omega=}\ \Omega\mathbf{k}\text{, \ \ }\mathbf{\omega}=\omega 
\mathbf{k}
\end{equation*}
In this case (\ref{Tspace}) reads 
\begin{equation}
T=T^{\omega}+T^{\rho}+T^{\sigma}=\frac{1}{2}\frac{\Omega^{2}}{I}+\frac{1}{2}%
\dot{\rho}^{2}+T^{\sigma}   \label{Tplane}
\end{equation}
On the other hand, from the orthogonal decomposition of each $\mathbf{\dot{a}%
}_{j}$ into its rotational and radial component 
\begin{equation}
\mathbf{\dot{a}}_{j}=\omega_{j}(\mathbf{k\times a}_{j})+\frac{\dot{\rho}_{j}%
}{\rho_{j}}\mathbf{a}_{j}\text{ , \ \ \ where }\rho_{j}^{2}=I_{j}, 
\label{split3}
\end{equation}
the total kinetic energy also adds up to 
\begin{equation}
T=\frac{1}{2}\sum I_{i}\omega_{i}^{2}+\frac{1}{8}\sum\frac{\dot{I}_{j}^{2}}{%
I_{j}}   \label{T2}
\end{equation}
Therefore, by combining (\ref{Tplane}) and (\ref{T2}) the \textquotedblright
intricate\textquotedblright\ energy term $T^{\sigma}$, responsible for the
change of shape, is given by 
\begin{equation}
T^{\sigma}=\frac{1}{2}\sum I_{i}\omega_{i}^{2}+(\frac{1}{8}\sum\frac{\dot {I}%
_{j}^{2}}{I_{j}}-\frac{1}{2}\dot{\rho}^{2})-\frac{1}{2}\frac{\Omega^{2}}{I} 
\label{T3}
\end{equation}

Now, start from the above expression to express $T^{\sigma}$ purely in terms
of the individual moments of inertia $I_{j}$.

\begin{lemma}
Let $\omega_{j}$ be the (scalar) angular velocity of $\mathbf{a}_{j}$. Then 
\begin{equation}
\omega_{1}=\frac{1}{I}(I_{3}\dot{\alpha}_{2}-I_{2}\dot{\alpha}_{3})+\frac{%
\Omega}{I}\text{ \ etc. \ (cyclic permutation of indices)}   \label{angvel}
\end{equation}
\begin{equation}
\dot{\alpha}_{1}=\frac{1}{8m_{1}m_{2}m_{3}\Delta}(-2m_{1}\dot{I}_{1}+C_{3}%
\frac{\dot{I}_{2}}{I_{2}}+C_{2}\frac{\dot{I}_{3}}{I_{3}})\text{ \ etc.} 
\label{alfa1}
\end{equation}
\end{lemma}

\begin{proof}
Set $\theta_{j}$ to be the angle of $\mathbf{a}_{j}$ with respect to a
chosen reference direction in the plane. Then $\omega_{j}=$ $\dot{\theta}%
_{j} $ and 
\begin{equation*}
\alpha_{i}=\theta_{i+2}-\theta_{i+1},\text{ \ }\dot{\alpha}_{i}=\omega
_{i+2}-\omega_{i+1}\text{\ , i mod 3}
\end{equation*}
The total (scalar) angular momentum sums up to 
\begin{align*}
\Omega & =\sum m_{i}(\mathbf{a}_{j}\mathbf{\times\dot{a}}_{j})\cdot \mathbf{k%
}=\sum\Omega_{j}=\sum I_{j}\omega_{j} \\
& =\mathbf{X\times(\omega\times X)\cdot k}=I\omega
\end{align*}
Consequently, 
\begin{equation*}
I_{1}\omega_{1}=\Omega-I_{2}\omega_{2}-I_{3}\omega_{3}=\Omega-I_{2}\dot {%
\alpha}_{3}+I_{3}\dot{\alpha}_{2}-(I_{2}\omega_{1}+I_{3}\omega_{1}) 
\end{equation*}
and this proves formula (\ref{angvel}).

Next, by differentiating the Ceva-cosine formula (\ref{Ceva-cos}) for $%
\alpha_{1}$ with respect to $t$ and use the expression for $\sin\alpha_{1} $
from the Ceva-sine formula (\ref{Ceva-sine}) for $\alpha_{1}$, we obtain the
formula (\ref{alfa1}).
\end{proof}

Substitution of the expressions (\ref{alfa1}) into (\ref{angvel}) also
leads\ to the following formula involving only $I_{j}$'s, namely 
\begin{align}
\omega_{1} & =\frac{1}{8m_{1}m_{2}m_{3}\Delta I}[(C_{3}I_{3}-C_{2}I_{2})%
\frac{\dot{I}_{1}}{I_{1}}-(C_{1}+2m_{2}I_{3})\dot{I}_{2}  \label{omega2} \\
& +(C_{1}+2m_{3}I_{2})\dot{I}_{3}]+\frac{\Omega}{I}\text{ \ etc. }  \notag
\end{align}
and this completes the proof of Theorem C1.

Furthermore, using either (\ref{angvel}), (\ref{alfa1}) and the relation $%
\dot{\alpha}_{1}+\dot{\alpha}_{2}+\dot{\alpha}_{3}=0$, or using (\ref{omega2}%
) directly, we calculate 
\begin{align}
\ \frac{1}{2}\sum I_{i}\omega_{i}^{2}-\frac{\Omega^{2}}{2I} & =\frac{1}{2I}%
(I_{1}I_{2}\dot{\alpha}_{3}^{2}+I_{2}I_{3}\dot{\alpha}_{1}^{2}+I_{3}I_{1}%
\dot{\alpha}_{2}^{2})\   \label{Iomega} \\
& =\frac{1}{8I}\left( \sum_{\text{i mod 3}}(\frac{4I_{1}I_{2}I_{3}m_{i}}{Q}%
+I_{i}-I\ )\frac{\dot{I}_{i}^{2}}{I_{i}}+(2-\frac{4C_{i}I_{i}}{Q})\dot {I}%
_{i+1}\dot{I}_{i+2}\right) \   \notag
\end{align}
and finally by insertion into (\ref{T3}) we deduce the formula $\ $%
\begin{equation}
T^{\sigma}=\ \frac{1}{2IQ}\left( \sum_{\text{i mod 3}}m_{i}I_{i+1}I_{i+2}%
\dot{I}_{i}^{2}\ -C_{i}I_{i}\dot{I}_{i+1}\dot{I}_{i+2}\right)   \label{T4}
\end{equation}
Consequently, the metric (\ref{metric6}) on $\bar{M}$ may be written 
\begin{equation}
d\bar{s}^{2}=d\rho^{2}+\rho^{2}d\sigma^{2}\   \label{dsigma2a}
\end{equation}
where\ 
\begin{equation*}
d\sigma^{2}\ =\frac{1}{I^{2}Q}\left( \sum_{\text{i mod 3}%
}m_{i}I_{i+1}I_{i+2}dI_{i}^{2}-C_{i}I_{i}dI_{i+1}dI_{i+2}\right) 
\end{equation*}
is the induced metric on the shape space $M^{\ast}=(I=1)$. Indeed, the
metric expression $d\sigma^{2}$ is a tensor on $M^{\ast}$ since it is
invariant under scaling in $\bar{M}.$

On the other hand, on $M^{\ast}$ the relation $I_{1}+I_{2}+I_{3}=1$ implies $%
dI_{1}+dI_{2}+dI_{3}=0$, and therefore the above metric on $M^{\ast}$ can be
restated as 
\begin{equation}
d\sigma^{2}=\frac{1}{Q^{\ast}}\left\{ 
\begin{array}{c}
\left[ -I_{2}^{2}+(1-m_{2})I_{2}\right] dI_{1}^{2}+\left[
-I_{1}^{2}+(1-m_{1})I_{1}\right] dI_{2}^{2} \\ 
-\left[ 2I_{1}I_{2}-(1-m_{2})I_{1}-(1-m_{1})I_{2}+m_{3}\right] dI_{1}dI_{2}%
\end{array}
\right\}   \label{dsigma2b}
\end{equation}
where $Q^{\ast}$ denotes the restriction of $Q$ to $M^{\ast}$ and we have
used the mass normalization $m_{1}+m_{2}+m_{3}=1.$

\begin{lemma}
The area form of $(M^{\ast},d\sigma^{2})$ is 
\begin{equation*}
dA=\frac{1}{2\sqrt{Q^{\ast}}}dI_{1}\wedge dI_{2}
\end{equation*}
\end{lemma}

\begin{proof}
As usual, the area form expresses as 
\begin{equation*}
dA=\sqrt{D}dI_{1}\wedge dI_{2}
\end{equation*}
where 
\begin{equation*}
D=\frac{1}{Q^{\ast2}}\ \left\{ 
\begin{array}{c}
I_{1}I_{2}(1-m_{1}-I_{1})(1-m_{2}-I_{2}) \\ 
-\frac{1}{4}\left[ 2I_{1}I_{2}-(1-m_{2})I_{1}-(1-m_{1})I_{2}+m_{3}\right]
^{2}%
\end{array}
\right\} =\frac{1}{4Q^{\ast}}
\end{equation*}
is the determinant of the metric (\ref{dsigma2b}).
\end{proof}

Finally, we turn to the kinematic 1-forms (\ref{1forms}) on $\bar{M}$, whose
definition is suggested by the expressions (\ref{omega2}) for the individual
angular velocities. Regarded as 1-forms on $M^{\ast}$ they are related to
the area form by 
\begin{align*}
d\Theta_{1} & =dI_{3}\wedge d\alpha_{2}-dI_{2}\wedge d\alpha_{3} \\
& =\frac{1}{2\sqrt{Q^{\ast}}}\left\{ 
\begin{array}{c}
dI_{3}\wedge(-2m_{2}dI_{2}+\frac{C_{1}}{I_{3}}dI_{3}+\frac{C_{3}}{I_{1}}%
dI_{1}) \\ 
-dI_{2}\wedge(-2m_{3}dI_{3}+\frac{C_{2}}{I_{1}}dI_{1}+\frac{C_{1}}{I_{2}}%
dI_{2})%
\end{array}
\right\} \\
& =\frac{1}{2\sqrt{Q^{\ast}}}(\frac{C_{2}+C_{3}+2(m_{2}+m_{3})I_{1}}{I_{1}}%
)dI_{1}\wedge dI_{2} \\
& =\frac{1}{\sqrt{Q^{\ast}}}dI_{1}\wedge dI_{2}=2dA
\end{align*}
This proves formula (\ref{2form}) and, as observed in Section 2.2.3, this
also completes the proof of Theorem C2.

\subsubsection{ A purely kinematic proof of Theorem A}

From the metric expression (\ref{dsigma2a}) it follows that $\bar{M}$ is a
Riemannian cone over the shape space $(M^{\ast},d\sigma^{2})$, expressed in (%
\ref{Mstar}) as the union of two isometric disks along their common boundary
circle $E^{\ast}$. Both disks are parametrized by the region $Q^{\ast}\geq0$
in the $(I_{1},I_{2})$-plane, where 
\begin{equation}
Q^{\ast}(I_{1},I_{2})=Q(I_{1},I_{2},1-I_{1}-I_{2})\text{, \ \ }0\leq
I_{i}\leq1   \label{Qstar}
\end{equation}
is the quadratic form (\ref{Qform}) with $I_{3}=1-I_{1}-I_{2}.$ In the
following we will describe our original calculations in \cite{1995} leading
to the discovery of the \emph{universal sphericality .}

At first glance, the mass distribution $\left\{ m_{i}\right\} $ is
intricately involved in the formula (\ref{dsigma2b}) of $d\sigma^{2}$, so we
will focus attention on the mass dependent quadratic form $Q^{\ast}$. The
major step of the proof is, in fact, the algebraic approach of seeking
better coordinates by transforming the metric tensor $d\sigma^{2}$ into a
simpler one. The geometric proof using the Hopf bundle (see Section 3.2.3
below) is, in fact, our \emph{second} proof.

Intuitively, one expects that optimal simplicity and maximal symmetry is
achieved by a suitable affine transformation of the $(I_{1},I_{2})$-plane
which transforms the region $Q^{\ast}\geq0$ into the unit disk and makes the
metric more "transparent". This simple idea was, indeed, the key leading to
such a remarkable coordinate transformation.

As indicated in Figure 2, $Q^{\ast}=0$ defines an ellipse which is tangent
to the triple of lines given by $I_{1}=0,I_{2}=0$ and $I_{3}=1-I_{1}-I_{2}=0$%
. It is easy to see that its center of symmetry is the point $%
(m_{1}^{\ast},m_{2}^{\ast})$, so we first set 
\begin{equation}
\tilde{I}_{1}=I_{1}-m_{1}^{\ast},\text{ \ }\tilde{I}_{2}=I_{2}-m_{2}^{\ast } 
\label{I-tilde}
\end{equation}
and obtain 
\begin{equation*}
Q^{\ast}=m_{1}m_{2}m_{3}-(1-m_{2})^{2}\tilde{I}_{1}^{2}-(1-m_{1})^{2}%
\tilde
{I}_{2}^{2}-2(m_{3}-m_{1}m_{2})\tilde{I}_{1}\tilde{I}_{2}
\end{equation*}
This suggests a rotation through the angle 
\begin{equation*}
\psi_{0}=\frac{1}{2}\tan^{-1}\frac{2(m_{1}m_{2}-m_{3})}{%
(m_{1}-m_{2})(1+m_{3})}
\end{equation*}
and new coordinates $\tilde{x},\tilde{y}$ defined by 
\begin{equation*}
\tilde{I}_{1}=\tilde{x}\cos\psi_{0}-\tilde{y}\text{ }\sin\psi_{0}\text{ , \ }%
\tilde{I}_{2}=\tilde{x}\sin\psi_{0}+\tilde{y}\text{ }\cos\psi_{0}\text{ }
\end{equation*}
Then 
\begin{equation}
Q^{\ast}=m_{1}m_{2}m_{3}-\mu_{1}\tilde{x}^{2}-\mu_{2}\tilde{y}^{2} 
\label{Qstar1}
\end{equation}
where 
\begin{align*}
\mu_{1} & =\frac{1}{2}((1-m_{1})^{2}+(1-m_{2})^{2})+\frac{1}{2}%
(m_{1}-m_{2})(1+m_{3})\cos2\psi_{0} \\
& +(m_{3}-m_{1}m_{2})\sin2\psi_{0} \\
\mu_{2} & =\frac{1}{2}((1-m_{1})^{2}+(1-m_{2})^{2})-\frac{1}{2}%
(m_{1}-m_{2})(1+m_{3})\cos2\psi_{0} \\
& -(m_{3}-m_{1}m_{2})\sin2\psi_{0}
\end{align*}
and we notice the identity 
\begin{equation*}
(-(1-m_{2})^{2}+(1-m_{1})^{2})\sin2\psi_{0}+2(m_{3}-m_{1}m_{2})\cos2%
\psi_{0}=0 
\end{equation*}
Thus, by setting 
\begin{equation*}
\tilde{x}=\sqrt{\frac{m_{1}m_{2}m_{3}}{\mu_{1}}}x\text{ , \ \ }\tilde{y}=%
\sqrt{\frac{m_{1}m_{2}m_{3}}{\mu_{2}}}y 
\end{equation*}
the expression (\ref{Qstar1}) transforms to 
\begin{equation}
Q^{\ast}=m_{1}m_{2}m_{3}(1-x^{2}-y^{2})   \label{Qstar2}
\end{equation}
Therefore, the following combined transformation 
\begin{align}
I_{1} & =\cos\psi_{0}\sqrt{\frac{m_{1}m_{2}m_{3}}{\mu_{1}}}x-\sin\psi _{0}%
\sqrt{\frac{m_{1}m_{2}m_{3}}{\mu_{2}}}y+m_{1}^{\ast}  \label{comb} \\
I_{2} & =\sin\psi_{0}\sqrt{\frac{m_{1}m_{2}m_{3}}{\mu_{1}}}x+\cos\psi _{0}%
\sqrt{\frac{m_{1}m_{2}m_{3}}{\mu_{2}}}y+m_{2}^{\ast}  \notag
\end{align}
will transform the formula (\ref{dsigma2b}) of $d\sigma^{2}$ into 
\begin{equation}
d\sigma^{2}=\frac{1}{4}\frac{(1-y^{2})dx^{2}+(1-x^{2})dy^{2}+2xydxdy}{%
1-x^{2}-y^{2}}   \label{dsigma3}
\end{equation}
From here, we simply set 
\begin{equation}
x=\sin\varphi\cos\theta\text{ , \ \ }y=\sin\varphi\sin\theta 
\label{spherical}
\end{equation}
which will transform (\ref{dsigma3}) into the metric (\ref{metric7}). This
proves that 
\begin{equation*}
(M^{\ast},d\sigma^{2})\simeq S^{2}(1/2). 
\end{equation*}

\subsubsection{The Hopf fibration and a geometric proof of Theorem A}

The moduli space $\bar{M}$ is, by definition, an $SO(3)$-orbit space with
the induced differential structure, and according to (\ref{Mbar3}) it is
also the orbit space $\ $%
\begin{equation}
\bar{M}=M/SO(3)\simeq\mathbb{R}^{4}/SO(2)\approx\mathbb{R}^{3}=C(S^{2}) 
\label{Mbar4}
\end{equation}
of the orthogonal transformation group $(SO(2),\mathbb{R}^{4}).$ As a
quotient of a Riemannian space by a compact group of isometries $\bar{M}$
has the induced \emph{orbital distance metric} which measures the distance
between orbits in $\mathbb{R}^{4}$ (or $M)$.

Let $S^{3}=S^{3}(1)$ $\subset$ $\mathbb{R}^{4}$ be the unit sphere and
recall the well known classical Hopf fibration, which in the above metric
setting reads 
\begin{equation}
SO(2)\rightarrow S^{3}\rightarrow S^{3}/SO(2)=\mathbb{C}P^{1}\simeq
S^{2}(1/2),   \label{Hopf}
\end{equation}
where the projection is a Riemannian submersion and the quotient space is
the round 2-sphere of radius 1/2. Combined with (\ref{Mbar4}) we have an
isometry 
\begin{equation*}
\bar{M}\simeq\mathbb{R}^{4}/SO(2)=C(S^{3}/SO(2))=C(S^{2}(1/2)) 
\end{equation*}
of Riemannian cones over the 2-sphere $M^{\ast}\simeq S^{2}(1/2).$ The cone $%
\bar{M}$ is homeomorphic to $\mathbb{R}^{3}$, but they are only
diffeomorphic away from the cone vertex (or base point $O$) which
corresponds to the origin $0\in\mathbb{R}^{3}$.

Finally, to complete the proof of Theorem A it remains to observe that the
above orbital distance metric actually coincides with the kinematically
defined one. The two metrics are, for example, determined by the kinetic
energy they associate to \textquotedblright motions\textquotedblright\ in $%
\bar{M}$. These are the image curves of virtual m-triangle motions $\mathbf{X%
}(t)$, which can always be chosen with vanishing angular momentum, namely
they are \emph{horizontal} (cf. Section 2.1.1). These motions are planar,
say $\mathbf{X}(t)$ is a curve in $\mathbb{R}^{4}\subset M_{0}$. Horizontal
curves are those perpendicular to the $SO(2)$-orbits, and at a point $%
\mathbf{X}\neq0$ the horizontal tangent vectors constitute the subspace $%
\mathcal{H}(\mathbf{X})\simeq\mathbb{R}^{3}$ consisting of all $\mathbf{Y}$
such that $\mathbf{X\times Y}=0$.

Now, the orbital distance metric on $\bar{M}$ is defined by demanding the
projection $\pi:\mathbb{R}^{4}\rightarrow\mathbb{R}^{4}/SO(2)$ $=\bar{M}$ to
be a Riemannian submersion, that is, that the tangent map $d\pi$ takes $%
\mathcal{H}(\mathbf{X})$ isometrically to the tangent space of $\bar{M}$ at $%
\pi(\mathbf{X})$. Equivalently, the kinetic energy associated to a moduli
curve is the same as the kinetic energy of a horizontal lifting. On the
other hand, the kinematic metric (\ref{dsbar}) also associates to a moduli
curve the kinetic energy $\bar{T}$ of a lifting with vanishing angular
momentum. Consequently, the two metrics on $\bar{M}$ are identical.

\subsection{Linear motions of m-triangles}

According to Newton's inertia law, in a center of mass reference frame and
in the absence of forces, the trajectory of the three-body system in the
Euclidean configuration space $M_{0}$ will be a geodesic, namely a \emph{%
linear motion} 
\begin{equation}
\delta(t)=(1-t)\delta_{1}+t\delta_{2}   \label{linear}
\end{equation}
where $\delta_{1}=(\mathbf{a}_{1}\mathbf{,a}_{2}\mathbf{,a}_{3}),\delta
_{2}=(\mathbf{b}_{1}\mathbf{,b}_{2}\mathbf{,b}_{3})$ are appropriate
m-triangles. Such motions are also characterized by having constant velocity 
$(\delta_{2}-\delta_{1})$, and the motion (\ref{linear}) has constant
angular momentum 
\begin{equation}
\mathbf{\Omega}=\delta_{1}\times\delta_{2}   \label{omega}
\end{equation}
Moreover, twice the action integral (\ref{J10}) of the motion from $\delta
_{1}$ to $\delta_{2}$ is the squared distance 
\begin{equation}
2\tint Tdt=\left\vert \delta_{1}-\delta_{2}\right\vert ^{2}=\tsum
m_{i}\left\vert \mathbf{a}_{i}\mathbf{-b}_{i}\right\vert ^{2} 
\label{action1}
\end{equation}

On the other hand, it is clear that the two m-triangles $\delta_{i}$ lie in
a common plane if the vector (\ref{omega}) is zero, namely the linear motion
(\ref{linear}) has vanishing angular momentum. Then the motions (\ref{linear}%
) provide a useful tool in analyzing the kinematic geometry of m-triangles
since their moduli curves are exactly the geodesics in the moduli space $(%
\bar
{M},d\bar{s}^{2})$. Their shape curves will be arcs along great
circles (geodesics) on the round sphere $M^{\ast}=$ $S^{2}$.

Let us collect some simple facts, assuming $\delta_{1}$ and $\delta_{2}$ are
m-triangles in the xy-plane (with normal vector $\mathbf{k}$) and $\mathbf{%
\Omega}=0.$

\begin{example}
If $\delta_{1}$ and $\delta_{2}$ have the same orientation, then the
distance between their congruence classes $\bar{\delta}_{i}$ in $\bar{M}$ is 
\begin{equation}
dist(\bar{\delta}_{1},\bar{\delta}_{2})\ =\left\vert \delta_{1}-\delta
_{2}\right\vert   \label{dist}
\end{equation}
\end{example}

\begin{example}
Consider two congruence classes $\bar{\delta}_{1}$, $\bar{\delta}_{2}$ in $%
\bar{M}$ whose shapes $\delta_{i}^{\ast}$ are different and not antipodal
points on $S^{2}$. It is not difficult to see that representative
m-triangles $\delta_{i}$ can be chosen in exactly two ways, modulo a
rotation of the xy-plane. The two choices are $(\delta_{1},\delta_{2})$ and $%
(\delta _{1},-\delta_{2})$ for suitable $\delta_{1}$ and $\delta_{2}. $

The shape curves of the corresponding linear motions (\ref{linear}), for $%
0\leq t\leq1,$ are the two geodesic arcs $\Gamma_{\pm}^{\ast}$ between $%
\delta_{1}^{\ast}$ and $\delta_{2}^{\ast}$ whose union is a great circle.
Each of the shape curves extends (as $t\rightarrow\pm\infty)$ to the whole
circle, minus the limit point $(\delta_{1}\mp\delta_{2})^{\ast} $ as $\left|
t\right| \rightarrow\infty$, which lies on the opposite arc $\Gamma_{\mp
}^{\ast}$.

We also remark that the relative position of $\delta_{1}$ and $\delta_{2}$
can be calculated from the line integrals of the kinematic 1-forms $%
\Theta_{i}$ along the geodesic arc $\Gamma_{\pm}^{\ast}$, see Theorem C1 and
(\ref{1forms}).
\end{example}

For easy reference, the shape of the three types of binary collisions are
the following three points on $E^{\ast}$%
\begin{equation}
\mathfrak{b}_{ij}:I_{i}=\frac{m_{i}m_{k}}{1-m_{k}},\text{ \ \ }I_{j}=\frac{%
m_{j}m_{k}}{1-m_{k}},\text{ \ \ }I_{k}\ =(1-m_{k})   \label{binary}
\end{equation}
where $\mathfrak{b}_{ij}=\mathfrak{b}_{ji}$ represents an m-triangle $%
\mathbf{(a}_{1}\mathbf{,a}_{2}\mathbf{,a}_{3}\mathbf{)}$\textbf{\ }with $%
\mathbf{a}_{i}=\mathbf{a}_{j}$ and $I=1$, and $\left\{ i,j,k\right\}
=\left\{ 1,2,3\right\} $. There are two more points on the sphere $M^{\ast}$
which are of kinematic importance, namely the north pole and south pole $%
\left\{ \mathcal{N}\text{\textsc{,}\QTR{cal}{S}}\right\} $. Each pole is, of
course, the \emph{geometric center} of the corresponding pole $%
M_{\pm}^{\ast} $, and we refer to Corollary \ref{poles} for an intrinsic
geometric characterization of their shape.

\begin{lemma}
The north pole (resp. south pole) is the shape of the positively (resp.
negatively) oriented m-triangle whose normalized individual moments of
inertia equal the dual masses, namely 
\begin{equation}
\mathcal{N}\text{ (or \textsc{S)}}:I_{j}=m_{j}^{\ast}=\frac{1}{2}(1-m_{j})%
\text{, \ }j=1,2,3   \label{geocenter}
\end{equation}
\end{lemma}

\begin{proof}
Let $\delta_{0}=(\mathbf{a}_{1}\mathbf{,a}_{2}\mathbf{,a}_{3})$ be an
m-triangle with $I_{j}$ as in (\ref{geocenter}). It suffices to show that
the three points (\ref{binary}) on the equator $E^{\ast}$ have the same
distance in $M^{\ast}$ to the point $\delta_{0}^{\ast}$ $\in M_{\pm}^{\ast}$
. Let $\delta_{1}=\mathbf{(b}_{1}\mathbf{,b}_{2}\mathbf{,b}_{3}\mathbf{)}$
be the (unit size) m-triangle with $\mathbf{b}_{1}=\sqrt{2}\mathbf{a}_{1}$
and $\mathbf{b}_{2}=\mathbf{b}_{3}=-\frac{m_{1}}{1-m_{1}}\mathbf{b}_{1}$. It
is easily checked that $\mathbf{\Omega}$ $=0$ in (\ref{omega}), that is, the
linear motion between $\delta_{0}$ and $\delta_{1}$ has vanishing angular
momentum.

The image of $\delta_{1}$ in $\bar{M}$ is the point $\delta_{1}^{\ast }=%
\mathfrak{b}_{23}$ on $E^{\ast}$, and according to (\ref{dist}) 
\begin{equation}
\left\vert \delta_{0}-\delta_{1}\right\vert =\sqrt{\sum m_{j}\left\vert 
\mathbf{a}_{j}\mathbf{-b}_{j}\right\vert ^{2}}=\sqrt{2-\sqrt{2}}=2\sin \frac{%
\pi}{8}   \label{dist2}
\end{equation}
equals the distance between $\delta_{0}^{\ast}$ and $\delta_{1}^{\ast}$ in $%
\bar{M}$. Therefore, their (spherical) distance in $M^{\ast}$ is equal to $%
\pi/4$. On the other hand, it is clear from the above calculation (or by
symmetry) that similar choices of $\delta_{1}$ with $\delta_{1}^{\ast }=%
\mathfrak{b}_{12}$ or $\mathfrak{b}_{31}$ lead to the same distance $\pi/4.$
\end{proof}

\subsection{Eigenvalues and eigenframe of the inertia tensor}

The bilinear form $B_{\mathbf{X}}$ defined by (\ref{B}) is identical to the
well known \emph{inertia tensor} in classical mechanics, for the special
case of an m-triangle $\mathbf{X}$ viewed as a rigid body.\ This is useful
in the kinematic study of non-planary motions of m-triangles. The geometric
interpretation of the quadratic form is that it calculates the moment of
inertia $I_{\mathbf{\omega}}$ of the body with respect to the central axis
through the vector $\mathbf{\omega}$, hence also the rotational kinetic
energy due to the angular velocity $\mathbf{\omega}$, namely 
\begin{equation*}
2T^{\omega}=B_{\mathbf{X}}(\mathbf{\omega,\omega})=\left\vert \mathbf{\omega
\times X}\right\vert ^{2}=\left\vert \mathbf{\omega}\right\vert ^{2}I_{%
\mathbf{\omega}}
\end{equation*}
By an \emph{eigenframe} of $\mathbf{X}$ we mean an orthonormal basis in
3-space consisting of eigenvectors (i.e., along the principal axes) of $B_{%
\mathbf{X}}$, whose \emph{eigenvalues} are the associated moments of
inertia. We will use the following notation for the eigenvalues and
associated eigenframe, 
\begin{equation}
\lambda_{1}\leq\lambda_{2}\leq\lambda_{3}\text{\ }\longleftrightarrow (%
\mathbf{u}_{1},\mathbf{u}_{2},\mathbf{n)},   \label{eigen}
\end{equation}
where $\mathbf{u}_{1}$ and $\mathbf{u}_{2}$ span the plane $\Pi(\mathbf{X)}$
if the triangle is nondegenerate, whereas in the collinear case $\lambda
_{1}=0$ and $\left\{ \mathbf{u}_{2},\mathbf{n}\right\} $ can be any
orthonormal basis of the normal plane $\Pi(\mathbf{X)}^{\perp}$.

\begin{lemma}
\label{eigen3}The eigenvalues of $B_{\mathbf{X}}$ are related by 
\begin{equation*}
\lambda_{1}+\lambda_{2}=\lambda_{3}=I\text{, \ \ \ \ }\lambda_{1}\lambda
_{2}=4m_{1}m_{2}m_{3}\Delta^{2}, 
\end{equation*}
and consequently 
\begin{equation}
\lambda_{i}=\frac{1}{2}(I\pm\sqrt{I^{2}-16m_{1}m_{2}m_{3}\Delta^{2}})\text{,
\ }i=1,2   \label{lambda}
\end{equation}
\end{lemma}

\begin{proof}
We consider the case that $\mathbf{X}$ is nondegenerate, and then $I$ is the
moment of inertia with respect to the normal direction. Let $\left\{ \mathbf{%
u}_{1}\mathbf{,u}_{2}\right\} $ be an orthonormal frame of\ $\Pi(\mathbf{X)} 
$ consisting of eigenvectors of $B_{\mathbf{X}}$ , namely 
\begin{equation*}
B_{\mathbf{X}}(\mathbf{u}_{1}\mathbf{,u}_{1})=\lambda_{1},\text{ \ }B_{%
\mathbf{X}}(\mathbf{u}_{2}\mathbf{,u}_{2})=\lambda_{2},\text{ \ \ }B_{%
\mathbf{X}}(\mathbf{u}_{1}\mathbf{,u}_{2})=0 
\end{equation*}
Then 
\begin{align*}
\lambda_{1}+\lambda_{2} & =\left\vert \mathbf{u}_{1}\mathbf{\times X}%
\right\vert ^{2}+\left\vert \mathbf{u}_{2}\mathbf{\times X}\right\vert
^{2}=\sum m_{j}(2\left\vert \mathbf{a}_{j}\right\vert ^{2}-(\mathbf{u}_{1}%
\mathbf{\cdot a}_{j})^{2}-(\mathbf{u}_{2}\mathbf{\cdot a}_{j})^{2}) \\
& =\sum m_{j}\left\vert \mathbf{a}_{j}\right\vert ^{2}=I
\end{align*}

Set 
\begin{equation}
\mathbf{a}_{1}=a_{11}\mathbf{u}_{1}+a_{12}\mathbf{u}_{2},\text{ \ \ }\mathbf{%
a}_{2}=a_{21}\mathbf{u}_{1}+a_{22}\mathbf{u}_{2}   \label{expanda1}
\end{equation}
Then on the one hand 
\begin{align}
B_{\mathbf{X}}(\mathbf{a}_{1},\mathbf{a}_{1})B_{\mathbf{X}}(\mathbf{a}_{2},%
\mathbf{a}_{2})-B_{\mathbf{X}}(\mathbf{a}_{1},\mathbf{a}_{2})^{2} & =
\label{BB} \\
\left| \mathbf{a}_{1}\times\mathbf{X}\right| ^{2}\cdot\left| \mathbf{a}%
_{2}\times\mathbf{X}\right| ^{2}-\left[ (\mathbf{a}_{1}\times\mathbf{X)\cdot 
}(\mathbf{a}_{2}\times\mathbf{X)}\right] ^{2}\mathbf{\ } & =\left| 
\begin{array}{cc}
a_{11} & a_{12} \\ 
a_{21} & a_{22}%
\end{array}
\right| ^{2}\lambda_{1}\lambda_{2}  \notag
\end{align}
where by the Ceva-area law (\ref{area}) 
\begin{equation*}
\left| 
\begin{array}{cc}
a_{11} & a_{12} \\ 
a_{21} & a_{22}%
\end{array}
\right| ^{2}=\left| \mathbf{a}_{1}\times\mathbf{a}_{2}\right|
^{2}=4m_{3}^{2}\Delta^{2}, 
\end{equation*}
and on the other hand, 
\begin{align}
B_{\mathbf{X}}(\mathbf{a}_{1},\mathbf{a}_{1}) & =m_{2}\left| \mathbf{a}%
_{1}\times\mathbf{a}_{2}\right| ^{2}+m_{3}\left| \mathbf{a}_{1}\times\mathbf{%
a}_{3}\right| ^{2}=4m_{2}m_{3}(m_{2}+m_{3})\Delta ^{2}  \notag \\
\text{{}}B_{\mathbf{X}}(\mathbf{a}_{2},\mathbf{a}_{2}) &
=4m_{1}m_{3}(m_{1}+m_{3})\Delta^{2}  \label{BB3} \\
\text{{}}B_{\mathbf{X}}(\mathbf{a}_{1},\mathbf{a}_{2}) &
=-4m_{1}m_{2}m_{3}\Delta^{2}\text{\ \ \ \ \ }  \notag
\end{align}
When\ the expressions (\ref{BB3}) are substituted into (\ref{BB}) we obtain 
\begin{equation*}
\lambda_{1}\lambda_{2}=4m_{1}m_{2}m_{3}\Delta^{2}
\end{equation*}
and then formula (\ref{lambda}) follows.
\end{proof}

\begin{corollary}
\label{poles}The poles (\ref{geocenter}) are the shapes uniquely
characterized by any of the two equivalent conditions :

(i) $\lambda_{1}=\lambda_{2}$ (i.e. ''umbilical'' shape)

(ii) the m-triangle attains the maximal area 
\begin{equation}
\Delta_{\max}=\frac{I}{4\sqrt{m_{1}m_{2}m_{3}}},   \label{areamax}
\end{equation}
among all m-triangles with the same moment of inertia $I$.
\end{corollary}

\begin{proof}
Clearly, $\lambda_{1}=\lambda_{2}$ if and only if the area $\Delta$ is given
by the formula of (\ref{areamax}). On the other hand, let us maximize the
area function $Q=C_{1}C_{2}+C_{2}C_{3}+C_{3}C_{1}$ (cf. (\ref{Isum}), (\ref%
{Qform})), using Lagrange's multiplier method subject to the constraint $I=1$%
. It follows that 
\begin{equation*}
C_{i}=-m_{i}m_{i}^{\ast}\ +m_{j}m_{j}^{\ast}+m_{k}m_{k}^{\ast}
\end{equation*}
or equivalently $I_{j}=m_{j}^{\ast}$, by (\ref{notation2}).
\end{proof}

The following result will also be useful. Briefly, it says that a linear
motion whose shape curve is a meridian arc from a pole to the equator, has a
constant eigenframe.

\begin{lemma}
\label{frame}Let $\delta_{0}=(\mathbf{a}_{1}\mathbf{,a}_{2}\mathbf{,a}_{3})$
be an m-triangle with the shape of a pole (\ref{geocenter}), and let $%
\left\{ \mathbf{u}_{1}\mathbf{,u}_{2}\right\} $ be an orthonormal frame of
the plane $\Pi(\delta_{0}).$ Moreover, let $\delta_{1}=\mathbf{(b}_{1},%
\mathbf{b}_{2}\mathbf{,b}_{3})$ be a degenerate m-triangle satisfying $%
\delta_{0}\times\delta_{1}=0$ and $\mathbf{u}_{2}\cdot\mathbf{b}_{j}=0$ for
all $j$. Then 
\begin{equation*}
B_{t}(\mathbf{u}_{1}\mathbf{,u}_{2})=0\text{ \ }
\end{equation*}
holds along the linear motion $\delta_{t}=\mathbf{(}1-t)\delta_{0}+t\delta
_{1}$, where $B_{t}$ is the inertia tensor of $\delta_{t}$, cf. (\ref{B}).
Hence, $\left\{ \mathbf{u}_{1}\mathbf{,u}_{2}\right\} $ is an eigenframe for 
$\delta_{t}$ for each $t$.
\end{lemma}

\begin{proof}
From the above Corollary it follows that 
\begin{equation*}
B_{0}(\mathbf{u}_{1}\mathbf{,u}_{2})=B_{1}(\mathbf{u}_{1}\mathbf{,u}_{2})=0 
\end{equation*}
Moreover, by the assumptions 
\begin{align*}
(\mathbf{u}_{2}\times\delta_{0})\cdot(\mathbf{u}_{1}\times\delta_{1}) &
=\tsum \limits_{j}m_{j}(\mathbf{u}_{2}\times\mathbf{a}_{j})\cdot(\mathbf{u}%
_{1}\times \mathbf{b}_{j}) \\
& =\tsum \limits_{j}m_{j}\left[ (\mathbf{u}_{1}\cdot\mathbf{u}_{2})(\mathbf{a%
}_{j}\cdot \mathbf{b}_{j})-(\mathbf{u}_{2}\cdot\mathbf{b}_{j})(\mathbf{u}%
_{1}\cdot\mathbf{a}_{j}\right] =0
\end{align*}
Therefore \qquad\qquad\qquad\qquad\qquad\qquad\qquad%
\begin{align*}
B_{t}(\mathbf{u}_{1}\mathbf{,u}_{2}) & =(\mathbf{u}_{1}\times\lbrack \mathbf{%
(}1-t)\delta_{0}+t\delta_{1}])\cdot(\mathbf{u}_{2}\times \lbrack\mathbf{(}%
1-t)\delta_{0}+t\delta_{1}]) \\
& =(1-t)^{2}B_{0}(\mathbf{u}_{1}\mathbf{,u}_{2})+t^{2}B_{1}(\mathbf{u}_{1}%
\mathbf{,u}_{2}) \\
& +(1-t)t[(\mathbf{u}_{1}\times\delta_{0})\cdot(\mathbf{u}_{2}\times
\delta_{1})\pm(\mathbf{u}_{2}\times\delta_{0})\cdot(\mathbf{u}_{1}\times
\delta_{1})] \\
& =(1-t)t(\mathbf{u}_{1}\times\mathbf{u}_{2})\cdot(\delta_{0}\times\delta
_{1})=0
\end{align*}
\end{proof}

\section{The spherical representation of shape space $M^{\ast}$}

By the \emph{spherical representation} we refer to an identification of $%
M^{\ast}$ with a round 2-sphere, with a distinguished (northern) hemisphere $%
M_{+}^{\ast}$ whose natural orientation induces the positive orientation of
the equator $E^{\ast}$ and hence the (eastward) direction of increasing
longitude. We also assume the (cyclic) ordering $\mathfrak{b}_{23}$, $%
\mathfrak{b}_{31},\mathfrak{b}_{12}$ of the three binary collision points
(lying on $E^{\ast}$) is in the positive direction. Finally, the
correspondence should represent the kinematic geometry and hence is an
isometry 
\begin{equation}
M^{\ast}\rightarrow S^{2}(1/2)   \label{corr1}
\end{equation}
which identifies each shape $\delta^{\ast}$ with a specific point on the
sphere. We will develop methods enabling us to express the spherical
coordinates in terms of intrinsic invariants of $\delta^{\ast}$, and
conversely.

Let $(r,\theta)$ denote \emph{polar coordinates} on $S^{2}(1/2)$, where $r$
is the \emph{polar distance }which measures the spherical distance from $%
\delta^{\ast}$ to the north pole $\mathcal{N}\in M_{+}^{\ast}$ and $\theta$
is the longitude angle. For convenience, we also introduce spherical
coordinates $(\varphi,\theta)$, where the angle $\varphi=2r$ is the \emph{%
colatitude} with $\varphi=0$ at the north pole. In these coordinates the
Riemannian metric of the sphere $M^{\ast}$ expresses as 
\begin{align}
d\sigma^{2} & =dr^{2}+\frac{1}{4}\sin^{2}(2r)d\theta^{2}=\frac{1}{4}%
(d\varphi^{2}+\sin^{2}\varphi\text{ }d\theta^{2})  \label{metric7} \\
0 & \leq r\leq\frac{\pi}{2}\text{, \ }0\leq\varphi\leq\pi\text{, \ }%
0\leq\theta\leq2\pi  \notag
\end{align}

\begin{remark}
\label{convention}The choice of the zero meridian $\theta=0$ is a matter of
convenience, and until further notice our convention is that it passes
through $\mathfrak{b}_{23}$. Indeed, only longitude differences $(\theta
-\theta^{\prime})$ is an intrinsic property of shapes of m-triangles, see
Section 4.2 and 4.3.
\end{remark}

\subsection{Geometric interpretation of the polar distance r}

Let $\delta$ be a nonzero positively oriented m-triangle, that is, $r\leq
\pi/4$. We will investigate the relationship between the polar distance $r$
and the geometric invariants of $\delta$. Let $\delta_{1}^{\ast}$ be the
intersection point between the equator $E^{\ast}$ and the meridian passing
through $\delta^{\ast}$. Choose unit size representatives 
\begin{equation*}
\delta_{0}=(\mathbf{a}_{1,}\mathbf{a}_{2}\mathbf{,a}_{3})\text{, \ \ }%
\delta_{1}=(\mathbf{b}_{1,}\mathbf{b}_{2}\mathbf{,b}_{3}) 
\end{equation*}
of the pole $\mathcal{N}$ and $\delta_{1}^{\ast}$ with $\delta_{0}\times
\delta_{1}=0$, and observe that the above meridian is the shape curve of the
linear motion $\delta_{t}=(1-t)\delta_{0}+t\delta_{1}$, $0\leq t\leq1$.
Henceforth, let $t$ be the unique value such that $\delta_{t}^{\ast}=%
\delta^{\ast}$.

Let $C$ be the cone surface (cf. also Definition \ref{cone}) in $\bar{M}$
spanned by the rays through points on the above meridian between $\mathcal{N}
$ and $\delta_{1}^{\ast}$. $C$ is isometric to a Euclidean sector of angular
width $\pi/4$, see Figure 3. The cord distance between $\mathcal{N}$ and $%
\delta_{1}^{\ast}$ is the number in (\ref{dist2}). On the other hand, the
cord distance between $\mathcal{N}$ and $\bar{\delta}_{t}$ can be computed
in two different ways, namely 
\begin{equation*}
dist(\mathcal{N},\bar{\delta}_{t})=2t\sin\frac{\pi}{8}=\sin\frac{\pi}{8}-\cos%
\frac{\pi}{8}\tan(\frac{\pi}{8}-r) 
\end{equation*}
Hence, 
\begin{equation}
t=\frac{1}{2}(1-\cot\frac{\pi}{8}\tan(\frac{\pi}{8}-r))   \label{t}
\end{equation}
and moreover, 
\begin{equation}
dist(O,\bar{\delta}_{t})=\frac{\cos\frac{\pi}{8}}{\cos(\frac{\pi}{8}-r)}. 
\label{dist3}
\end{equation}

Next, let us compute the eigenvalues of $B_{t}=B_{\delta_{t}}$, namely

\begin{equation*}
\lambda_{1}^{\prime}=B_{t}(\mathbf{u}_{1}\mathbf{,u}_{1})\text{, \ \ }%
\lambda_{2}^{\prime}=B_{t}(\mathbf{u}_{2}\mathbf{,u}_{2})\text{,}
\end{equation*}
where by Lemma \ref{frame} we have chosen an orthonormal frame $\left\{ 
\mathbf{u}_{1},\mathbf{u}_{2}\right\} $ of the plane $\Pi(\delta_{0})$ with $%
\mathbf{u}_{2}\cdot\mathbf{b}_{j}=0$ for all $j$.\ It follows that 
\begin{align*}
\lambda_{1}^{\prime} & =B_{t}(\mathbf{u}_{1}\mathbf{,u}_{1})=\left\vert 
\mathbf{u}_{1}\times\delta_{t}\right\vert ^{2}=\left\vert \mathbf{u}%
_{1}\times(1-t)\delta_{0}\right\vert ^{2} \\
& =(1-t)^{2}B_{0}(\mathbf{u}_{1},\mathbf{u}_{1})=\frac{1}{2}(1-t)^{2}
\end{align*}
Therefore, since $\delta^{\ast}$ and $\bar{\delta}_{t}$ differ by the
scaling factor (\ref{dist3}), the eigenvalues of $B_{\delta^{\ast}}$ are 
\begin{equation}
\lambda_{1}=\frac{\cos\frac{\pi}{8}}{\cos(\frac{\pi}{8}-r)}%
\lambda_{1}^{\prime}=\frac{1}{2}(1-t)^{2}\frac{\cos\frac{\pi}{8}}{\cos(\frac{%
\pi}{8}-r)}   \label{lambda1}
\end{equation}
and $\lambda_{2}=1-\lambda_{1}.$

Finally, we can use (\ref{t}), (\ref{lambda1}) and the identity 
\begin{equation*}
\lambda_{1}\lambda_{2}=4m\,_{1}m_{2}m_{3}\Delta^{2}
\end{equation*}
to solve for $r$ as a function of the area $\Delta$ of $\delta^{\ast}$. We
state the final result as follows :

\begin{lemma}
The polar distance $r$ for an arbitrary given positively oriented m-triangle 
$\delta$ is given by the formula 
\begin{equation}
\cos(2r)=4\sqrt{m_{1}m_{2}m_{3}}\frac{\Delta}{I}   \label{cos2r}
\end{equation}
where $\tsum m_{i}=1$ and $\Delta$ $($resp. $I)$ is the area (resp. moment
of inertia) of $\delta.$
\end{lemma}

Equivalently, by (\ref{lambda}) there is the formula 
\begin{equation}
\sin2r=\frac{1}{I}\left\vert \lambda_{1}-\lambda_{2}\right\vert 
\label{sin2r}
\end{equation}

\subsection{Geometric interpretation of the longitude angle $\protect\theta$}

Let $\delta_{0}$ be an oriented m-triangle whose shape $\delta_{0}^{\ast}$
is the pole $\mathcal{N}$ or \QTR{cal}{S} (i.e . $r=0$ or $\pi/2)$. Recall
from Lemma \ref{frame}, it is possible to deform $\delta_{0}$, through a
linear motion with zero angular momentum, to the shape of any given
degenerate m-triangle. Then the shape curve will be the meridian from the
pole to a point $\delta_{1}^{\ast}$ on the equator circle $E^{\ast}$.
Moreover, the line spanned by the final configuration $\delta_{1}$ is
uniquely determined by $\delta_{0}$, and there is a constant eigenframe
throughout the deformation.

Now, let us consider two points $\delta_{1}^{\ast},$ $\delta_{2}^{\ast}$ on $%
E^{\ast}$ and seek an interpretation of their spherical distance in $M^{\ast
}=S^{2}(1/2)$.

\begin{theorem}
\label{ang1}Let $\delta_{0}=(\mathbf{a}_{1}\mathbf{,a}_{2}\mathbf{,a}_{3})$
be an m-triangle, of maximal area for a fixed moment of inertia, and let $%
\delta_{1},\delta_{2}$ be degenerate (but nonzero) m-triangles satisfying
the vanishing angular momentum condition 
\begin{equation*}
\delta_{0}\times\delta_{1}=\delta_{0}\times\delta_{2}=0 
\end{equation*}
for the linear motions from $\delta_{0}$ to $\delta_{i},i=1,2$. Then the
angle $\psi$ between the lines spanned by $\delta_{1}$ and $\delta_{2}$ is
equal to the distance between the associated points $\delta_{1}^{\ast}$ and $%
\delta _{2}^{\ast}$ in the shape space $M^{\ast}$, namely 
\begin{equation*}
\psi=\frac{1}{2}\left| \theta_{2}-\theta_{1}\right| 
\end{equation*}
\end{theorem}

\begin{proof}
We may assume all m-triangles are confined to the xy-plane, the shape $%
\delta_{0}^{\ast}$ is the north pole and $\delta_{i}^{\ast}$ has longitude
angle $\theta_{i},i=1,2$. Consider the piecewise linear motion whose
associated shape curve is the spherical triangle $D$ in $M^{\ast}$ with
vertices $\delta_{1}^{\ast},\delta_{2}^{\ast},\delta_{0}^{\ast}$. Starting
from $\delta_{1}$, the motion passes successively through the m-triangles $%
\delta_{1,}\delta_{2}^{\prime},\delta_{0},\delta_{1}^{\prime}$. Here $%
\delta_{2}^{\prime}$ is congruent to $\delta_{2}$ and is situated in the
same line as $\delta_{1}$, whereas $\delta_{1}^{\prime}$ is congruent to $%
\delta_{1}$ but is actually situated in a line making the angle $\psi$ with
the original line.

Next, let us apply the Gauss-Bonnet formula (\ref{GB}) to the region $D$,
thus obtaining an equality between twice the area of $D$ and the angle $\psi$%
. Finally, we simply combine this with the fact that, as a geodesic triangle
on the sphere of radius 1/2, the area of $D$ equals one quarter of its angle
at $\delta_{0}^{\ast}$, namely the angle $\left|
\theta_{2}-\theta_{1}\right| $.
\end{proof}

In general, it turns out that the longitude angle $\theta$ of an m-triangle
is determined by the relative position and size of the eigenframe and
normalized area $\Delta/I$ respectively. To make this relationship precise,
let $\delta=(\mathbf{a}_{1}\mathbf{,a}_{2}\mathbf{,a}_{3})$ be a
non-degenerate, positively oriented m-triangle in the xy-plane with normal
vector 
\begin{equation*}
\frac{\mathbf{a}_{1}\times\mathbf{a}_{2}}{\left\vert \mathbf{a}_{1}\times%
\mathbf{a}_{2}\right\vert }=\mathbf{k}\text{ }
\end{equation*}
and assume the shape $\delta^{\ast}$ is not the pole $\mathcal{N}$.

\begin{theorem}
\label{ang2}Let $\left\{ \mathbf{u}_{1}\mathbf{,u}_{2}\right\} $ be an
eigenframe of $\delta$, where $\mathbf{u}_{1}$ is the eigenvector of the
inertia tensor $B_{\delta}$ associated with the smallest eigenvalue $%
\lambda_{1}$, and let $\psi_{i}\ $ be the (oriented) angle from $\mathbf{a}%
_{1}$ to $\mathbf{u}_{i}$. Then the following identity \ 
\begin{equation}
\tan\frac{\theta}{2}=-\frac{1+\sin\varphi}{\cos\varphi}\tan\psi_{1}\ =\frac{%
1+\sin\varphi}{\cos\varphi}\cot\psi_{2}\   \label{angle2}
\end{equation}
relates the angle $\psi_{i}$ to the spherical coordinates $(\varphi,\theta)$
of the shape $\delta^{\ast}$ on the 2-sphere $M^{\ast}$, where $\varphi$ is
the colatitude and $\theta$ is the longitude (eastward, with $\theta=0$ at $%
\mathfrak{b}_{23}$).
\end{theorem}

\begin{remark}
The above formula holds for any choice of eigenframe since $\psi_{1}$
changes by $\pm\pi$ if $\mathbf{u}_{1}$ is replaced by $-\mathbf{u}_{1}$.
\end{remark}

\begin{proof}
The first step of the proof is to derive a formula which expresses $\psi_{1}$
solely in terms of intrinsic invariants of the m-triangle, together with a
simple recipe for calculating this angle. Then, by applying the Gauss-Bonnet
formula (cf. Theorem C2) we shall deduce formula (\ref{angle2}).

Since $\psi_{2}=\psi_{1}\pm\pi/2$ we need only prove the first identity in (%
\ref{angle2}). First of all, in order to have the angle $\psi_{1}$ uniquely
defined we must specify the choice of eigenframe. Namely, let $\left\{ 
\mathbf{u}_{1}\mathbf{,u}_{2}\right\} $ be a positive frame and hence $%
\mathbf{u}_{1}\mathbf{\times u}_{2}=\mathbf{k}$, and moreover, we assume $%
\mathbf{u}_{1}$ chosen so that 
\begin{equation}
-\frac{\pi}{2}\leq\psi_{1}<\frac{\pi}{2}   \label{range2}
\end{equation}

Let $\left\{ \mathbf{e}_{1}\mathbf{,e}_{2}\right\} $ be the orthonormal
frame derived from $\left\{ \mathbf{a}_{1}\mathbf{,a}_{2}\right\} $ by the
Gram-Schmidt algorithm, with $\mathbf{e}_{1}=\mathbf{a}_{1}/\left| \mathbf{a}%
_{1}\right| $. Using the expressions (\ref{BB3}) it is not difficult to
deduce 
\begin{align}
A & =B_{\delta}(\mathbf{e}_{1}\mathbf{,e}_{1})=4m_{1}m_{2}m_{3}(m_{2}+m_{3})%
\frac{\Delta^{2}}{I_{1}}  \notag \\
B & =B_{\delta}(\mathbf{e}_{1}\mathbf{,e}%
_{2})=((m_{2}+m_{3})C_{3}-2m_{1}m_{2}I_{1})\frac{\Delta}{I_{1}}  \label{ABC}
\\
C & =B_{\delta}(\mathbf{e}_{2}\mathbf{,e}_{2})=I-A  \notag
\end{align}
By writing 
\begin{align}
\mathbf{u}_{1} & =\cos\psi_{1}\mathbf{e}_{1}+\sin\psi_{1}\mathbf{e}_{2}
\label{u1u2} \\
\mathbf{u}_{2} & =-\sin\psi_{1}\mathbf{e}_{1}+\cos\psi_{1}\mathbf{e}_{2} 
\notag
\end{align}
and inserting these expressions into $B_{\delta}(\mathbf{u}_{1}\mathbf{,u}%
_{2})=0$, we deduce the formula 
\begin{equation}
\tan2\psi_{1}=\frac{2B}{A-C}=\frac{2B}{2A-I}   \label{angle1}
\end{equation}
and similarly 
\begin{equation}
\lambda_{1}-\lambda_{2}=B_{\delta}(\mathbf{u}_{1}\mathbf{,u}_{1})-B_{\delta
}(\mathbf{u}_{2}\mathbf{,u}_{2})=(A-C)\cos2\psi_{1}+2B\sin2\psi_{1} 
\label{l-diff}
\end{equation}

How is $\psi_{1}$ determined from (\ref{angle1}) and (\ref{l-diff})? Assume
first $B=0$. With the normalization $\tsum m_{1}=1,\tsum I_{i}=1$, this
happens when 
\begin{equation*}
I_{1}(m_{3}-m_{1}m_{2})+I_{2}(1-m_{1})^{2}=(1-m_{1})m_{3}
\end{equation*}
For example, with uniform mass distribution this holds for the isosceles
triangle with $\left| \mathbf{a}_{2}\right| =\left| \mathbf{a}_{3}\right| $.
Note that $\left\{ \mathbf{e}_{1}\mathbf{,e}_{2}\right\} $ is also a
positive eigenframe, and the eigenvalues $\lambda_{i}$ equals $A$ and $C$.
Since $\lambda_{1}<\lambda_{2}$, by assumption, we deduce from (\ref{l-diff}%
) that $A>C$ implies $\psi_{1}=\frac{\pi}{2}$, and $A<C$ implies $%
\psi_{1}=0. $

Next, assume $B\neq0$. By combining (\ref{angle1}) and (\ref{l-diff}) we
eliminate $\cos2\psi_{1}$ and obtain the expression 
\begin{equation}
\lambda_{1}-\lambda_{2}=\frac{(A-C)^{2}+4B^{2}}{2B}\sin2\psi_{1} 
\label{eigendiff}
\end{equation}
Consequently, $\sin2\psi_{1}$ has the opposite sign of $B$, namely 
\begin{equation}
0<\psi_{1}<\frac{\pi}{2}\text{ \ if }B<0\text{, \ \ }-\frac{\pi}{2}<\psi
_{1}<0\text{ if }B>0\text{ \ \ }   \label{angle}
\end{equation}

Finally, we turn to the proof of formula (\ref{angle2}). On the sphere $%
M^{\ast}$, let $\delta_{1}^{\ast}$ be the intersection point of $E^{\ast}$
and the meridian passing through $\delta^{\ast}$, and consider the spherical
triangle on $M^{\ast}$ with vertices $\mathfrak{b}_{23},\delta_{1}^{\ast
},\delta^{\ast}$, whose area is denoted by $\bar{\Delta}$. The right angle
at the vertex $\delta_{1}^{\ast}$ has adjacent edges of length 
\begin{equation*}
s=\frac{\left\vert \theta\right\vert }{2}\leq\frac{\pi}{2}\text{ \ and }%
\frac{\pi}{4}-r, 
\end{equation*}
and by applying the spherical sine law and area formula to the magnified
triangle on $S^{2}(1)$ with area $\tilde{\Delta}=4\bar{\Delta}$, we have 
\begin{align*}
\tan\frac{\tilde{\Delta}}{2} & =\frac{\sin2s\sin(\frac{\pi}{2}-2r)}{%
1+\cos2s+\cos(\frac{\pi}{2}-2r)+\cos2s\cos(\frac{\pi}{2}-2r)} \\
& =\frac{\sin2s}{1+\cos2s}\frac{\sin(\frac{\pi}{2}-2r)}{1+\cos(\frac{\pi}{2}%
-2r)}=\tan s\tan(\frac{\pi}{4}-r)
\end{align*}

On the other hand, applying the Gauss-Bonnet formula (Theorem C2) to the
triangle in Figure 4, it is not difficult to see that the total rotation of
the vector $\mathbf{a}_{1}$ is through the angle $\psi_{1}$, hence 
\begin{equation*}
\pm\psi_{1}=2\bar{\Delta}=\frac{\tilde{\Delta}}{2}
\end{equation*}
and consequently 
\begin{equation}
\tan\psi_{1}=\pm\tan\frac{\theta}{2}\tan(\frac{\pi}{4}-r)   \label{angle3}
\end{equation}

We claim that the sign to be used in (\ref{angle3}) is -1. This can be seen
by considering the situation where $\delta_{1}^{\ast}$ lies between $%
\mathfrak{b}_{23}$ and $\mathfrak{b}_{31}$, by observing that $\mathfrak{%
\psi }_{1}$ decreases as $\theta$ increases (i.e. when $\delta_{1}^{\ast}$
approaches $\mathfrak{b}_{31})$. This completes the proof of formula (\ref%
{angle2}).
\end{proof}

\subsection{Intrinsic form of the spherical representation}

We will focus attention on the \textquotedblright inverse\textquotedblright\
of the correspondence (\ref{corr1}), namely 
\begin{equation*}
(\varphi,\theta)\rightarrow(I_{1},I_{2},I_{3}),\sum I_{j}=1 
\end{equation*}
The first step in this direction was, in fact, our kinematic proof of
Theorem A (cf. Section 3.2.2). Namely, by substituting (\ref{spherical})\
into (\ref{comb}) and considering the special case of $m_{3}=m_{1}m_{2}$, it
is easy to verify that the expressions (\ref{comb}) simplify to 
\begin{equation}
I_{1}=m_{1}^{\ast}(1+\sin\varphi\cos\theta)\text{, \ }I_{2}=m_{2}^{\ast
}(1+\sin\varphi\sin\theta)   \label{intrins}
\end{equation}
These formulas are, indeed, a special case of a general intrinsic
description of the spherical representation, purely in terms of geometric
concept.

The \emph{longitude distance }between $\delta_{1}^{\ast}=(\varphi_{1},%
\theta_{1})$ and $\delta_{2}^{\ast}=(\varphi_{2},\theta_{2})$ is, by
definition, the angle $\left\vert \theta_{1}-\theta_{2}\right\vert $ mod $%
2\pi$. Then it is easy to check that (\ref{intrins}) can be stated as 
\begin{equation}
I_{i}=m_{i}^{\ast}(1+\sin\varphi\cos\tilde{\theta}_{i})\text{, \ }i=1,2,3 
\label{intrins2}
\end{equation}
where $\tilde{\theta}_{1}$ (resp. $\tilde{\theta}_{2}$ or $\tilde{\theta}%
_{3})$ is the longitude distance between $\delta^{\ast}=(\varphi,\theta)$
and the binary collision point $\mathfrak{b}_{23}$ (resp. $\mathfrak{b}_{31}$
or $\mathfrak{b}_{12}$)$.$

On the other hand, consider the three distance functions on $%
M^{\ast}=S^{2}(1/2)$%
\begin{equation}
\sigma_{i}=\sigma_{i}(\delta^{\ast})=dist(\delta^{\ast},\mathfrak{b}%
_{i+1,i+2})\text{ \ (}i\text{ }\func{mod}3)   \label{dist1}
\end{equation}
which measure the (spherical) distances to the points $\mathfrak{b}_{ij}$.
By the spherical cosine law applied to the triangle with vertices $\mathcal{N%
},\mathfrak{b}_{ij},\delta^{\ast}$, it follows that 
\begin{equation*}
\cos2\sigma_{i}=\sin\varphi\cos\tilde{\theta}_{i}\text{, \ }i=1,2,3 
\end{equation*}
and consequently (\ref{intrins2}) has the invariant form 
\begin{equation}
I_{i}=m_{i}^{\ast}(1+\cos2\sigma_{i})\text{, \ }i=1,2,3   \label{intrins3}
\end{equation}
This may be stated as 
\begin{equation}
\cos2\sigma_{i}=\frac{I_{i}}{m_{i}^{\ast}}-1=\frac{1}{m_{i}^{\ast}}\tilde {I}%
_{i}\text{,\ \ cf. (\ref{I-tilde}) }   \label{intrins5}
\end{equation}
or equivalently 
\begin{equation}
\sigma_{i}=\arccos\sqrt{\frac{I_{i}}{1-m_{i}}}\text{ \ \ \ }(\sum I_{i}=1) 
\label{intrins6}
\end{equation}

Thus, in order to establish (\ref{intrins3}) or (\ref{intrins2}) as a
general formula it suffices to verify formula (\ref{intrins6}) in general.
Again, the basic idea we use is to construct a suitable linear motion in $%
M_{0}$, as we did in Chapter 3, namely we consider the linear motion whose
shape curve is the (shortest) geodesic from $\delta^{\ast}$ to $\mathfrak{b}%
_{ij}$. We shall calculate the length of this curve in the following way.

Let $\delta=(\mathbf{a}_{1},\mathbf{a}_{2},\mathbf{a}_{3})$ be a given
m-triangle, normalized with $I=1$, and consider the linear motion of
vanishing angular momentum 
\begin{equation*}
\delta(t)=(1-t)\delta+t(\sqrt{\frac{1-m_{1}}{I_{1}}}\mathbf{a}_{1},-\frac{%
m_{1}}{\sqrt{(1-m_{1})I_{1}}}\mathbf{a}_{1},-\frac{m_{1}}{\sqrt{%
(1-m_{1})I_{1}}}\mathbf{a}_{1}) 
\end{equation*}
between\ $\delta=\delta(0)$ and the normalized m-triangle $%
\delta_{1}=\delta(1)$ with the shape $\delta_{1}^{\ast}=\mathfrak{b}_{23}$.
Its shape curve is the desired geodesic.

The length $\left\vert \delta-\delta_{1}\right\vert $ of the segment in $%
M_{0}$ from $\delta$ to $\delta_{1}$ is also the length $\bar{\sigma}_{1}$
of the chord in $\bar{M}$ between $\delta^{\ast}$ and $\delta_{1}^{\ast}$,
see (\ref{dist}) and Figure 4. By applying the Ceva-cosine law (\ref%
{Ceva-cos}) to the calculation of inner products $\mathbf{a}_{i}\cdot\mathbf{%
a}_{j}$ we arrive at the expression 
\begin{equation*}
\bar{\sigma}_{1}^{2}=\left\vert \delta-\delta_{1}\right\vert ^{2}=2(1-\sqrt{%
\frac{I_{1}}{1-m_{1}}}) 
\end{equation*}
and consequently 
\begin{equation*}
\sigma_{1}=2\arcsin\frac{\bar{\sigma}_{1}}{2}=\arccos\sqrt{\frac{I_{1}}{%
1-m_{1}}}
\end{equation*}

\subsection{The reduced Newton's equation in spherical coordinates}

Let us utilize the structure of $\bar{M}$ as a cone over the 2-sphere to
express the reduced Newton's equation of Theorem E1 in terms of the
spherical coordinate system $(\rho,\varphi,\theta)$, where $\rho=\sqrt{I}$
measures the distance from the base point $O$ (cone vertex). The
relationship between coordinate functions $\left\{ I_{j}\right\} ,\left\{
r_{ij}\right\} $ and $\left\{ \rho,\varphi,\theta\right\} $ is expressed by (%
\ref{r/I}) and (\ref{intrins2}), thus enabling us to transform the equation
to a system purely in terms of $\rho,\varphi,\theta$. This change of
variable is, however, rather messy, but an equivalent system can be worked
out in several ways. For example, we obtain the following system of three 2.
order equations 
\begin{align}
(i)\text{ \ }0 & =\ddot{\rho}+\frac{\dot{\rho}^{2}}{\rho}-\frac{1}{\rho }%
(U+2h)  \notag \\
(ii)\text{ \ }0 & =\text{\ }\ddot{\varphi}+2\frac{\dot{\rho}}{\rho}\dot{%
\varphi}-\frac{1}{2}\sin(2\varphi)\dot{\theta}^{2}-\frac{4}{\rho^{2}}\frac{%
\partial U}{\partial\varphi}  \label{Newton2} \\
(iii)\text{ \ \ }0 & =\ddot{\theta}+2\frac{\dot{\rho}}{\rho}\dot{\theta }%
+2\cot(\varphi)\dot{\varphi}\dot{\theta}-\frac{4}{\rho^{2}}\frac{1}{\sin
^{2}\varphi}\frac{\partial U}{\partial\theta}  \notag
\end{align}
valid for planary three-body motions with a fixed energy level $h$. The
angular momentum $\Omega$ constant is not implicit in these equations since
it is an integration constant defined by the initial value problem. In fact,
we have also the equations

\begin{equation*}
T-U=h\text{, \ \ \ \ }\ddot{I}=2T+2h\text{,}
\end{equation*}
namely the energy equation and the Lagrange-Jacobi equation (cf. (\ref{L-J}%
)). The latter is precisely equation (i) in (\ref{Newton2}), and the energy
integral 
\begin{equation*}
U+h-\frac{\Omega^{2}}{2\rho^{2}}=\frac{1}{2}\dot{\rho}^{2}+\frac{\rho^{2}}{8}%
(\dot{\varphi}^{2}+(\sin\varphi)^{2}\text{ }\dot{\theta}^{2}) 
\end{equation*}
makes any of the two equations (ii) or (iii) superfluous.

\subsection{Ceva-type relations in the spherical representation}

The classical Ceva theorem tells us that the lines from the vertices to the
center of mass of an m-triangle $\delta$ divide the triangle into
subtriangles whose areas are in the proportion 
\begin{equation*}
\Delta_{1}:\Delta_{2}:\Delta_{3}=m_{1}:m_{2}:m_{3}
\end{equation*}
On the other hand, a point $\delta^{\ast}$ on a hemisphere $M_{\pm}^{\ast}$
divides it into three spherical triangles with areas $A_{i}$, with the
common vertex $\delta^{\ast}$ and the binary collision points $\mathfrak{b}%
_{12}$, $\mathfrak{b}_{23}$, $\mathfrak{b}_{31}$ as the other vertices, cf.
Figure 5. In this way, various (normalized) geometric invariants of $\delta$
such as sides, areas, angles (cf. Figure 1) have their spherical "dual"
counterparts, although the dual quantity may be of a different type. There
are, for example, the three central angles $\alpha_{i}$ (resp. $\pi_{i})$ of 
$\delta$ (resp. at the shape point $\delta^{\ast}$) with $%
\sum\alpha_{i}=\tsum \pi_{i}=2\pi$. According to the following lemma, the
areas $A_{i}$ are dual to the angles $\alpha_{i}$, and later we also show
the areas $\Delta_{i}$ are dual to certain sides of the spherical triangles.

\begin{lemma}
Let $\delta=(\mathbf{a}_{1}\mathbf{,a}_{2}\mathbf{,a}_{3})$ be an m-triangle
with central angle $\alpha_{i}$ opposite to the vector $\mathbf{a}_{i}$.
Then the area of the spherical triangle in $M^{\ast}$ with vertices $%
\mathfrak{b}_{12},\delta^{\ast},\mathfrak{b}_{31}$ equals 
\begin{equation*}
A_{1}=\frac{1}{2}(\pi-\alpha_{1}) 
\end{equation*}
\end{lemma}

\begin{proof}
Let $L_{2},L_{3}$ be the lines spanned by the vectors $\mathbf{a}_{2}$ and $%
\mathbf{a}_{3}$ respectively. There is an obvious piecewise linear motion
with $\mathbf{\Omega}=0$ which starts at $\delta$ and collapses the triangle
to a degenerate configuration of shape $\mathfrak{b}_{31}$ along $L_{2}$,
and continues along $L_{2}$ until the shape of $\mathfrak{b}_{12}$ is
reached. This motion keeps the direction of $\mathbf{a}_{2}$ unaltered.

On the other hand, the linear motion with $\mathbf{\Omega}=0$ which
collapses $\delta$ to a configuration of shape $\mathfrak{b}_{12}$ along $%
L_{3}$, will rotate $\mathbf{a}_{2}$ to a vector along $L_{3}$ which lies
opposite to $\mathbf{a}_{3}$. Hence, the total change of position when $%
\delta$ is deformed according to the above piecewise linear motion whose
shape curve encloses the spherical triangle, is equal to the angle $\pi-$ $%
\alpha_{1}$. Finally, by the Gauss-Bonnet formula, this is twice the area $%
A_{1}$ of the triangle.
\end{proof}

Next, we turn to the mutual distances, that is, the sides of the m-triangle $%
\delta$ 
\begin{equation*}
s_{1}=r_{23}=\left\vert \mathbf{a}_{2}-\mathbf{a}_{3}\right\vert \text{ \
etc. \ \ }\ 
\end{equation*}
and ask for their spherical counterpart, namely the spherical distances $%
\sigma_{i}$ from $\delta^{\ast}$ to the binary collision points. The
quantities $s_{i}$ have, indeed, a nice geometric interpretation in the
vector algebra representation described below, see (\ref{side4}). But first,
by combining (\ref{r/I}) and (\ref{intrins5}), the identity 
\begin{equation}
s_{i}^{2}=\frac{1-m_{i}-I_{i}}{\hat{m}_{i}}=\frac{1-m_{i}}{\hat{m}_{i}}(1-%
\frac{I_{i}}{1-m_{i}})=\frac{1-m_{i}}{\hat{m}_{i}}\sin^{2}\sigma _{i} 
\label{side2}
\end{equation}
holds, where we have assumed normalization $I=1$ and (as usual) $\sum
m_{i}=1 $, cf. also (\ref{mass2}) for notation. By summation over $i$ the
condition $I=1$ reads 
\begin{equation}
1=\sum\hat{m}_{i}s_{i}^{2}=\sum(1-m_{i})\sin^{2}\sigma_{i}\text{ \ \ or \ }%
\sum m_{i}^{\ast}\cos2\sigma_{i}=0   \label{side3}
\end{equation}
where the first identity is just the normalized version of Lagrange's
formula (\ref{Isum}).

Now, let $\delta_{0}$ be an m-triangle whose shape is a pole $%
\delta_{0}^{\ast}=$ $\mathcal{N}$ or \QTR{cal}{S}, and let 
\begin{equation}
\left\{ \alpha_{1},\alpha_{2},\alpha_{3}\right\} \text{, \ \ }\left\{
\beta_{1},\beta_{2},\beta_{3}\right\} \text{\ }   \label{triple2}
\end{equation}
denote the central angles $\alpha_{i}$ of $\delta_{0}$ and the central
angles $\beta_{i}=\pi_{i}$ at $\delta_{0}^{\ast}$, respectively, see Figure
5. In particular, $\beta_{1}$ is the longitude distance between $\mathfrak{b}%
_{31}$ and $\mathfrak{b}_{12}$, or equivalently, $\beta_{1}/2$ is their
distance in $M^{\ast}$. The relationship between the two triples in (\ref%
{triple2}) follows from the above lemma, namely 
\begin{equation}
\beta_{i}=4A_{i}=2\pi-2\alpha_{i}   \label{central}
\end{equation}

On the other hand, the triple of angles and the (normalized) mass
distribution uniquely determine each other. In one direction, we apply the
Ceva-cosine law to $\delta_{0}$ and obtain 
\begin{equation}
\cos\alpha_{1}=\frac{-\sqrt{m_{2}m_{3}}}{\sqrt{1-m_{2}}\sqrt{1-m_{3}}}\text{%
, \ }\cos\beta_{1}=\frac{m_{2}m_{3}-m_{1}}{(1-m_{2})(1-m_{3})}\text{, \ etc.}
\label{angles}
\end{equation}
In particular, the angles are in the range 
\begin{equation}
\frac{\pi}{2}<\alpha_{i}<\pi\text{ , \ \ \ }0<\beta_{i}<\pi   \label{range}
\end{equation}
We also note that $\cos\beta_{i}$ can be calculated by applying a distance
function $\sigma_{j},j\neq i$, see (\ref{dist1}). For example, $\cos\beta
_{1}=\cos(2\sigma_{3}(\mathfrak{b}_{31}))$, and by (\ref{intrins5}) and the
fact that $\mathfrak{b}_{31}$ has $I_{3}=\frac{m_{2}m_{3}}{1-m_{2}}$, we
deduce again the above expression for $\cos\beta_{1}$.

In the other direction, we would like to know which angles in the range (\ref%
{range}) are actually realizable for some mass distribution. The condition
on the angles is 
\begin{equation}
2\sin\beta_{i}<\sum\sin\beta_{j}\text{, \ for each }i   \label{realize}
\end{equation}
and the corresponding (normalized) mass distribution is defined by 
\begin{equation}
m_{1}=1-\frac{2\sin\beta_{1}}{\sum\sin\beta_{j}}=\frac{1+\sum\cos\beta_{j}}{%
-1+\sum\cos\beta_{j}-2\cos\beta_{1}}\text{ \ , \ \ etc.}   \label{mass}
\end{equation}
We omit the simple proof of this, remarking that the realizability condition
(\ref{realize}) also has a nice geometric interpretation. Namely, consider
the three binary collision points $\mathfrak{b}_{12},\mathfrak{b}_{23},%
\mathfrak{b}_{31}$ as the vertices of a triangle in the Euclidean disk (of
radius $1/2$) with the equator circle $E^{\ast}$ as boundary. In general,
let us call a triangle \emph{central }if its circumcenter $O$ lies in its
interior. Then the realizability condition simply says that the above
triangle must be central. For another property of this triangle we also
refer to Lemma \ref{dual}.

\subsection{The vector algebra representation of the kinematic geometry}

Since the kinematic study of m-triangles and their motions essentially
involves spherical geometry, we are naturally led to the vector algebra in
the Euclidean 3-space $\mathbb{R}^{3}$, where inner products and
determinants are the basic invariants. Therefore, it is sometimes convenient
to represent $M^{\ast}$ by the sphere $S^{2}(1)$ of unit vectors in $\mathbb{%
R}^{3}$ and hence its cone $\bar{M}$ becomes the whole Euclidean space
3-space. Thus we introduce the \emph{vector algebra representation} of the
moduli space $\bar{M}$ by constructing the following transformation between
Riemannian cones 
\begin{align}
\bar{M} & =C(S^{2}(1/2))\overset{\Psi}{\rightarrow}C(S^{2}(1))=\mathbb{R}^{3}
\label{cones} \\
\Psi & :(\rho,r,\theta)\rightarrow(\rho^{2},2r,\theta)=(I,\varphi ,\theta) 
\notag
\end{align}
which magnifies $M^{\ast}$ to a sphere of radius 1 and squares the distance
to the origin. It is a diffeomorphism away from the base point (or origin) $%
O $, namely the class of the triple collision $\rho=0$.

In (\ref{cones}), $(I,\varphi,\theta)$ are the usual spherical coordinates
in 3-space associated with the Euclidean coordinates $(x,y,z)$, that is, 
\begin{equation*}
x=I\sin\varphi\cos\theta\text{, \ }y=I\sin\varphi\sin\theta,\text{ \ }%
z=I\cos\varphi\text{\ }
\end{equation*}
where $0\leq\varphi\leq\pi,$ $0\leq\theta\leq2\pi$. The kinematic metric (%
\ref{dsbar1}) on $\bar{M}$, expressed as a metric on $\mathbb{R}^{3}$, now
becomes the following conformal modification of the Euclidean metric, namely 
\begin{equation}
d\bar{s}^{2}=d\rho^{2}+\rho^{2}d\sigma^{2}=\frac{dx^{2}+dy^{2}+dz^{2}}{4%
\sqrt{x^{2}+y^{2}+z^{2}}}\   \label{dsbar3}
\end{equation}

A variable point on $M^{\ast}$ will be represented by a unit vector 
\begin{equation*}
\delta^{\ast}\rightarrow\ \mathbf{p}=(x,y,z)\in S^{2}(1)\subset\mathbb{R}%
^{3}\ 
\end{equation*}
and we fix the following notation and location of binary collision points
(cf. (\ref{angles}) and Figure 6) 
\begin{align}
\mathfrak{b}_{23} & \rightarrow\mathbf{\hat{b}}_{1}=(1,0,0)  \notag \\
\mathfrak{b}_{31} & \rightarrow\mathbf{\hat{b}}_{2}=(\cos\beta_{3},\sin
\beta_{3},0)  \label{binary1} \\
\mathfrak{b}_{12} & \rightarrow\mathbf{\hat{b}}_{3}=(\cos\beta_{2},-\sin%
\beta_{2},0)  \notag
\end{align}
Moreover, (\ref{intrins5}) now reads 
\begin{equation}
\ \mathbf{p\cdot\hat{b}}_{i}=\cos2\sigma_{i}=\ \frac{1}{m_{i}^{\ast}}\tilde {%
I}_{i}=\left\{ 
\begin{array}{cc}
x & i=1 \\ 
x\cos\beta_{3}+y\sin\beta_{3} & i=2 \\ 
x\cos\beta_{2}-y\sin\beta_{2} & i=3%
\end{array}
\right.   \label{inner2}
\end{equation}

Recall that $2\sigma_{i}$ is the spherical distance between $\mathbf{p}$ and 
$\mathbf{\hat{b}}_{i}$; hence, by (\ref{side2}) there is the simple formula 
\begin{equation}
s_{i}=\frac{1}{2}\sqrt{\frac{1-m_{i}}{\hat{m}_{i}}}\left\vert \mathbf{p-%
\hat {b}}_{i}\right\vert   \label{side4}
\end{equation}
which expresses the three mutual distances $s_{i}$ for the normalized
m-triangle $\delta$ as the Euclidean distances (modulo a fixed factor) from $%
\mathbf{p\in}S^{2}$ to the three fixed points $\mathbf{\hat{b}}_{i}$ lying
on the circle $x^{2}+y^{2}=1,z=0.$

\begin{lemma}
\label{dual}There is a unique mass distribution $(m_{1}^{\prime},m_{2}^{%
\prime},m_{3}^{\prime})$ such that $(\mathbf{\hat{b}}_{1},\mathbf{\hat{b}}%
_{2},\mathbf{\hat{b}}_{3})$ becomes an m-triangle with center of mass at
origin, that is, 
\begin{equation*}
m_{1}^{\prime}\mathbf{\hat{b}}_{1}+m_{2}^{\prime}\mathbf{\hat{b}}%
_{2}+m_{3}^{\prime}\mathbf{\hat{b}}_{3}=0\text{,}
\end{equation*}
namely the dual masses $m_{i}^{\prime}=m_{i}^{\ast}$, cf. (\ref{mass2}).
\end{lemma}

\begin{proof}
Since the triangle is central the origin lies in its interior, so there are
barycentric coordinates $q_{i}>0$, unique up to a common multiple, such that 
$\sum q_{i}\mathbf{\hat{b}}_{i}=0$. On the other hand, by combining (\ref%
{side3}) and (\ref{inner2}), $\sum m_{i}^{\ast}\mathbf{\hat{b}}_{i}\cdot%
\mathbf{p}\ =0$ holds for all $\mathbf{p}$ and consequently $\sum
m_{i}^{\ast}\mathbf{\hat{b}}_{i}=0$.
\end{proof}

\subsection{An integral formula for the distance function on $M^{\ast}$}

The kinematic Riemannian metric $d\sigma^{2}$, expressed in terms of
coordinates $I_{1},I_{2}$ as in (\ref{dsigma2b}), may be viewed as the
infinitesimal version of an integral formula for the distance function $%
\sigma(p,p^{\prime})$ on each hemisphere $M_{\pm}^{\ast}$ of $%
M^{\ast}=S^{2}(1/2)$. Our calculation of such an integral formula will be
based upon a special type of coordinates; namely, we choose the points $%
\left\{ \mathfrak{b}_{23},\mathfrak{b}_{31}\right\} $ as a \emph{bipolar
system} whose associated distance functions $\left\{
\sigma_{1},\sigma_{2}\right\} $ constitute a coordinate system on each
hemisphere $M_{\pm}^{\ast}$. In our final formula, however, we shall express
the distance in terms of $I_{1},I_{2}$, and more simply in terms of their
translates $\tilde{I}_{i}=I_{i}-m_{i}^{\ast}$.

To this end, it is convenient to apply the above vector algebra
representation, where we use the unit sphere $S^{2}(1)$ representation of $%
M^{\ast}$ rather than the sphere $S^{2}(1/2)$, and the collision points $%
\mathfrak{b}_{23},\mathfrak{b}_{31}$ and variable points $p,p^{\prime}$ are
replaced by the unit vectors $\mathbf{\hat{b}}_{1},\mathbf{\hat{b}}_{2},%
\mathbf{p},\mathbf{p}^{\prime}$, respectively. Thus, $\sigma(p,p^{\prime })$
is half of the spherical distance between $\mathbf{p}$ and $\mathbf{p}%
^{\prime}$ on the unit sphere.

We shall reduce the calculation of the distance function $\sigma$ to a
simple vector algebra involving determinants and Lagrange's formula : 
\begin{equation}
\det(\mathbf{\hat{b}}_{1},\mathbf{\hat{b}}_{2},\mathbf{p})\det(\mathbf{\hat {%
b}}_{1},\mathbf{\hat{b}}_{2},\mathbf{p}^{\prime})=\left\vert 
\begin{array}{ccc}
1 & \mathbf{\hat{b}}_{1}\cdot\mathbf{\hat{b}}_{2} & \mathbf{\hat{b}}_{1}\cdot%
\mathbf{p}^{\prime} \\ 
\mathbf{\hat{b}}_{2}\cdot\mathbf{\hat{b}}_{1} & 1 & \mathbf{\hat{b}}_{2}\cdot%
\mathbf{p}^{\prime} \\ 
\mathbf{p}\cdot\mathbf{\hat{b}}_{1} & \mathbf{p}\cdot\mathbf{\hat{b}}_{2} & 
\mathbf{p}\cdot\mathbf{p}^{\prime}%
\end{array}
\right\vert   \label{Lagr}
\end{equation}
By writing the left side as

\begin{equation*}
\left[ \det(\mathbf{\hat{b}}_{1},\mathbf{\hat{b}}_{2},\mathbf{p})^{2}\det(%
\mathbf{\hat{b}}_{1},\mathbf{\hat{b}}_{2},\mathbf{p}^{\prime})^{2}\right]
^{1/2}
\end{equation*}
and applying Lagrange's formula to each square, the right side of (\ref{Lagr}%
) equals the product 
\begin{equation}
\left\vert 
\begin{array}{ccc}
1 & \mathbf{\hat{b}}_{1}\cdot\mathbf{\hat{b}}_{2} & \mathbf{\hat{b}}_{1}\cdot%
\mathbf{p} \\ 
\mathbf{\hat{b}}_{2}\cdot\mathbf{\hat{b}}_{1} & 1 & \mathbf{\hat{b}}_{2}\cdot%
\mathbf{p} \\ 
\mathbf{p}\cdot\mathbf{\hat{b}}_{1} & \mathbf{p}\cdot\mathbf{\hat{b}}_{2} & 1%
\end{array}
\right\vert ^{1/2}\left\vert 
\begin{array}{ccc}
1 & \mathbf{\hat{b}}_{1}\cdot\mathbf{\hat{b}}_{2} & \mathbf{\hat{b}}_{1}\cdot%
\mathbf{p}^{\prime} \\ 
\mathbf{\hat{b}}_{2}\cdot\mathbf{\hat{b}}_{1} & 1 & \mathbf{\hat{b}}_{2}\cdot%
\mathbf{p}^{\prime} \\ 
\mathbf{p}^{\prime}\cdot\mathbf{\hat{b}}_{1} & \mathbf{p}^{\prime}\cdot%
\mathbf{\hat{b}}_{2} & 1%
\end{array}
\right\vert ^{1/2}   \label{Lagr2}
\end{equation}
and this gives us an identity where $\mathbf{\hat{b}}_{1}\cdot\mathbf{\hat{b}%
}_{2}=\cos\beta_{3}$ is a constant, the inner product 
\begin{equation*}
\mathbf{p}\cdot\mathbf{p}^{\prime}=\cos2\sigma(p,p^{\prime}) 
\end{equation*}
appears linearly, and the other non-constant entries $\mathbf{\hat{b}}%
_{i}\cdot\mathbf{p}=\cos2\sigma_{i}$, $\mathbf{\hat{b}}_{i}\cdot \mathbf{p}%
^{\prime}=\cos2\sigma_{i}^{\prime}$ in the determinants are of the type (\ref%
{intrins5}) (or (\ref{inner2})).

Let $D,D^{\prime}$ be the determinants in (\ref{Lagr2}). Solving the above
determinant identity with respect to $\mathbf{p}\cdot\mathbf{p}^{\prime}$
gives 
\begin{equation}
\cos2\sigma(p,p^{\prime})=\frac{1}{\sin^{2}\beta_{3}}(\sqrt{DD^{\prime}}+F) 
\label{dist4}
\end{equation}
where $F$ is a bilinear form of both vectors $(\tilde{I}_{1},\tilde{I}_{2})$
and $(\tilde{I}_{1}^{\prime},\tilde{I}_{2}^{\prime})$, and moreover, $F=0$
and $D^{\prime}=\sin^{2}\beta_{3}$ when $p^{\prime}$ is the pole \textsc{P}
of the hemisphere. The latter observation implies 
\begin{equation*}
\cos2\sigma(p\mathbf{,}\text{\textsc{P}})=\cos2r=4\Delta\sqrt{m_{1}m_{2}m_{3}%
}=\frac{\sqrt{D}}{\sin\beta_{3}}\text{, \ \ \ cf. (\ref{cos2r}) }
\end{equation*}
Consequently, by the Ceva-Heron formula (\ref{C-H}) 
\begin{equation*}
D=\frac{\sin^{2}\beta_{3}}{m_{1}m_{2}m_{3}}Q^{\ast}=\frac{\sin^{2}\beta_{3}}{%
m_{1}m_{2}m_{3}}(m_{1}m_{2}m_{3}-Q_{0}^{\ast}(\tilde{I}_{1},\tilde{I}_{2})) 
\end{equation*}
where $Q^{\ast}=Q|_{M^{\ast}}$ is the restriction of the quadratic form (\ref%
{Qform}), and 
\begin{equation}
Q_{0}^{\ast}=(1-m_{2})^{2}\tilde{I}_{1}^{2}+(1-m_{1})^{2}\tilde{I}%
_{2}^{2}+2(m_{3}-m_{1}m_{2})\tilde{I}_{1}\tilde{I}_{2}   \label{Q0star}
\end{equation}

On the other hand, taking $p=p^{\prime}$\textbf{\ }in (\ref{dist4}) gives 
\begin{equation*}
\frac{F}{\sin^{2}\beta_{3}}=\frac{m_{1}m_{2}m_{3}-Q^{\ast}}{m_{1}m_{2}m_{3}}=%
\frac{Q_{0}^{\ast}}{m_{1}m_{2}m_{3}}\text{ ,}
\end{equation*}
and therefore, $F$ as a function of $(\tilde{I}_{1},\tilde{I}_{2},\tilde
{I}%
_{1}^{\prime},\tilde{I}_{2}^{\prime})$, is a constant times the polarization
of $Q_{0}^{\ast}$ in (\ref{Q0star}). This establishes the general spherical
distance formula 
\begin{equation}
\sigma(p,p^{\prime})=\frac{1}{2}\arccos(\frac{\sqrt{Q^{\ast}Q^{\prime\ast}}+%
\frac{1}{2}Pol(Q_{0}^{\ast})}{m_{1}m_{2}m_{3}})   \label{dist5}
\end{equation}
where 
\begin{equation*}
\frac{1}{2}Pol(Q_{0}^{\ast})=(1-m_{2})^{2}\tilde{I}_{1}\tilde{I}_{1}^{\prime
}+(1-m_{1})^{2}\tilde{I}_{2}\tilde{I}_{2}^{\prime}+(m_{3}-m_{1}m_{2})(\tilde{%
I}_{1}\tilde{I}_{2}^{\prime}+\tilde{I}_{1}^{\prime}\tilde{I}_{2}) 
\end{equation*}

\begin{remark}
By spherical trigonometry it is easy to see that the polarization term in (%
\ref{dist5}) can be expressed in polar coordinates as 
\begin{equation*}
\sin\varphi\sin\varphi^{\prime}\cos(\theta-\theta^{\prime}) 
\end{equation*}
\end{remark}

\section{Motions of m-triangles with conserved angular momentum}

In the previous chapters we have investigated the kinematic quantities,
their general relationships, and the resulting kinematic identities are
valid for any motion of m-triangles, with no explicit assumption on the
invariance of the angular momentum vector $\mathbf{\Omega}$. In dynamics,
however, the motion is governed by a potential function and one is primarily
interested in the trajectories of the equations of motion, which is a second
order ODE. In these cases the invariance of $\mathbf{\Omega}$ is generally
seen as a consequence of (rotational) symmetry properties of the potential
function, as in the Newtonian case discussed in the introductory chapter.
From this viewpoint, the linear motions (cf. Section 3.3) are the
trajectories in the trivial case of a constant potential function, but still
we have found them to be useful in our survey of the kinematic geometry of
the moduli space $\bar
{M}$.

In this chapter we turn to the study of virtual m-triangle motions, in full
generality except with the explicit assumption that the angular momentum is
conserved. Our aim is also to complete the proofs of the Main Theorems B, D,
E1, E2 and F stated in Section 2.2.

\begin{remark}
\label{Weier} It is a classical result, dating (at least) back to
Weierstrass, that a 3-body motion with $\mathbf{\Omega}=0$ must be planar.
There is a simple and purely kinematic proof of this fact. Namely, if $(%
\mathbf{X}(t),\mathbf{n}(t))$ is a motion of oriented m-triangles and $%
\mathbf{X}=(\mathbf{a}_{1}\mathbf{,a}_{2}\mathbf{,a}_{3})$ is nondegenerate,
then $\mathbf{\dot{n}}$ $=0$ if and only if $q_{1}=q_{2}=0$, where $q_{i}=%
\mathbf{n\cdot\dot{a}}_{i}$. Therefore, the proof follows from the identity 
\begin{equation*}
\pm(\mathbf{a}_{1}\times\mathbf{a}_{2})\times\mathbf{\Omega}=\tilde{q}_{1}%
\mathbf{a}_{1}+\tilde{q}_{2}\mathbf{a}_{2}, 
\end{equation*}
where for $\left\{ i,j\right\} =\left\{ 1,2\right\} $, $\tilde{q}_{i}$ is a
linear combination of $q_{i}$ and $q_{j}$ with coefficients $2\Delta
m_{i}(1-m_{j})$ and $2\Delta m_{i}m_{j}$, respectively.
\end{remark}

\subsection{Moving eigenframe and intrinsic decomposition of velocities}

Consider an m-triangle motion $\mathbf{X}(t)=(\mathbf{a}_{1}\mathbf{,a}_{2}%
\mathbf{,a}_{3})$ with a (continuous) eigenframe $\mathfrak{F}(t)=(\mathbf{u}%
_{1}\mathbf{,u}_{2}\mathbf{,n})$ of its inertia tensor $B_{\mathbf{X}}$ (\ref%
{B}). For convenience, let us assume $\mathbf{X}(t)$ is nondegenerate (say,
for some time interval) and hence $\mathbf{X}$ spans a plane $\Pi(\mathbf{X}%
)=lin\left\{ \mathbf{u}_{1}\mathbf{,u}_{2}\right\} $. Any vector $\mathbf{v}$
in 3-space has an orthogonal splitting, $\mathbf{v=v}^{\tau}+\mathbf{v}%
^{\eta}$, where $\mathbf{v}^{\tau}$ is the \emph{tangential} component lying%
\emph{\ }in the plane $\Pi(\mathbf{X})$ and $\mathbf{v}^{\eta}$ is the \emph{%
normal} component.

We will combine the splitting (\ref{Xdot}) of the velocity of $\mathbf{X}(t)$
with its tangential and normal decomposition, namely\ we decompose the
individual velocities $\mathbf{\dot{a}}_{i}$ into their tangential and
normal parts and write 
\begin{equation}
\mathbf{\dot{X}}=\mathbf{\dot{X}}^{\tau}+\mathbf{\dot{X}}^{\eta}=(\mathbf{%
\dot{a}}_{1}^{\tau},\mathbf{\dot{a}}_{2}^{\tau},\mathbf{\dot{a}}%
_{3}^{\tau})+(\mathbf{\dot{a}}_{1}^{\eta},\mathbf{\dot{a}}_{2}^{\eta },%
\mathbf{\dot{a}}_{3}^{\eta})   \label{split2}
\end{equation}
Recall the roles of the (instantaneous) angular velocity vector $\mathbf{%
\omega=\omega(}t\mathbf{)}$ and the angular momentum $\mathbf{\Omega }$,
which by the inertia operator (\ref{inert-op}) essentially determine each
other. They are responsible for the purely rotational (or rigid) motion of
the m-triangle, and in accordance with (\ref{split2}) the rotational
velocity\ has the splitting 
\begin{equation}
\mathbf{\dot{X}}^{\omega}=\mathbf{\omega\times X=\omega}^{\eta}\mathbf{%
\times X}+\mathbf{\omega}^{\tau}\mathbf{\times X\ =\mathbf{(\dot{X}}%
^{\omega})^{\tau }}+\mathbf{(\dot{X}}^{\omega})^{\eta}   \label{rot}
\end{equation}

\begin{lemma}
The normal velocity component of the m-triangle motion is purely rotational,
that is, 
\begin{equation*}
\mathbf{\dot{X}}^{\eta}=\mathbf{(\dot{X}}^{\omega})^{\eta}=\mathbf{\omega }%
^{\tau}\mathbf{\times X}
\end{equation*}
where $\mathbf{\omega}^{\tau}$ is the tangential component of $\mathbf{%
\omega }$, and moreover, 
\begin{equation*}
\mathbf{\dot{n}}=\mathbf{\omega\times n}=\mathbf{\omega}^{\tau}\mathbf{%
\times n}
\end{equation*}
\end{lemma}

\begin{proof}
Consider the 2-parameter family of degenerate m-triangles 
\begin{equation*}
(\lambda\mathbf{u}_{1}+\mu\mathbf{u}_{2})\times\mathbf{X}=(c_{1}\mathbf{n,c}%
_{2}\mathbf{n,}c_{3}\mathbf{n)}
\end{equation*}
where the constants $c_{i}=c_{i}(\lambda,\mu)$ are linear combinations of $%
\lambda$ and $\mu$ with the relation $\sum m_{i}c_{i}=0$. The last identity
also expresses the range of the linear transformation $(\lambda,\mu
)\rightarrow(c_{1},c_{2},c_{3})$.

On the other hand, for certain constants $k_{i}$ 
\begin{equation*}
\mathbf{\dot{X}}^{\eta}=(k_{1}\mathbf{n,}k_{2}\mathbf{n,}k_{3}\mathbf{n)}%
\text{, \ }\sum m_{i}k_{i}=0, 
\end{equation*}
and consequently the system of equations $c_{i}=k_{i},i=1,2,3,$ has a unique
solution $(\lambda,\mu)$, that is, there is a unique vector $\mathbf{\omega }%
^{\prime}$ such that $\mathbf{\dot{X}}^{\eta}=\mathbf{\omega}^{\prime}\times%
\mathbf{X}$. Clearly, $\mathbf{\omega}^{\prime}$ is just the tangential
component $\mathbf{\omega}^{\tau}$ of $\mathbf{\omega}$. Finally, $\mathbf{n}
$ is a multiple of $\mathbf{a}_{1}\times\mathbf{a}_{2}$ and $\mathbf{\dot{n}}
$ is a tangential vector, so it is a simple vector algebra calculation to
verify that $\mathbf{\dot{n}}=\mathbf{\omega}^{\tau}\mathbf{\times n.}$
\end{proof}

It follows that the tangential velocity version of (\ref{Xdot}) reads 
\begin{equation*}
\mathbf{\dot{X}}^{\tau}=\mathbf{\omega}^{\eta}\mathbf{\times X}+\mathbf{%
\dot
{X}}^{h}
\end{equation*}
and, in particular, the horizontal velocity of an m-triangle motion is
always tangential, whereas the rotational velocity (\ref{rot}) in general
has a tangential and normal component.

Now, we turn to the angular momentum $\ $ 
\begin{equation*}
\mathbf{\Omega=\mathbf{X}\times\mathbf{\dot{X}}^{\omega}=X}\times (\mathbf{%
\omega\times X)}
\end{equation*}
\ which we assume is a fixed vector along the z-axis, say, and the expansion 
$\ $ 
\begin{equation}
\ \mathbf{\Omega}\text{\ }=\Omega\mathbf{k}=\mathbf{\Omega}^{\tau }+\mathbf{%
\Omega}^{\eta}=(g_{1}\mathbf{u}_{1}+g_{2}\mathbf{u}_{2})+g_{3}\mathbf{n}%
\text{ }\   \label{expans1}
\end{equation}
defines its (time dependent) coordinate vector $(g_{1},g_{2},g_{3})$
relative to the moving frame $\mathfrak{F}(t)$. The inner product of $%
\mathbf{\Omega}$ with a vector $\mathbf{v}$ may be written as 
\begin{equation*}
\mathbf{\Omega\cdot v=(X}\times\mathbf{\omega)\times X\cdot v}=\mathbf{(X}%
\times\mathbf{\omega)\cdot(X}\times\mathbf{v)}=B_{\mathbf{X}}(\mathbf{\omega
,v)}
\end{equation*}
Hence, by letting $\mathbf{v=v}_{i}$ be any of the vectors from $\mathfrak{F}
$, 
\begin{equation*}
g_{i}=\mathbf{\Omega\cdot v}_{i}=B_{\mathbf{X}}(\mathbf{\omega,v}%
_{i})=\lambda_{i}\mathbf{\omega\cdot v}_{i}\text{, \ }i=1,2,3, 
\end{equation*}
and we obtain the expansion 
\begin{equation}
\mathbf{\omega=\omega}^{\tau}+\mathbf{\omega}^{\eta}=(\frac{g_{1}}{\lambda
_{1}}\mathbf{u}_{1}+\frac{g_{2}}{\lambda_{2}}\mathbf{u}_{2})+\frac{g_{3}}{I}%
\mathbf{n},   \label{expans2}
\end{equation}
In particular, the rotational kinetic energy can be expressed as 
\begin{equation}
T^{\omega}=\frac{1}{2}\left\vert \mathbf{\dot{X}}^{\omega}\right\vert ^{2}=%
\frac{1}{2}B_{\mathbf{X}}(\mathbf{\omega,\omega)}=\frac{1}{2}(\frac {%
g_{1}^{2}}{\lambda_{1}}+\frac{g_{2}^{2}}{\lambda_{2}}+\frac{g_{3}^{2}}{I}) 
\label{T-rot}
\end{equation}
with tangential and normal parts 
\begin{equation*}
(T^{\omega})^{\tau}=\frac{1}{2}\frac{g_{3}^{2}}{I},\text{ \ \ \ }(T^{\omega
})^{\eta}=\frac{1}{2}(\frac{g_{1}^{2}}{\lambda_{1}}+\frac{g_{2}^{2}}{%
\lambda_{2}}) 
\end{equation*}

Now, let us also have a closer look at the individual velocities in (\ref%
{split2}) and their splitting, namely 
\begin{equation}
\mathbf{\dot{a}}_{i}=\mathbf{\dot{a}}_{i}^{\tau}\ +\mathbf{\dot{a}}%
_{i}^{\eta }=\left( \omega_{i}(\mathbf{n\times a}_{i})+\frac{\dot{\rho}_{i}}{%
\rho_{i}}\mathbf{a}_{i}\right) +(\mathbf{\omega}^{\tau}\times\mathbf{a}_{i})%
\text{, \ cf. (\ref{split3})}   \label{split5}
\end{equation}
where $\omega_{i}$ is the scalar tangential angular velocity of $\mathbf{a}%
_{i}$. We claim that 
\begin{equation}
\omega_{i}=\omega_{i}^{0}+\frac{g_{3}}{I}\text{ ,\ \ cf. Remark }\ref{RemC1} 
\label{split4}
\end{equation}
where $\omega_{i}^{0}$ is the scalar angular velocity in the case of
vanishing angular momentum, and a formula is given in Theorem C1, see (\ref%
{kin2}).\ The proof of (\ref{split4}) is really the same as in Section
3.2.1, if we only consider velocity components in the plane $\Pi(\mathbf{X})$
and replace $\mathbf{\Omega}$ by its normal component $\mathbf{\Omega}%
^{\eta} $. The identities in Section 3.2.1, in fact, expresses kinematic
relationships valid at each moment $t$, and there is no need to assume $%
\mathbf{\Omega}$ is a constant.

Finally, according to (\ref{split5}) the individual kinetic energy terms
expresses as 
\begin{equation}
T_{i}=T_{i}^{\tau}+T_{i}^{\eta}=(\frac{1}{2}\omega_{i}^{2}I_{i}+\frac{\dot {I%
}_{i}^{2}}{8I_{i}})+\frac{1}{2}m_{i}\left\vert \mathbf{\omega}^{\tau}\times%
\mathbf{a}_{i}\right\vert ^{2}\text{ }   \label{split6}
\end{equation}
and hence they depend, in fact, only on the moduli curve $\bar{\Gamma}(t)$
of the motion and the moving frame coordinates $g_{i}$ of the angular
momentum.

\subsection{Final proof of the Main Theorems D, B, E1,E2}

\subsubsection{The Euler equations and proof of Theorem D}

Consider the horizontal (i.e. with vanishing angular momentum) m-triangle
motion $\mathbf{X}^{h}(t)=(\mathbf{b}_{1}\mathbf{,b}_{2},\mathbf{b}_{3})$,
with the same moduli curve $\bar{\Gamma}(t)$ as $\mathbf{X}(t)$ and with
moving eigenframe 
\begin{equation*}
\mathfrak{F}^{h}(t)=(\mathbf{u}_{1}^{h}(t)\mathbf{,u}_{2}^{h}(t)\mathbf{,n(}%
t_{0})), 
\end{equation*}
subject to the initial conditions 
\begin{equation}
\mathbf{X}^{h}(t_{0})=\mathbf{X}(t_{0})\text{, \ \ }\mathfrak{F}^{h}(t_{0})=%
\mathfrak{F}(t_{0})   \label{initial}
\end{equation}
The existence of this motion is the statement of Theorem B in the simple
case of vanishing angular momentum (cf. Section 2.2.2).

Let $\Gamma^{\ast}(t)=(\varphi(t),\theta(t))$ be the associated shape curve
on $M^{\ast}=S^{2}$, expressed in the usual spherical coordinates, and
consider the two nearby points 
\begin{equation*}
\mathbf{p}=(\varphi(t_{0}),\theta(t_{0})\text{, \ }\mathbf{p}^{\prime
}=(\varphi(t_{0}+\Delta t),\theta(t_{0}+\Delta t)) 
\end{equation*}
with the longitude difference $\Delta\theta$. The meridians through $\mathbf{%
p}$ and $\mathbf{p}^{\prime}$ intersect the equator circle $E^{\ast}$ in the
points $\mathbf{e}$ and $\mathbf{e}^{\prime}$ respectively, and there is the
piecewise geodesic closed path 
\begin{equation*}
\mathbf{p\rightarrow e\rightarrow e}^{\prime}\mathbf{\rightarrow p}^{\prime
}\rightarrow\mathbf{p}
\end{equation*}
enclosing the shaded region as indicated in Figure 7, whose area (on the
unit sphere) is 
\begin{equation*}
\Delta A\equiv\cos\varphi\text{ }\Delta\theta\text{ \ }
\end{equation*}
modulo higher orders of $\Delta\theta$.

It follows from Theorem C2 and Theorem \ref{ang1} applied to the above path,
with a piecewise linear motion in the plane perpendicular to $\mathbf{n(}%
t_{0})$, that 
\begin{align}
\mathbf{u}_{1}^{h}(t_{0}+\Delta t) & \equiv\cos(\Delta\psi)\mathbf{u}%
_{1}(t_{0})+\sin(\Delta\psi)\mathbf{u}_{2}(t_{0})  \label{u-sharp} \\
\mathbf{u}_{2}^{h}(t_{0}+\Delta t) & \equiv-\sin(\Delta\psi)\mathbf{u}%
_{1}(t_{0})+\cos(\Delta\psi)\mathbf{u}_{2}(t_{0})  \notag
\end{align}
modulo higher orders of $\Delta\theta$, where 
\begin{equation}
\Delta\psi=\frac{1}{2}\cos\varphi\text{ }\Delta\theta   \label{angle5}
\end{equation}
Consequently, we infer from (\ref{u-sharp}) and (\ref{angle5}) 
\begin{equation}
\mathbf{\dot{u}}_{1}^{h}(t_{0})=(\frac{1}{2}\dot{\theta}\cos\varphi \text{)}%
\mathbf{u}_{2}|_{t_{0}}\text{, \ }\mathbf{\dot{u}}_{2}^{h}(t_{0})=(-\frac{1}{%
2}\dot{\theta}\cos\varphi\text{)}\mathbf{u}_{1}|_{t_{0}}   \label{u-sharp1}
\end{equation}

Next, consider the following intrinsic frame version of (\ref{Xdot})

\begin{equation}
\mathbf{\dot{u}}_{i}=\mathbf{\omega\times u}_{i}+\mathbf{\dot{u}}%
_{i}^{h},i=1,2\text{; \ \ \ }\mathbf{\dot{n}=\omega\times n,}   \label{dot}
\end{equation}
By taking the inner product with $\mathbf{\Omega}$ on both sides of these
identities, using (\ref{expans1}) and (\ref{expans2}), we perform the
following calculations : 
\begin{align*}
\mathbf{\Omega\cdot\dot{u}}_{1}(t_{0}) & =\mathbf{\Omega\cdot}(\mathbf{%
\omega\times u}_{1}+\mathbf{\dot{u}}_{1}^{h})|_{t_{0}}\  \\
& =\mathbf{\Omega\cdot}\left( \mathbf{(}\frac{g_{1}}{\lambda_{1}}\mathbf{u}%
_{1}+\frac{g_{2}}{\lambda_{2}}\mathbf{u}_{2}+\frac{g_{3}}{I}\mathbf{n)\times
u}_{1}+(\frac{1}{2}\dot{\theta}\cos\varphi\text{)}\mathbf{u}_{2}\right)
|_{t_{0}}\  \\
& =\mathbf{\Omega\cdot}\left( \mathbf{-}\frac{g_{2}}{\lambda_{2}}\mathbf{n+(}%
\frac{g_{3}}{I}+\frac{1}{2}\dot{\theta}\cos\varphi)\mathbf{u}_{2}\right)
|_{t_{0}}\mathbf{\ } \\
& =g_{2}\left( (\frac{1}{I}-\frac{1}{\lambda_{2}})g_{3}+\frac{1}{2}%
\cos\varphi\text{ }\dot{\theta}\right) |_{t_{0}}\ 
\end{align*}
\begin{align*}
\mathbf{\Omega\cdot\dot{u}}_{2}(t_{0}) & =\mathbf{\Omega\cdot}(\mathbf{%
\omega\times u}_{2}+\mathbf{\dot{u}}_{2}^{h})|_{t_{0}}\  \\
& =\mathbf{\Omega\cdot}\left( \mathbf{(}\frac{g_{1}}{\lambda_{1}}\mathbf{u}%
_{1}+\frac{g_{2}}{\lambda_{2}}\mathbf{u}_{2}+\frac{g_{3}}{I}\mathbf{n)\times
u}_{2}-(\frac{1}{2}\dot{\theta}\cos\varphi\text{)}\mathbf{u}_{1}\right)
|_{t_{0}} \\
& =\mathbf{\Omega\cdot}\left( \frac{g_{1}}{\lambda_{1}}\mathbf{n+(-}\frac{%
g_{3}}{I}-\frac{1}{2}\dot{\theta}\cos\varphi)\mathbf{u}_{1}\right) |_{t_{0}}%
\mathbf{\ } \\
& =g_{1}\left( (\frac{1}{\lambda_{1}}-\frac{1}{I})g_{3}-\frac{1}{2}%
\cos\varphi\text{ }\dot{\theta}\right) |_{t_{0}}
\end{align*}
\begin{align*}
\mathbf{\Omega\cdot\dot{n}} & =\mathbf{\Omega\cdot(\omega\times
n)=\Omega\cdot}\left( \mathbf{(}\frac{g_{1}}{\lambda_{1}}\mathbf{u}_{1}+%
\frac{g_{2}}{\lambda_{2}}\mathbf{u}_{2}+\frac{g_{3}}{I}\mathbf{n)\times n}%
\right) |_{t_{0}} \\
& =\mathbf{\Omega\cdot}\left( \mathbf{-}\frac{g_{1}}{\lambda_{1}}\mathbf{u}%
_{2}+\frac{g_{2}}{\lambda_{2}}\mathbf{u}_{1}\right) |_{t_{0}}=(\frac{1}{%
\lambda_{2}}-\frac{1}{\lambda_{1}})g_{1}g_{2}|_{t_{0}}
\end{align*}
Since $\mathbf{\Omega}$ is a constant vector and time $t_{0}$ is arbitrary,
the above three identities amount precisely to the ODE (\ref{Euler}), and
this completes the proof of Theorem D.

Finally, we turn to the precession angle $\chi(t)$ which records the motion
of the normal vector $\mathbf{n}$ around the z-axis, that is, the fixed $%
\mathbf{\Omega}$-axis. For example, using spherical coordinates $(\tilde{%
\varphi},\tilde{\theta})$ on the unit sphere in Euclidean 3-space, with $%
\tilde{\varphi}=0$ at the north pole $\mathbf{k}$ and $\chi(t)=$ $\tilde{%
\theta}(t)$, it is a simple exercise to deduce the first equality in (\ref%
{prec}), and by substituting $\mathbf{\dot{n}}=$ $\mathbf{\omega\times n}$
the second expression in (\ref{prec}) follows by calculating cross products
in the frame $\mathfrak{F}$.

\subsubsection{The lifting problem and proof of Theorem B}

To complete the proof in the general case $\mathbf{\Omega}\neq0$, let us
also choose a horizontal lifting $\mathbf{X}^{h}(t)$ of \emph{\ }$\bar{\Gamma%
}(t)$. Then it follows from the identity (\ref{Xdot}) that the lifting $%
\Gamma(t)$ is the motion $\mathbf{X}(t)=(\mathbf{a}_{1}(t),\mathbf{a}_{2}(t),%
\mathbf{a}_{3}(t))$ determined by the following initial value problem$%
\mathbf{\ }$%
\begin{equation}
\frac{d}{dt}\mathbf{X}=(\mathbf{\omega}\times\mathbf{X)+}\frac{d}{dt}\mathbf{%
X}^{h}\text{, \ \ \ }\mathbf{X(}t_{0})=\Gamma(t_{0})   \label{initial2}
\end{equation}
Here the vector $\mathbf{\omega}$ is a function of $\mathbf{X}$ and the
constant vector $\mathbf{\Omega}$, namely for a fixed and nondegenerate $%
\mathbf{X}$, $\mathbf{\omega}$ is found by inverting the inertia operator on
3-space : 
\begin{equation*}
\mathbb{I}_{\mathbf{X}}:\mathbf{\omega\rightarrow X\times(\omega}\times%
\mathbf{X)=\Omega}
\end{equation*}
The matrix of this operator is 
\begin{equation*}
B=\left\vert \mathbf{X}\right\vert ^{2}Id-ADA^{t}\sim
diag(\lambda_{1},\lambda_{2},\lambda_{3}) 
\end{equation*}
where $Id$ is the identity, the vectors $\mathbf{a}_{i}$ are the columns of $%
A$, and 
\begin{equation*}
D=diag(m_{1},m_{2},m_{3}) 
\end{equation*}
The eigenvalues $\lambda_{i}$ of $B$ are listed in Section 2.2.4, and one of
them vanishes when $\mathbf{X}\neq0$ is degenerate. However, in that case it
is easy to check that the indeterminacy of $\mathbf{\omega}$ is a summand
along the line $\Pi(\mathbf{X})$ and consequently the summand $\mathbf{%
\omega }\times\mathbf{X}$ in (\ref{initial2}) is still well defined as a
function of $\mathbf{X}$ and $\mathbf{\Omega}$.

For another proof of Theorem B, more directly related to the construction of
a position curve $\gamma(t)$ in $SO(3)$, we use either Theorem C1 or D. The
first theorem applies to planary motions and calculates a position curve $%
\gamma(t)$ in $SO(2)$, recording the rotation of the vectors $\mathbf{a}_{i}$%
. In the non-planary case the position of the m-triangle is represented by
the position of its moving eigenframe $\mathfrak{F}$, and the latter is
determined by the coordinate vector $(g_{1},g_{2},g_{3})$ of $\mathbf{\Omega}
$ relative to $\mathfrak{F}$ together with the precession angle $\chi$ (of
the normal vector $\mathbf{n}$) in the "invariant" plane perpendicular to $%
\mathbf{\Omega}$. The four functions $g_{i}$ and $\chi$ are the solution of
an initial value problem depending only on the moduli curve $\bar{\Gamma}(t)$%
, as explained by Theorem D and formula (\ref{prec}).

\subsubsection{Geometric reduction of Newton's equation and proof of Theorem
E1 and E2}

To derive the reduced Newton's equations from the Newton's equations (\ref%
{Newton1}), we differentiate the kinematic quantities $I_{i}$ up to second
order, for example, $\dot{I}_{1}=2m_{1}\mathbf{a}_{1}\cdot \mathbf{\dot{a}}%
_{1}$ and 
\begin{align}
\ddot{I}_{1} & =2m_{1}\left\vert \mathbf{\dot{a}}_{1}\right\vert ^{2}+2m_{1}%
\mathbf{a}_{1}\cdot\mathbf{\ddot{a}}_{1}  \notag \\
& =4T_{1}+2m_{1}\mathbf{a}_{1}\cdot\left( \frac{m_{2}}{r_{12}^{3}}(\mathbf{a}%
_{2}-\mathbf{a}_{1})+\frac{m_{3}}{r_{13}^{3}}(\mathbf{a}_{3}-\mathbf{a}%
_{1})\right)   \label{2diff}
\end{align}
Then we use the Ceva-cosine law (\ref{Ceva-cos}), stated in the form 
\begin{equation*}
\mathbf{a}_{i}\cdot\mathbf{a}_{j}=\frac{-1}{2m_{i}m_{j}}C_{k}=\frac{1}{%
2m_{i}m_{j}}(m_{k}I_{k}-m_{i}I_{i}-m_{j}I_{j}), 
\end{equation*}
to replace all inner products in (\ref{2diff}) by linear combinations of the 
$I_{i}^{\prime}s$. This procedure leads to the differential equations (\ref%
{redu1}).

In the above differential equations the individual kinetic energies $T_{i}$
are crucial terms, and their actual splitting (\ref{split6}) distinguishes
the two cases of planary and non-planary 3-body motions $\Gamma(t)$. Of
course, the actual case is also decided by the initial data $\Gamma(t_{0}),%
\dot {\Gamma}(t_{0})$. However, it is not decided by $\Gamma(t_{0})$ and the
angular momentum vector, unless $\Gamma(t_{0})$ is nondegenerate. Clearly,
the statements of Theorem E1 and E2 must be modified if they should also
cover the case where the initial configuration $\Gamma(t_{0})$ is collinear.

First, observe that a planary three-body motion is characterized by having
all normal kinetic energies $T_{i}^{\eta}=0$, and then its moduli curve $%
\bar{\Gamma}(t)$ is a solution of the $\Omega$-reduced equations (\ref{redu1}%
) with $T_{i}=T_{i}^{\tau}$ given by the first summand in (\ref{split6}).

Conversely, let $\bar{\gamma}(t)$ be a moduli curve which is a solution of
this ODE, for a given value of $\Omega$. By Theorem B there is a unique
lifting $\mathbf{X}(t)$, namely a virtual motion in the xy-plane, with
angular momentum $\Omega\mathbf{k}$ and specified initial position $\mathbf{X%
}(t_{0})$. From this knowledge we may calculate the initial velocity 
\begin{equation*}
\mathbf{\dot{X}}(t_{0})=\mathbf{\omega}(t_{0})\times\mathbf{X}(t_{0})+%
\mathbf{\dot{X}}^{h}(t_{0})\text{, \ \ }\mathbf{\omega}(t_{0})=\frac{\Omega 
}{I(t_{0})}\mathbf{k,}
\end{equation*}
since the horizontal velocity $\mathbf{\dot{X}}^{h}(t_{0})$ is determined by 
$\frac{d}{dt}\bar{\gamma}(t_{0})$. On the other hand, Newton's equations (%
\ref{Newton1}) also has a unique solution $\Gamma(t)$ with the above initial
conditions, namely $\Gamma(t_{0})=\mathbf{X}(t_{0})$ and $\dot{\Gamma}%
(t_{0})=\mathbf{\dot{X}}(t_{0})$, and clearly the moduli curve of $\Gamma(t)$
is a solution of the $\Omega$-reduced ODE. By uniqueness of the lifting we
conclude that $\Gamma(t)=\mathbf{X}(t)$ for all $t$, and this completes the
proof of Theorem E1.

Next, we turn to the general case,\ described by Theorem E2. The kinetic
energies $T_{i}$ in the $\Omega$-reduced ODE have the general form (\ref%
{split6}), and we \underline{claim} they depend only on the moduli curve and
the functions $g_{i}$. This clearly holds for the tangential summand $%
T_{i}^{\tau}$, whose expression is even independent of $g_{1}$ and $g_{2}$.
On the other hand, the normal summand is, say, for $i=1$: 
\begin{align}
T_{1}^{\eta} & =\frac{1}{2}m_{1}\left\vert \mathbf{\omega}^{\tau}\times%
\mathbf{a}_{1}\right\vert ^{2}=\frac{1}{2}m_{1}\left\vert (\frac{g_{1}}{%
\lambda_{1}}\mathbf{u}_{1}+\frac{g_{2}}{\lambda_{2}}\mathbf{u}%
_{2})\times(\cos\psi_{1}\mathbf{u}_{1}+\sin\psi_{1}\mathbf{u}%
_{2})\right\vert ^{2}  \notag \\
& =\frac{1}{2}m_{1}(\frac{g_{1}}{\lambda_{1}}\sin\psi_{1}-\frac{g_{2}}{%
\lambda_{2}}\cos\psi_{1})^{2}   \label{kin3}
\end{align}
where $\psi_{1}$ is the angle between $\mathbf{u}_{1}$ and $\mathbf{a}_{1}$,
satisfying 
\begin{equation}
\tan\psi_{1}=\frac{\cos\varphi}{1+\sin\varphi}\tan\frac{\theta}{2},\text{ \
cf. Theorem \ref{ang2} }   \label{ang6}
\end{equation}
In this formula $\theta$ is the longitude angle measured from the binary
collision point $\mathbf{\hat{b}}_{1}$ and is increasing in the direction
towards $\mathbf{\hat{b}}_{2}$. It follows (e.g. by symmetry) that one
obtains the corresponding formula for $T_{2}^{\eta}$ and $T_{3}^{\eta}$ from
(\ref{kin3}) when $\psi_{1}$ is replaced by the corresponding angle $%
\psi_{2} $ and $\psi_{3}$, determined by the same formula (\ref{ang6}) with $%
\theta$ measured from $\mathbf{\hat{b}}_{2}$ or $\mathbf{\hat{b}}_{3}$,
respectively. This proves the above claim.

\begin{problem}
It is an interesting task to simplify the expression for the energies $%
T_{i}^{\eta}$, and to express them in terms of coordinates $I_{j}$ as in the
case of $T_{i}^{\tau}$. But we leave the topic here.
\end{problem}

Thus, in the general case our \textquotedblright reduced\textquotedblright\
ODE actually consists of the reduced Newton's equations (\ref{redu1})
together with the Euler equations (\ref{Euler}). The initial data will be a
given nondegenerate configuration $\mathbf{X}(t_{0})$ and angular momentum
vector $\mathbf{\Omega}$, and as before, this determines the initial
velocity $\mathbf{\dot{X}}(t_{0})$ and hence the motion $\mathbf{X}(t)$ is
unique (by Newton's equation (\ref{Newton1})).

However, to avoid ambiguity in the initial value problem for the "reduced"
ODE we must also specify the initial orientation (i.e. the normal vector $%
\mathbf{n(}t_{0})$) which decides whether the shape curve starts out on the
upper or lower hemisphere of $M^{\ast}$. With this choice the initial
eigenframe $\left\{ \mathbf{u}_{1}\mathbf{,u}_{2}\mathbf{,n}\right\}
|_{t=t_{0}},$ and hence also the initial values $g_{i}(t_{0})$, will be
unique relative to a fixed convention, say, the angle $\psi_{1}$ between $%
\mathbf{u}_{1}$ and $\mathbf{a}_{1}$ is (initially) in the range (\ref%
{range2}).\ Now, the remaining part of the proof of Theorem E2 is similar to
the proof of Theorem E1.

\subsection{Geometric reduction of the least action principles}

Recall from Section 1.2 the two classical least action principles\ with the
action integral $J_{1}$ and $J_{2}$, respectively. The underlying geometric
structure naturally associated to the former is \emph{Riemannian} while that
of the latter is \emph{symplectic}. Thus the two types of least action
principles are radically different in their basic geometric setting,
although both of them characterize the same motion. Indeed, it is easy to
verify that the Euler-Lagrange equations of both variational principles are
equivalent to the Newton's equation of motion (\ref{Newton1}).

However, the classical approach to the three-body problem is mainly based on 
$J_{2}$, namely the least action principle of Hamilton and the
Hamilton-Jacobi theory, in the framework of canonical transformations and
symplectic geometry. On the other hand, the kinematic geometry of
m-triangles is, on the other hand, more naturally associated with the least
action principle of Euler-Lagrange-Jacobi and the action integral $J_{1}$,
and it is in this geometric framework that we have established many basic
results, such as the universal sphericality, the kinematic Gauss-Bonnet
formula (and geometric phase), kinematic moving frames and the generalized
Euler equations.

\subsubsection{Proof of Theorem F}

We consider virtual 3-body motions in the xy-plane, represented by motions $%
\delta(t)$ of oriented m-triangles with $\mathbf{k}$ as their common normal
vector, and hence a (nondegenerate) m-triangle $\delta=(\mathbf{a}_{1}%
\mathbf{,a}_{2}\mathbf{,a}_{3})$ is positively oriented if $\mathbf{a}_{1}%
\mathbf{\times a}_{2}$ points in the direction of $\mathbf{k}$. The
corresponding configuration space is $\mathbb{R}^{4}$, see (\ref{4-bundle}),
and $\mathbb{R}^{4}/SO(2)=\bar{M}$ is the full moduli space. The $SO(2)$%
-orbit $\bar{\delta}$ of an oriented m-triangle $\delta$ will be regarded
both as a point in $\bar{M}$ and as a subset (congruence class) of $\mathbb{R%
}^{4}$. In the sequel, all motions in $\mathbb{R}^{4}$ are also assumed to
have a constant angular momentum.

The set of differentiable ($C^{1}$-smooth) curves in\ $\bar{M}$ from $\bar{%
\delta}$ to $\bar{\delta}_{1}$ is denoted $\mathfrak{\bar{P}}_{\bar{\delta},%
\bar{\delta}_{1}}$, and similarly $\mathfrak{P}_{\delta ,\bar{\delta}_{1}}$%
denotes the set of differentiable curves in $\mathbb{R}^{4}$ which start at $%
\delta$ and terminate at the orbit $\bar{\delta}_{1}$. For a fixed angular
momentum $\Omega$ the kinetic energy 
\begin{equation*}
T=T_{\Omega}=\bar{T}+T^{\omega}=\bar{T}+\frac{\Omega^{2}}{2I}, 
\end{equation*}
the potential function $U$, the Lagrange function $L_{\Omega}=T_{\Omega}+U$
and total energy $h=T_{\Omega}-U$, are functions which are also defined at
the level of $\bar{M}$. Therefore, for fixed value of $\Omega,h$ or time
interval $\left[ 0,t_{1}\right] $, we may consider corresponding subsets of $%
\mathfrak{\bar{P}}_{\bar{\delta},\bar{\delta}_{1}}$ 
\begin{equation}
\text{ }\mathfrak{\bar{P}}_{\bar{\delta},\bar{\delta}_{1}}(h)\text{, \ }%
\mathfrak{\bar{P}}_{\bar{\delta},\bar{\delta}_{1}}(\left[ t_{0},t_{1}\right]
)   \label{sets}
\end{equation}
with the obvious meaning, and similarly subsets of $\mathfrak{P}_{\delta ,%
\bar{\delta}_{1}}$ 
\begin{equation}
\mathfrak{P}_{\delta,\bar{\delta}_{1}}(\Omega),\text{ \ }\mathfrak{P}%
_{\delta,\bar{\delta}_{1}}(\Omega,h),\text{ }\mathfrak{P}_{\delta,\bar{%
\delta }_{1}}(\Omega,\left[ t_{0},t_{1}\right] )   \label{sets2}
\end{equation}
Clearly, the solution curves of Newton's equation belong to sets of type (%
\ref{sets2}).

We can also define the reduced action integrals 
\begin{equation}
\bar{J}_{1,\Omega}=\int T_{\Omega}dt\text{ , \ }\bar{J}_{2,\Omega}\text{\ }%
=\int L_{\Omega}dt   \label{actionredu}
\end{equation}
acting on moduli curves, and there is the following commutative diagram

\begin{equation}
\begin{array}{ccc}
\mathfrak{\bar{P}}_{\bar{\delta},\bar{\delta}_{1}} & \overset{\pi_{\Omega}}{%
\longleftarrow} & \mathfrak{P}_{\delta,\bar{\delta}_{1}}(\Omega) \\ 
\bar{J}_{i,\Omega}\searrow &  & \swarrow J_{i,\Omega} \\ 
& \mathbb{R} & 
\end{array}
\label{triang}
\end{equation}
where the map $\pi_{\Omega}$ takes a curve $\Gamma(t)$ to its moduli curve $%
\bar{\Gamma}(t)$ and $J_{i,\Omega}$ is the restriction of $J_{i}$ to the
subspace of curves with fixed angular momentum $\Omega$. Moreover, for $%
\delta$ a nondegenerate m-triangle the map $\pi_{\Omega}$ is, in fact, a
bijection due to the unique lifting property described by Theorem B.

Now, let us turn to the proof of Theorem F, which we restate as follows :

\begin{theorem}
The solution curves of the planary $\Omega$-reduced Newton's equation can be
characterized as the extremal curves of $\bar{J}_{1,\Omega}$ (resp. $\bar
{J%
}_{2,\Omega})$ restricted to the sets 
\begin{equation*}
\mathfrak{\bar{P}}_{\bar{\delta},\bar{\delta}_{1}}(h),\text{\ resp. }%
\mathfrak{\bar{P}}_{\bar{\delta},\bar{\delta}_{1}}(\left[ t_{0},t_{1}\right]
)\text{\ }
\end{equation*}
of moduli curves, with fixed energy $h$ or time interval $\left[ t_{0},t_{1}%
\right] $, respectively.
\end{theorem}

\begin{proof}
First of all, extremal curves of $\bar{J}_{1,\Omega}$ (resp. $\bar
{J}%
_{2,\Omega}$) are solutions of the associated Euler-Lagrange equations, and
one checks that these are second order ODE whose solution curves are (as
usual) uniquely determined by their initial position and velocity.

On the other hand, the $\Omega$-reduced Newton's equation is also a second
order ODE whose solution curves are uniquely determined by their initial
position and velocity. Therefore, to show that the Euler-Lagrange equations
and the $\Omega$-reduced Newton's equation have the same solutions it
suffices to verify this locally. More precisely, it suffices to show that
any small segment of a solution curve of the $\Omega$-reduced Newton's
equation is also an extremal curve of $\bar{J}_{1,\Omega}$ (resp. $\bar{J}%
_{2,\Omega}$).

Let $\bar{\Gamma}$ be a small segment from $\bar{\delta}$ to $\bar{\delta}%
_{1}$ of a solution curve of the $\Omega$-reduced Newton's equation. We may
assume the points are sufficiently close to ensure that $\bar{\Gamma}$ is
the only segment (of a solution) linking them.

Choose $\delta$ in the orbit $\bar{\delta}$. Since the actions in (\ref%
{actionredu}) are always nonnegative and the orbits $\bar{\delta}$ and $\bar{%
\delta}_{1}$ are sufficiently close, there exists a $J_{1}$-minimizing
(resp. $J_{2}$-minimizing) curve $\Gamma$ in $\mathbb{R}^{4}$ between $\delta
$ and the orbit $\bar{\delta}_{1}$, say $\delta_{1}$ is its end point. Then $%
\Gamma$ is a solution of Newton's equation and it must, in fact, be the
lifting of $\bar{\Gamma}$. Moreover, it is the unique curve in $\mathfrak{P}%
_{\delta,\bar{\delta}_{1}}(\Omega)$ with minimal action integral of $J_{1} $
(resp. $J_{2}$).

Consequently, $\Gamma$ is, of course, also a small segment of an extremal
curve of $J_{i,\Omega}$ and therefore by (\ref{triang}) its moduli curve $%
\bar{\Gamma}$ is a small segment of an extremal curve of $\bar{J}_{i,\Omega} 
$.
\end{proof}

\begin{remark}
Theorem F does not extend to the case of general three-body motions. The
reason is that the kinetic energy of a motion $\delta(t)$ does not depend
only on the moduli curve and the angular momentum vector $\mathbf{\Omega}$,
but also on the instantaneous configuration $\delta(t)$. Hence, one cannot
proceed as above, since it is not clear what should be the appropriate
action $\bar
{J}_{i,\Omega}$ at the moduli space level.
\end{remark}

\section{The Newtonian potential function}

In this chapter our primary task is to analyze the Newtonian potential
function (\ref{U1}) and its crucial dependence on the mass distribution. The
function is naturally defined at moduli space level, 
\begin{equation}
U=\sum\limits_{i=1}^{3}\frac{\hat{m}_{i}}{s_{i}}=\sum\limits_{i=1}^{3}\frac{%
\hat{m}_{i}^{3/2}}{\sqrt{(1-m_{i})I-I_{i}}}   \label{U2}
\end{equation}
where in each half-space $\bar{M}_{\pm}=\mathbb{R}_{\pm}^{3}$ the two
triples $(I_{1},I_{2},I_{3})$ and $(s_{1},s_{2},s_{3})$, related by (\ref%
{r/I}), are natural coordinate systems. Certainly, $U$ has the simplest
possible form\ when expressed by the mutual distances $s_{i}$. Even so,
sometimes it is also convenient to use Euclidean coordinates or their
associated spherical coordinates $(I,\varphi,\theta)$, where $I=\rho^{2}=%
\sqrt{x^{2}+y^{2}+z^{2}\text{, }}$as explained in Section 4.5.

Let $U^{\ast}$ be the restriction of $U$ to the \textquotedblright
unit\textquotedblright\ sphere $M^{\ast}=S^{2}=(I=1)$. As a function of the
coordinates $I_{j}$ (resp. $s_{j})$ $U$ is homogeneous of degree $-\frac{1}{2%
}$ (resp. $-1)$, namely 
\begin{equation*}
U(I_{1},I_{2},I_{3})=\frac{1}{\rho}U^{\ast}(\frac{I_{1}}{I},\frac{I_{2}}{I},%
\frac{I_{3}}{I})=\frac{1}{\rho}U^{\ast}(\delta^{\ast})=\frac{1}{\rho }%
U^{\ast}(\varphi,\theta) 
\end{equation*}
where $\delta^{\ast}$ $\longleftrightarrow(\varphi,\theta)$ represents the
shape of an m-triangle. For the sake of convenience, the formula for $%
U^{\ast }$ in terms of spherical coordinates is 
\begin{equation}
U^{\ast}=\sum_{i=1}^{3}\frac{\hat{m}_{i}^{3/2}(m_{i}^{\ast})^{-1/2}}{\sqrt{%
1-\sin\varphi\cos(\theta-\theta_{i})}}\text{ }   \label{Ustar}
\end{equation}
where $\theta_{1},\theta_{2},\theta_{3}$ are the longitude angles of the
binary collision points $\mathfrak{b}_{23},\mathfrak{b}_{31},\mathfrak{b}%
_{12}$, respectively.$\mathbf{\ }$This follows by substituting the
expression (\ref{intrins2}) with $\tilde{\theta}_{i}=$ $\theta-\theta_{i}$
into (\ref{U2}). In particular, with the convention that $\theta_{1}=0$ made
in Remark \ref{convention}, we must use 
\begin{equation*}
(\theta_{1},\theta_{2},\theta_{3})=(0,\beta_{3},-\beta_{2}) 
\end{equation*}
where the angles $\beta_{i}$, described in (\ref{triple2}) - (\ref{angles}),
measure the longitude differences between the points $\mathfrak{b}_{kl}$.

The analysis of $U$ trivially reduces to that of the restriction $U^{\ast
}=U\left\vert _{M^{\ast}}\right. $. In fact, by symmetry it suffices to
investigate $U^{\ast}$ on the closed upper hemisphere, $0\leq\varphi\leq\pi
/2$. $U^{\ast}$ has no maximum points since $U^{\ast}$ tends to $\infty$ at
the singular points $\mathfrak{b}_{kl}$ (which are poles of $U^{\ast}$). On
the other hand, it is a classical result, dating (at least) back to
Lagrange, that $U^{\ast}$ has a unique minimum value at the shape of a
regular triangle, namely (for $\rho=1)$%
\begin{equation}
s_{1}=s_{2}=s_{3}=\frac{1}{\sqrt{\hat{m}}},\text{ \ \ or \ \ }I_{i}=1-m_{i}-%
\frac{\hat{m}_{i}}{\hat{m}}\text{\ }   \label{phys}
\end{equation}
This defines a unique point $\mathfrak{p}_{0}^{\pm}$ on each hemisphere $%
M_{\pm}^{\ast}$ which we also refer to as the \emph{physical center }(as
opposed to the poles which are the \emph{geometric center}). It is easy to
prove the above statement using the coordinates $s_{i}$ and Lagrange's
multiplier method subject to the constraint $1=I=\tsum m_{i}^{\ast}s_{i}^{2}$%
, cf. (\ref{side3}). Another proof follows from Lemma \ref{zero} below. The
spherical coordinates of the shape (\ref{phys}) is worked out in Section
8.8, cf. (\ref{phys1}), (\ref{phys2}).

\subsection{Vector algebra analysis of the Newtonian function}

The vector algebra representation of $\bar{M}$ in Section 4.5 is also a
convenient setting for the local analysis of $U^{\ast}$, namely the function
on the unit sphere of Euclidean 3-space defined by 
\begin{equation}
U^{\ast}(\mathbf{p})=\sum\limits_{i=1}^{3}U_{i}^{\ast}(\mathbf{p}%
)=\sum\limits_{i=1}^{3}\frac{k_{i}}{\left\vert \mathbf{p}-\mathbf{\hat{b}}%
_{i}\right\vert },\text{ \ \ \ }k_{i}=\frac{2\hat{m}_{i}^{3/2}}{\sqrt{1-m_{i}%
}}   \label{U3}
\end{equation}
where 
\begin{equation*}
\left\vert \mathbf{p}-\mathbf{\hat{b}}_{i}\right\vert =\frac{k_{i}}{\hat
{m}%
_{i}}s_{i}\text{ \ \ \ \ \ (cf. (\ref{side4}))}
\end{equation*}
is the Euclidean distance from $\mathbf{p}$ to the binary collision point $%
\mathbf{\hat{b}}_{i}$.

Fix a point $\mathbf{p}$ on the sphere with $z>0$, and consider nearby
points $\mathbf{p}^{\prime}=\mathbf{p+x}$ on the sphere, that is, $\mathbf{x}
$ is subject to the constraint 
\begin{equation}
2\mathbf{p\cdot x+x\cdot x}=0,   \label{cond}
\end{equation}
which in turn implies 
\begin{equation*}
\left\vert \mathbf{p}^{\prime}-\mathbf{\hat{b}}_{i}\right\vert
^{2}=\left\vert \mathbf{p-\hat{b}}_{i}\right\vert ^{2}-2\mathbf{\hat{b}}%
_{i}\cdot\mathbf{x}
\end{equation*}
Hence, there is the following expansion at $\mathbf{p}$%
\begin{equation}
U^{\ast}(\mathbf{p+x})=\sum\limits_{i=1}^{3}\frac{U_{i}^{\ast}(\mathbf{p})}{%
(1-z_{i})^{1/2}}=\sum\limits_{i=1}^{3}U_{i}^{\ast}(\mathbf{p})\tsum
\limits_{n=0}^{\infty}c_{n}z_{i}^{n}=\sum\limits_{n=0}^{\infty}F_{n}(\mathbf{%
p;x})   \label{series}
\end{equation}
where we use the notation 
\begin{align*}
z_{i} & =\frac{2\mathbf{\hat{b}}_{i}\cdot\mathbf{x}}{\left\vert \mathbf{p}-%
\mathbf{\hat{b}}_{i}\right\vert ^{2}}=(\frac{m_{i}^{\ast}m_{i}}{\bar{m}})%
\frac{\mathbf{\hat{b}}_{i}\cdot\mathbf{x}}{s_{i}^{2}} \\
c_{n} & =\frac{1\cdot3\cdot5\cdot\cdot\cdot(2n-1)}{2^{n}n!}\text{, \ }c_{0}=1
\end{align*}

\begin{remark}
It is easy to see that the convergence condition $\left\vert
z_{i}\right\vert <1$ for the series of $U_{i}^{\ast}(\mathbf{p+x)}$ in (\ref%
{series}) is equivalent to the condition 
\begin{equation*}
(\mathbf{p+x)\cdot\hat{b}}_{i}>2\mathbf{p\cdot\hat{b}}_{i}-1 
\end{equation*}
Geometrically, this means that $\mathbf{p+x}$ belongs to the hemispherical
cap $D_{i}$ centered at $\mathbf{\hat{b}}_{i}$ which is cut out by the plane
parallel to the tangent plane at $\mathbf{\hat{b}}_{i}$ and separated by the
distance $R_{i}=2(1-\mathbf{p}\cdot\mathbf{\hat{b}}_{i})$. In particular, $%
D_{i}$ is the whole hemisphere if $\mathbf{p}\cdot\mathbf{\hat{b}}_{i}\leq0$%
, and by Lemma \ref{dual}, we know $\mathbf{p}\cdot\mathbf{\hat{b}}_{i}\leq0$
holds for at least one $i$. The domain of convergence for the series of $%
U^{\ast}$ is the \textquotedblright polygonal\textquotedblright\ region $%
\cap D_{i}.$
\end{remark}

The zero order term of the expansion (\ref{series}) is, of course, $%
F_{0}=U^{\ast}(\mathbf{p)}$, and the first order term is 
\begin{equation}
F_{1}(\mathbf{p;x)=(}\frac{k_{1}\mathbf{\hat{b}}_{1}}{\left\vert \mathbf{p}-%
\mathbf{\hat{b}}_{1}\right\vert ^{3}}+\frac{k_{2}\mathbf{\hat{b}}_{2}}{%
\left\vert \mathbf{p}-\mathbf{\hat{b}}_{2}\right\vert ^{3}}+\frac {k_{3}%
\mathbf{\hat{b}}_{3}}{\left\vert \mathbf{p}-\mathbf{\hat{b}}_{3}\right\vert
^{3}})\cdot\mathbf{x}   \label{F1}
\end{equation}
which is essentially the gradient of $U^{\ast}$ at $\mathbf{p}$. To make
this precise, consider the following function from $S^{2}(1)$ to the
xy-plane 
\begin{equation}
\mathbf{p\rightarrow B(p)=}\sum\limits_{i=1}^{3}\frac{k_{i}\mathbf{\hat{b}}%
_{i}}{\left\vert \mathbf{p}-\mathbf{\hat{b}}_{i}\right\vert ^{3}}\in\mathbb{R%
}^{2}   \label{B0}
\end{equation}

\begin{lemma}
The gradient vector of $U^{\ast}$ at $\mathbf{p}$ is given by 
\begin{equation}
\nabla U^{\ast}(\mathbf{p)=B(p)-(B(p)\cdot p)p}   \label{B1}
\end{equation}
\end{lemma}

\begin{proof}
Since by (\ref{F1}) 
\begin{equation*}
\nabla U^{\ast}(\mathbf{p)\cdot x}=F_{1}(\mathbf{p;x)}=\mathbf{B(p)\cdot x}
\end{equation*}
and $\mathbf{x}$ is tangential to $\mathbf{p}$ in the limit as $\mathbf{x}%
\rightarrow0$, it follows that 
\begin{equation*}
\nabla U^{\ast}(\mathbf{p)\cdot t}=\mathbf{B(p)\cdot t}
\end{equation*}
holds for all tangent vectors $\mathbf{t}$. Hence, the tangent vector $%
\nabla U^{\ast}(\mathbf{p)}$ is the orthogonal projection of $\mathbf{B(p)}$
in the direction of $\mathbf{p}$.\ 
\end{proof}

\begin{lemma}
\label{zero}The zero points of $\mathbf{B}$ are the two critical points of $%
U^{\ast}$ outside the equator circle $z=0$, namely the physical center $%
\mathbf{\hat{p}}_{0}^{\pm}$ (cf. (\ref{phys})) on each hemisphere $z>0$ or $%
z<0$.
\end{lemma}

\begin{proof}
By Lemma \ref{dual}, $\mathbf{B(p)}=0$ if and only if for some constant $%
\lambda>0$, 
\begin{equation*}
m_{1}^{\ast}\frac{\left\vert \mathbf{p}-\mathbf{\hat{b}}_{1}\right\vert ^{3}%
}{k_{1}}=m_{2}^{\ast}\frac{\left\vert \mathbf{p}-\mathbf{\hat{b}}%
_{2}\right\vert ^{3}}{k_{2}}=m_{3}^{\ast}\frac{\left\vert \mathbf{p}-\mathbf{%
\hat{b}}_{3}\right\vert ^{3}}{k_{3}}=\lambda, 
\end{equation*}
and by (\ref{side4}), this is equivalent to $s_{1}^{3}=s_{2}^{3}=s_{3}^{3}=%
\frac{\lambda}{2}$, namely $s_{1}=s_{2}=s_{3}.$ In particular, $\mathbf{p}$
is a point with $z\neq0$, cf. (\ref{z0}). On the other hand, for $z\neq0$ it
is easy to see that $\mathbf{B(p)}=0$ if and only if $\nabla U^{\ast}(%
\mathbf{p)}=0$.
\end{proof}

The identity (\ref{B1}) also implies that the critical points of $U^{\ast}$
on the unit circle $z=0$ are the \textquotedblright
eigenvectors\textquotedblright\ of $\mathbf{B}$, in the sense that $\mathbf{%
B(p)}=\lambda\mathbf{p}$ for some $\lambda$, necessarily equal to $\mathbf{%
B(p)\cdot p}\neq0$. Clearly, $\mathbf{B}$ has a pole at $\mathbf{\hat{b}}%
_{i} $ and $\mathbf{B(p)/}\left\vert \mathbf{B(p)}\right\vert $ tends to $%
\mathbf{\hat{b}}_{i}$ as $\mathbf{p}$ tends to $\mathbf{\hat{b}}_{i}$. A
simple analysis of $\mathbf{B}$ will show there are exactly one solution $%
\mathbf{p} $ between each pair $\mathbf{\hat{b}}_{j},\mathbf{\hat
{b}}_{k}$
of poles. These are the so-called Euler points $\mathbf{\hat{e}}_{i},i=1,2,3$%
, and they are the saddle points of $U^{\ast}$on the 2-sphere. We omit the
proof of this well known fact.

\subsubsection{ Series expansion of $U^{\ast}$ at its minimum point}

Henceforth, we shall focus attention on the expansion (\ref{series}) at the
physical center $\mathbf{\hat{p}}_{0}=(\hat{x}_{,}\hat{y},\hat{z}),\hat{z}>0$%
, that is, $\mathbf{\hat{p}}_{0}$ is the minimum point of $U^{\ast}$ on the
hemisphere $z>0$. The coordinates of $\mathbf{\hat{p}}_{0}$ are the
following mass dependent constants 
\begin{subequations}
\begin{align}
\hat{x} & =\mathbf{\hat{p}}_{0}\cdot\mathbf{\hat{b}}_{1}=1-\frac{\bar{m}}{%
\hat{m}m_{1}^{\ast}m_{1}}  \label{x0} \\
\hat{y} & =\frac{\sqrt{\bar{m}}}{\hat{m}}(\frac{m_{2}-m_{3}}{m_{2}+m_{3}})
\label{y0} \\
\hat{z} & =\cos2r_{0}=4\Delta_{0}\sqrt{m_{1}m_{2}m_{3}}=\sqrt{3}\frac {\sqrt{%
\bar{m}}}{\hat{m}}   \label{z0}
\end{align}
where $\Delta_{0}$ is the area of the regular triangle with $I=1$. These
expressions follow from (\ref{cos2r}), (\ref{inner2}) and (\ref{phys}).
Note, for example, that $\hat{z}$ becomes arbitrarily small when some mass $%
m_{i}$ tends to zero, and $\hat{z}=1$, that is, $\mathbf{\hat{p}}_{0}$ is
the north pole $\mathcal{N}$, precisely when the masses are equal.
Concerning the sign of $\hat{y}=\pm\sqrt{1-\hat{x}^{2}-\hat{z}^{2}}$, we
used (\ref{inner2}) to check, for example, that $\hat{y}>0$ if $m_{2}>m_{3}$.

The constant term of the series (\ref{series}) is the minimum value 
\end{subequations}
\begin{equation}
F_{0}=U^{\ast}(\mathbf{\hat{p}}_{0})=\sum U_{i}^{\ast}(\mathbf{\hat{p}}%
_{0})=\sum\sqrt{\hat{m}}\hat{m}_{i}=\hat{m}^{3/2},\text{ }   \label{min}
\end{equation}
and $F_{1}=0$, of course. Moreover, by (\ref{phys}), $\hat{m}s_{i}^{2}=1$
holds for all $i$, and therefore the $n$-th order term can be written as 
\begin{equation}
F_{n}(\mathbf{\hat{p}}_{0};\mathbf{x})=\left( \frac{c_{n}\hat{m}^{n+\frac {1%
}{2}}}{\bar{m}^{n-1}}\right) \sum\limits_{i=1}^{3}\frac{1}{m_{i}}%
(m_{i}m_{i}^{\ast})^{n}(\mathbf{\hat{b}}_{i}\cdot\mathbf{x)}^{n}   \label{Fn}
\end{equation}

Now, let us turn to the local analysis of the series, namely $U^{\ast}$
developed as a power series in suitable coordinates around $\mathbf{\hat{p}}%
_{0}$. To this end, consider a positively oriented orthonormal frame of the
Euclidean 3-space of type \ 
\begin{equation}
\left\{ \mathbf{t}_{1},\mathbf{t}_{2},\mathbf{\hat{p}}_{0}\right\} \text{, \ 
}\mathbf{t}_{i}\in\Pi\text{ for }i=1,2   \label{frame2}
\end{equation}
where $\Pi$ is the tangent space of the sphere at $\mathbf{\hat{p}}_{0}$. By
condition (\ref{cond}), the components of $\mathbf{x=p-\hat{p}}_{0}$ must
satisfy 
\begin{equation}
\mathbf{x}=\xi\mathbf{t}_{1}+\eta\mathbf{t}_{2}+\zeta\mathbf{\hat{p}}_{0}%
\text{, \ \ }2\zeta+\xi^{2}+\eta^{2}+\zeta^{2}=0,   \label{cond2}
\end{equation}
and consequently the map 
\begin{equation*}
(\xi,\eta)\rightarrow(\xi,\eta,\zeta), 
\end{equation*}
where 
\begin{equation}
\zeta=-1+\sqrt{1-(\xi^{2}+\eta^{2})}=-\frac{1}{2}(\xi^{2}+\eta^{2})-\frac {1%
}{8}(\xi^{2}+\eta^{2})^{2}-...,   \label{param}
\end{equation}
is a parametrization of the region of the sphere lying above the plane
through the origin and parallel to $\Pi$. Geometrically, the unit disk of $%
\Pi$ is projected down to the sphere in the direction of the normal vector $%
\mathbf{\hat{p}}_{0}$.

It is natural to choose $\mathbf{t}_{1}$ tangent to the meridian through $%
\mathbf{\hat{p}}_{0}$. Indeed, the intrinsic nature of this condition will
lead to a series expansion whose coefficients are essentially symmetric
functions of the masses $m_{i}$. The projection in the xy-plane of such an
orthonormal basis $\left\{ \mathbf{t}_{1},\mathbf{t}_{2}\right\} $ of $\Pi$
is (up to sign) given by 
\begin{equation}
\mathbf{t}_{1}^{\prime}=\frac{\hat{z}}{\sqrt{1-\hat{z}^{2}}}(\hat{x},\hat {y}%
)\text{, \ }\mathbf{t}_{2}^{\prime}=\frac{1}{\sqrt{1-\hat{z}^{2}}}(-\hat {y},%
\hat{x})=\mathbf{t}_{2}   \label{frame3}
\end{equation}

In order to express the $n$-th term (\ref{Fn}) of the $U^{\ast}$-series in
terms of $\xi,\eta$, we proceed as follows. Expand the \textquotedblright
variables\textquotedblright\ in (\ref{Fn}) 
\begin{equation}
\mathbf{\hat{b}}_{i}\cdot\mathbf{x}=a_{i}^{\prime}\xi+b_{i}^{\prime}\eta
+c_{i}^{\prime}\zeta\text{, \ }i=1,2,3   \label{variable}
\end{equation}
where the three triples $(a_{i}^{\prime},b_{i}^{\prime},c_{i}^{\prime})$ of
coefficients are specific functions of the parameters $m_{i}$ determined by 
\begin{equation}
a_{i}^{\prime}=\mathbf{\hat{b}}_{i}\cdot\mathbf{t}_{1}=\mathbf{\hat{b}}%
_{i}\cdot\mathbf{t}_{1}^{\prime}\text{, \ }b_{i}^{\prime}=\mathbf{\hat{b}}%
_{i}\cdot\mathbf{t}_{2}\text{, \ }c_{i}^{\prime}=\mathbf{\hat{b}}_{i}\cdot%
\mathbf{\hat{p}}_{0}   \label{coeff}
\end{equation}
For example, by (\ref{binary1}) and (\ref{frame3}), 
\begin{equation*}
a_{1}^{\prime}=\frac{\hat{x}\hat{z}}{\sqrt{1-\hat{z}^{2}}}\text{, \ }%
b_{1}^{\prime}=\frac{-\hat{y}}{\sqrt{1-\hat{z}^{2}}}\text{ , \ }%
c_{1}^{\prime }=\hat{x}, 
\end{equation*}
where $\hat{x},\hat{y},\hat{z}$ are the known functions in (\ref{x0}) - (\ref%
{z0}), and the three triples in (\ref{coeff}) permute covariantly with the
parameters $m_{i}$.

In view of (\ref{Fn}), it is slightly more convenient to replace (\ref%
{variable}) by 
\begin{equation*}
(m_{i}m_{i}^{\ast})\mathbf{\hat{b}}_{i}\cdot\mathbf{x}=a_{i}\xi+b_{i}%
\eta+c_{i}\zeta\text{, \ }i=1,2,3 
\end{equation*}
and hence we calculate the modified coefficients, namely 
\begin{align}
a_{i} & =\frac{\sqrt{3\bar{m}}}{\sqrt{\hat{m}^{2}-3\bar{m}}}%
(m_{i}m_{i}^{\ast}-\frac{\bar{m}}{\hat{m}}),  \notag \\
b_{i} & =\frac{-\sqrt{\bar{m}}m_{i}}{2\sqrt{\hat{m}^{2}-3\bar{m}}}%
(m_{i+1}-m_{i+2})\text{ \ \ \ \ }(i\text{ }\func{mod}3)   \label{coeff1} \\
c_{i} & =m_{i}m_{i}^{\ast}-\frac{\bar{m}}{\hat{m}}  \notag
\end{align}
Thus, with the coordinate system $(\xi,\eta)$, we finally arrive at the
following \emph{symmetrization} of the power series in (\ref{series} ) : 
\begin{equation}
U^{\ast}(\xi,\eta)=\sum\limits_{n=0}^{\infty}K_{n}\sum\limits_{i=1}^{3}\frac{%
1}{m_{i}}(a_{i}\xi+b_{i}\eta+c_{i}\zeta)^{n}=\sum\limits_{n=0}^{\infty
}K_{n}\sum\limits_{j+k=n}\left( 
\begin{array}{c}
n \\ 
j%
\end{array}
\right) A_{j,k}\xi^{j}\eta^{k}   \label{series1}
\end{equation}
where 
\begin{equation*}
K_{n}=\frac{1\cdot3\cdot5\cdot\cdot\cdot(2n-1)}{2^{n}n!}\frac{\hat{m}^{n+%
\frac{1}{2}}}{\bar{m}^{n-1}}
\end{equation*}
and $A_{j,k}$, $k$ even, are symmetric functions of the masses $m_{i}$,
whereas $A_{j,k}$ is alternating symmetric when $k$ is odd.

\begin{remark}
The Newton sums $S_{k}=\sum m_{i}^{k}$ are, of course, polynomials of $%
\hat
{m}$ and $\bar{m}$, cf. (\ref{symm}). For example, 
\begin{equation*}
S_{4}=1-4\hat{m}+4\bar{m}+2\hat{m}^{2}
\end{equation*}
In terms of the $S_{k}$ it is rather straightforward to obtain explicit
expressions for the above symmetric functions $A_{j,k}$. For $k$ odd, the
alternating function $A_{j,k}$ is a product of a symmetric function and the
basic alternating function 
\begin{align}
\mathfrak{A} & =(m_{1}-m_{2})(m_{2}-m_{3})(m_{3}-m_{1})  \label{alt1} \\
& =\sum_{i\func{mod}3}m_{i}(1-m_{i})(m_{i+1}-m_{i+2})  \notag
\end{align}
\end{remark}

\subsubsection{The quadratic term of $U^{\ast}$}

We shall work out explicitly the quadratic term of the function $U^{\ast}$
expanded at the physical center $\mathbf{\hat{p}}_{0}$, namely the $n=2$
term of (\ref{series}) or (\ref{series1}) 
\begin{equation}
F_{2}=\frac{3\hat{m}^{5/2}}{8\bar{m}}(A\xi^{2}+2B\xi\eta+C\eta^{2})=\kappa(%
\tilde{\lambda}_{1}\tilde{\xi}^{2}+\tilde{\lambda}_{2}\tilde{\eta }^{2}) 
\label{F2}
\end{equation}
where $(\tilde{\xi},\tilde{\eta})$ is the coordinate system of a
diagonalizing frame $\left\{ \mathbf{\tilde{t}}_{1},\mathbf{\tilde{t}}%
_{2}\right\} $ of the tangent plane $\Pi$ at $\mathbf{\hat{p}}_{0}$.

First, let us determine the coefficients $A=A_{2,0},B=A_{1,1},C=A_{0,2}$ in (%
\ref{series1}) as symmetric or alternating functions of the symbols $m_{i}$.
By using (\ref{alt1}), the identities 
\begin{equation*}
\sum m_{i}(1-m_{i})^{2}=\hat{m}+3\bar{m}\text{, \ }\sum
m_{i}(m_{i+1}-m_{i+2})^{2}=\hat{m}-9\bar{m}
\end{equation*}
and the expressions (\ref{coeff1}), we find 
\begin{align*}
A & =\sum\frac{1}{m_{i}}a_{i}^{2}=\frac{1}{4}\frac{3\bar{m}}{\hat{m}}(\frac{%
\hat{m}^{2}+3\hat{m}\bar{m}-4\bar{m}}{\hat{m}^{2}-3\bar{m}}) \\
C & =\sum\frac{1}{m_{i}}b_{i}^{2}=\frac{1}{4}\frac{\bar{m}(\hat{m}-9\bar {m})%
}{\hat{m}^{2}-3\bar{m}} \\
B & =\sum\frac{1}{m_{i}}a_{i}b_{i}=\frac{1}{4}(\frac{-\sqrt{3}\bar{m}}{\hat{m%
}^{2}-3\bar{m}})\mathfrak{A}
\end{align*}
Hence, the eigenvalues of the quadratic in (\ref{F2}) are determined from
the equations 
\begin{equation*}
\tilde{\lambda}_{1}+\tilde{\lambda}_{2}=A+C=\frac{\bar{m}}{\hat{m}}\text{, \
\ }\tilde{\lambda}_{1}\tilde{\lambda}_{2}=AC-B^{2}=\frac{3}{4}\frac{\bar
{m}%
^{2}}{\hat{m}}, 
\end{equation*}
which yield 
\begin{equation}
\tilde{\lambda}_{i}=\frac{\bar{m}}{2\hat{m}}(1\pm\sqrt{1-3\hat{m}}) 
\label{eig}
\end{equation}
Consequently, 
\begin{equation}
F_{2}=\frac{3}{16}\hat{m}^{3/2}\left( \left( 1\pm\mu\right) \tilde{\xi}%
^{2}+(1\mp\mu)\tilde{\eta}^{2}\right) ,\text{ \ \ }\mu=\sqrt{1-3\hat{m}}%
\text{,}   \label{F2x}
\end{equation}
where $(\tilde{\xi},\tilde{\eta})$ are coordinates with respect to the
eigenvectors 
\begin{equation*}
\mathbf{\tilde{t}}_{1}=\cos\tilde{\alpha}\text{ }\mathbf{t}_{1}+\sin \tilde{%
\alpha}\text{ }\mathbf{t}_{2}\text{, \ \ }\mathbf{\tilde{t}}_{2}=-\sin\tilde{%
\alpha}\text{ }\mathbf{t}_{1}+\cos\tilde{\alpha}\text{ }\mathbf{t}_{2}
\end{equation*}
obtained by rotating the intrinsic frame $\left\{ \mathbf{t}_{1},\mathbf{t}%
_{2}\right\} $ of the plane $\Pi_{0}$.

Similar to (\ref{angle1}), the angle $\tilde{\alpha}$ is given by 
\begin{equation}
\tan2\tilde{\alpha}=\frac{2B}{A-C}=\frac{-\sqrt{3}\hat{m}%
(m_{1}-m_{2})(m_{2}-m_{3})(m_{3}-m_{1})}{\hat{m}^{2}+9\bar{m}\hat{m}-6\bar{m}%
}   \label{angle4}
\end{equation}
Then it also follows from (\ref{eigendiff}) that the largest eigenvalue $%
\tilde{\lambda}_{1}\sim1+\mu$ corresponds to the vector $\mathbf{\tilde{t}}%
_{1}$ if and only if $\sin2\tilde{\alpha}$ and $B$ have the same sign.

\begin{remark}
In the simplest case of uniform mass distribution, $m_{i}=1/3$, the analysis
of $U^{\ast}$ is much simpler than in the general case. In this case, where
the geometrical and physical center coincide, some of the above expressions
such as (\ref{angle4}), are of indeterminate type. However, if $%
m_{i}=m_{j}\neq m_{k}$ then $\tan2\tilde{\alpha}=0$, and hence the frame $%
\left\{ \mathbf{t}_{1},\mathbf{t}_{2}\right\} $ is already diagonalizing.
\end{remark}

\section{A geometric setting for the study of triple collisions}

Recall the well known fact, stated by Weierstrass and proved by Sundman (cf. 
\cite{Sund1}, \cite{Sund2}), that three-body motions leading to triple
collision must have vanishing angular momentum, $\mathbf{\Omega}=0$, and
consequently they are also planary. Thus we shall focus attention on planary
virtual motions $\mathbf{X}(t)$, namely curves in the configuration space $%
M_{0}\simeq\mathbb{R}^{4}$, with zero angular momentum, and we continue to
use the vector algebra representation (cf. Section 4.5) of the moduli space $%
\bar{M}=M_{0}/SO(2)=\mathbb{R}^{3}$, where the (equator) xy-plane $\bar
{E}=%
\mathbb{R}^{2}$ represents congruence classes of eclipse (i.e. collinear)
configurations.

The total kinetic energy $T=\bar{T}$ can be expressed as a positive definite
quadratic differential form on\ $\bar{M}\smallsetminus\left\{ O\right\} $,
namely the kinematic Riemannian metric 
\begin{equation}
d\bar{s}^{2}=2Tdt^{2}=d\rho^{2}+\rho^{2}d\sigma^{2}=d\rho^{2}+\frac{\rho^{2}%
}{4}ds^{2}\ \text{, \ cf. (\ref{dsbar3})}   \label{metric3}
\end{equation}
which describes $\bar{M}\ $as the Riemannian cone over the shape space $%
M^{\ast}=S^{2}(1/2)$, and 
\begin{equation}
ds^{2}=4d\sigma^{2}=d\varphi^{2}+(\sin^{2}\varphi)d\theta^{2} 
\label{metric8}
\end{equation}
is the metric of the magnified sphere $S^{2}(1).$

Following Jacobi, we introduce the following conformal modification of the
metric (\ref{metric3}) for each energy level $h$, namely 
\begin{equation}
d\bar{s}_{h}^{2}=(U+h)d\bar{s}^{2},\text{ \ \ \ }   \label{metric5}
\end{equation}
which we refer to as the \emph{physical metric}, and transform the action
integral $\bar{J}_{1,0}$ of (\ref{actionredu}) into the arc-length integral
in the Riemannian space 
\begin{equation}
(\bar{M}_{h},d\bar{s}_{h}^{2}):\text{ }\bar{M}_{h}=\left\{ \mathbf{p}\in 
\bar{M};U(\mathbf{p})+h\geq0\right\}   \label{Mbarh}
\end{equation}
Consequently, the trajectories of three-body motions with total energy $h$
are mapped to curves in $\bar{M}$ which are geodesics in the space $(\bar{M}%
_{h},d\bar{s}_{h}^{2})$. Notice that reflection in the equator plane $\bar{E}
$ $\subset\bar{M}$, that is, the transformation $\varphi\rightarrow\pi
-\varphi$, restricts to an involutive isometry of the Riemannian space (\ref%
{Mbarh}) with the eclipse subspace $\bar{E}_{h}$ $=\bar{E}$ $\cap\bar
{M}%
_{h}$ as fixed point set, and hence this is a totally geodesic submanifold
of $\bar{M}_{h}$. In particular, a geodesic curve in $(\bar{M}_{h},d\bar
{s}%
_{h}^{2})$ is transversal to $\bar{E}_{h}$, unless it lies entirely in $\bar{%
E}_{h}$. Moreover, for a simple geometrical reason, a shortest geodesic in $(%
\bar{M}_{h},d\bar{s}_{h}^{2})$ between a point outside $\bar{E}_{h}$ and the
origin $O$ cannot have any intermediate intersection with $\bar{E}_{h}$.

Finally, we note that Newton's equation (\ref{Newton1}) has a 1-parameter
group of space-time scaling symmetries $\left\{ g_{s}\right\} $, where $%
g_{s} $ sends a solution $\mathbf{X}(t)$ to a solution 
\begin{equation}
\mathbf{Y}(t)=e^{2s/3}\mathbf{X}(e^{-s}t)   \label{scale}
\end{equation}
and changes the energy level from $h$ to $e^{-2s/3}h$. Hence, all the
Riemannian structures in (\ref{Mbarh}) with energy $h$ of the same sign are
mutually homothetic, and consequently there are essentially only three
distinct cases, namely when the total energy $h$ is negative, zero or
positive.

\subsection{Geodesic rays and distance estimates}

Clearly, for $h\geq0$ the variety $\bar{M}_{h}$ is the whole moduli space $%
\bar{M}$, whereas for $h<0$ it is the star-shaped union of all ray segments 
\begin{equation}
\left[ O,\frac{U^{\ast}(\mathbf{p})}{\left| h\right| }\mathbf{p}\right] 
\text{, }\mathbf{p}\in M^{\ast}   \label{segment}
\end{equation}
from the origin $O$ to the point where the ray through $\mathbf{p}$
intersects the boundary $\partial\bar{M}_{h}$, that is, the level surface $%
U=-h$. By definition, the physical metric (\ref{metric5}) vanishes on the
boundary, meaning that the distance between any two boundary points is zero.

For $h<0$ the length of any ray segment (\ref{segment}) is 
\begin{equation}
L_{h}(\mathbf{p})=\int d\bar{s}_{h}=\int\limits_{0}^{\frac{U^{\ast}(\mathbf{p%
})}{\left\vert h\right\vert }}\sqrt{\frac{U^{\ast}(\mathbf{p})}{\rho}+h}%
\text{ }d\rho\text{,}   \label{length}
\end{equation}
and therefore there is a unique pair of shortest length 
\begin{equation}
L_{h}=L_{h}(\mathbf{\hat{p}}_{0}^{\pm})=\int\limits_{0}^{\frac{\mu_{0}}{%
\left\vert h\right\vert }}\sqrt{\frac{\mu_{0}}{\rho}+h}\text{ }d\rho 
\label{length2}
\end{equation}
where the points $\mathbf{\hat{p}}_{0}^{\pm}$ on the hemispheres $z>0$ and $%
z<0$ represent the shape of a regular triangle and hence $U^{\ast}$ has the
minimal value 
\begin{equation*}
\mu_{0}=U^{\ast}(\mathbf{\hat{p}}_{0}^{\pm})=\hat{m}^{3/2}\text{ \ , cf. (%
\ref{min}) }
\end{equation*}

The following is a useful fact in Riemannian geometry which follows from
general analysis of the first variation of arc-length.

\begin{lemma}
\label{compare} Let $ds^{2}$ and $d\tilde{s}^{2}$ be two Riemannian metrics
on a given manifold such that$\ $%
\begin{equation*}
d\tilde{s}^{2}=f^{2}ds^{2}
\end{equation*}
where $f$ is a smooth and positive function, that is, $d\tilde{s}^{2}$ is a
conformal modification of $ds^{2}.$ Let $\Gamma$ be a $C^{2}$-smooth curve
and let $\mathbf{n}$ denote a normal vector at a given point on $\Gamma.$
Then the geodesic curvatures of $\Gamma$ in (the direction of $\mathbf{n)}$
with respect to the two metrics are related by 
\begin{equation*}
\mathcal{\tilde{K}(}\mathbf{n)}=\mathcal{K(}\mathbf{n)-}\frac{d}{d\mathbf{n}}%
\ln f 
\end{equation*}
\end{lemma}

We shall apply the lemma to the kinematic and physical metric, namely the
metrics $d\bar{s}^{2}$ and $d\bar{s}_{h}^{2}$ on $\bar{M}$, cf. (\ref%
{metric5}). Thus, a moduli curve $\bar{\Gamma}$ is a geodesic with respect
to $d\bar
{s}_{h}^{2}$ if and only if its geodesic curvature $\mathcal{%
\tilde{K}(}\mathbf{n)}$ with respect to $d\bar{s}_{h}^{2}$ vanishes for all $%
\mathbf{n}$, or equivalently 
\begin{equation}
\mathcal{K(}\mathbf{n)}=\frac{1}{2}\frac{d}{d\mathbf{n}}\ln(U+h) 
\label{curv}
\end{equation}
where $\mathcal{K(}\mathbf{n)}$ is the geodesic curvature in the normal
direction $\mathbf{n}$, with respect to $d\bar{s}^{2}$.

The simplest type of 3-body motions are the \emph{shape invariant }ones,
that is, the shape curve is a single point on $M^{\ast}=S^{2}$ and hence the
moduli curve is confined to a ray emanating from $O$ in the cone $\bar{M}$.
Rays are, of course, geodesics with respect to the kinematic metric $d\bar{s}%
^{2}$, consequently a ray through $\mathbf{p}\in S^{2}$ (or a ray segment (%
\ref{segment}) if $h<0)$ is also a geodesic of the metric $d\bar{s}_{h}^{2}$
if and only if the normal derivative vanishes in all directions $\mathbf{n}$
normal to the ray, that is, 
\begin{equation*}
\frac{d}{d\mathbf{n}}\ln(\frac{1}{\rho}U^{\ast}+h)=0 
\end{equation*}
This condition is independent of the radial coordinate $\rho$, and for $%
\rho=1$ the vectors $\mathbf{n}$ span the tangent plane of $S^{2}$ at the
point $\mathbf{p}$. Consequently, the solutions are the five critical points 
$\mathbf{p}$ of $U^{\ast}$, namely the three saddle points (called Euler
points) $\mathbf{\hat{e}}_{i}$ on the equator circle $\rho=1$ in the
xy-plane, and the pair $\mathbf{\hat{p}}_{0}^{\pm}$ of minimumspoints (also
called Lagrange points). Thus, there are altogether exactly five geodesic
rays (or ray segments) in $(\bar{M}_{h},d\bar{s}_{h}^{2})$.

\begin{lemma}
For $h<0$, the two ray segments $\left[ O,\frac{\mu_{0}}{\left\vert
h\right\vert }\mathbf{\hat{p}}_{0}^{\pm}\right] $ are the unique shortest
geodesic curves in $(\bar{M}_{h},d\bar{s}_{h}^{2})$ linking a boundary point
and the base point $O$ (ignoring curve pieces of zero length along $\partial%
\bar{M}_{h})$. In particular, the distance from $\partial\bar{M}_{h}$ to $O$
is the number $L_{h}$ in (\ref{length2}).
\end{lemma}

\begin{proof}
In ($\bar{M},d\bar{s}^{2})$, let $B_{h}$ be the geodesic ball of radius $%
\frac{\mu_{0}}{\left\vert h\right\vert }$ centered at $O$. It lies inside $%
\bar{M}_{h}$ and touches $\partial\bar{M}_{h}$ at the two points $\frac {%
\mu_{0}}{\left\vert h\right\vert }\mathbf{\hat{p}}_{0}^{\pm}$. If $\ \bar{%
\Gamma}$ is any curve between $O$ and a point $\mathbf{q}$ on $\partial\bar{M%
}_{h}$, let $\bar{\Gamma}_{1}$ be the portion of $\bar{\Gamma}$ between $O$
and the first point $\mathbf{q}_{1}$ on $\partial\bar{M}_{h}.$ From the
calculation 
\begin{equation*}
L(\bar{\Gamma})\geq L(\bar{\Gamma}_{1})=\int\sqrt{h+U}d\bar{s}\geq\int \sqrt{%
h+\frac{\mu_{0}}{\rho}}d\bar{s}\geq\int_{0}^{\frac{\mu_{0}}{\left\vert
h\right\vert }}\sqrt{h+\frac{\mu_{0}}{\rho}}d\rho=L_{h}
\end{equation*}
it is clear that the two ray segments of length $L_{h}$ are, indeed, the
shortest curves, and they are unique (modulo a portion along the boundary).
\end{proof}

\begin{remark}
For $h\geq0$, the geodesic rays through $\mathbf{\hat{p}}_{0}^{\pm}$ are
still length minimizing, whereas for any $h$ this fails for the three
geodesic rays (or segments) in the eclipse plane $\bar{E}$.
\end{remark}

\subsection{Existence of triple collision motions with minimal action}

We shall combine the above differential geometric setting, Theorem F and
Hilbert's direct method to study the \emph{existence problem }of three-body
motions leading to triple collision, starting from a given non-degenerate
m-triangle and with \emph{minimal} action integral (\ref{J10}), say. When
this problem is pushed down to the level of $\bar{M}$, at a given energy
level $h$, it can be reduced to the problem of existence of a shortest
geodesic, with respect to the metric $d\bar{s}_{h}^{2}$, between a given
point in $\bar
{M}_{h}-\bar{E}$ $\ $and the base point $O$.

\begin{remark}
\label{Kepler} For any energy $h$ there are Newtonian\ motions, with the
constant shape of a regular triangle or an Euler configuration, through
which the m-triangle shrinks homothetically to a triple collision in finite
time. To find the time parametrization of such a three-body motion, in fact,
amounts to solve a two-body (or Kepler) problem, and this leads to the
classical solutions found by Lagrange and Euler, see \cite{Euler}, \cite%
{Lagrange}, \cite{S-M}. For these motions minimal action is achieved for the
Lagrange motions (regular triangle), but not for the Euler motions (which
are collinear).
\end{remark}

More generally, let us first consider the case $h<0$, and define the variety 
\begin{equation}
\bar{D}_{h}=\left\{ \mathbf{p}\in\bar{M}_{h};d(\mathbf{p},O)\leq d(\mathbf{p}%
,\partial\bar{M}_{h})+L_{h}\right\}   \label{Dhbar}
\end{equation}
where $d(\mathbf{p},O)$ (resp. $d(\mathbf{p},\partial\bar{M}_{h})$) is the
distance between $O$ and $\mathbf{p}$ (resp. $\partial\bar{M}_{h})$ in $\bar{%
M}_{h}$ with the metric $d\bar{s}_{h}^{2}$. The interior $D_{h}$ (resp.
boundary $\partial\bar{D}_{h}$) of $\bar{D}_{h}$ is defined by strict
inequality (resp. equality) in (\ref{Dhbar}).

\begin{remark}
The two surfaces $\partial\bar{M}_{h}$ and $\partial\bar{D}_{h}$ are
interesting geometric objects in the study of triple collision orbits. They
touch each other at the two points $\frac{\mu_{0}}{\left\vert h\right\vert }%
\mathbf{\hat{p}}_{0}^{\pm}$ on $\partial\bar{M}_{h}$ closest to the cone
vertex $O$.
\end{remark}

We will prove the following existence result :

\begin{theorem}
(cf. \cite{1994}, Theorem 5) Starting from a given oriented m-triangle $%
\delta$ whose congruence class $\bar{\delta}$ belongs to $\bar{D}_{h}$, $h<0 
$, there is a three-body motion with total energy $h$ and minimal action
integral which leads to a triple collision.
\end{theorem}

\begin{proof}
As a consequence of Theorem F, the proof reduces to the existence of a curve
with minimal length in $\bar{M}_{h}$ linking $\bar{\delta}$ to $O$. For $%
\bar{\delta}$ in $D_{h}$ such a curve is necessarily a geodesic.

Assume first $\bar{\delta}\in D_{h}$, and let $\left\{ \bar{\Gamma}%
_{i}\right\} $ be a sequence of curves in $\bar{M}_{h}$ between $\bar{\delta 
}$ and $O$ whose lengths satisfy 
\begin{equation*}
L(\bar{\Gamma}_{i})<d(\bar{\delta},\partial\bar{M}_{h})+L_{h}\text{, \ \ }%
\lim_{i\rightarrow\infty}L(\bar{\Gamma}_{i})=d(\bar{\delta},O) 
\end{equation*}
In particular, each $\bar{\Gamma}_{i}$ is disjoint from the boundary $%
\partial\bar{D}_{h}$.

Let us divide $\bar{\Gamma}_{i}$ into $m2^{i}$ segments of equal length and
replace each segment by the unique shortest geodesic between its end points.
Then it is quite straightforward to apply the direct method of Hilbert to
find a suitable subsequence of $\left\{ \bar{\Gamma}_{i}\right\} $ with a
limiting curve $\bar{\Gamma}$, and this is necessarily a geodesic curve in $%
\bar{M}_{h}$ between $\bar{\delta}$ and $O$ with minimal length $L(\bar{%
\Gamma})=d(\bar{\delta},O)$.

Assume next $\bar{\delta}\in$\ $\partial\bar{D}_{h}$, and let $\left\{ \bar{%
\delta}_{k}\right\} $ be as sequence of points in $D_{h}$ with $\bar{\delta}$
as its limit. Moreover, let $\left\{ \bar{\Gamma}_{k}\right\} $ be a
sequence of curves, where $\bar{\Gamma}_{k}$ is a shortest geodesic between $%
\bar{\delta}_{k}$ and $O$, that is, $L(\bar{\Gamma}_{k})=d(\bar{\delta}%
_{k},O)$. It follows that 
\begin{equation}
\lim_{k\rightarrow\infty}d(\bar{\delta}_{k},O)=d(\bar{\delta},O)   \label{L}
\end{equation}
and it is not difficult so see that there is a suitable subsequence of $%
\left\{ \bar{\Gamma}_{k}\right\} $ with a limiting curve $\bar{\Gamma}$
whose length is the limit (\ref{L}), and moreover, $\bar{\Gamma}$ is a
geodesic between $\bar{\delta}$ and $O$.
\end{proof}

Finally, we consider the case $h\geq0$, namely when $\bar{M}_{h}=\bar{M}$,
and then we have the following analogue of the above theorem .

\begin{theorem}
(cf. \cite{1994}, Theorem 5') Starting from a given oriented m-triangle $%
\delta$, for a given energy level $h\geq0$ there is always a three-body
motion leading to triple collision and with minimal action integral.
\end{theorem}

\begin{proof}
This is similar to the case $\bar{\delta}\in D_{h}$ of the previous proof,
and the application of Hilbert's direct method will give the existence of
the curve we seek.
\end{proof}

\begin{remark}
The direct method of Hilbert can, of course, also be applied to study the
existence problem \ of a geodesic curve $\bar{\Gamma}$ realizing the minimal
distance between two given points $\bar{\delta}_{1}$ and $\bar{\delta}_{2}$
of $(\bar{M}_{h},d\bar{s}_{h}^{2}).$ Then, by Theorem B, there are liftings $%
\Gamma$ of $\bar{\Gamma}$ which are planary three-body motions (with
specified angular momentum) starting from a given oriented m-triangle $%
\delta_{1}$ belonging to the congruence class $\bar{\delta}_{1}$. However,
the end point configuration $\delta_{2}$ of $\Gamma$ is already determined
by $\bar{\Gamma}$ and $\delta_{1}$, according to Theorem C2. Consequently,
only three-body motions with specific relative positions of their initial
and terminal m-triangles $\delta_{i}$ can have minimal action integrals. In
the above two theorems there is no such relative position constraint since
the triple collision configuration $\delta_{2}=O$ consists of a single
congruence class.
\end{remark}

\subsection{The uniqueness problem for triple collision motions with minimal
action}

In view of the above existence theorems, it is natural to investigate the
following uniqueness problem for motions starting from a given configuration
at a sufficiently large energy level.

\begin{problem}
To a given non-degenerate oriented m-triangle $\delta$ and energy level $h$
above a lower bound, say $h\geq\mathfrak{h}(\bar{\delta})$, is there a
unique three-body motion from $\delta$ leading to a triple collision with
minimal action integral ?
\end{problem}

This problem requires a considerable amount of in-depth analysis of the
geodesic equation of $(\bar{M}_{h},d\bar{s}_{h}^{2})$, and here we shall
leave it as an open problem.

However, to facilitate future analytical studies of the above problem and
related problems we shall discuss a geometric reduction technique which
reflects some useful feature of the Riemannian structure of $\bar{M}%
=C(M^{\ast})$ as a cone over the subspace $M^{\ast}=(\rho=1)$. In $\bar{M}$
the integral curves of the vector field $\frac{\partial}{\partial\rho}$ are
the rays emanating from $O$, and they define the \emph{radial }(i.e. a
natural "vertical") direction at every point $\neq O$.

The above problem is, indeed, simple and has an optimal solution in the
special case mentioned in Remark \ref{Kepler}, namely for the shape
invariant motions of Lagrange type. A 3-body motion is \emph{shape invariant 
}if its moduli curve $\bar{\Gamma}$ is confined to a ray, that is, the
associated shape curve $\Gamma^{\ast}$ is a single point. In fact, if a
Newtonian motion is shape invariant over some time interval of length $>0,$
then for all time $\Gamma^{\ast}$ is a single point (necessarily a critical
point of $U^{\ast}$). However, in view of Remark \ref{Kepler}, the collinear
solutions with the shape of an Euler point are not even action minimizing.

In general, let $\bar{\Gamma}$ be a smooth curve in $\bar{M}-\left\{
O\right\} $. Since the associated shape curve $\Gamma^{\ast}$ is the radial
projection of $\bar{\Gamma}$ onto the transversal subspace $M^{\ast}$, $%
\Gamma^{\ast}$ will be smooth as long as $\bar{\Gamma}$ is transversal to
the radial direction, whereas a \emph{cusp }may occur at points where this
fails. Hence, in the long run the typical shape curves of 3-body motions are
rather piecewise smooth, but still they can be parametrized by arc-length.
Moreover, unless $\Gamma^{\ast}$ is a single point, $\bar{\Gamma}$ may also
be parametrized by the arc-length parameter of $\Gamma^{\ast}$.

\begin{definition}
\label{cone}Let $\bar{\Gamma}$ be a curve in the moduli space $\bar{M}$ and
let $\Gamma^{\ast}$ be the associated shape curve. The cone consisting of
all rays emanating from $O$ and passing through points on $\bar{\Gamma}$ (or 
$\Gamma^{\ast})$ is called the \emph{cone surface}\textit{\ of }$\bar{\Gamma}
$ (or $\Gamma^{\ast})$, and it is denoted either $C(\bar{\Gamma})$ or $C($ $%
\Gamma^{\ast})$.
\end{definition}

We assume $\bar{\Gamma}$ (and hence also $\Gamma^{\ast}$) has a given
orientation. Since the metric on $M^{\ast}=S^{2}(1/2)$ is denoted by $%
d\sigma^{2}$ (cf. e.g. \ref{dsbar1}), $\sigma$ also denotes the arc-length
parameter of $\Gamma^{\ast}$. For $\sigma$ ranging over some interval $\left[
\sigma_{0},\sigma_{1}\right] $, the corresponding surface $C(\Gamma^{\ast})$
is immersed in $(\bar{M},d\bar{s}^{2})$ with the induced kinematic metric 
\begin{equation}
d\bar{s}^{2}|_{C(\Gamma^{\ast})}=d\rho^{2}+\rho^{2}d\sigma^{2}\text{; \ \ \ }%
\sigma_{0}\leq\vartheta\leq\sigma_{1},   \label{metric}
\end{equation}
and hence it is isometric to a flat Euclidean sector of angular width $%
\sigma_{1}-\sigma_{0}$, with $(\rho,\sigma)$ as polar coordinates centered
at the origin $O$.

The moduli space $\bar{M}\approx\mathbb{R}^{3}$ has the standard (right
handed) orientation and, in particular, the 2-sphere $M^{\ast}$ has the
induced orientation with $\frac{\partial}{\partial\rho}$ as positive normal
vector field. The surface $C(\Gamma^{\ast})$ is naturally oriented with the
positive orthonormal frame 
\begin{equation}
\left\{ \frac{\partial}{\partial\rho},\frac{1}{\rho}\frac{\partial}{%
\partial\sigma}\right\} ,   \label{stationary}
\end{equation}
and we choose its normal vector field $\mathbf{\nu}$ so that $\mathbf{\nu}$
followed by the frame (\ref{stationary}) is a positive orthonormal frame in $%
\bar{M}$.

In (\ref{metric}) the curve $\Gamma^{\ast}$ becomes the circular arc of
radius $\rho=1$, whereas the (original) moduli curve $\bar{\Gamma}$ appears
as a radial deformation of $\Gamma^{\ast}$. When $\bar{s}$ in (\ref{metric}
is viewed as the arc-length parameter of 
\begin{equation*}
\bar{\Gamma}:\bar{s}\rightarrow(\rho(\bar{s}),\sigma(\bar{s})), 
\end{equation*}
(\ref{metric}) becomes an identity along the curve.

The extrinsic geometry of $C(\Gamma^{\ast})$ $\subset\bar{M}$ is completely
determined by the extrinsic geometry of $\Gamma^{\ast}\subset M^{\ast}$.
Indeed, the \emph{lines of curvature }are the two families of \emph{%
coordinate curves,} namely the rays ($\sigma$ constant) and the
\textquotedblright circles\textquotedblright\ ($\rho$ constant) in $%
C(\Gamma^{\ast}).$ The principal curvature of $C(\Gamma^{\ast})$ at a point $%
\bar{\delta}\ $is zero in the ray direction and is equal to $\mathcal{K}%
_{g}^{\ast}/\rho$ in the direction of $\frac{\partial}{\partial\sigma}$,
where $\mathcal{K}_{g}^{\ast}$ is the geodesic curvature of $\Gamma^{\ast}$
in $M^{\ast}$ at the corresponding point $\delta^{\ast}$.

Along the curve $\bar{\Gamma}$ we will also consider the positive
orthonormal moving frame $\left\{ \mathbf{\tau,\eta,\nu}\right\} $, where $%
\mathbf{\tau }$ is the unit tangent vector in the (chosen) positive
direction of $\bar{\Gamma}$, and hence the frame $\left\{ \mathbf{\tau,\eta}%
\right\} $ of $C(\Gamma^{\ast})$ differs from the stationary frame (\ref%
{stationary}) by a certain rotation angle $\alpha$. Namely, we define the 
\emph{(radial) inclination angle }$\alpha$ of $\bar{\Gamma}$ by wrtiting 
\begin{gather}
\mathbf{\tau}=\cos\alpha\frac{\partial}{\partial\rho}+\sin\alpha\frac{1}{%
\rho }\frac{\partial}{\partial\sigma},\text{ \ }\mathbf{\eta}=-\sin\alpha 
\frac{\partial}{\partial\rho}+\cos\alpha\frac{1}{\rho}\frac{\partial}{%
\partial\sigma}  \label{frame4} \\
\cos\alpha=\frac{d\rho}{d\bar{s}}\text{, \ }\sin\alpha=\rho\frac{d\sigma }{d%
\bar{s}}\text{\ , \ }\cot\alpha=\frac{1}{\rho}\frac{d\rho}{d\sigma}=\frac{d}{%
d\sigma}\ln\rho   \label{frame5}
\end{gather}
Briefly, $\alpha$ is the angle between the ray direction and the tangent
direction, and $0\leq\alpha\leq\pi$ since $\sin\alpha$ in (\ref{frame5}) is
not negative. The extreme values $\alpha=0,\pi$ occur when $\bar{\Gamma}$ is
not transversal to the radial direction, in which case $\alpha$ and $%
\Gamma^{\ast}$ (as functions of $\sigma$ or time) may encounter a
singularity, namely $\Gamma^{\ast}$ encounters a cusp.

\begin{remark}
The angle $\alpha(\sigma)$ and the radial distance $\rho(\sigma)$ are
mutually dependent according to (\ref{frame5}). For example, we have for $%
\rho (\sigma_{0})\neq0$ 
\begin{equation}
\rho(\sigma)=\rho(\sigma_{0})\exp(\int_{\sigma_{0}}^{\sigma}\cot\alpha
(\sigma)d\sigma)   \label{rho}
\end{equation}
\end{remark}

The geodesic condition for a curve $\bar{\Gamma}$ in $(\bar{M}_{h},d\bar
{s}%
_{h}^{2})$ is evidently equivalent to two identities of type (\ref{curv}),
namely for two linearly independent normal vectors $\mathbf{n}$ to $\bar{%
\Gamma}$. Thus, we shall consider the two cases 
\begin{align}
(i)\text{ }\mathbf{n} & =\mathbf{\eta}\text{ : tangential to }C(\bar{\Gamma }%
)\text{ \ \ }  \label{cases} \\
(ii)\text{ }\mathbf{n} & =\mathbf{\nu}\text{ : perpendicular to }C(\bar{%
\Gamma})  \notag
\end{align}

The first case amounts to the characterization of $\bar{\Gamma}$ as a
geodesic in the (truncated) cone surface $C(\bar{\Gamma})\cap\bar{M}_{h}$
with the metric $d\bar{s}_{h}^{2}$, as follows :

\begin{lemma}
Let $u(\sigma)$ be the restriction of the potential function $U^{\ast}$
along the shape curve $\Gamma^{\ast}(\sigma)$. Then the geodesic equation
for the moduli curve $\bar{\Gamma}$ in the cone surface $C(\bar{\Gamma})$
with the metric 
\begin{equation*}
d\bar{s}_{h}^{2}=(\frac{1}{\rho}U^{\ast}+h)(d\rho^{2}+\rho^{2}d\sigma^{2}) 
\end{equation*}
is equivalent to the equation 
\begin{equation}
\frac{d\alpha}{d\bar{s}}+\frac{d\sigma}{d\bar{s}}=\frac{1}{2\rho}\left(
\sin\alpha\frac{u(\sigma)}{u(\sigma)+h\rho}+\cos\alpha\frac{%
u^{\prime}(\sigma)}{u(\sigma)+h\rho}\right)   \label{geo1}
\end{equation}
\end{lemma}

\begin{proof}
Let $\mathbf{\eta}$ \ be the normal vector in (\ref{frame4}). Then the
geodesic condition is by (\ref{curv}) 
\begin{equation}
\mathcal{K}(\mathbf{\eta)=}\frac{1}{2}\frac{d}{d\mathbf{\eta}}\ln (\frac{%
u(\sigma)}{\rho}+h)   \label{geo2}
\end{equation}
where $\mathcal{K}(\mathbf{\eta)}$ is the (geodesic) curvature of $\bar {%
\Gamma}$ in the Euclidean sector (\ref{metric}). However, in a Euclidean
plane it is easy to see that $\mathcal{K}(\mathbf{\eta)}$ can be expressed
as $d\zeta/d\bar{s}$, where $\zeta$ is the angle between a fixed reference
ray $\sigma=0$ (say, the positive x-axis) and the tangent line, in fact, $%
\zeta=\alpha+\sigma$, see Figure 8. Finally, calculation of the normal
derivative on the right side of the identity (\ref{geo2}), using the
orthonormal frame (\ref{stationary}), leads to the formula (\ref{geo1}).
\end{proof}

In the second case of (\ref{cases}) the geodesic condition is the identity (%
\ref{curv}) with $\mathbf{n}$ equal to the normal $\mathbf{\nu}$ of the
surface. In this case $\mathcal{K}(\mathbf{\nu})$ equals the normal
sectional curvature of $C(\bar{\Gamma})$ in the direction of $\bar{\Gamma}$,
namely the value $\Pi(\mathbf{\tau,\tau)}$ of the second fundamental form.
The latter has the frame (\ref{stationary}) as eigenvectors, with
eigenvalues $0$ and $\mathcal{K}_{g}^{\ast}/\rho$ respectively, and hence by
(\ref{frame4}) and Euler's classical formula for the decomposition of normal
geodesic curvature 
\begin{equation}
\mathcal{K}_{g}^{\ast}\sin^{2}\alpha=\rho\mathcal{K}(\mathbf{\nu})=\frac{%
\rho }{2}\frac{d}{d\mathbf{\nu}}\ln(\frac{1}{\rho}U^{\ast}+h) 
\label{knormal2}
\end{equation}

As a summary we now state the following theorem, valid as long as the
quantities involved are well defined.

\begin{theorem}
\label{1994-Th6}(cf. \cite{1994}, Theorem 6) \ In the moduli space $\bar{M}$
with the kinematic Riemannian metric $d\bar{s}^{2}$, let $\bar{\Gamma}$ be
the oriented moduli curve of a three-body motion with total energy $h$ and
vanishing angular momentum, and let $\Gamma^{\ast}$ be the corresponding
shape curve on the sphere $M^{\ast}=S^{2}(1/2)$ with unit tangent (resp.
normal) vector $\mathbf{\tau}^{\ast}$ (resp.\ $\mathbf{\nu}^{\ast}$) so that 
$\left\{ \mathbf{\tau}^{\ast},\mathbf{\nu}^{\ast}\right\} $ is a positive
frame on the sphere. Then $\bar{\Gamma}$ $=(\rho,\Gamma^{\ast})$ can be
characterized as a solution of the following system of ODE 
\begin{align}
(i)\text{ } & \text{: }\frac{d\alpha}{d\sigma}=-1+\frac{1}{2}\frac {u(\sigma)%
}{u(\sigma)+h\rho}\left( 1+\cot\alpha\frac{d}{d\mathbf{\tau}^{\ast }}%
\ln(U^{\ast})\right)  \label{ODE1} \\
(ii)\text{ } & \text{: }\mathcal{K}_{g}^{\ast}\sin^{2}\alpha=\ \frac{1}{2}%
\frac{u(\sigma)}{u(\sigma)+h\rho}\ \frac{d}{d\mathbf{\nu}^{\ast}}%
\ln(U^{\ast})\newline
\notag
\end{align}
\newline
where $\mathcal{K}_{g}^{\ast}$ is the geodesic curvature of $\Gamma^{\ast}$
in $M^{\ast}$, $\sigma$ is the arc-length parameter of $\Gamma^{\ast}$, $%
U^{\ast}$ is the restriction of $U$ to $M^{\ast}$ and $u(\sigma)$ is its
further restriction along $\Gamma^{\ast}$, and $\alpha \in\lbrack0,\pi]$ is
the angle between the (outgoing) ray direction and $\bar{\Gamma}$ in $\bar{M}
$.
\end{theorem}

\begin{proof}
By using the expression for $\sin\alpha$ in (\ref{frame5}), equation (\ref%
{geo1}) can be stated as 
\begin{equation*}
\frac{d\alpha}{d\bar{s}}=(-1+\frac{1}{2}\frac{u(\sigma)}{u(\sigma)+h\rho })%
\frac{d\sigma}{d\bar{s}}+\frac{1}{2\rho}\cos\alpha\frac{u^{\prime}(\sigma )}{%
u(\sigma)+h\rho}
\end{equation*}
When we replace the arc-length parameter $\bar{s}$ of $\bar{\Gamma}$ by $%
\sigma$, using a formula from (\ref{frame5}), this equation reads 
\begin{equation*}
\frac{d\alpha}{d\sigma}=(-1+\frac{1}{2}\frac{u(\sigma)}{u(\sigma)+h\rho })+%
\frac{1}{2}\cot\alpha\frac{u^{\prime}(\sigma)}{u(\sigma)+h\rho}, 
\end{equation*}
and by viewing $u^{\prime}(\sigma)/u(\sigma)$ as the tangential derivative
of $\ln(U^{\ast})$ we obtain the first equation (\ref{ODE1}).

The second equation of (\ref{ODE1}) is merely a reformulation of (\ref%
{knormal2}), whose right side may be expressed as 
\begin{equation*}
\frac{1}{2}\frac{d}{d\mathbf{\nu}^{\ast}}\ln(\frac{1}{\rho}U^{\ast}+h)=\frac{%
2^{-1}}{u(\sigma)+h\rho}\frac{d}{d\mathbf{\nu}^{\ast}}U^{\ast}
\end{equation*}
Here we use the fact that the normal vector $\mathbf{\nu}$ \ along $\bar{%
\Gamma}$ may be identified with the scaling of $\mathbf{\nu}^{\ast}$ by the
factor $1/\rho$, that is, $\mathbf{\nu=\nu}^{\ast}/\rho$, and moreover,
differentiation in the direction of $\mathbf{\nu}$ \ commutes with the
scaling.
\end{proof}

\begin{remark}
\label{invariance}The above system (\ref{ODE1}) is easily seen to be scaling
invariant. Namely, when the size function $\rho$ is multiplied by a fixed
constant $k>0$, the energy level changes as $h\rightarrow h/k$ and hence the
product\ $h\rho$ stays invariant. In particular, since the energy level $h=0$
is invariant with respect to scaling of solutions, the explicit dependence
on $\rho$ in (\ref{ODE1}) disappears in this case. Moreover, the angle $%
\alpha$, geometrically interpreted in (\ref{frame4}) as the inclination
angle of the moduli curve $\bar{\Gamma}$, is a neat scaling invariant which
together with $\Gamma^{\ast}$ represents $\bar{\Gamma}$ uniquely up to
scaling. In fact, $\rho$ is generally obtained from $\alpha$ by
\textquotedblright quadrature\textquotedblright\ along $\Gamma^{\ast}$, cf. (%
\ref{rho}). This explains the following result.
\end{remark}

\begin{corollary}
Let the pair $(\alpha,\Gamma^{\ast})$ represent the moduli curve $\bar{%
\Gamma }$ of a three-body motion with vanishing angular momentum and
vanishing total energy, where the shape curve $\Gamma^{\ast}$ is not a
single point and is viewed as a curve on the standard sphere $S^{2}$ of
radius 1. Then $(\alpha,\Gamma^{\ast})$ is a solution of the following
system of ODE 
\begin{equation}
\left\{ 
\begin{array}{c}
\frac{d\alpha}{ds}=-\frac{1}{4}+\frac{1}{2}\cot\alpha\frac{d}{d\mathbf{\tau }%
^{\ast}}\ln(U^{\ast})\medskip \\ 
\mathcal{K}_{g}^{\ast}\sin^{2}\alpha=\ \frac{1}{2}\frac{d}{d\mathbf{\nu}%
^{\ast}}\ln(U^{\ast})\ \medskip%
\end{array}
\right.   \label{ODE}
\end{equation}
where $s=2\sigma$ is the arc-length parameter of $\Gamma^{\ast}$ on $S^{2}$
and $\mathcal{K}_{g}^{\ast}$ is its geodesic curvature. Moreover, a solution 
$(\alpha,\Gamma^{\ast})$ can only encounter a singularity (cusp) when $%
\alpha=0$ or $\pi$, or when $\Gamma^{\ast}$ reaches a collision point.
\end{corollary}

We are particularly interested\ in applying the system (\ref{ODE1}) or (\ref%
{ODE}) to the study of triple collision motions. \ A triple collision is
simply expressed by the condition $\rho=0$, but this singular event is not
explicitly visible in (\ref{ODE}) since the variable $\rho$ is eliminated.
However, the term $h\rho$ in (\ref{ODE1}) also disappears when $\rho
\rightarrow0$, so the two systems should behave ''similarly'' in the limit.
Hence, the system (\ref{ODE}) is likely to be significant also when $h\neq0$.

One of the major results of Sundman and Siegel in their work on the local
analysis of triple collisions prove the existence of both a \emph{limiting
shape}, necessarily a critical point of $U^{\ast}$, and a \emph{limiting
position}, cf. \cite{Sund1}, \cite{Sund2}, \cite{Siegel1}, \cite{Siegel2}.
The existence of a limiting position is the statement that the 3-body motion 
$\Gamma(t)$ has a \textquotedblright size normalized\textquotedblright\
limit, 
\begin{equation}
\frac{\Gamma(t)}{\left\vert \Gamma(t)\right\vert }\rightarrow\delta , 
\label{limitpos}
\end{equation}
at the configuration space level, and we shall express the statement
concerning the limiting shape (due to Sundman) by saying the pair $%
(\alpha,\Gamma^{\ast})$ approaches a specific pair ($\hat{\alpha}%
,\delta^{\ast})$, namely 
\begin{equation}
\hat{\alpha}\in\left\{ 0,\pi\right\} =\partial\lbrack0,\pi]\text{ , \ }%
\delta^{\ast}\in\left\{ \mathbf{\hat{e}}_{1},\mathbf{\hat{e}}_{2},\mathbf{%
\hat{e}}_{3},\mathbf{\hat{p}}_{0}^{+},\mathbf{\hat{p}}_{0}^{-}\right\}
\subset S^{2}   \label{limit}
\end{equation}
It is also known (cf. e.g. Siegel-Moser\cite{S-M}, p. 89) that an Euler
point $\mathbf{\hat{e}}_{i}$ can only be the limiting shape of a triple
collision motion confined to a fixed line. (However, $\mathbf{\hat{e}}_{i}$
may well be the limiting shape of a non-collinear motion as $%
t\rightarrow\pm\infty$ ).

The two \textquotedblright boundary\textquotedblright\ values of $\alpha$ in
(\ref{limit}) actually distinguish between the two events \emph{triple
explosion }and \emph{triple collision,} as follows : $\alpha=0$ when $\bar{%
\Gamma}$ starts (or \textquotedblright explodes\textquotedblright) out from
the cone vertex $O$ of $\bar{M}$ , and $\lim\alpha=\pi$ when $\bar {\Gamma}$
is oriented towards $O\ $and terminates with a \textquotedblright total
collapse\textquotedblright. Anyhow, we are free to run a three-body motion
in either directions, and the associated initial value problem for (\ref%
{ODE1}) or (\ref{ODE}) is (a priori) of singular type in the above case
since $\sin\alpha=0$. In fact, a solution $(\alpha,\Gamma^{\ast})$ of (\ref%
{ODE}) may also encounter another type of singularity (called cusp) when $%
\sin\alpha=0$, but with $\rho\neq0$. In Chapter 8 these events and related
problems will be further investigated in selected testing cases.

\section{Case study of triple collision motions with zero energy}

\subsection{The basic setting and statement of Theorem G}

Due to the simplicity of the system (\ref{ODE}), the special case of
vanishing total energy, $h=0$, lends itself as the simplest testing case of
three-body motions leading to a triple collision. We shall investigate this
case more carefully, and for convenience, let us also restrict ourselves to
the case of equal masses, $m_{i}=1/3$, which largely simplifies the series
expansions of the potential function and its derivatives.

We shall address the triple collision problem as an initial value problem,
namely as a \emph{triple explosion, }although we usually write
\textquotedblright triple collision\textquotedblright\ motions. Let us first
recall the so-called \emph{Lagrange-Jacobi} equation which is the result of
differentiating $I=\rho^{2}$ twice with respect to time $t$, using the
homogeneity of $U$ and conservation of energy $h=T-U$, namely in our case 
\begin{equation}
\frac{d^{2}}{dt^{2}}I=2(T+h)=2T>0   \label{L-J}
\end{equation}
It follows that $I$ is a nonnegative convex function of time $t$, and
starting from a triple collision (say, $I(0)=0)$ it is strictly increasing
and tends to $\infty$ as $t\rightarrow\infty$.

Following the setup from Section 7.3, we seek a description of the moduli
curves of triple collision motions, valid for some appropriate time interval 
$\left[ 0,t_{1}\right] $. For this purpose it is convenient to use the
coordinates $(\rho,\varphi,\theta)$ in the cone $\bar{M}$ $=C(M^{\ast})$,
where as before $\rho=\sqrt{I}$ and $(\varphi,\theta)$ are spherical
coordinates on the unit sphere $S^{2}=M^{\ast}$ centered at the north pole $%
\mathcal{N}$. In this setting a moduli curve $\bar{\Gamma}$ and the
associated shape curve $\Gamma^{\ast}$ have coordinate representations 
\begin{equation}
\bar{\Gamma}(s)=(\rho(s),\varphi(s),\theta(s))\text{, \ }\Gamma^{\ast
}(s)=(\varphi(s),\theta(s)),   \label{A0}
\end{equation}
where $s=s(t)$ is the arc-length parameter $s=s(t)$ of $\Gamma^{\ast}$ and
is an increasing function of time $t$. Here we must exclude, of course, the
well understood shape invariant motions, namely the trivial case that $%
\Gamma ^{\ast}$ is a single point (in which case $\rho(t)$ is the solution
of\ a 1-dimensional Kepler problem).

Thus, we seek a description of all those moduli curves $\bar{\Gamma}(s)$
emanating from a triple collision, at $s=0$ say. According to Remark \ref%
{invariance} it suffices to consider the class of $\bar{\Gamma}$ modulo
scaling, represented by the pair $(\alpha,\Gamma^{\ast})$ where $\alpha
(s)\in\lbrack0,\pi]$ is the (radial) inclination angle of $\bar{\Gamma}$. In
fact, recall from (\ref{rho}) that the size function $\rho$ of $\bar{\Gamma}$
is recovered from $(\alpha,\Gamma^{\ast})$ by the general quadrature formula 
\begin{equation}
\rho(s)=\rho(s_{0})e^{\frac{1}{2}\int_{s_{0}}^{s}\cot(\alpha)ds}\text{, \ \ }%
\rho(s_{0})\neq0   \label{rho3}
\end{equation}

\begin{remark}
Using the parameter $s$ rather than time $t$ is, of course, crucial for our
geometric approach below. The relationship between $s$ and $t$ is follows
from the kinematic metric, using e.g. (\ref{metric}), (\ref{frame5}), (\ref%
{rho3}), (\ref{A14}). Namely, in the case of zero angular momentum there are
the identities $\ $%
\begin{equation*}
\ 2(U+h)dt^{2}=2Tdt^{2}=d\bar{s}^{2}=d\rho^{2}+\frac{\rho^{2}}{4}%
ds^{2}=(\cos^{2}\alpha)d\bar{s}^{2}+\frac{\rho^{2}}{4}ds^{2}, 
\end{equation*}
from which we deduce the relationship 
\begin{equation}
dt=\frac{\rho(s)\ }{2^{3/2}\sin\alpha(s)\sqrt{u(s)/\rho(s)+h}}ds\text{ \ \ \ 
}   \label{dsdt}
\end{equation}
where $u(s)=$ $U^{\ast}(\Gamma^{\ast}(s))$. Moreover, by switching over to $%
t $ it is, in fact, not difficult to see that $\rho$ (at any time $t_{0})$
can be determined solely from the time parametrized shape curve $%
\Gamma^{\ast}(t) $ and the normal derivative of $U^{\ast}$ (near $t=t_{0})$.
\end{remark}

Now, resuming the assumption $h=0$, our approach is to determine the above
pairs $(\alpha,\Gamma^{\ast})$ by solving the system

\begin{equation}
ODE^{\ast}:\left\{ 
\begin{array}{c}
\frac{d\alpha}{ds}=-\frac{1}{4}+\frac{1}{2}\cot\alpha\frac{d}{d\mathbf{\tau }%
^{\ast}}\ln(U^{\ast})\medskip \\ 
\mathcal{K}_{g}^{\ast}\sin^{2}\alpha=\ \frac{1}{2}\frac{d}{d\mathbf{\nu}%
^{\ast}}\ln(U^{\ast})\ \medskip \\ 
1=(\frac{d\varphi}{ds})^{2}+(\sin^{2}\varphi)(\frac{d\theta}{ds})^{2}\medskip%
\end{array}
\right.   \label{ODE*}
\end{equation}
as an appropriate initial value problem which represents a triple collision,
see (\ref{solutions}). The system (\ref{ODE*}) is a copy of (\ref{ODE})
since the third equation merely expresses the constraint that $\Gamma^{\ast}$
is a curve on the unit sphere. We will refer to the first and second
equation of (\ref{ODE*}) as the \emph{inclination }and \emph{curvature }%
equation respectively. The first one relates the growth of the inclination
angle $\alpha$ (of the moduli curve $\bar{\Gamma}$) with the tangential
derivative of $\ln(U^{\ast})$ along $\Gamma^{\ast}$, whereas the second one
- which is of order two- relates $\alpha$ to the geodesic curvature of $%
\Gamma^{\ast}$ and the normal derivative of $\ln(U^{\ast})$ along $%
\Gamma^{\ast}$.

The following theorem summarizes the main result of this chapter. It
describes the family $\mathfrak{S}(\mathbf{\hat{p}}_{0})$ of all shape
curves $\Gamma^{\ast}(s)$ representing triple collision motions, with the
limiting shape of $\mathbf{\hat{p}}_{0}$ at the collision.

\textbf{Theorem G}$_{1}$\emph{\textbf{\ }In the case of uniform mass
distribution and zero total energy, consider the family }$\mathfrak{S}(%
\mathbf{\hat{p}}_{0})$ \emph{of arc-length parametrized shape curves }$%
\Gamma^{\ast}(s),s\geq0$\emph{,} \emph{which emanate from the north pole }$%
\mathbf{\hat{p}}_{0}=$ $\Gamma^{\ast}(0)$ \emph{of the 2-sphere }$S^{2}$ 
\emph{and\ represent 3-body motions with a triple collision at }$s=0\emph{.}$%
\emph{\ This family has the following properties :}

(i) \ \ \emph{There is a unique curve }$\Gamma_{\theta_{0}}^{\ast}$ \emph{%
for each initial longitude direction }$\theta_{0}$\emph{. }

(ii) \ \emph{The family is invariant under the induced action of the
dihedral isometry group }$\mathfrak{D}_{3}$ \emph{of }$S^{2}$\emph{\ which
fixes }$\mathbf{\hat{p}}_{0}$ \emph{and permutes the three Euler points }$%
\mathbf{\hat{e}}_{i}$.\emph{\ In particular, }$\Gamma_{\theta_{0}+2\pi
/3}^{\ast}$\emph{\ is obtained from }$\Gamma_{\theta_{0}}^{\ast}$\emph{\ by
rotating the sphere, }$\theta\rightarrow\theta+2\pi/3.$\emph{\ }

(iii) \emph{Each curve stays within a sector of angular width }$\pi /3$\emph{%
\ and bounded by meridians representing the shape of isosceles triangles, at
least until the first eclipse (i.e. crossing the equator circle). }

(iv)\emph{\ } \emph{Each curve extends analytically through }$s=0$\emph{\
and }$\Gamma_{\theta_{0}+\pi}^{\ast}(s)=\Gamma_{\theta_{0}}^{\ast}(-s) $%
\emph{, and it has no singularity before the first eclipse}.

(v) \emph{\ For each }$\theta_{0}$\emph{\ the associated inclination angle
function }$\alpha_{\theta_{0}}(s)$\emph{\ is the unique analytic solution of
(\ref{ODE*}) along }$\Gamma_{\theta_{0}}^{\ast}$\emph{, with the singular
initial condition }$\alpha(0)=0,\alpha^{\prime}(0)>0.$\emph{\ Moreover, }$%
\alpha_{\theta_{0}+\pi}(s)=-\alpha_{\theta_{0}}(-s).$\emph{\ }

(vi) \emph{The sign of the curvature of the curves in }$\mathfrak{S}(\mathbf{%
\hat{p}}_{0})$\emph{\ depends only on the sector, and in neigboring sectors
the sign is opposite.}

\begin{remark}
In the case of rectilinear three-body motions, the corresponding sets $%
\mathfrak{S}(\mathbf{\hat{e}}_{i})$ are obviously \textquotedblright
congruent\textquotedblright, each consisting of the pair $%
\Gamma_{\pm}^{\ast} $ of arcs of the equator circle, in opposite directions
and starting at the Euler point $\mathbf{\hat{e}}_{i}$. The associated
inclination angle function $\alpha_{\pm}(s)$ is defined by a unique analytic
function $\alpha(s)$ so that $\alpha_{\pm}(s)=\alpha(s)$ for $s\geq0$, and $%
\alpha(-s)=-\alpha(s)$. We refer to Section 8.4.
\end{remark}

We also refer to Section 8.6.2 for more information concerning the geometric
behavior of the curves in\ $\mathfrak{S}(\mathbf{\hat{p}}_{0})$. Indeed, we
are actually close to a stronger version of Theorem G$_{1}$, but the proof
needs more elaboration, so we formulate the following conjecture as an open
problem. \ 

\begin{conjecture}
\label{conject}The different triple collison shape curves $\Gamma^{\ast
}(s),s\geq0$, intersect the equator circle the first time at different
points, and moreover, each point on the circle is reached by a unique curve.
The curves do not intersect each other, except possibly after the first
eclipse.\textbf{\ }
\end{conjecture}

\begin{corollary}
Under the current hypothesis of uniform mass distribution and zero total
energy, consider the "moduli space" consisting of all three-body motions in
3-space which start from a triple explosion at time $t=0$, and moreover, the
motion is neither rectilinear nor shape invariant (i.e. homographic). This
space can be naturally identified with the manifold 
\begin{equation*}
\mathbb{\ }SO(3)\times SO(2)\times\mathbb{R}^{+}
\end{equation*}
In particular, the "moduli space" for those triple collision three-body
motions confined to a fixed plane is 
\begin{equation*}
O(2)\times SO(2)\times\mathbb{R}^{+}
\end{equation*}
\end{corollary}

Indeed, starting with the space $\mathfrak{S}(\mathbf{\hat{p}}_{0})\simeq
SO(2)$ of shape curves described in Theorem G$_{1}$, $SO(2)\times \mathbb{R}%
^{+}$ is the space of their associated curves $\bar{\Gamma}(s)$ in the
moduli space $\bar{M}$ since the size function $\rho(s)$ can be scaled by
any positive number $\lambda\in\mathbb{R}^{+}$ without affecting the shape
curve. Next, the possible liftings $\Gamma(s)$ of $\bar{\Gamma}(s)$ to
three-body motions with $\Gamma(0)=0$ are distinguished by the normalized
limit $\delta$ in (\ref{limitpos}), which is an oriented, regular m-triangle 
$\Delta$ of unit size. We may identify the various positions of such an
oriented triangle with the rotation group $SO(3)$ which measures its
\textquotedblright deviation\textquotedblright\ from a fixed \ reference
position. In terms of the fibration 
\begin{equation*}
O(2)\rightarrow SO(3)\rightarrow SO(3)/O(2)\simeq\mathbb{R}P^{2}
\end{equation*}
we can say that the projective plane $\mathbb{R}P^{2}$ represents the
choices of 2-planes containing $\Delta$ (and the motion), whereas $O(2)$
represents the possible positions of an oriented regular triangle in a given
plane. However, we mention that there is no global \textquotedblright
field\textquotedblright\ of reference positions (or gauge) for all the
planes, since this would imply the fibration is trivial, which is certainly
not true.

\subsection{Analysis of the potential function for equal masses}

In this chapter we shall choose the zero meridian $\theta=0$ for the polar
coordinate system $(\varphi,\theta)$ of $S^{2}$ different from the
convention in Remark \ref{convention}. Namely, the three binary collision
points $\mathbf{\hat{b}}_{i},i=1,2,3$, which are now equally spaced along on
the equator circle $\varphi=\pi/2$, will have the longitude angles 
\begin{equation}
\theta_{1}=-\frac{\pi}{3},\text{ \ }\theta_{2}=\frac{\pi}{3},\text{ \ }%
\theta_{3}=\pi\text{ \ \ \ (cf. Figure 9)}   \label{A2.1}
\end{equation}
Note, for example, the antipodal point of $\mathbf{\hat{b}}_{i}$ is the
Euler point $\mathbf{\hat{e}}_{i}$, and now $\theta=0$ at $\mathbf{\hat{e}}%
_{3}$. It is also convenient to use negative values of $\varphi$, with the
usual interpretation so that $(\varphi,\theta)$ and $(-\varphi,\theta+\pi)$
is the same point on the 2-sphere. This is consistent with our trigonometric
formulae for $U^{\ast}(\varphi,\theta)$ below, see (\ref{A3.5}) and (\ref%
{P123}).$\qquad$

For convenience, let us normalize $U^{\ast}$ by a constant factor to make
its minimum value $U^{\ast}(\mathbf{\hat{p}}_{0})=1$. In fact, scaling of $%
U^{\ast}$ has no effect on the system (\ref{ODE*}). Thus, for $\mathbf{p\in }%
S^{2}$ 
\begin{equation}
U^{\ast}(\mathbf{p})=\frac{\sqrt{2}}{3}\sum_{i=1}^{3}\frac{1}{\left\vert 
\mathbf{p-\hat{b}}_{i}\right\vert }=\frac{1}{3}\sum_{i=1}^{3}\frac {1}{%
\mathbf{(}1-z_{i})^{1/2}},\text{ \ }   \label{A3}
\end{equation}
where (for any mass distribution, indeed) 
\begin{equation}
\left\vert \mathbf{p-\hat{b}}_{i}\right\vert =\ \sqrt{2}\mathbf{(}%
1-z_{i})^{1/2}\text{, \ with\ }z_{i}=\sin\varphi\cos(\theta-\theta _{i})%
\text{,}   \label{A3.5}
\end{equation}
is the usual Euclidean distance from $\mathbf{p}$ to the binary collision
point $\mathbf{\hat{b}}_{i}$ and $\theta_{i}$ is the longitude angle of $%
\mathbf{\hat{b}}_{i}$, cf. (\ref{side4}), (\ref{Ustar}), (\ref{U3}), \ref%
{A2.1}.

Consequences of the invariance of $U^{\ast}$ with respect to permutation of
the points $\mathbf{\hat{b}}_{i}$ will be analyzed and exploited later (cf.
Section 8.3.1). At the algebraic level, however, the following
symmetrization technique will facilitate the analysis of $U^{\ast}$ and
related series expansions. For each integer $k\geq0$, define 
\begin{equation*}
S_{k}=\sum\limits_{i=1}^{3}\cos^{k}(\theta-\theta_{i}) 
\end{equation*}
and write 
\begin{equation*}
f(x)=\prod\limits_{i=1}^{3}(x-\cos(\theta-\theta_{i}))=x^{3}-\frac{3}{4}x+%
\frac{1}{4}\cos(3\theta) 
\end{equation*}
Logarithmic differentiation of $f(x)$ leads to the formal identity 
\begin{equation*}
\left( \sum\limits_{k=0}^{\infty}\frac{S_{k}}{x^{k+1}}\right) f(x)=f^{\prime
}(x)=3x^{2}-\frac{3}{4}
\end{equation*}
from which we deduce the recursive formula 
\begin{equation}
S_{k+3}=\frac{3}{4}S_{k+1}-\frac{1}{4}\cos(3\theta)S_{k},\text{ \ }k\geq0 
\label{A5}
\end{equation}
For convenience, the first few $S_{k}$ are listed as follows : 
\begin{align*}
S_{0} & =3,\text{ \ }S_{1}=0,\text{ \ }S_{2}=\frac{3}{2},\text{ \ }S_{3}=-%
\frac{3}{4}\cos(3\theta),\text{ \ }S_{4}=\frac{9}{8} \\
S_{5} & =-\frac{15}{16}\cos(3\theta),\text{ \ }S_{6}=\frac{27}{32}+\frac {3}{%
16}\cos^{2}(3\theta),\text{ \ }S_{7}=-\frac{63}{64}\cos(3\theta)
\end{align*}

It follows, for example, that $S_{k}$ as a polynomial in $\cos(3\theta)$ has
all its nonzero coefficients positive (resp. negative) when $k$ is even
(resp. $k$ is odd). By expanding $U^{\ast}$ as a sum of binomial series in
the variables $z_{i}$ and using the identity 
\begin{equation*}
\sum\limits_{i=1}^{3}z_{i}^{k}=(\sin^{k}\varphi)S_{k}, 
\end{equation*}
we arrive at the following trigonometric series 
\begin{align}
U^{\ast} & =U^{\ast}(\varphi,\theta)=\frac{1}{3}\sum\limits_{k=0}^{\infty }%
\binom{-\frac{1}{2}}{k}(-1)^{k}(\sin^{k}\varphi)S_{k}  \label{A6} \\
& =1+\frac{3}{16}\sin^{2}\varphi-\frac{5}{64}(\sin^{3}\varphi)(\cos
3\theta)+...+  \notag
\end{align}

For later use we also introduce the following functions on the sphere 
\begin{equation}
F(\mathbf{p})=\frac{\sqrt{2}}{3}\sum\limits_{i=1}^{3}\frac{\sin(\theta
-\theta_{i})}{\left\vert \mathbf{p}-\mathbf{\hat{b}}_{i}\right\vert ^{3}},%
\text{ \ \ }G(\mathbf{p})=\frac{\sqrt{2}}{3}\sum\limits_{i=1}^{3}\frac {%
\cos(\theta-\theta_{i})}{\left\vert \mathbf{p}-\mathbf{\hat{b}}%
_{i}\right\vert ^{3}}   \label{A7}
\end{equation}
It follows that 
\begin{equation}
\frac{\partial U^{\ast}}{\partial\theta}=-F\sin\varphi,\text{ \ \ }\frac{%
\partial U^{\ast}}{\partial\varphi}=G\cos\varphi   \label{A8}
\end{equation}

\begin{remark}
\label{F,G} It is easy to check that $F$ vanishes precisely along the six
meridians 
\begin{equation}
\Gamma_{k}^{\ast}:\theta=k\frac{\pi}{3},\text{ \ \ \ }0\leq k\leq 5 
\label{merid}
\end{equation}
passing through either a binary collision point $\mathbf{\hat{b}}_{i}$ or an
Euler point $\mathbf{\hat{e}}_{i}$ $(=-\mathbf{\hat{b}}_{i})$. Hence, $F$
changes its sign by crossing these meridians, but on the other hand, the
function $G$ is positive (except undefined at $\mathbf{\hat{p}}_{0}$). For
example, $F$ is negative for $0<\theta<\pi/3$, and in this sector the
gradient flow of $U^{\ast}$ is depicted in Figure 9.
\end{remark}

\subsection{Reduction, regularity and singularity}

Here we shall describe a finite group acting on moduli curves and, in
particular, it is a symmetry group of the space of solutions $s\rightarrow
(\alpha(s),\Gamma^{\ast}(s))$ of (\ref{ODE*}). Moreover, regularity and
singularity aspects of the solutions we seek are also briefly discussed.

\subsubsection{Discrete symmetries and reduction}

In addition to time translation and space-time scaling symmetries which
transform solutions of the general 3-body problem as in (\ref{scale})),
there is also an additional symmetry group of order $4$ which we denote by 
\begin{equation}
\mathfrak{D}_{1}\times\mathbb{Z}_{2}=\left\langle \bar{\sigma},\bar{\tau }%
\right\rangle \simeq\mathbb{Z}_{2}\times\mathbb{Z}_{2}   \label{4-group}
\end{equation}
The involution $\bar{\tau}$ represents reversal of time, $t\rightarrow-t$ ,
and its induced action on oriented curves in the moduli space is expressed
by 
\begin{equation}
\bar{\tau}:(s,\alpha)\rightarrow(\tilde{s},\tilde{\alpha})=(-s,\pi -\alpha)%
\text{\ \ \ \ (reversal of direction)\ }   \label{reverse}
\end{equation}
which takes a solution $(\alpha(s),\Gamma^{\ast}(s))$ of the system (\ref%
{ODE*}) on the interval $s_{1}<s<s_{2}$ to the ''reverse''\ solution in the
opposite direction and defined on the interval $-s_{2}<\tilde{s}<-s_{1}$(or
any translation of this interval). The other involution $\bar{\sigma}$ is a
purely geometric symmetry, arising from the reversal of orientation of
m-triangles. At the moduli space level, the latter is the reflection of $%
\bar{M}\simeq\mathbb{R}^{3}$ in the (equator) xy-plane, that is, the map $%
\varphi\rightarrow\pi-\varphi$ in the coordinates $(\rho,\varphi,\theta)$.

On the other hand, under the present assumption of equal masses, there is
also the (order 6) dihedral isometry group $\mathfrak{D}_{3}\subset O(2)$ of 
$S^{2}$ which fixes the poles $\mathbf{\hat{p}}_{0}^{\pm}$ and permutes the
Euler points $\mathbf{\hat{e}}_{i}$. It is a symmetry group of the 3-body
problem since it leaves $U^{\ast}$ invariant, cf. Section 8.2. The action is
generated by the rotation $\theta\rightarrow\theta+2\pi/3$ and the
reflection $\theta\rightarrow-\theta$, and altogether we have a symmetry
group of order 24, 
\begin{equation}
\mathfrak{G}=\mathfrak{D}_{6}\times\mathbb{Z}_{2}=(\mathfrak{D}_{3}\times%
\mathfrak{D}_{1})\times\mathbb{Z}_{2}   \label{24-group}
\end{equation}
where we also regard $\mathfrak{D}_{6}\subset O(3)$ as a dihedral isometry
group of $S^{2}$ generated by reflections. Thus, the action of $\mathfrak{D}%
_{6}$ divides the sphere into $12$ congruent spherical triangles called $%
\emph{chambers}$, and we choose one of them to be our \emph{fundamental
chamber}, namely the following geodesic triangle on the upper hemisphere \ 
\begin{equation}
\mathfrak{C}_{0}=\left\{ (\varphi,\theta)\in S^{2};\text{ }0\leq\varphi \leq%
\frac{\pi}{2},\text{ }0\leq\theta\leq\frac{\pi}{3}\right\}   \label{A4}
\end{equation}
(cf. Figure 9) with the vertices 
\begin{equation}
\mathbf{\hat{p}}_{0}=(\varphi=0)\text{, \ }\mathbf{\hat{e}}_{3}=(\varphi
=\pi/2,\text{ }\theta=0)\text{, \ }\mathbf{\hat{b}}_{2}=(\varphi=\pi/2,\text{
}\theta=\pi/3)   \label{vertices}
\end{equation}

In particular, the action of $\mathfrak{D}_{6}$ on solutions $(\alpha
,\Gamma^{\ast})$ of (\ref{ODE*}) reduces the study of solutions to the study
of \textquotedblright solution segments\textquotedblright\ $\Gamma^{\ast}$
inside $\mathfrak{C}_{0}$. In particular, we may restrict the study of
triple collision solutions $\Gamma^{\ast}$ to those emanating from the
vertex $\mathbf{\hat{p}}_{0}$ or $\mathbf{\hat{e}}_{3}$ with initial
direction leading into the chamber $\mathfrak{C}_{0}$. The natural first
step of this program is to look for solutions whose curvature equation in (%
\ref{ODE*}) is trivially satisfied, and Section 8.4 is devoted to this
preliminary study.

\subsubsection{Cusps and other singularities \qquad}

It is well known that a 3-body motion $\Gamma(t)$ can be \textquotedblright
regularized\textquotedblright\ through a binary collision (cf. \cite{S-M}),
and the only "real" singularity must be a triple collision. Away from
collisions the moduli curve $\bar{\Gamma}$ and the shape curve\ $%
\Gamma^{\ast }$ are also analytic functions when we parametrize by time $t$
or the arc-length $\bar{s}$ of $\bar{\Gamma}$. However, we also want to
parametrize by the arc-length $s\ $of $\Gamma^{\ast}$, and then
singularities may occur at specific instants where the time derivative $\dot{%
s}(t)\geq0$ vanishes. Although this type of "singularity" is rather
artificial, it has geometric significance which explains the possible cusps
of the embedded curve $\Gamma^{\ast}$ on the 2-sphere. These are also
singularities of the system (\ref{ODE*}), and in this subsection we shall
discuss them in some detail.

Since $\frac{d\bar{s}}{dt}>0$ for all $t$, the event $\dot{s}(t_{1})=0$ is
equivalent to the condition $\alpha(t_{1})=0$ or $\pi$. In this case, $\frac{%
d}{dt}\Gamma^{\ast}(t_{1})=0$ and we say $\mathbf{p=}$ $\Gamma^{\ast
}(t_{1}) $ is a \emph{halting point }for the shape curve.\emph{\ }%
Geometrically, this can be a singular point for $\Gamma^{\ast}$, namely it
is the type of singularity that may occur when a regular curve $\bar{\psi}%
(t) $ in 3-space with vertical tangent at $t=t_{1}$ is projected to a curve $%
\psi^{\ast}(t)$ in the xy-plane, cf. also (\ref{A0}). On the other hand,
when $\Gamma^{\ast}$ is parametrized by $s$ and $s_{1}=s(t_{1})$, the unit
tangent vector $\frac{d}{ds}\Gamma^{\ast}(s_{1})$ still exists (as a
one-sided limit) at the halting point $\mathbf{p}$. Moreover, the
(one-sided) geodesic curvature $\mathcal{K}_{g}^{\ast}$ of $\Gamma^{\ast}$
will be bounded near $\mathbf{p}$. In fact, for a \textquotedblright
thin\textquotedblright\ region $0<\rho_{1}\leq\rho\leq\rho_{1}+\delta\rho$
of the moduli space $(\bar
{M},d\bar{s}^{2})$ with the kinematic metric (%
\ref{metric3}),$\ $the projection to the shape space $(M^{\ast},ds^{2})$ may
be viewed as a Riemannian submersion modulo an almost constant scaling.
Restricting to the above region, the curvature of the moduli curve $\bar{%
\Gamma}$ is certainly bounded, and its image curve in $M^{\ast}$ will also
have bounded curvature.

Now, consider a pair $(\alpha(s),\Gamma^{\ast}(s))$ which is a solution of
the system (\ref{ODE*}). The pair is \emph{regular }on the interval $%
(s_{1},s_{2})$ if the three functions $\alpha(s),\varphi(s),\theta(s)$ are
analytic and $\alpha(s)\neq0,\pi$, and a \emph{singularity} is encountered
at $s=s_{i}$ if we cannot extend the functions regularly beyond this point.
In that case it follows from (\ref{ODE*}) that either $\sin\alpha(s_{i})=0$,
in which case we call $\mathbf{p}=\Gamma^{\ast}(s_{i})$ a \emph{cusp}, or $%
\mathbf{p}$ is a binary collision point $\mathbf{\hat{b}}_{j}$ (in which
case $\alpha (s_{i})=\pi/2$ and $\alpha^{\prime}(s_{i})=\infty)$. On the
interval $(s_{1},s_{2})$ the growth of $\alpha(s)$ is governed by the
inclination angle equation (cf. (\ref{ODE*})) 
\begin{equation}
\frac{d\alpha}{ds}=-\frac{1}{4}+\frac{1}{2}\cot\alpha(s)D(s)   \label{eq1}
\end{equation}
whose dependence on the shape curve is solely through the tangential
logarithmic derivative 
\begin{equation}
D(s)=\nabla(\ln U^{\ast})\cdot\mathbf{\tau}^{\ast}=\frac{u^{\prime}(s)}{u(s)}%
\text{, \ \ \ }u(s)=U^{\ast}(\Gamma^{\ast}(s))   \label{D(s)}
\end{equation}

Assume there is a halting point at $s=s_{i}$, say $\alpha(s_{i})=0$ and
write $\mathbf{p}=\Gamma^{\ast}(s_{i})$. If\ $D(s_{i})\neq0$, then by (\ref%
{eq1}) $\alpha^{\prime}(s_{i})=\pm\infty$ and $\mathbf{p}$ is a cusp. On the
other hand, if $D(s_{i})=0$ and $\mathbf{p}$ is not a critical point of $%
U^{\ast}$, then the curvature (i.e. second) equation of (\ref{ODE*}), whose
right side is nonzero at $s=s_{i}$, would force the geodesic curvature of $%
\Gamma^{\ast}$ to become infinitely large towards $\mathbf{p}$. However, as
observed above, such a behavior of the shape curve is not possible. Hence, $%
D(s_{i})=0$ is only possible when the halting point $\mathbf{p}$ of $%
\Gamma^{\ast}$ is also a critical point of $U^{\ast}$.

Finally, assume the halting point $\mathbf{p}$ is also a critical point of $%
U^{\ast}$. For $s$ close to $s_{i}$ we have $\cot\alpha$ $\sim1/\alpha$, and
the local behavior of $\alpha(s)$ near $s=s_{i}$ is largely governed by
equation (\ref{eq1}), from which we can show $\alpha^{\prime}(s)$ is bounded
in a neighborhood of $s_{i}$. Choose some $s_{0}$ so that $s_{1}<s_{0}<s_{2}$
and consider the two cases depending on whether $s$ is approaching $s_{i}$
from above or below : 
\begin{align}
i)\text{ \ }s_{i} & =s_{1}:\text{ }\int_{s_{0}}^{s_{1}}\cot\alpha(s)\text{ }%
ds\approx\int_{s_{0}}^{s_{1}}\frac{ds}{\alpha(s)}=-\infty  \label{int1} \\
ii)\text{ \ }s_{i} & =s_{2}:\text{ }\int_{s_{0}}^{s_{2}}\cot\alpha(s)\text{ }%
ds=\infty   \label{int2}
\end{align}
In particular, in case i) an associated 3-body motion must necessarily
encounter a triple collision at $s=s_{1}$ since formula (\ref{rho3}) implies 
$\rho(s_{1})=0$, whereas in case ii) $\rho(s_{2})=\infty$ and $\mathbf{%
p=\Gamma}^{\ast}(s_{2})$ is the limiting shape of the 3-body motion as $%
t\rightarrow\infty$. In contrast to this, for a cusp singularity with $%
0<\rho(s_{i})<\infty$, we have $\alpha(s)\rightarrow0$ and $\alpha^{\prime
}(s)\rightarrow\pm\infty$ as $s\rightarrow s_{i}$, in such a way that the
integrals (\ref{int1}) or (\ref{int2}) will converge. For example, this
would be the case if $\alpha(s)\sim k(s-s_{i})^{p}$ for some constant $k$
and $p<1$.

We claim, in fact, that $(\alpha(s),\Gamma^{\ast}(s))$ is regularizable at
the above halting point $\mathbf{p=}$ $\mathbf{\Gamma}^{\ast}(s_{i})$ which
is also a critical point of $U^{\ast}$, that is, as a solution of (\ref{ODE*}%
) the functions can be extended analytically beyond $s_{i}$. Indeed, once
the derivative $\alpha^{\prime}(s_{i})$ exists, the power series
developments at $s=s_{i}$ of the functions $\alpha(s),\varphi(s),\theta(s)$
are recursively determined from the system (\ref{ODE*}). The recursive
scheme is worked out in detail in Section 8.4 for the case (\ref{int1}) with
a triple collision at the north pole $\mathbf{\hat{p}}_{0}=\mathbf{\Gamma}%
^{\ast}(s_{1})$, and case ii) is similar.

The collinear type of triple collision is at an Euler point such as $\mathbf{%
p=}$ $\mathbf{\hat{e}}_{3}$, in which case $\Gamma^{\ast}(s)$ moves along
the equator. In particular, $u(s)$ in (\ref{D(s)}) has a minimum at $s=s_{1}$%
. In fact, approaching $\mathbf{\hat{e}}_{3}$ from other directions (not
tangential to the equator) would force $\alpha^{\prime}(s)$ to become
complex valued. It is remarkable that $\alpha^{\prime}(s_{1})=a_{0}$ (resp. $%
b_{0})$ turns out to be the same constant for all possible triple collision
curves emanating from $\mathbf{\hat{p}}_{0}^{\pm}$ (resp. $\mathbf{\hat{e}}%
_{i}$), see (\ref{a0b0}).

In order to describe analytically the regularization of the triple collision
motions it is convenient to extend the domain of the angle $\alpha$ to
negative values as well. Indeed, $-\alpha$ should be identified with $%
\pi-\alpha$, and therefore we introduce the $\alpha$\emph{-circle } 
\begin{equation}
\frac{\left[ 0,\pi\right] }{(0\sim\pi)}\simeq S^{1}\subset\mathbb{C}%
:\alpha\rightarrow z_{\alpha}=e^{2i\alpha}   \label{circle}
\end{equation}
as the new domain for $\alpha$. Then the continuous motion $z_{\alpha}(s)$
on the circle illustrates the qualitative behavior of the solution $(\alpha
(s),\mathbf{\Gamma}^{\ast}(s))$ and hence also the moduli curve $\bar{\Gamma 
}(s)$ of an associated 3-body motion. For example, $\rho(s)$ is increasing
(resp. decreasing) with $s$ when $z_{\alpha}$ lies on the upper (resp.
lower) semicircle. A halting point is characterized by $z_{\alpha}=1$, and
it represents either a cusp, a triple collision ($\rho\rightarrow0$), or an
escape $(\rho\rightarrow\infty)$.

The only way $z_{\alpha}(s)$ enters the other semicircle is at $z_{\alpha}=1$%
, but for $\rho\rightarrow0$ or $\infty$, since at a cusp $z_{\alpha}$
\textquotedblright bounces back\textquotedblright\ on the same semicircle.
Similarly, $z_{\alpha}$ reaches the value $-1$ at a binary collision point $%
\mathbf{\hat{b}}_{i}$, but again $z_{\alpha}$ \textquotedblright bounces
back\textquotedblright\ (if the moduli curve is continued via regularization
with $\rho$ increasing ).

\begin{summary}
The solution $(z_{\alpha}(s),\Gamma^{\ast}(s))$ has a singularity at $%
s=s_{i} $ if either i) the pair takes the value $(1,\mathbf{p)}$ where $%
\mathbf{p}$ does not belong to the set 
\begin{equation}
\left\{ \pm\mathbf{\hat{p}}_{0},\mathbf{\hat{e}}_{1},\mathbf{\hat{e}}_{2},%
\mathbf{\hat{e}}_{3}\right\} \cup\left\{ \mathbf{\hat{b}}_{1},\mathbf{\hat{b}%
}_{2},\mathbf{\hat{b}}_{3}\right\} ,   \label{8 points}
\end{equation}
or ii) it takes the value $(-1,\mathbf{\hat{b}}_{i}),i>0$. The singularity
is a cusp (resp. a binary collision) in the first (resp. second) case.
\end{summary}

\subsubsection{The initial value problem for triple collision solutions}

For later reference we introduce the set $\mathfrak{S}(\mathbf{p})$ of all
shape curves $\Gamma^{\ast}(s),s\geq0$, representing a triple collision
motion with the limiting shape $\mathbf{p}=\Gamma^{\ast}(0)$ at the
collision. Here $\mathbf{p}$ can be any of the five points of the first
subset in (\ref{8 points}). In fact, each $\Gamma^{\ast}(s)$ is associated
with a unique (inclination angle) function $\alpha(s)$ so that the pair $%
(\alpha ,\Gamma^{\ast})$ is a solution of the system $ODE^{\ast}$ in (\ref%
{ODE*}) with the additional and singular initial condition $\ $ 
\begin{equation}
\text{\ \ }i)\text{ }\alpha(0)=0\text{, \ \ }ii)\text{\ }\alpha^{\prime
}(0)\geq0   \label{condition}
\end{equation}
In particular, for a fixed $\Gamma^{\ast}(s)$, $\alpha(s)$ is the unique
solution of (\ref{ODE*}) and (\ref{condition}). Thus we may as well consider
the totality of pairs $(\alpha,\Gamma^{\ast})$ and define 
\begin{equation}
\mathfrak{S}(\mathbf{p})=\left\{ (\alpha(s),\Gamma^{\ast}(s))\text{; }%
s\geq0,\Gamma^{\ast}(0)=\mathbf{p,}\text{ }\alpha(0)=0,\alpha^{\prime}(0)%
\geq0\right\}   \label{solutions}
\end{equation}
as the solutions of a specific initial value problem for the system $%
ODE^{\ast}$, as indicated.

For the calculation of the sets (\ref{solutions}) we may assume (by
symmetry) the initial shape $\mathbf{p}$ belongs to the fundamental chamber $%
\mathfrak{C}_{0}$, namely $\mathbf{p}$ is either its vertex $\mathbf{\hat{e}}%
_{3}$\textbf{\ }or $\mathbf{\hat{p}}_{0}$, see (\ref{vertices}). Observe
that the set (\ref{solutions}) has the induced \emph{symmetry group} $%
\mathfrak{G}_{\mathbf{p}}$ which is the \textquotedblright
isotropy\textquotedblright\ subgroup at $\mathbf{p}$ of $\mathfrak{G}$ (cf. (%
\ref{24-group})) 
\begin{equation}
\mathfrak{G}_{\mathbf{\hat{e}}_{3}}=\left\{ 1,\bar{r}\right\} \times\left\{
1,\bar{\tau}\right\} \simeq\mathfrak{D}_{1}\times\mathbb{Z}_{2}\text{, \ \ }%
\mathfrak{G}_{\mathbf{\hat{p}}_{0}}=\mathfrak{D}_{3}\times\mathbb{Z}_{2}%
\text{\ }   \label{iso}
\end{equation}
where $\bar{r}\in$ $\mathfrak{D}_{3}$ is the reflection $\theta\rightarrow
-\theta$.

\begin{remark}
Contrary to the above, the initial value problem for (\ref{ODE*}) at a cusp
or binary collision is not well defined since we would have $\alpha^{\prime
}(s_{i})=\pm\infty$. Therefore, one cannot continue a solution $(\alpha
(s),\Gamma^{\ast}(s))$ across the singularity using only the system $%
ODE^{\ast}$. To circumvent the problem one should, for example, turn to the
moduli curve $\bar{\Gamma}$ itself, which is regular in any case. In Section
8.4.2 below we shall use Newton's equation of motion more directly to
develop in time a specific solution through several cusps.
\end{remark}

\subsection{Isosceles and collinear triple collision motions}

In this section we take the opportunity to illustrate the above approach
applied to the \textquotedblright simplest\textquotedblright\ type of triple
collision motions apart from the shape invariant ones. We shall also supply
with numerical calculations, for comparison reasons and illustration of
examples only.

Namely, consider the possibility that the shape curve $\Gamma^{\ast}$ of a
triple collision motion is confined to a great circle on the sphere, that
is, $\mathcal{K}_{g}^{\ast}(s)$ $=0$ for all $s.$ Then the curvature
equation in (\ref{ODE*}) forces the normal derivative of $U^{\ast}$ along $%
\Gamma^{\ast}$ to vanish. Hence, by (\ref{A8}) and Remark \ref{F,G}, $%
\Gamma^{\ast}$ \textquotedblright moves\textquotedblright\ either on the
equator circle ($\varphi=\pi/2$) or on one of the six meridians (\ref{merid}%
) passing through an Euler point $\mathbf{\hat{e}}_{i}$ or a binary
collision point $\mathbf{\hat{b}}_{i}$. These meridians represent the shape
of isosceles triangles, so the associated 3-body motions are either of
collinear type or isosceles triangle type.

By the symmetry reduction explained in Section 8.3.1 it suffices to consider
three separate cases, namely $\Gamma^{\ast}$ is (initially) confined to one
of the boundary arcs of the fundamental chamber (\ref{A4}). We list the
starting point (at the triple collision) and a choice of arc-length
parameter $s$ to be used (up to the first cusp or binary collision point) in
each case : 
\begin{align}
(1)\text{ \ }\mathbf{\hat{p}}_{0} & :\theta=0,\text{ \ \ \ }s=\varphi \geq0%
\text{ \ }  \notag \\
(2)\text{ }\ \mathbf{\hat{e}}_{3} & :\varphi=\pi/2,\text{ }s=\theta \geq0
\label{triple3} \\
(3)\text{ \ }\mathbf{\hat{p}}_{0} & :\theta=\pi/3,\text{ }s=\varphi \geq0 
\notag
\end{align}

\subsubsection{The inclination angle $\protect\alpha$ and its ODE}

We shall investigate the first equation of (\ref{ODE*}) 
\begin{equation}
\frac{d\alpha}{ds}=-\frac{1}{4}+\frac{1}{2}\cot(\alpha)D_{i}(s)\text{ , }%
i=1,2,3   \label{diff1}
\end{equation}
for each of the three cases (\ref{triple3}), where $D_{i}(s)$ is calculated
from the appropriate expression of $U^{\ast}=U^{\ast}(\varphi,\theta)=U^{%
\ast }(\varphi,\theta+2\pi/3)$, namely (cf. (\ref{A3.5})) 
\begin{align}
\text{\ \ }U^{\ast}(\varphi,0) & =\ \frac{1}{3}(\frac{2}{\sqrt{1-\frac{1}{2}%
\sin\varphi}}+\frac{1}{\sqrt{1+\sin\varphi}})\text{, \ }  \label{P123} \\
U^{\ast}(\pi/2,\theta) & =\frac{1}{3}(\frac{1}{\sqrt{1-\cos(\theta-\frac {\pi%
}{3})}}+\frac{1}{\sqrt{1-\cos(\theta+\frac{\pi}{3})}}+\frac{1}{\sqrt{%
1+\cos\theta}})  \notag \\
U^{\ast}(\varphi,\pi/3) & =U^{\ast}(-\varphi,0)=U^{\ast}(\varphi ,\pi) 
\notag
\end{align}
As will be demonstrated below, there is a unique solution of the initial
value problem (\ref{condition}).

The derivatives $D_{i}$ are the following analytic functions expanded at the
point $\mathbf{\hat{p}}_{0}$ in case (1) and (3), and $\mathbf{\hat{e}}_{3}$
in case (2) : 
\begin{align}
D_{1}(\varphi) & =\frac{\frac{d}{d\varphi}U^{\ast}(\varphi,0)}{U^{\ast
}(\varphi,0)}=\varphi(\frac{3}{8}-\frac{15}{64}\varphi+\frac{23}{256}%
\varphi^{2}-....)  \notag \\
D_{2}(\theta) & =\frac{\frac{d}{d\theta}U^{\ast}(\pi/2,\theta)}{%
U^{\ast}(\pi/2,\theta)}=\theta(\frac{29}{20}+\frac{1801}{1200}\theta^{2}+%
\frac {17569}{12000}\theta^{4}+...)  \label{D12} \\
D_{3}(\varphi) & =\frac{\frac{d}{d\varphi}U^{\ast}(\varphi,\pi/3)}{U^{\ast
}(\varphi,\pi/3)}=\varphi(\frac{3}{8}+\frac{15}{64}\varphi+\frac{23}{256}%
\varphi^{2}+...)  \notag
\end{align}
It follows that $D_{1}(\varphi)=-D_{3}(-\varphi)$ and hence case (3) of (\ref%
{triple3}) can be subsumed under case (1) by using the range $\varphi<0$ and
moreover, $\alpha<0$ interpreted in accordance with (\ref{circle}). In fact,
if $\alpha_{i}(\varphi)$ is the solution of (\ref{diff1}) for $i=1,3$, then $%
\alpha_{3}(\varphi)=-\alpha_{1}(-\varphi)$. Hence, we need only consider the
cases (1) and (2) of (\ref{diff1}).

To investigate the nature of the singularity $\alpha=0,$ let us first
approximate the functions $D_{1}(\varphi)$, $D_{2}(\theta)$ and 
\begin{equation*}
\cot\alpha=\alpha^{-1}-\frac{1}{3}\alpha-\frac{1}{45}\alpha^{3}+..., 
\end{equation*}
by their first term $\frac{3}{8}\varphi$, $\frac{29}{20}\theta$ and $%
\alpha^{-1}$ respectively. Then the initial value problem $\alpha(0)=0$ for
the simplified version of (\ref{diff1}) has the two straight line solutions 
\begin{align}
(1) & :\alpha=a_{0}\varphi,\text{ \ \ \ }a_{0}=\frac{\pm\sqrt{13}-1}{8}
\label{lines} \\
(2) & :\alpha=b_{0}\theta,\text{ \ \ \ }b_{0}=\frac{\pm\frac{1}{5}\sqrt{1185}%
-1}{8}  \notag
\end{align}
found by solving a second order polynomial, namely 
\begin{equation}
a_{0}=-\frac{1}{4}+\frac{3}{16}\frac{1}{a_{0}}\text{, \ \ }b_{0}=-\frac{1}{4}%
+\frac{29}{40}\frac{1}{b_{0}}   \label{lines1}
\end{equation}
However, the extended initial value condition (\ref{condition}) demands $%
\alpha$ to be initially increasing and hence selects the positive solution
in (\ref{lines}) as the leading coefficient for the two types of triple
collision, namely 
\begin{align}
\text{Lagrange type }\text{: } & a_{0}=\frac{\sqrt{13}-1}{8}\approx 0.326
\label{a0b0} \\
\text{Euler type } & \text{:}\text{ }b_{0}=\frac{\frac{1}{5}\sqrt{1185}-1}{8}%
\approx0.736  \notag
\end{align}

Returning to the original equation (\ref{diff1}) we can determine
recursively the power series expansion of $\alpha$,

\begin{align}
(1)\text{ \ }\alpha & =a_{0}\varphi(1+a_{1}\varphi+a_{2}\varphi ^{2}+...)
\label{alfaseries} \\
(2)\text{ \ }\alpha & =b_{0}\theta(1+b_{1}\theta+b_{2}\theta^{2}+...)  \notag
\end{align}
which depends solely on the initial oefficients $a_{0},b_{0}$ in (\ref{a0b0}%
). For convenience we list (with a few decimals only) the first terms of the
expansions : 
\begin{align*}
(1)\text{ \ }\alpha & \approx\varphi(0.3257-0.0955\varphi+0.0129\varphi
^{2}-0.0233\varphi^{3}+...) \\
(2)\text{ \ }\alpha & =\theta(0.7356+0.1941\theta^{2}+0.0487\theta^{4}+...)
\end{align*}
As indicated, in case (1) the series is alternating and in case (2) $%
\alpha(s)$ is an odd function since $b_{1}=b_{3}=b_{5}=...=0$.

The above series can be easily developed and used with high accuracy for
small $\varphi$ (or $\theta)$. However, we shall only use it to calculate an
initial value of $\alpha$ for some small $\varphi$ (or $\theta$) and then
solve equation (\ref{diff1}) by standard numerical procedures, on the
maximal interval bounded by the first singularity in each direction. Namely,
in case (1) we find 
\begin{align}
\lim_{\varphi\rightarrow-\pi/2}\alpha(\varphi) & =-\pi/2\text{, \ }%
\lim_{\varphi\rightarrow-\pi/2}\alpha^{\prime}(\varphi)=\infty  \label{lim}
\\
\lim_{\varphi\rightarrow\varphi_{1}}\alpha(\varphi) & =0\text{, \ }%
\lim_{\varphi\rightarrow\varphi_{1}}\alpha^{\prime}(\varphi)=-\infty\text{,
\ \ \ \ }\varphi_{1}\approx1.876\approx107.5^{\circ}  \notag
\end{align}
On the interval $[0,\varphi_{1}]$, $\alpha(\varphi)$ increases up to its
maximum at $\varphi\approx1.2$ and is thereafter decreasing, see Figure 10.

On the other hand, if we rotate by $180^{\circ}$ the graph of $\alpha
(\varphi)$ over the interval $[-\pi/2,0]$, then we obtain the graph of $%
\alpha(\varphi)$ on $\left[ 0,\pi/2\right] $ for case (3) of (\ref{triple3}%
). Finally, the graph of $\alpha(\theta)$ in case (2) on the interval $%
[0,\pi/3]$ is quite similar to that of case (3), see Figure 11. Its graph
over $[-\pi/3,\pi/3]$ is symmetric with respect to the origin.

As calculated above, the shape curve $\Gamma^{\ast}$ along the meridian $%
\theta=0$ reaches the first cusp at $\varphi=\varphi_{1}$, which is beyond $%
\mathbf{\hat{e}}_{3}$. At the cusp the motion $\Gamma^{\ast}(s)$ changes its
direction and continues northward and across $\mathbf{\hat{e}}_{3}$. See
Section 8.4.2 and the time development of this motion.

Here is a brief summary of the analysis of the preliminary cases (\ref%
{triple3}) and the solution sets (\ref{solutions}) to which they belong :

\begin{itemize}
\item The set $\mathfrak{S}(\mathbf{\hat{e}}_{3})$ has only two solutions $%
(\alpha_{+},\Gamma_{+}^{\ast}),(\alpha_{-},\Gamma_{-}^{\ast})$, and they are
equivalent modulo the isometric reflection $\bar{r}:\theta\rightarrow-\theta$
belonging to $\mathfrak{D}_{3}\cap\mathfrak{G}_{\mathbf{\hat{e}}_{3}}$. Let $%
\Gamma^{\ast}(\theta),\theta\in(-\pi/3,\pi/3)$, be the arc-length
parametrization of the equator circle from $\mathbf{\hat{b}}_{1}$ to $%
\mathbf{\hat{b}}_{2}$. There is an analytic function $\alpha(\theta)$ so
that the pair $(\alpha,\Gamma^{\ast})$ is a solution of (\ref{ODE*}) and
moreover, 
\begin{align*}
\alpha_{-}(s) & =-\alpha(-s)\text{ for }s=-\theta\geq0\text{; \ }\alpha
_{+}(s)=\alpha(s)\text{ for }s=\theta\geq0 \\
\Gamma_{-}^{\ast}(s) & =\Gamma^{\ast}(-s)\text{ for }s=-\theta\geq0\text{; \ 
}\Gamma_{+}^{\ast}(s)=\Gamma^{\ast}(s)\text{ for }s=\theta\geq0
\end{align*}

\item The set $\mathfrak{S}(\mathbf{\hat{p}}_{0})$ has a unique solution $%
(\alpha_{0},\Gamma_{0}^{\ast})$ and $(\alpha_{\pi/3},\Gamma_{\pi/3}^{\ast})$
in case (1) and (3) of (\ref{triple3}), respectively. Consider also the
solution 
\begin{equation*}
\bar{\mu}(\alpha_{\pi/3},\Gamma_{\pi/3}^{\ast})=(\alpha_{\pi},\Gamma_{\pi
}^{\ast})\in\mathfrak{S}(\mathbf{\hat{p}}_{0}) 
\end{equation*}
along the meridian $\theta=\pi$, obtained by applying the rotation $\bar{\mu 
}\in\mathfrak{D}_{3}:$ $\theta\rightarrow\theta+2\pi/3$, as indicated, and
let $\Gamma^{\ast}(\varphi),\varphi\in(-\pi/2,\pi/2)$, be the arc-length
parametrization of the half-circle $(\mathbf{\hat{b}}_{3}\rightarrow \mathbf{%
\hat{p}}_{0}\rightarrow$ $\mathbf{\hat{e}}_{3})$. There is an analytic
function $\alpha(\varphi)$ so that the pair $(\alpha,\Gamma^{\ast})$ is a
solution of (\ref{ODE*}) and moreover, 
\begin{align*}
\alpha_{\pi}(s) & =-\alpha(-s)\text{ for }s=-\varphi\geq0\text{; \ }%
\alpha_{0}(s)=\alpha(s)\text{ for }s=\varphi\geq0 \\
\Gamma_{\pi}^{\ast}(s) & =\Gamma^{\ast}(-s)\text{ for }s=-\varphi \geq0\text{%
;\ }\Gamma_{0}^{\ast}(s)=\Gamma^{\ast}(s)\text{ for }s=\varphi\geq0
\end{align*}
\end{itemize}

\subsubsection{ Time dependence and Newton's equation}

We shall choose case (1) of (\ref{triple3}) and compare the above approach
using the system (\ref{ODE*}) with the time parametrized motion $\Gamma(t)$
using Newton's equation (\ref{Newton1}). Namely, in the xy-plane we consider
the motion of three point masses of mass $1/3$, symmetric with respect to
the y-axis and with position vectors 
\begin{equation*}
\mathbf{a}_{1}=(-x,y),\mathbf{a}_{2}=(x,y),\mathbf{a}_{3}=(0,-2y) 
\end{equation*}
Newton's equation (\ref{Newton1}) reads 
\begin{equation}
\ddot{x}=-\frac{1}{12}\frac{\left\vert x\right\vert }{x^{3}}-\frac{1}{3}%
\frac{x}{(x^{2}+9y^{2})^{3/2}},\text{ \ \ }\ddot{y}=-\frac{y}{%
(x^{2}+9y^{2})^{3/2}}   \label{Newton}
\end{equation}
\qquad As initial condition at time $t=0$, assume $y=0$ (i.e. $\Gamma^{\ast}$
is at the Euler point $\mathbf{\hat{e}}_{3}$) and moment of inertia $%
I=\rho^{2}=1$. Moreover, let $\beta$ denote the oriented angle from the
positive x-axis to the initial velocity vector $\mathbf{\dot{a}}_{2}$, whose
length is denoted $v$. Assuming the total energy $h=T-U$ vanishes, the
initial condition now reads 
\begin{equation}
(x,y)|_{t=0}=(\sqrt{\frac{3}{2}},0),\text{ \ }(\dot{x},\dot{y}%
)|_{t=0}=v(\cos\beta,\sin\beta)\text{, \ }v=\sqrt{\frac{\frac{5}{6}\sqrt{%
\frac{2}{3}}}{1+2\sin^{2}\beta}}   \label{t=0}
\end{equation}
Finally, let us assume $0<\beta<\pi$, which means the shape curve $%
\Gamma^{\ast}$ is heading southwards from $\mathbf{\hat{e}}_{3}$ for small $%
t>0.$

The above data specify a 1-parameter family (parametrized by $\beta)$ of
isosceles 3-body motions $\Gamma(t)=(x(t),y(t))$ with total energy $h=0$,
with normalized size $I=1$ and collinear shape $\mathbf{\hat{e}}_{3}$ at
time $t=0.$ We shall use the equation (\ref{Newton}) to investigate the time
dependence of the various geometric and kinematic quantities of the motion,
such as $I,T,\varphi,\alpha$, where 
\begin{align}
I & =\frac{2}{3}x^{2}+2y^{2}\text{, \ }\cos\varphi=\pm\frac{4}{3\sqrt{3}}%
\frac{\Delta}{I}=\frac{-4}{\sqrt{3}}\frac{xy}{\frac{2}{3}x^{2}+2y^{2}}
\label{quantities} \\
T & =U=\frac{1}{9}(\frac{2}{\left\vert x\right\vert }+\frac{2}{\sqrt {%
x^{2}+9y^{2}}})=\frac{1}{2}(\dot{\rho}^{2}+\frac{\rho^{2}}{4}\dot{\varphi }%
^{2})  \notag
\end{align}

By definition of the inclination angle $\alpha$, 
\begin{equation}
\cos\alpha=\frac{\frac{\partial}{\partial\rho}\cdot\frac{d}{dt}\bar{\Gamma }%
(t)}{\left| \frac{d}{dt}\bar{\Gamma}(t)\right| }=\frac{\dot{\rho}}{\sqrt{2T}}%
=\frac{\frac{2}{3}x\dot{x}+2y\dot{y}}{\sqrt{2IU}},   \label{cosalfa}
\end{equation}
and denoting the angle at $t=0$ by $\alpha_{0}$ we deduce 
\begin{equation}
\cos\alpha_{0}=\frac{3}{\sqrt{5}}(\frac{2}{3})^{1/4}v\cos\beta=\frac{\cos
\beta}{\sqrt{1+2\sin^{2}\beta}}   \label{alfa0}
\end{equation}
Thus, the correspondence $\beta\longleftrightarrow\alpha_{0}$ is a bijection
of the interval $(0,\pi)$ such that $\pi-\beta$ corresponds to $\pi-\alpha
_{0}$.

We claim there are exactly two values of $\beta$ leading to a triple
collision motion, namely the pair $\beta_{0}$ and $\pi-\beta_{0}$ for some $%
\beta _{0}<\pi/2$. This choice of $\beta=\beta_{0}$ yields $\dot{\rho}(0)>0$%
, and hence the triple collision occurred in the past, namely at some
negative time $t_{0}<0$, with the shape curve $\Gamma^{\ast}$ at the north
pole $\mathbf{\hat{p}}_{0}$. Similarly, using $\beta=\pi-\beta_{0}$ the
triple collision is reached at time $-t_{0}>0$ with $\Gamma^{\ast}$ at the
south pole.

The angle $\beta_{0}$ is calculated using the formula (\ref{alfa0}), where $%
\alpha_{0}=\alpha(\pi/2)$ and $\alpha(\varphi)$ is the solution of the
equation 
\begin{equation}
\frac{d\alpha}{d\varphi}=-1/4+\frac{\varphi}{2}\cot(\alpha)(\frac{3}{8}-%
\frac{15}{64}\varphi+\frac{23}{256}\varphi^{2}-....)   \label{diff2}
\end{equation}
with initial condition $\alpha(0)=0$, cf. case (1) of (\ref{diff1}) and (\ref%
{D12}). The solution found by the approach in Section 8.4.1, is
approximately 
\begin{equation*}
\alpha_{0}=\alpha(\pi/2)\approx0.18673...\text{, \ \ \ \ }%
\beta_{0}=0.10865.... 
\end{equation*}
Thus, we know the initial data (\ref{t=0}) corresponding to triple collision
motions. As a test, by running the system (\ref{Newton}) backwards in time
one will find that the triple collision occurs approximately at $%
t_{0}=-1.0228...$

It is also interesting to follow the shape curve $\Gamma^{\ast}$ of $%
\Gamma(t)$ for $t>0$, for example, using the ratio $y(t)/x(t)$ or
calculating $\varphi(t)$ directly from (\ref{quantities}). The solution $%
\Gamma(t)$ can be continued in time $t$ through the cusps since they are not
singularities for Newton's equation, and only truncation errors or numerical
instability may invalidate the calculation in the long run. At $%
t_{1}\approx10.4$, $y/x$ is maximal and $\varphi(t_{1})=\varphi_{1}$ (cf. (%
\ref{lim})) is the colatitude of the first cusp - here $\alpha$ $=0$ and $%
\Gamma^{\ast}(t)$ turns northward. After passing $\mathbf{\hat{e}}_{3}$
there is a second cusp where $\Gamma^{\ast}$ turns southward again and
crosses $\mathbf{\hat{e}}_{3}$, but the next cusp is closer to $\mathbf{\hat{%
e}}_{3}$, and so on. Concerning the possible asymptotic behavior of $%
\Gamma^{\ast}(t)$, at triple collision or as $t\rightarrow\infty$, see also
Section 8.6.3.

Finally, we consider the time dependent size function $\rho(t)=\sqrt{I(t)}\ $%
and compare it with the integral formula (\ref{rho3}). We have, by
assumption, $\rho=1$ at time $t=0$, and Newton's equation (\ref{Newton})
yields, for example, 
\begin{equation*}
\rho(-0.5)\approx0.634726\text{, \ \ }\varphi(-0.5)=\hat{\varphi}%
\approx1.381793 
\end{equation*}
On the other hand, numerical integration of the solution $\alpha(\varphi)$
of (\ref{diff2}) yields 
\begin{equation}
\rho|_{\varphi=\hat{\varphi}}=e^{\frac{1}{2}\int_{1}^{\hat{\varphi}%
}\cot\alpha d\varphi}\approx0.634725   \label{int}
\end{equation}
Alternatively, let us also evaluate this integral using the time
parametrized function $\alpha(t)$ calculated by (\ref{cosalfa}) and
developed via Newton's equation. Thus we change the variable $\varphi$ in (%
\ref{int}) to $t$ using (\ref{quantities}), namely 
\begin{equation*}
d\varphi=\varphi^{\prime}(t)dt=(\frac{d}{dt}\arccos(-\frac{4x(t)y(t)}{\sqrt {%
3}(\frac{2}{3}x(t)^{2}+2y(t)^{2})}))dt, 
\end{equation*}
and then numerical integration similar to the case (\ref{int}) yields 
\begin{equation*}
\rho(-0.5)=e^{\frac{1}{2}\int_{0}^{-0.5}\cot(\alpha)\varphi^{\prime}(t)dt}%
\approx0.634726 
\end{equation*}

\subsection{Analytic uniqueness of triple collision motions \qquad}

In this section we turn to the full family $\mathfrak{S}\mathbb{(}\mathbf{%
\hat{p}}_{0})$ of triple collision solutions $(\alpha,\Gamma^{\ast})$ of the
system $ODE^{\ast}$, with $\Gamma^{\ast}$ starting out from $\mathbf{\hat{p}}%
_{0}$. Due to the symmetry group $\mathfrak{D}_{3}$ we may assume the shape
curve $\Gamma^{\ast}$ enters the fundamental chamber $\mathfrak{C}_{0}$,
which limits the initial direction $\theta_{0}$ to the range $[0,\pi/3]$.
Therefore, in terms of spherical coordinates $(\varphi ,\theta)$ the
appropriate and complete initial condition (\ref{solutions}) now reads 
\begin{equation*}
\alpha(0)=0,\text{\ }\alpha^{\prime}(0)\geq0;\text{ }\varphi(0)=0,\text{ }%
\theta(0)=\theta_{0}\text{, }0\leq\theta_{0}\leq\pi/3 
\end{equation*}

The border cases $\theta_{0}=0,\pi/3$ are the cases (1) and (3) of (\ref%
{triple3}) already investigated in Section 8.4, and now it is natural to
generalize the procedure used there to the whole range of angles $\theta_{0}$%
. This time, however, the strength of the curvature equation of (\ref{ODE*})
must be fully utilized. At this point, we assume (tentatively) that the
functions $\alpha,\varphi,\theta$ have power series expansions at $\mathbf{%
\hat{p}}_{0}$, necessarily of type 
\begin{align}
\varphi & =s(c_{0}+c_{1}s+c_{2}s^{2}+c_{3}s^{3}+...+)  \notag \\
\theta & =\theta_{0}+s(d_{0}+d_{1}s+d_{2}s^{2}+...+),\text{ \ \ }0\leq
\theta_{0}\leq\frac{\pi}{3},  \label{A9} \\
\alpha & =a_{0}s(1+a_{1}s+a_{2}s^{2}+...+)  \notag
\end{align}
Recall from Section 8.1, we have excluded the trivial case of constant
shape, and hence $\alpha$ does not vanish identically.To justify the
notation in the third line, it will be demonstrated below that the leading
term is $a_{0}s$ with $a_{0}\neq0$.

By considering the leading coefficients of the series for $\varphi$ and $%
\theta$ the third equation of $ODE^{\ast}$ implies 
\begin{equation}
c_{0}=1,\text{ \ }c_{1}=0,\text{ \ }c_{2}=-\frac{1}{6}d_{0}^{2}\leq 0 
\label{A10}
\end{equation}
and clearly 
\begin{equation}
(\sin\varphi)\theta^{\prime}=\varepsilon(1-\varphi^{\prime2})^{1/2}=%
\varepsilon(-6c_{2}s^{2}+...+)^{1/2}\text{, \ \ }\varepsilon=\pm1 
\label{A11}
\end{equation}
The calculation of $a_{0}$ in the expansion $\alpha=a_{0}s+...$ is really
the same as in Section 8.4.1 and gives the same value (\ref{a0b0})
independent of $\theta_{0}$. To see this, we write for clarity the first
equation of $ODE^{\ast}$ as 
\begin{equation}
\alpha(2\alpha^{\prime}+\frac{1}{2})=(\alpha\cot\alpha)D(s),   \label{A13}
\end{equation}
where 
\begin{equation*}
\alpha\cot\alpha=1-\frac{1}{3}\alpha^{2}-\frac{1}{45}\alpha^{4}-\frac{2}{945}%
\alpha^{6}-\frac{1}{4725}\alpha^{8}+O(\alpha^{10}), 
\end{equation*}
and the potential function (\ref{A6}) and its logarithmic derivative along $%
\Gamma^{\ast}$ have the expansions 
\begin{equation}
u(s)=U^{\ast}(\Gamma^{\ast}(s))=\sum\limits_{i=0}^{\infty}u_{i}s^{i}=1+\frac{%
3}{16}s^{2}-\frac{5}{64}(\cos3\theta_{0})s^{3}+....   \label{A14}
\end{equation}
\begin{align}
D(s) & =\frac{d}{ds}\ln(u)=\frac{u^{\prime}(s)}{u(s)}=\frac{1}{u}(\frac{%
\partial U^{\ast}}{\partial\varphi}\varphi^{\prime}+\frac{\partial U^{\ast}}{%
\partial\theta}\theta^{\prime})  \notag \\
& =s\sum\limits_{i=0}^{\infty}\mu_{i}s^{i}=s(\frac{3}{8}-(\frac{15}{64}%
\cos3\theta_{0})s+...)   \label{A15}
\end{align}
In particular, the first order term in (\ref{A15}) is independent of $%
\theta_{0}$ and the leading terms of (\ref{A13}) yield the single condition 
\begin{equation}
4a_{0}^{2}+a_{0}-3/4=0\text{ , with positive root\ : }a_{0}=\frac{\sqrt{13}-1%
}{8}   \label{roots}
\end{equation}
The identity (\ref{A13}) also provides recursive relations for the
calculation of $a_{k},k>0,$ expressed in terms of the coefficients $%
\mu_{i},i\leq k,$ see (\ref{A26}) below.

\subsubsection{The method of undetermined coefficients}

The proof of Theorem G$_{1}$, concerning the existence and uniqueness of the
curves, is based upon formal power series substitution for the three
functions (\ref{A9}) involved in $ODE^{\ast}$. This leads to a recursive
procedure - the method of \emph{undetermined coefficients} - which is
consistent and determines successively the higher order coefficients in
terms of $c_{2}$, $d_{0}$ and $\theta_{0}$. By (\ref{A10}) $c_{2}$ is
already determined by $d_{0}$, and at the final stage we shall find that $%
d_{0}$ is actually determined by $\theta_{0}$. Consequently, the expansions
in (\ref{A9}) are, indeed, determined by the initial angle $\theta_{0}$
alone and hence $\theta_{0}$ parametrizes the whole solution set $\mathfrak{S%
}(\mathbf{\hat
{p}}_{0}).$

On the 2-sphere there is the positive, orthonormal frame $\{\frac{\partial }{%
\partial\varphi},\frac{1}{\sin\varphi}\frac{\partial}{\partial\theta}\}$
associated with the coordinates $\varphi,\theta$. Along the oriented shape
curve $\Gamma^{\ast}$ we also have the positive, orthonormal moving frame $\{%
\mathbf{\tau}^{\ast},\mathbf{\nu}^{\ast}\}$, where $\mathbf{\tau}^{\ast}$ is
the tangent vector. The latter frame differs from the stationary frame by a
rotation angle $\beta$, namely in analogy with (\ref{frame4}) - (\ref{frame5}%
), 
\begin{gather}
\mathbf{\tau}^{\ast}=\cos\beta\frac{\partial}{\partial\varphi}+\frac{\sin
\beta}{\sin\varphi}\frac{\partial}{\partial\theta}\text{, \ }\mathbf{\nu }%
^{\ast}=-\sin\beta\frac{\partial}{\partial\varphi}+\frac{\cos\beta}{%
\sin\varphi}\frac{\partial}{\partial\theta}  \label{A17} \\
\cos\beta=\varphi^{\prime}\text{, \ }\sin\beta=(\sin\varphi)\theta^{\prime
}=\varepsilon(1-\varphi^{\prime}{}^{2})^{1/2}\text{, cf.\ (\ref{A11}) \ }\ 
\label{A17a}
\end{gather}

Now, we turn to the second equation of (\ref{ODE*}), namely the curvature
equation written as an identity 
\begin{equation}
LHS=RHS   \label{A18}
\end{equation}
between the left hand and right hand side \ 
\begin{align}
LHS & =(1-\cos2\alpha)\mathcal{K}_{g}^{\ast}U^{\ast}  \label{A19} \\
RHS & =U^{\ast}(\nabla U^{\ast}\cdot\mathbf{\nu}^{\ast})=-\sin\beta \frac{%
\partial U^{\ast}}{\partial\varphi}+\frac{\cos\beta}{\sin\varphi}\frac{%
\partial U^{\ast}}{\partial\theta}  \notag
\end{align}
In $LHS$ the geodesic curvature term decomposes as 
\begin{equation}
\mathcal{K}_{g}^{\ast}=\frac{d\beta}{ds}+\cos\varphi\frac{d\theta}{ds}%
=-\varepsilon(1-\varphi^{\prime2})^{-1/2}\varphi^{\prime\prime}+(\cos%
\varphi)\theta^{\prime}   \label{A20}
\end{equation}
where $\beta^{\prime}$ is calculated using (\ref{A17}). We have actually $%
\varepsilon=1$, see the Remark below.

By substituting the first expression for $\sin\beta$ in (\ref{A17}) into $%
RHS $ and comparing the leading terms of the expansions of $LHS$ and $RHS$,
it follows that either $c_{2}\neq0$ or all $c_{i}=0$ for $i$ $>1$, and that $%
c_{2}=0$ implies $\sin3\theta_{0}=0$. Thus, $c_{2}=0$ means $\theta_{0}=0 $
or $\pi/3$, that is, the two meridian solutions of isosceles triangle type
already discussed in Section 8.4.

Henceforth, we shall assume $c_{2}\neq0.$ Comparison of the leading terms
(of order 2) in (\ref{A18}) yields 
\begin{equation*}
4a_{0}^{2}d_{0}=-\frac{3}{8}d_{0}+\frac{15}{64}\sin3\theta_{0}
\end{equation*}
which combined with (\ref{A10}) gives 
\begin{equation}
d_{0}=\frac{15}{16}\frac{\sin3\theta_{0}}{(16a_{0}^{2}+\frac{3}{2})},\text{
\ \ \ }c_{2}=-\frac{75}{512}\frac{\sin^{2}3\theta_{0}}{(16a_{0}^{2}+\frac {3%
}{2})^{2}}\   \label{A21}
\end{equation}
with the approximate values 
\begin{equation*}
d_{0}\approx0.293\sin3\theta_{0},\text{ \ }c_{2}\approx-0.014\sin^{2}3%
\theta_{0}
\end{equation*}

\begin{remark}
\label{epsilon}In particular, in (\ref{A11}) we have $\varepsilon
=sgn(d_{0})=1$, and the expressions in (\ref{A21}) are, in fact, valid in
the whole closed chamber $\mathfrak{C}_{0}:$ $0\leq\theta_{0}\leq\pi/3$.
However, $\varepsilon=\pm1$ actually changes sign across the border meridian
of two neigboring chambers.
\end{remark}

In view of (\ref{A17}), (\ref{A20}) we also need the expansions 
\begin{align*}
(1-\varphi^{\prime2})^{1/2} & =d_{0}s(1+b_{1}s+b_{2}s^{2}+...) \\
(1-\varphi^{\prime2})^{-1/2} & =\frac{1}{d_{0}s}(1+\bar{b}_{1}s+\bar{b}%
_{2}s^{2}+...)
\end{align*}
where by writing $\tilde{c}_{k}=c_{k}/c_{2}$ for $k\geq3$ we have by simple
inspection $\ $%
\begin{align}
b_{n} & =\frac{n+3}{6}\tilde{c}_{n+2}+B_{n}(\tilde{c}_{3},...,\tilde {c}%
_{n+1})\text{, \ }n\geq1  \label{A22} \\
\bar{b}_{n} & =-\frac{n+3}{6\ }\tilde{c}_{n+2}+\bar{B}_{n}(\tilde{c}_{3},...,%
\tilde{c}_{n+1})\text{, \ }n\geq1  \notag
\end{align}
where $B_{n}$ and $\bar{B}_{n}$ are polynomials and $B_{1}=$ $\bar{B}_{1}=0$.

For simplicity, let $P(y_{1},y_{2},...)$ denote any (unspecified) polynomial
in the variables $y_{i}$, except that $y_{1}=\theta_{0}$ means it is
polynomial in $\sin3\theta_{0}$ and $\cos3\theta_{0}$. Using the notation 
\begin{align}
\sin\varphi & =s(1+g_{2}s^{2}+g_{3}s^{3}+...),\text{ \ \ \ \ \ \ \ }%
g_{k}=c_{k}+P(c_{2},...,c_{k-1})  \label{A23} \\
\cos\varphi & =1-\frac{1}{2}s^{2}+h_{4}s^{4}+h_{5}s^{5}+...,\text{ }%
h_{k}=P(c_{2},...,c_{k-2})  \notag
\end{align}
we derive from (\ref{A11}) the following formula for $d_{n}\ $%
\begin{equation}
(n+1)d_{n}=d_{0}b_{n}-(d_{0}g_{n}+2d_{1}g_{n-1}+...+(n-1)d_{n-2}g_{2}),\text{
}n\geq1,   \label{A24}
\end{equation}
and from (\ref{A20}) we calculate the curvature expansion 
\begin{equation*}
\mathcal{K}_{g}^{\ast}=2d_{0}(1+k_{1}s+k_{2}s^{2}+...) 
\end{equation*}
where 
\begin{align}
2d_{0}k_{n} & =\left( d_{0}\bar{b}_{n}+(n+1)d_{n}-(n+3)(n+2)\frac{c_{n+2}}{%
d_{0}}\right)  \label{A25} \\
& -\frac{1}{d_{0}}\sum_{k=3}^{n+1}(k+1)kc_{k}\bar{b}_{n+2-k}+%
\sum_{k=0}^{n-1}(k+1)d_{k}h_{n-k}  \notag
\end{align}

Next, let us have a closer look at the coefficients of $u(s)$ and $u^{\prime
}(s)/u(s)$, using (\ref{A6}), (\ref{A14}), (\ref{A15}), and also at the
recursive generation of the coefficients of $\alpha(s)$ using (\ref{A13}).
It follows that they are of type 
\begin{align}
u_{n} & =P(\theta_{0},c_{2},...,c_{n-2},d_{0},...,d_{n-4})\text{, \ \ }n\geq4
\notag \\
\mu_{n} & =P(\theta_{0},c_{2},...,c_{n},d_{0},...,d_{n-2})\text{, \ \ \ }%
n\geq2  \label{A26} \\
a_{n} & =P(a_{1},a_{2},...,a_{n-1},\mu_{1},...,\mu_{n})\text{, \ \ }n\geq1 
\notag
\end{align}
For example, 
\begin{equation*}
a_{1}=\frac{2\mu_{1}}{12a_{0}^{2}+a_{0}}=\frac{8\mu_{1}}{10-\sqrt{13}}=-%
\frac{15/8\ }{10-\sqrt{13}}\cos3\theta_{0}
\end{equation*}

For convenience, write 
\begin{equation}
1-\cos2\alpha=2a_{0}^{2}s^{2}(1+A_{1}s+A_{2}s^{2}+...),   \label{A27}
\end{equation}
where 
\begin{equation*}
A_{1}=2a_{1},\text{ \ }A_{2}=2a_{2}+a_{1}^{2}-\frac{1}{3}a_{0}^{2},\text{ \
. . . , }A_{n}=P(a_{1},a_{2},...,a_{n}), 
\end{equation*}
and thus we arrive at the following presentation of $LHS$ as a product of
series 
\begin{equation}
LHS=4a_{0}^{2}d_{0}s^{2}(1+A_{1}s+....)(1+k_{1}s+....)(1+u_{1}s+....) 
\label{A28}
\end{equation}

From the structure of $U^{\ast}$ as a trigonometric series (\ref{A6}) in the
variables $\sin\varphi$ and $\cos3\theta$, we can write 
\begin{equation}
\frac{\partial U^{\ast}}{\partial\varphi}=(\sin\varphi\cos\varphi)R_{1}\text{%
, \ }\frac{\partial U^{\ast}}{\partial\theta}=(\sin^{3}\varphi\sin3\theta
)R_{2}   \label{A29}
\end{equation}
where 
\begin{equation*}
R_{1}=\frac{3}{8}-\frac{15}{64}\sin\varphi\cos3\theta+...\text{, \ \ \ }%
R_{2}=\frac{15}{64}+\frac{945}{4096}\sin^{2}\varphi+.... 
\end{equation*}
are again polynomial series in the variables $\sin\varphi,\cos3\theta$.
Hence, by substituting the series of $\sin\varphi,\cos\varphi,\theta^{%
\prime},\varphi^{\prime},\cos3\theta,\sin3\theta\ $into the expression 
\begin{equation}
RHS=(\sin^{2}\varphi)\left( -\theta^{\prime}\cos\varphi R_{1}+\varphi
^{\prime}\sin3\theta R_{2}\right)   \label{A31}
\end{equation}
equation (\ref{A18}) renders a recursive procedure for the calculation of $%
c_{n}$ and $d_{n-2}$, $n\geq2,$ starting from $c_{2}$ and $d_{0}$ (\ref{A21}%
).

\begin{lemma}
\bigskip The coefficients $c_{m}$ and $d_{m-2}$ can be expressed as 
\begin{equation}
c_{m}=P(\theta_{0})c_{2}\text{, \ \ \ }d_{m-2}=P(\theta_{0})d_{0};\text{ \ }%
m\geq2   \label{A32}
\end{equation}
where $P(\theta_{0})$ denotes some polynomial of $\sin3\theta_{0}$ and $%
\cos3\theta_{0}$ (generally different for each coefficient).
\end{lemma}

\begin{proof}
The first step is to compare terms of order 3 in (\ref{A18}), which renders
the identity 
\begin{equation*}
4a_{0}^{2}d_{0}(A_{1}+k_{1}+u_{1})=-\frac{3}{4}d_{1}+\frac{15}{64}%
d_{0}\cos3\theta_{0}
\end{equation*}
where 
\begin{equation*}
A_{1}=2a_{1},\text{ \ }k_{1}=\frac{c_{3}}{c_{2}},\text{ \ }u_{1}=0,\text{ \ }%
d_{1}=\frac{d_{0}c_{3}}{3c_{2}}
\end{equation*}
Consequently, 
\begin{align}
c_{3} & =\frac{15(16a_{0}+1)}{4(16a_{0}^{2}+1)(12a_{0}+1)}(\cos3\theta
_{0})c_{2}  \label{A33} \\
d_{1} & =\frac{15(16a_{0}+1)}{12(16a_{0}^{2}+1)(12a_{0}+1)}(\cos3\theta
_{0})d_{0}  \notag
\end{align}

We proceed by induction and assume that (\ref{A32}) holds for $m$ in the
range $2\leq m\leq n$. \ By (\ref{A22}) - (\ref{A27}), we infer that $%
b_{m-2},\bar{b}_{m-2},k_{m-2},g_{m,}h_{m},u_{m},\mu_{m}$ and $a_{m}$ are all
of type $P(\theta_{0})$ for $m\leq n$. Furthermore, 
\begin{align}
nd_{n-1} & =\left( \frac{n+2}{6c_{2}}c_{n+1}+P(\theta_{0})\right) d_{0}\text{
\ \ }  \label{A34} \\
k_{n-1} & =\frac{(n+1)(n+2)}{12c_{2}}c_{n+1}+P(\theta_{0})  \notag
\end{align}

Consider the terms of order $n+1$ in equation (\ref{A18}). By equating the
coefficients of $s^{n+1}$ in $LHS$ and $RHS$ we deduce 
\begin{equation}
4a_{0}^{2}d_{0}k_{n-1}+P(\theta_{0})d_{0}=-\frac{3}{8}nd_{n-1}+P(\theta
_{0})d_{0}   \label{A35}
\end{equation}
Here, $k_{n-1}$ and $d_{n-1}$ are the only coefficients depending on $%
c_{n+1} $, and by substituting their expressions from (\ref{A34}) into the
identity (\ref{A35}), we deduce the identity 
\begin{equation*}
\left( \frac{4a_{0}^{2}(n+1)(n+2)}{12}+\frac{n+2}{16}\right)
c_{n+1}=P(\theta_{0})c_{2}, 
\end{equation*}
and consequently, 
\begin{equation*}
c_{n+1}=P(\theta_{0})c_{2},\ \ \ d_{n-1}=P(\theta_{0})d_{0}
\end{equation*}
This completes the induction step, and hence (\ref{A32}) holds for all $%
m\geq2$.
\end{proof}

This settles the existence and uniqueness question for the series expansions
(\ref{A9}), for each initial longitude angle $\theta_{0}$. Their radius of
convergence is certainly positive (e.g. by an inductive argument showing the
coefficients are bounded), but we shall not try to estimate the radius here.
Clearly, for $\theta_{0}=k\pi/3$ the radius is at most $\pi/2$.

\subsubsection{Symmetries of the solution set $\mathfrak{S}(\mathbf{\hat{p}}%
_{0})$}

In the previous subsection it was established that the triple collision
solution set (\ref{solutions}) for $\mathbf{p}=\mathbf{\hat{p}}_{0}$ is
naturally parametrized by angles $\theta_{0}$, namely

\begin{equation}
\mathfrak{S}(\mathbf{\hat{p}}_{0})=\left\{
(\alpha_{\theta_{0}},\Gamma_{\theta_{0}}^{\ast});0\leq\theta_{0}<2\pi\right%
\}   \label{solu}
\end{equation}
is in 1-1 correspondence with points on a circle and therefore inherits
\textquotedblright symmetries\textquotedblright\ of a circle. However, the
actual symmetry group should act with orbits representing the various
\textquotedblright species' or \textquotedblright congruence"\ classes of
solutions, and moreover, knowledge of each class suffices to generate all
solutions by a straightforward transformation procedure. We contend that $%
\mathfrak{G}_{\mathbf{\hat{p}}_{0}}$defined in (\ref{iso}) is, in fact, the
appropriate group.

First of all, $\mathfrak{G}_{\mathbf{\hat{p}}_{0}}$ contains the group $%
\mathfrak{D}_{3}$ which acts on the 2-sphere and represents the purely
geometric symmetries. On the other hand, we have also seen that each
solution curve $\Gamma^{\ast}=\Gamma_{\theta_{0}}^{\ast}$ corresponds to
three analytic functions $(\alpha(s),\varphi(s),\theta(s))$ in a
neighborhood of $s=0$, and for $s<0$ these functions also describe a motion
approaching a triple collision as $s\rightarrow0^{-}$. Hence, by inverting
its direction we should obtain a triple collision motion emanating with
initial longitude angle $\theta_{0}+\pi$, which by uniqueness must be the
solution in (\ref{solu}) labelled by $\theta_{0}+\pi$. Consequently, in
agreement with the summary of Section 8.4.1, for each \textquotedblright
antipodal\textquotedblright\ pair $(\mathfrak{\alpha}_{\theta_{0}},\Gamma_{%
\theta_{0}}^{\ast}),(\mathfrak{\alpha }_{\theta_{0}+\pi},\Gamma_{\theta_{0}+%
\pi}^{\ast})$ in (\ref{solu}) there is an analytic curve $\Gamma^{\ast}(s)$
passing through $\mathbf{\hat{p}}_{0}$ and an analytic function $\alpha(s)$
so that $(\alpha(s),\Gamma^{\ast}(s))$ is a solution of the system (\ref%
{ODE*}), and moreover, 
\begin{align*}
\alpha_{\theta_{0}+\pi}(s) & =-\alpha(-s)\text{ and\ }\alpha_{%
\theta_{0}}(s)=\alpha(s)\text{ for }s\geq0 \\
\Gamma_{\theta_{0}+\pi}^{\ast}(s) & =\Gamma^{\ast}(-s)\text{ and \ }%
\Gamma_{\theta_{0}}^{\ast}(s)=\Gamma^{\ast}(s)\text{ for }s\geq0
\end{align*}

In particular, the inversion operator $\bar{\tau}$ applied to $(\alpha
(s),\Gamma^{\ast}(s))$ induces an involution 
\begin{equation*}
\bar{\tau}:(\mathfrak{\alpha}_{\theta_{0}},\Gamma_{\theta_{0}}^{\ast
})\rightarrow(\mathfrak{\alpha}_{\theta_{0}+\pi},\Gamma_{\theta_{0}+\pi}^{%
\ast}) 
\end{equation*}
of the set $\mathfrak{S}(\mathbf{\hat{p}}_{0})$ which commutes with the
action of $\mathfrak{D}_{3}$, and together they generate the dihedral \emph{%
symmetry group } 
\begin{equation}
\mathfrak{G}_{\mathbf{\hat{p}}_{0}}=\mathfrak{D}_{3}\times\left\{ 1,\bar {%
\tau}\right\} \text{\ }\simeq\mathfrak{D}_{6}
\end{equation}
which may be viewed as an \textquotedblright isotropy\textquotedblright\
subgroup of $\mathfrak{G}$ in (\ref{24-group}). In effect, this divides the
fundamental chamber $\mathfrak{C}_{0}$ (\ref{A4}) in two sectors of angular
width $\pi/6$, say 
\begin{equation*}
\mathfrak{\tilde{C}}_{0}:0\leq\theta_{0}\leq\pi/6 
\end{equation*}
is our \emph{reduced} fundamental chamber. However, we remark that $%
\mathfrak{G}_{\mathbf{\hat{p}}_{0}}$ does not act on $S^{2}$, so $\mathfrak{%
\bar{C}}_{0}$ is not a fundamental domain in the geometric sense. But
solutions starting out in this region suffice to generate the whole solution
set (\ref{solu}) using analytic continuation.

The power series developments (\ref{A9}) of the three functions $\alpha
(s),\varphi(s),\theta(s)$, where $\Gamma_{\theta_{0}}^{\ast}(s)=(\varphi
(s),\theta(s))$ is the shape curve with initial direction $\theta_{0}$ at
the north pole, also exhibit a specific symmetry pattern which reflects the $%
\mathfrak{D}_{6}$-symmetry of their coefficients. For example, consider the
\textquotedblright reflection\textquotedblright\ in $\mathfrak{G}_{\mathbf{%
\hat{p}}_{0}}$ 
\begin{equation*}
\theta_{0}\rightarrow\pi/3-\theta_{0}
\end{equation*}
which divides $\mathfrak{C}_{0}$ into two reduced chambers, and write 
\begin{equation*}
X=\cos3\theta_{0},\text{ \ \ }Y=\sin3\theta_{0}
\end{equation*}
\begin{equation*}
a_{k}=\mathfrak{\ }\text{\textsc{A}}_{k}(X,Y)\text{, \ \ \ \ }c_{k}=\text{%
\textsc{C}}_{k}(X,Y)c_{2},\text{ \ \ \ \ }d_{k}=\text{\textsc{D}}%
_{k}(X,Y)d_{0}
\end{equation*}
where \textsc{A}$_{k},$\textsc{C}$_{k},$\textsc{D}$_{k}$ are polynomials of
two variables (not unique, of course, since $X,Y$ are algebraic dependent).
Let $a_{k},c_{k},d_{k}$ and $\bar{a}_{k},\bar{c}_{k},\bar{d}_{k}$ be the
coefficients of the solutions $(\alpha,\varphi,\theta),(\bar{\alpha},\bar{%
\varphi},\bar{\theta})$ corresponding to initial angles $\theta_{0}$ and $%
\pi/3-\theta_{0}$, respectively. Then we have 
\begin{equation*}
\left\{ 
\begin{array}{c}
a_{k}=\bar{a}_{k}\text{, }c_{k}=\bar{c}_{k}\text{, \ }d_{k}=\bar{d}_{k}\text{%
, \ \ \ \ \ \ \ \ \ \ for\ }k\text{ even} \\ 
a_{k}=-\bar{a}_{k}\text{, \ }c_{k}=-\bar{c}_{k}\text{, \ }d_{k}=-\bar{d}_{k}%
\text{, \ for\ }k\text{ odd}%
\end{array}
\right. 
\end{equation*}
Equivalently, as functions of $X$ the polynomials \textsc{A}$_{k},$\textsc{C}%
$_{k},$\textsc{D}$_{k}$ are odd (resp. even) functions for $k$ odd (resp.
even). This is due to the fact that $Y$ is invariant whereas $X$ changes
sign under the substitution $\theta_{0}\rightarrow\pi/3-\theta_{0}$.

\subsubsection{Symbolic manipulations and numerical calculation of power
series}

The recursive scheme used in Section 8.5.1 will generate all higher order
coefficients as polynomials of $Y=\sin3\theta_{0}$ and $X=\cos3\theta_{0}$,
but the explicit calculations involve a substantial amount of symbolic
manipulations. For example, various types of algebraic operations, together
with composition, are applied to power series.

In principle, calculations involving elementary functions of power series,
such as $\sin(\sum p_{i}x^{i})$, can be reduced to symbolic manipulations on
power series of the type 
\begin{equation*}
(p_{0}+p_{1}x+p_{2}x^{2}+...)^{n}=\sum\limits_{k=0}^{\infty}P_{k}^{(n)}x^{k}
\end{equation*}
where the n-th \emph{multinomial polynomial }$P_{k}^{(n)}$ records the
n-partitions and associated multinomial coefficients which can be calculated
recursively with some effort.

On the other hand, available computer software developed for symbolic
computation have built-in procedures which effectively generate the
intermediate power series expansions as well as recursive formulas. We have
employed such symbolic software for the calculation\footnote{%
The symbolic and numerical calculations were performed in 1995 by Chee-Whye
Chin, an undergraduate student at U.C. Berkeley, using \emph{Mathematica }%
software.\emph{\ }} of $a_{k},c_{k},d_{k}$ in (\ref{A9}), for small $k$, see
also Section 8.7. For convenience, we list the first of them below (omitting
the already known $c_{2},d_{0},a_{0})$, and we remark that the exact (or
symbolic) expressions are growing fast in complexity as $k$ increases : 
\begin{align}
\text{\ \ }c_{3} & =c_{2}\frac{15(10+\sqrt{13})}{116}(\cos3\theta _{0})
\label{A36} \\
c_{4} & =c_{2}\left( \frac{2040762505+136353812\sqrt{13}}{2891425280}+\frac{%
20984375(113+20\sqrt{13})}{2891425280}(\cos6\theta_{0})\right)  \notag \\
d_{1} & =d_{0}\frac{5(10+\sqrt{13})}{116}(\cos3\theta_{0})  \notag \\
d_{2} & =d_{0}\left( \frac{2(19733316+84347\sqrt{13})}{216856896}+\frac{%
302175(113+20\sqrt{13})}{216856896}(\cos6\theta_{0})\right)  \notag \\
a_{1} & =\frac{-5(10+\sqrt{13})}{232}(\cos3\theta_{0})  \notag \\
a_{2} & =\frac{3(28004+4175\sqrt{13})}{2745024}+\frac{-25(644+47\sqrt{13})}{%
2745024}(\cos6\theta_{0})  \notag \\
& +\frac{4350(9+\sqrt{13})}{2745024}\sqrt{113+20\sqrt{13}}%
(\sin^{2}3\theta_{0})  \notag
\end{align}

\subsection{Global behavior of the shape of triple collision motions}

First we shall investigate the curvature properties of the flow consisting
of the gradient lines of the potential function $U^{\ast}$ on the sphere $%
S^{2}(1)$. This information will be related to the curvature properties of
the "flow" consisting of those curves $\Gamma^{\ast}$ belonging to the set (%
\ref{solu}), that is, the triple collision shape curves emanating from the
north pole $\mathbf{\hat{p}}_{0}$.

\subsubsection{Differential geometry of the gradient flow of $U^{\ast}$}

We start with the following elementary result about curves on the unit
sphere $S^{2}$ in Euclidean 3-space.

\begin{lemma}
\label{triple4}Let $t\rightarrow\mathbf{p}(t)$ be a parametrized curve on $%
S^{2}$. Then its geodesic curvature is given by the following triple product 
\begin{equation}
K_{g}(t)=(\frac{dt}{ds})^{3}\mathbf{p\times\dot{p}}\cdot\mathbf{\ddot{p}}=%
\mathbf{p\times p}^{\prime}\cdot\mathbf{p}^{\prime\prime}   \label{A37}
\end{equation}
where (as usual) $s$ is arc-length, $\mathbf{\dot{p}=}\frac{d}{dt}\mathbf{p}$
and $\mathbf{p}^{\prime}=\frac{d}{ds}\mathbf{p}$.
\end{lemma}

\begin{proof}
The unit tangent vector $\mathbf{\tau}^{\ast}=\mathbf{p}^{\prime}$ points in
the positive direction of the curve, and $\mathbf{p}^{\prime\prime}$ is the
curvature vector in 3-space. With $\mathbf{\nu}^{\ast}=\mathbf{p\times\tau }%
^{\ast}$ as the normal vector field along the curve, $\left\{ \mathbf{\tau }%
^{\ast}\mathbf{,\nu}^{\ast}\right\} $ is a positively oriented frame of the
sphere. By definition, the geodesic curvature vector in the sphere is the
orthogonal projection of $\mathbf{p}^{\prime\prime}$ into the tangent plane,
namely 
\begin{equation*}
\mathbf{p}^{\prime\prime}-(\mathbf{p}^{\prime\prime}\cdot\mathbf{p})\mathbf{p%
}=K_{g}\mathbf{\nu}^{\ast}
\end{equation*}
where the coefficient $K_{g}$ is the (scalar) geodesic curvature. Clearly, $%
K_{g}$ equals the triple product in (\ref{A37}).
\end{proof}

Next, we turn to the gradient field $\nabla U^{\ast}$ on the sphere, whose
integral curves will be referred to as the \emph{gradient lines }of $%
U^{\ast} $. Their geodesic curvature will be denoted by $K_{g}$. Until
further notice there is no restriction on the mass distribution, and we use
the expressions (\ref{U3}), (\ref{B0}) and (\ref{B1}) for $U^{\ast}$, the
vector function $\mathbf{B(p)}$ and the gradient $\nabla U^{\ast}$ field,
respectively.

\begin{lemma}
The geodesic curvature of the gradient line of $U^{\ast}$ passing through $%
\mathbf{p}$ is given by the following triple product 
\begin{equation}
K_{g}(\mathbf{p})=\frac{3}{\left\vert \nabla U^{\ast}(\mathbf{p})\right\vert
^{3}}\mathbf{B}\times\mathbf{C}\cdot\mathbf{p}   \label{A38}
\end{equation}
where $\mathbf{B=B(p)}$ and 
\begin{equation}
\mathbf{C}=\mathbf{C}(\mathbf{p})=\sum\limits_{i=1}^{3}\frac{k_{i}}{%
\left\vert \mathbf{p}-\mathbf{\hat{b}}_{i}\right\vert ^{5}}\left[ \mathbf{%
p\times B}\cdot(\mathbf{p\times\hat{b}}_{i})\right] \mathbf{\hat{b}}_{i} 
\label{A39}
\end{equation}
\end{lemma}

\begin{proof}
Consider a gradient line parametrized by arc-length, $s\rightarrow \mathbf{p}%
(s)$. We can write 
\begin{equation*}
\nabla U^{\ast}(\mathbf{p})=f(s)\mathbf{p}^{\prime}(s) 
\end{equation*}
where $f(s)=\left\vert \nabla U^{\ast}(\mathbf{p(}s\mathbf{)})\right\vert $,
and hence by (\ref{B1}) 
\begin{equation*}
\mathbf{p}^{\prime}=\frac{\mathbf{B-(B}\cdot\mathbf{p)p}}{\left\vert \nabla
U^{\ast}\right\vert }
\end{equation*}
On the other hand, $0=\mathbf{p\times p}^{\prime}\cdot\nabla U^{\ast }=%
\mathbf{p\times p}^{\prime}\cdot\mathbf{B}$ and by differentiation 
\begin{equation}
\mathbf{p\times p}^{\prime\prime}\cdot\mathbf{B}=-\mathbf{p\times p}^{\prime
}\cdot\frac{d}{ds}\mathbf{B}   \label{A40}
\end{equation}
Hence, by substituting the expression 
\begin{equation*}
\mathbf{B}=\left\vert \nabla U^{\ast}\right\vert \mathbf{p}^{\prime }+(%
\mathbf{B}\cdot\mathbf{p)p}
\end{equation*}
into the left side of (\ref{A40}), we obtain 
\begin{equation}
\left\vert \nabla U^{\ast}\right\vert \mathbf{p\times p}^{\prime}\cdot%
\mathbf{p}^{\prime\prime}=\mathbf{p\times p}^{\prime}\cdot\frac{d}{ds}%
\mathbf{B}=\frac{\mathbf{p}\times\mathbf{B}}{\left\vert \nabla U^{\ast
}\right\vert }\cdot\frac{d}{ds}\mathbf{B}   \label{A41}
\end{equation}

Finally, we calculate from (\ref{B0}) 
\begin{equation*}
\frac{d}{ds}\mathbf{B}=3\sum_{i=1}^{3}k_{i}\frac{(\mathbf{\hat{b}}_{i}\cdot%
\mathbf{p}^{\prime})}{\left\vert \mathbf{p-\hat{b}}_{i}\right\vert ^{5}}%
\mathbf{\hat{b}}_{i}=\frac{3}{\left\vert \nabla U^{\ast}\right\vert }\sum
k_{i}\frac{(\mathbf{\hat{b}}_{i}\times\mathbf{p)\cdot(B}\times\mathbf{p)}}{%
\left\vert \mathbf{p-\hat{b}}_{i}\right\vert ^{5}}\mathbf{\hat{b}}_{i}
\end{equation*}
and by substituting this into (\ref{A41}) it follows from Lemma \ref{triple4}
\begin{equation*}
K_{g}=\mathbf{p\times p}^{\prime}\cdot\mathbf{p}^{\prime\prime}=\frac {3}{%
\left\vert \nabla U^{\ast}\right\vert ^{3}}\sum k_{i}\frac{(\mathbf{p}\times%
\mathbf{B)\cdot(p}\times\mathbf{\hat{b}}_{i}\mathbf{)}}{\left\vert \mathbf{p-%
\hat{b}}_{i}\right\vert ^{5}}(\mathbf{p}\times\mathbf{B}\cdot\mathbf{\hat{b}}%
_{i}) 
\end{equation*}
This expression can be rewritten as (\ref{A38}).
\end{proof}

\begin{problem}
\label{curvK}Regarding $K_{g}$ as a function on $S^{2}$, determine the
curves defined by the condition $K_{g}(\mathbf{p})=0.$
\end{problem}

\begin{remark}
Note that $\mathbf{B(p)}$ and $\mathbf{C(p)}$ are vectors in the xy-plane in
the Euclidean model $\bar{M}$ $=\mathbb{R}^{3}$, and the function $K_{g}$ on
the sphere $S^{2}:x^{2}+y^{2}+z^{2}=1$ is undefined precisely at the
critical points of $U^{\ast}$, namely for $z\geq0$ these are the points $%
\mathbf{\hat
{b}}_{i},0\leq i\leq3$, and the minimumspoint (physical
center) $\mathbf{\hat
{p}}_{0}^{+}$. From the triple product formula (\ref%
{A38}) it follows that $K_{g}(\mathbf{p})$ vanishes on the eclipse circle ($%
z=0)$, whereas for $z>0$ $K_{g}(\mathbf{p})$ vanishes if and only if $%
\mathbf{B(p)}$ and $\mathbf{C(p)}$ are linearly dependent.
\end{remark}

Henceforth, we shall retain our assumption of uniform mass distribution, and
a deeper understanding of the function $K_{g}$ will be achieved. In fact, in
this case the above problem has a simple solution, as explained at the end
of this subsection.

By assumption, 
\begin{equation*}
m_{i}=\frac{1}{3},\text{ \ }k_{i}=\frac{\sqrt{2}}{3},\text{ \ }\mathbf{%
\hat
{b}}_{i}\cdot\mathbf{\hat{b}}_{j}=-\frac{1}{2}\text{ \ \ \ }(i,j\geq1%
\text{ and }i\neq j) 
\end{equation*}
and we introduce the three distance functions 
\begin{equation}
\delta_{i}=\delta_{i}(\mathbf{p})=\left\vert \mathbf{p-\hat{b}}%
_{i}\right\vert =\sqrt{2}\sqrt{1-\mathbf{p\cdot\hat{b}}_{i}},\text{ }i\geq1,%
\text{\ \ \ cf. Section 6.1}   \label{A42}
\end{equation}
which are algebraically related by 
\begin{equation}
\delta_{1}^{2}+\delta_{2}^{2}+\delta_{3}^{2}=6,   \label{A42.1}
\end{equation}
due to the identity $\sum\mathbf{p}\cdot\mathbf{\hat{b}}_{i}=\mathbf{p}\cdot$
$\sum\mathbf{\hat{b}}_{i}=0$ and 
\begin{equation}
\mathbf{p\cdot\hat{b}}_{i}=1-\frac{1}{2}\delta_{i}^{2}   \label{A42.2}
\end{equation}

Let us write 
\begin{align}
\mathbf{B(p)} & =\frac{\sqrt{2}}{3}\sum_{i=1}^{3}e_{i}\mathbf{\hat{b}}_{i}=%
\frac{\sqrt{2}}{3}\left( (e_{1}-e_{3})\mathbf{\hat{b}}_{1}+(e_{2}-e_{3})%
\mathbf{\hat{b}}_{2}\right)  \label{A43} \\
\mathbf{C(p)} & =\frac{1}{9}\sum_{i=1}^{3}f_{i}\mathbf{\hat{b}}_{i}=\frac {1%
}{9}\left( (f_{1}-f_{3})\mathbf{\hat{b}}_{1}+(f_{2}-f_{3})\mathbf{\hat{b}}%
_{2}\right)   \label{A43.5}
\end{align}
where 
\begin{equation}
e_{i}=\frac{1}{\delta_{i}^{3}},\text{ \ }f_{i}=\frac{3\sqrt{2}}{%
\delta_{i}^{5}}(\mathbf{p\times B(p))}\cdot\mathbf{(p\times\hat{b}}_{i}) 
\label{A44}
\end{equation}
\qquad To simplify our notation we denote products (monomials) of the
functions $\delta_{i}$ by 
\begin{equation}
\delta_{a,b,c}=\delta_{1}^{a}\delta_{2}^{b}\delta_{3}^{c},   \label{A45}
\end{equation}
where $a,b,c$ are nonnegative integers, and the \emph{alternating}
polynomial generated by the monomial (\ref{A45}) is 
\begin{equation}
A_{a,b,c}=\left\vert 
\begin{array}{ccc}
\delta_{1}^{a} & \delta_{2}^{a} & \delta_{3}^{a} \\ 
\delta_{1}^{b} & \delta_{2}^{b} & \delta_{3}^{b} \\ 
\delta_{1}^{c} & \delta_{2}^{c} & \delta_{3}^{c}%
\end{array}
\right\vert =\sum_{\sigma\in S_{3}}sgn(\sigma)\delta_{a,b,c}^{\sigma } 
\label{A46}
\end{equation}
where $S_{3}$ is the permutation group of $\left\{ \delta_{1},\delta
_{2},\delta_{3}\right\} $ acting on monomials in the obvious way. In
particular, the \emph{basic} \emph{alternating} polynomial is 
\begin{equation}
A=A_{0,1,2}=\left\vert 
\begin{array}{ccc}
1 & 1 & 1 \\ 
\delta_{1} & \delta_{2} & \delta_{3} \\ 
\delta_{1}^{2} & \delta_{2}^{2} & \delta_{3}^{2}%
\end{array}
\right\vert
=(\delta_{1}-\delta_{2})(\delta_{2}-\delta_{3})(\delta_{3}-\delta_{1}) 
\label{A47}
\end{equation}
On the other hand, the \emph{symmetric} function generated by\ $\delta
_{a,b,c}$, where we may assume $a\geq b\geq c$, is the smallest $S_{3}$%
-invariant sum 
\begin{equation}
S_{a,b,c}=\sum\delta_{i}^{a}\delta_{j}^{b}\delta_{k}^{c}=\sum_{\sigma\in
S_{3}}\delta_{a,b,c}^{\sigma}   \label{A48}
\end{equation}
containing $\delta_{a,b,c}$. In particular, $S_{2,0,0}=6,$ by (\ref{A42.1}).

Note that $S_{a,b,c}$ is unchanged, whereas $A_{a,b,c}$ may change sign,
when $a,b,c$ are permuted. The alternating polynomials can be decomposed as
a product of the basic alternating function (\ref{A47}) and a symmetric
function, for example 
\begin{align}
A_{0,2,4} & =(S_{2,1,0}+2S_{1,1,1})A  \notag \\
A_{0,1,5} & =(S_{3,0,0}+S_{2,1,0}+S_{1,1,1})A  \label{A49} \\
A_{0,1,6} & =(S_{4,0,0}+S_{3,1,0}+S_{2,2,0}+S_{2,1,1})A  \notag \\
A_{0,3,6} & =(S_{4,2,0}+S_{4,1,1}+2S_{3,2,1}+3S_{2,2,2})A  \notag
\end{align}

The induced action of $S_{3}$ on the triples $\left\{
e_{1},e_{2},e_{3}\right\} $ and $\left\{ f_{1},f_{2},f_{3}\right\} $ is
covariant with the action on $\left\{
\delta_{1},\delta_{2},\delta_{3}\right\} $. Certainly, the above
coefficients $e_{i}$ in (\ref{A44}) are simple rational functions of the $%
\delta_{j}^{\prime}s$, and now we prove a similar statement for the $%
f_{i}^{\prime}s$, as follows.

\begin{lemma}
As a rational function of $\delta_{1},\delta_{2},\delta_{3}$%
\begin{align*}
f_{1} & =\frac{1}{\delta_{6,6,6}}(-3\delta_{1,3,6}-3\delta_{1,6,3}+2%
\delta_{0,6,6}+\delta_{3,6,3}+\delta_{3,3,6}+\delta_{1,5,6} \\
+ & \delta_{1,6,5}-\frac{1}{2}\delta_{2,6,6}-\frac{1}{2}\delta_{3,5,6}-\frac{%
1}{2}\delta_{3,6,5}),
\end{align*}
and $f_{2}$ (resp. $f_{3})$ is obtained from $f_{1}$ (resp. $f_{2})$ by
cyclic permutation, $a\rightarrow b\rightarrow c\rightarrow a$, of the
indices of each monomial $\delta_{a,b,c}.$
\end{lemma}

\begin{proof}
Use the identities (\ref{A42.2}) and substitute the expression (\ref{A43})
for $\mathbf{B}$ into the formula (\ref{A44}) for $f_{1}$, namely 
\begin{equation*}
f_{1}=\frac{3\sqrt{2}}{\delta_{1}^{5}}\left( \mathbf{B\cdot\hat{b}}_{1}-(%
\mathbf{p\cdot\hat{b}}_{1})(\mathbf{B\cdot p)}\right) 
\end{equation*}
Then one obtains the above rational expression for $f_{1}$ by
straightforward calculations, and by symmetry it is also clear that $f_{2}$
and $f_{3}$ are obtained from $f_{1}$ as claimed.
\end{proof}

Now, turning to the formula (\ref{A38}) for the curvature function $K_{g}$
and inserting the expressions (\ref{A43}), (\ref{A43.5}), we write 
\begin{equation*}
\mathbf{B(p)}\times\mathbf{C(p)\cdot k=}\frac{\sqrt{6}}{54}\tilde{A}, 
\end{equation*}
where $\mathbf{k}$ is the unit normal vector of the xy-plane and $\tilde{A}$
is, by definition, the function 
\begin{equation}
\tilde{A}(\delta_{1},\delta_{2},%
\delta_{3})=(e_{1}f_{2}-e_{2}f_{1})+(e_{2}f_{3}-e_{3}f_{2})+(e_{3}f_{1}-e_{1}f_{3}) 
\label{A50}
\end{equation}
When the products $e_{i}f_{j}$ are calculated using the above lemma, for
example 
\begin{align*}
e_{1}f_{2} & =\frac{1}{\delta_{6,6,6}}(-3\delta_{3,1,3}-3\delta
_{0,1,6}+2\delta_{3,0,6}+\delta_{3,3,3}+\delta_{3,1,5}+\delta_{0,3,6} \\
& +\delta_{2,1,6}-\frac{1}{2}\delta_{3,2,6}-\frac{1}{2}\delta_{3,3,5}-\frac{1%
}{2}\delta_{2,3,6}),
\end{align*}
the expression (\ref{A50}) may be written as 
\begin{equation}
\tilde{A}=-\frac{1}{\delta_{6,6,6}}%
(3A_{0,1,6}+A_{0,3,6}+A_{1,3,5}+A_{1,2,6})   \label{A51}
\end{equation}
Finally, substitution of expressions from (\ref{A49}) into (\ref{A51})
yields 
\begin{equation*}
\tilde{A}(\delta_{1},\delta_{2},\delta_{3})=-\tilde{S}(\delta_{1},\delta
_{2},\delta_{3})A 
\end{equation*}
where 
\begin{align*}
\tilde{S}(\delta_{1},\delta_{2},\delta_{3}) & =\frac{1}{(S_{1,1,1})^{6}}%
[3(S_{4,0,0}+S_{3,1,0}+S_{2,2,0}+S_{2,1,1})+S_{4,2,0}+S_{4,1,1} \\
&
+2S_{3,2,1}+3(S_{1,1,1})^{2}+2S_{1,1,1}S_{2,1,0}+3(S_{1,1,1})^{2}+S_{1,1,1}S_{3,0,0}]
\end{align*}
In summary, we have established the following proposition, where the factor $%
\tilde{S}$ of $K_{g}$ is always positive !

\begin{proposition}
\label{K}The geodesic curvature function of the gradient lines of $U^{\ast}$
on the unit sphere $S^{2}$ is given by the product 
\begin{equation}
K_{g}(\mathbf{p})=-\frac{\sqrt{6}\text{ }z\text{ }\tilde{S}}{18\left| \nabla
U^{\ast}(\mathbf{p})\right| ^{3}}A\text{, \ \ \ }\mathbf{p}=(x,y,z),\text{ }%
z\geq0   \label{A52}
\end{equation}
\end{proposition}

Observe that the $\mathfrak{D}_{3}$-chambers of the (upper) hemisphere of $%
S^{2}$ are defined by inequalities $\delta_{i}\leq\delta_{j}\leq\delta_{k}$,
and therefore $A$, and hence $K_{g}$ as well, has constant sign in each
chamber. For example, the fundamental chamber (\ref{A4}) is given by 
\begin{equation*}
\mathfrak{C}_{0}:\delta_{2}\leq\delta_{1}\leq\delta_{3}
\end{equation*}
and here $K_{g}\geq0$. The meridians which are the walls of the $\mathfrak{D}%
_{3}$-chambers are defined by relations of type $\delta_{i}=\delta_{j}$;
these are the zero set of the function $A$. Together with the equator circle
they are the curves where $K_{g}$ vanishes (or is undefined), and this also
solves Problem \ref{curvK} (in the case of uniform mass distribution).

It is easy to visualize the gradient flow on the 2-sphere. For example, in
the interior of the spherical triangle $\mathfrak{C}_{0}$ the flow has the
vertex $\mathbf{\hat{p}}_{0}$ as source and converges towards the vertex $%
\mathbf{\hat{b}}_{2}$, with positive curvature everywhere. The flow in $%
\mathfrak{C}_{0}$ is illustrated in Figure 9.

\subsubsection{Geometry of the triple collision shape curves}

It is possible to draw qualitative information about the family $\mathfrak{S}%
(\mathbf{\hat{p}}_{0})$ of shape curves by relating it with the geometry of
the gradient flow of $U^{\ast}$. By symmetry it suffices to consider those
curves $\Gamma_{\theta_{0}}^{\ast}$ starting out in the chamber $\mathfrak{C}%
_{0}$, that is, $0\leq\theta_{0}\leq\frac{\pi}{3}$, and we observe that the
boundary meridians $\theta_{0}=0$ and $\theta_{0}=\pi/3$, emanating from the
north pole $\mathbf{\hat{p}}_{0}$ towards the equator, are themselves both
shape curves and gradient lines. So, the question is what one can say about
the shape curves in the interior of $\mathfrak{C}_{0}$?

A rough description goes as follows. In $\mathfrak{C}_{0}$ there are three
different \textquotedblright flows\textquotedblright\ of curves emanating
from $\mathbf{\hat{p}}_{0}$, namely the shape curves, the gradient lines and
the meridians ($\theta$ constant). At each point $\mathbf{p}%
=\Gamma^{\ast}(s_{1}),s_{1}>0$, the shape curve $\Gamma^{\ast}(s),s>s_{1}$,
is \textquotedblright trapped\textquotedblright\ between the gradient line\
and the meridian through $\mathbf{p}$. Being positively curved one may
imagine the shape curves arising by gradually bending the meridians towards
the gradient lines by means of a "force" field directed eastward, see Figure
12.

To be more precise, we shall focus on five properties as stated below. For
this purpose we introduce two angular functions $\beta(s),\gamma(s)$ as
follows. Namely, $\beta$ is the angle between the meridian and $%
\Gamma^{\ast} $, as defined in (\ref{A17}). It is the oriented angle from $%
\frac{\partial }{\partial\varphi}$ to the velocity vector $\mathbf{\tau}%
^{\ast}=\frac{d}{ds}\Gamma^{\ast}$, and $\gamma$ is the angle from $\frac{%
\partial}{\partial\varphi}$ to the gradient vector 
\begin{align*}
\nabla U^{\ast} & =\frac{\partial U^{\ast}}{\partial\varphi}\frac{\partial }{%
\partial\varphi}+\frac{1}{\sin^{2}\varphi}\frac{\partial U^{\ast}}{%
\partial\theta}\frac{\partial}{\partial\theta} \\
& =\left\vert \nabla U^{\ast}\right\vert (\cos\gamma\frac{\partial}{%
\partial\varphi}+\sin\gamma\frac{1}{\sin\varphi}\frac{\partial}{%
\partial\theta})
\end{align*}
We also recall the role of the inclination angle $\alpha(s)$, which is not
directly related to the geometry of the spherical curve $\Gamma^{\ast}$
itself, but to the associated moduli curve $\bar{\Gamma}$. However, the
curvature equation from (\ref{ODE*}) can now be stated as 
\begin{equation}
2U^{\ast}\mathcal{K}_{g}^{\ast}\sin^{2}\alpha=\left\vert \nabla\mathbf{(}%
U^{\ast})\right\vert \cos(\frac{\pi}{2}+\beta-\gamma)   \label{curv3}
\end{equation}
and hence relates all three angles with the curvature of $\Gamma^{\ast}$.

Now, we contend that the following properties are valid for $\Gamma^{\ast}=$ 
$\Gamma_{\theta_{0}}^{\ast},0<\theta_{0}<\frac{\pi}{3}$, at least until $%
\Gamma^{\ast}$ leaves $\mathfrak{C}_{0}$ the first time (but not necessarily
later) :

\begin{itemize}
\item (i) $0<\alpha<\frac{\pi}{2}$, for $s>0.$ In particular, the curve $%
s\rightarrow$ $\Gamma^{\ast}(s)$ has no cusp singularity for $s<\pi/2.$

\item (ii) The spherical coordinates\ $\varphi(s),\theta(s)$ of $%
\Gamma^{\ast }(s)$ are strictly increasing functions of $s.$

\item (iii) $0<\beta\leq\gamma<\frac{\pi}{2}$, for $s>0.$

\item (iv) The geodesic curvature $\mathcal{K}_{g}^{\ast}$ of $\Gamma^{\ast
}(s),s\geq0,$ is nonnegative$.$

\item (v)$\ \Gamma^{\ast}$ leaves the chamber $\mathfrak{C}_{0}$ by crossing
its boundary arc $(\mathbf{\hat{e}}_{3},\mathbf{\hat{b}}_{2})$ on the
equator circle. \ 
\end{itemize}

In order to verify these statements one may proceed as follows. First, note
that $0<\gamma<\pi/2$ follows from the fact, due to (\ref{A8}) and Remark %
\ref{F,G}, that $\frac{\partial U^{\ast}}{\partial\varphi}>0$ and $\frac{%
\partial U^{\ast}}{\partial\theta}>0$ inside $\mathfrak{C}_{0}$, and
moreover,\ the gradient lines emanate from $\mathbf{\hat{p}}_{0}$ with $%
\gamma=0$ and approach the binary collision point $\mathbf{\hat{b}}_{2}$
with $\gamma=\pi/2$ in the limit. This also explains why property (v)
follows from (iii), and using (\ref{A17} and (\ref{curv3}) we also readily
deduce properties (ii) and (iv) from (iii). Thus, we are left with the
statements (i) and (iii), and let us first establish property (iii) (using
property (i) if necessary).

Observe that $\beta\geq0$ by (\ref{A17}) and Remark \ref{epsilon}. But $%
\beta=0$ for some $s$ would imply $\theta^{\prime}=0,\varphi^{\prime}=1$,
and hence by (\ref{A20}) $\mathcal{K}_{g}^{\ast}(s)$ would vanish. However,
with $\mathbf{p}$ in the interior of $\mathfrak{C}_{0}$ we also have $\frac {%
\partial U^{\ast}}{\partial\theta}(\mathbf{p)}\neq0$ and then $\gamma>0$ in
the right side of the identity (\ref{curv3}). Consequently, $\mathcal{K}%
_{g}^{\ast}(s)\neq0$ and this contradiction shows $\beta>0$ must hold. Again
by (\ref{curv3}), $\mathcal{K}_{g}^{\ast}$ is positive as long as $%
\beta<\gamma$, and this certainly holds for small $s$ since 
\begin{equation*}
\mathcal{K}_{g}^{\ast}(0)=d_{0}=\frac{15}{16}\frac{\sin3\theta_{0}}{%
(16a_{0}^{2}+\frac{3}{2})}
\end{equation*}
We claim that $\beta\leq\gamma$ holds (at least) until $\Gamma^{\ast}(s)$
leaves the chamber. To see this, suppose we had $\beta=\gamma$ for $s=s_{1}$%
, $\beta<\gamma$ (resp. $\beta>\gamma)$ for $s<s_{1}$ (resp. $s>s_{1})$ and $%
s$ close to $s_{1}.$ Then $\Gamma^{\ast}$ would be tangent to the gradient
line at $\mathbf{p}=\Gamma^{\ast}(s_{1})$ and is (locally) lying on the
\textquotedblright upper\textquotedblright\ side of it, hence $\mathcal{K}%
_{g}^{\ast}(s_{1})\geq K_{g}(\mathbf{p})>0.$ This contradicts the fact that $%
\mathcal{K}_{g}^{\ast}(s_{1})=0$, by (\ref{curv3}).

\begin{remark}
\label{inc}Property (iii) implies that $U^{\ast}$ increases along the curve $%
\Gamma^{\ast}(s),s\geq0.$ In general, the event $\ \beta=\gamma$ means $%
\Gamma^{\ast}$ is perpendicular to the level curve of $U^{\ast}$, and by (%
\ref{curv3}) this can happen for two reasons, namely i) $\mathcal{K}%
_{g}^{\ast}$ vanishes or ii) $\Gamma^{\ast}$ reaches a cusp. We can rule out
the second case due to property (i), but only up to the first crossing of
the equator.
\end{remark}

Finally, we turn to property (i). From the relations 
\begin{equation*}
\ \text{\ \ }\frac{d\rho}{ds}\frac{ds}{dt}=\frac{d\rho}{dt}>0,\text{ \ }%
\frac{d\rho}{ds}=\rho(s)\cot\alpha,\text{ \ cf. (\ref{rho3}),}
\end{equation*}
we deduce $0\leq\alpha<\pi/2$. In fact, $\alpha=\pi/2$ (and $ds/dt=\infty)$
only at the vertex $\mathbf{\hat{b}}_{2}$, and our claim is that $\alpha=0$
only holds at $\mathbf{\hat{p}}_{0}=$ $\Gamma^{\ast}(0)$.

It is certainly evident from the numerical analysis of the shape curves (cf.
Table 1) that $\alpha>$ $0$ for $s>0$, at least until $\Gamma^{\ast}$ leaves 
$\mathfrak{C}_{0}$. Indeed, using numerical data and a continuity argument
we can establish the uniform lower bound $\alpha\geq0.18$ for $\pi/4\leq
\varphi\leq\pi/2$. However, we shall also explain an alternative and more
qualitative approach to settle the problem.

To show $\alpha>0$ holds, let assume the contrary and recall the geometric
arguments in the setting in Section 7.3, where we regarded $\Gamma^{\ast
}(\sigma)$, $\sigma=s/2$, as a curve on the sphere $S^{2}(1/2)$ and $%
C(\Gamma^{\ast})$ $\subset\bar{M}$ denotes the cone surface with the induced
Euclidean metric (\ref{metric}). The associated moduli curve\ $\bar
{s}%
\rightarrow$ $\bar{\Gamma}(\bar{s})$ lies in this surface, and assuming the
first cusp occurs at $\sigma=$ $\sigma_{1}$ we consider the Euclidean sector 
$0\leq\sigma\leq\sigma_{1}$ bounded by the rays $\sigma=0$ and $\sigma
=\sigma_{1}$. By our assumptions, there is a bijective correspondence $[0,%
\bar{s}_{1}]$ $\longleftrightarrow$ $[0,\sigma_{1}]$ between the arc-length
parametrizations of the moduli curve $\bar{\Gamma}$ and $\Gamma^{\ast}$, and 
$\alpha>0$ for $0<\bar{s}<\bar{s}_{1}$.

The curve $\bar{\Gamma}(\bar{s})$ starts out from the origin and its radial
distance $\rho$ is increasing. It has the positively oriented moving frame $%
\left\{ \mathbf{\tau,\eta}\right\} $ of (\ref{frame4}), where $\mathbf{\tau}%
\ $(resp. $\mathbf{\eta)}$\ is the unit tangent (resp. normal) vector.$%
\mathbf{\ }$By (\ref{geo2}) and a well known formula for the curvature of
curves in the Euclidean plane, the curvature of $\bar{\Gamma}$ in the above
sector can be expressed as 
\begin{equation}
\frac{d\zeta}{d\bar{s}}=\frac{1}{2}\frac{d}{d\mathbf{\eta}}(\ln U)\text{, \
\ \ }\zeta=\sigma+\alpha\text{ , cf. Figure 8}   \label{curv2}
\end{equation}
Towards the point $\bar{\Gamma}(\bar{s}_{1})$ the curve $\bar{\Gamma}$
becomes tangential to the boundary ray $\sigma=\sigma_{1}$, that is, the
angle $\alpha$ decreases to zero. Therefore $\frac{d\sigma}{d\bar{s}}$
vanishes, by (\ref{frame5}), and hence 
\begin{equation*}
\frac{d\zeta}{d\bar{s}}\leq0\text{ \ as }\bar{s}\rightarrow\bar{s}_{1}
\end{equation*}
Moreover, the frame $\left\{ \mathbf{\tau,\eta}\right\} $ approaches $%
\left\{ \frac{\mathbf{\partial}}{\partial\rho}\mathbf{,}\frac{1}{\rho_{1}}%
\frac{\partial}{\partial\sigma}\right\} $ as $\bar{s}\rightarrow\bar{s}_{1}$%
, so we also deduce 
\begin{equation*}
\frac{d}{d\mathbf{\eta}}(\ln U)\rightarrow\frac{u^{\prime}(\sigma)}{u(\sigma
)}|_{\sigma=\sigma_{1}}\leq0 
\end{equation*}
and hence $u(\sigma)=U^{\ast}(\Gamma^{\ast}(\sigma))$ is decreasing at $%
\sigma_{1}$, or possibly $u^{\prime}(\sigma_{1})=0$ and hence $\Gamma^{\ast}$
is tangential to the level curve of $U^{\ast}$ on the sphere. However, $%
U^{\ast}$ is actually increasing towards the point $\Gamma^{\ast}(\sigma
_{1})$ by Remark \ref{inc} (which applies here since there is no cusp for $%
\sigma<\sigma_{1})$. This contradiction rules out any occurrence of cusps
inside $\mathfrak{C}_{0}$.

\subsubsection{Final escape limiting behavior of the shape curves}

As an interesting example, recall the time parametrized meridian solution $%
\Gamma_{0}^{\ast}(t)$ from Section 8.4.2. The first cusp appears at $%
t_{1}\approx10.4$, with colatitude $\varphi_{1}\approx107.5^{\circ}$. The
next cusps occur roughly at times $t_{2}\approx435$, $t_{3}\approx162400$, $%
t_{4}\approx5^{.}10^{7}$, $t_{5}\approx1.24^{.}10^{9}$ (with due regard to
numerical instability) and they appear to be approaching $\mathbf{\hat{e}}%
_{3}$ as a final limit of the shape curve. The behavior of the curve
resembles a damped oscillation converging to its \textquotedblright
stability\textquotedblright\ point $\mathbf{\hat{e}}_{3}$ as $t\rightarrow
\infty$. Let us refer to this limiting behavior as \emph{irregular}, namely
the limit shape $\mathbf{p}$ is reached through converging cusps and $%
\alpha^{\prime}(s)$ has no limit, as in the above example. Such a behavior
at the final escape at infinity is, however, not necessarily related to the
fact that the shape curve is a triple collision curve in the other direction.

Thus, one may consider more generally the limiting behavior of moduli curves 
$\bar{\Gamma}(t)$ of three-body motions as $t\rightarrow\infty$, assuming
the limit shape $\mathbf{p}$ exists. In the irregular case, however, we do
not claim that $\mathbf{p}$ is necessarily a central configuration (that is,
a critical point of $U^{\ast})$, although this is rather likely. On the
other hand, if $\mathbf{p}$ is a central configuration, we claim that it is
an Euler points $\mathbf{\hat{e}}_{i}$, and moreover, $\Gamma^{\ast}$ is not
confined to the equator circle.

We define the final limiting behavior to be \emph{regular }if the final
shape $\mathbf{p}=\Gamma^{\ast}(s_{2})$ is a central configuration, where $%
\alpha(s)>0$ for $s_{2}-\epsilon<s<s_{2}$ and $\alpha(s_{2})=0$. Moreover, $%
\Gamma^{\ast}$ is confined to the equator circle (and hence the 3-body
motion is collinear) if $\mathbf{p=\hat{e}}_{i}$. As in the case of triple
collisions, a useful tool in the study of such limiting behavior is again
the system ODE$^{\ast}$ (\ref{ODE*}), whose solutions are pairs $(\alpha
(s),\Gamma^{\ast}(s))$. Then the fact that $\rho\rightarrow\infty$ as $%
s\rightarrow s_{2}$ is expressed by the divergence of the integral (\ref%
{int2}), and moreover, the system (\ref{ODE*}) itself imposes the condition
that\textbf{\ }$\mathbf{p}$ must be a critical point of $U^{\ast}$.

A closer study of the above regular solutions $(\alpha(s),\Gamma^{\ast}(s))$
near the final limit may proceed in the same way as we studied triple
collision shape curves in Section 8.5. For convenience, let us translate the
arc-length parameter, $s$ $\rightarrow s-s_{2}$, and consider the power
series expansions of $\alpha(s)$ and $\Gamma^{\ast}(s)$ at $s=0$. The
calculations are similar to the triple collision case worked out in Section
8.4.1 and 8.5.1, but this time $\alpha^{\prime}(s)$ converges to the
negative root of the polynomial in (\ref{lines1}), namely

\begin{equation}
a_{0}=-\frac{\sqrt{13}+1}{8}\approx-0.575\,69\text{, \ \ }b_{0}=-\frac {%
\frac{1}{5}\sqrt{1185}+1}{8}\approx-0.\,\allowbreak985\,6   \label{a0b0n}
\end{equation}
Thus, in the collinear (Euler) case with the limit shape $\Gamma^{\ast}(0)$ $%
=$ $\mathbf{\hat{e}}_{3}$, with $\Gamma^{\ast}(s)$ an arc-length
parametrization of the equator circle near $\mathbf{\hat{e}}_{3}$, the
differential equation (\ref{diff1}) in the case $i=2$ has a unique solution $%
\alpha(s)$ with $\alpha(0)=0,$ $\alpha^{\prime}(0)=b_{0}$.

Next, for the final limit shape of Lagrange type we consider the following
(singular) initial conditions 
\begin{equation*}
\Gamma^{\ast}(0)=\mathbf{\hat{p}}_{0}\text{; }\alpha(0)=0,\alpha^{\prime
}(0)=a_{0}\text{ (hence }\alpha(s)>0\text{ for }s<0) 
\end{equation*}
which define a family of analytic solutions $(\alpha(s),\Gamma^{\ast}(s))$
of the system (\ref{ODE*}). The calculations are similar to the triple
collision case, with $a_{0}$ equal to the positive number in (\ref{a0b0}),
worked out in Section 8.5.1. Therefore, we leave it to the reader to modify
these calculations and perhaps establish the same kind of analytic
uniqueness, namely that solutions are parametrized by the terminal angular
(longitude) direction $\theta_{0}$ of $\Gamma^{\ast}(s)$ at $\mathbf{\hat{p}}%
_{0}$, cf. (\ref{A9}). In particular, by reversing the direction of the
curve segment $\{\Gamma^{\ast}(s),s\geq0\}$ one obtains the shape curve with
terminal direction $\theta_{0}+\pi$.

\subsubsection{More about the asymtotic behavior at triple collision}

Finally, we turn to the asymtotic behavior, in terms of the time parameter $%
t $, of a triple collision taking place at $t=0$. Let $\bar{\Gamma}%
(t)=(\rho(t),\Gamma^{\ast}(t))$ be the moduli curve of such a motion, with $%
\Gamma^{\ast}(0)=\mathbf{\hat{p}}_{0}$ or $\mathbf{\hat{e}}_{i}$, and write $%
\mu=U^{\ast}(\Gamma^{\ast}(0))$. Then it is a classical result, dating back
to the work of Sundman and Siegel, that (for any energy level $h$ and mass
distribution) 
\begin{equation}
\rho(t)\sim\kappa t^{2/3}\text{ as\ }t\rightarrow0,\text{ \ \ \ }\kappa =(%
\frac{9}{2}\mu)^{1/3}   \label{asym1}
\end{equation}
and moreover, the total kinetic energy is asymtotically dominated by the
\textquotedblright change of size\textquotedblright, in the sense that 
\begin{equation}
T(t)\sim T^{\rho}=\frac{1}{2}\dot{\rho}(t)^{2}\sim\frac{2}{9}\kappa
^{2}t^{-2/3}\sim\frac{\mu}{\rho(t)}   \label{asym2}
\end{equation}
In particular, the residual kinetic energy 
\begin{equation}
T^{\sigma}=\frac{1}{8}\rho(t)^{2}(\frac{ds}{dt})^{2}=T-T^{\rho} 
\label{asym3}
\end{equation}
due to the \textquotedblright change of shape\textquotedblright\ must be of
lower order in $t$ than that of $T^{\rho}$, in the sense that $T^{\sigma
}/T^{\rho}\rightarrow0$. However, this does not exclude the possibility that 
$T^{\sigma}\rightarrow\infty$ as $t\rightarrow0$. We claim, however, that $%
T^{\sigma}\rightarrow0$, and we shall present the following (rather
heuristic) argument for this.

Using the relationship (\ref{rho3}) between $\rho(s$) and $\alpha(s)$ and
the series expansion of $\alpha(s)$ we obtain an expansion of type 
\begin{equation}
\rho(s)=\rho_{0}s^{p}(1+r_{1}s+r_{2}s^{2}+...)\text{, \ }p=1/2\alpha^{\prime
}(0)  \notag
\end{equation}
Hence, for suitable nonzero constants $\kappa_{0}$ and $\kappa_{s}$ 
\begin{equation*}
s(t)\sim\kappa_{0}\rho(t)^{1/p}=\kappa_{0}\rho(t)^{2\alpha^{\prime}(0)}\sim%
\kappa_{s}t^{4\alpha^{\prime}(0)/3}=\kappa_{s}t^{e}, 
\end{equation*}
where the exponent $e$ depends on the two types of triple collision, namely
by (\ref{a0b0}) 
\begin{align}
(i)\text{ }\Gamma^{\ast}(0) & =\mathbf{\hat{p}}_{0}:e=\frac{4}{3}a_{0}=\frac{%
1}{6}(\sqrt{13}-1)\approx0.\,\allowbreak434\,26\   \label{asym4} \\
(ii)\text{ }\Gamma^{\ast}(0) & =\mathbf{\hat{e}}_{i}:e=\frac{4}{3}b_{0}=%
\frac{1}{6}(\frac{1}{5}\sqrt{1185}-1)\approx0.\,\allowbreak 98079  \notag
\end{align}
Now, as is the case of $\rho(t)$, let us assume differentiation of $s(t)$
also commutes with taking asymtotic limit, namely $\dot{s}(t)\sim e\kappa
_{s}t^{e-1}$. Then by (\ref{asym3}) 
\begin{equation}
T^{\sigma}(t)\sim\frac{1}{8}(\kappa\kappa_{s}e)^{2}t^{4/3+2(e-1)}\
=\kappa_{\sigma}t^{\epsilon},\left\{ 
\begin{array}{c}
\text{Case (i) : }\epsilon\approx\allowbreak0.201\,85 \\ 
\text{Case (ii) : }\epsilon\approx\allowbreak1.\,\allowbreak294\,9%
\end{array}
\right.   \label{asym5}
\end{equation}

\subsection{Numerical solutions of triple collision motions}

We shall describe a modified approach to provide numerical $C^{1}$-data for
the 1-parameter family $\mathfrak{S(}\mathbf{\hat{p}}_{0})$ of shape curves
representing non-collinear triple collision motions, under the standing
assumption of equal masses $m_{i}$ $=1/3$ and zero total energy. Recall that
these curves $\Gamma^{\ast}(s)$ arise from solutions $(\alpha,\Gamma^{\ast})$
of the system (\ref{ODE*}) of ordinary differential equations, where $%
\Gamma^{\ast}=\Gamma_{\theta_{0}}^{\ast}$ starts from the north pole $%
\mathbf{\hat{p}}_{0}$ on the 2-sphere with initial (longitude) direction $%
\theta_{0}$, and the (radial inclination) angle $\alpha\geq0$ has the
initial value $\alpha=0$. The basic idea is to obtain numerical data close
to the initial point, by the analytical method, and use them as the initial
data for the remaining integration by means of a Runge-Kutta method. With
some more efforts we believe it is possible to settle Conjecture \ref%
{conject} by carefully combining numerical analysis and theory along these
lines. The following numerical analysis serves at least to illustrate the
geometry of those shape curves.

\subsubsection{Outline of a numerical approach}

As usual, $(\theta,\varphi)$ are the spherical coordinates of the 2-sphere $%
M^{\ast}\simeq S^{2}$, with $\varphi=0$ at the initial point $\mathbf{\hat
{%
p}}_{0}$. By symmetry (as explained earlier) we need only consider curves
whose initial direction $\theta_{0}$ lies in the interval $0<\theta_{0}<\pi
/3$, and moreover, we may as well use $\varphi$ to parametrize each curve
since $\varphi$ increases with the arc-length $s$. Thus, let 
\begin{equation*}
\Gamma_{\theta_{0}}^{\ast}=\left\{
(\varphi,\theta_{\theta_{0}}(\varphi)),\varphi\geq0;\lim_{\varphi%
\rightarrow0}\theta_{\theta_{0}}(\varphi)=\theta_{0}\right\} 
\end{equation*}
denote the shape curve with initial angle $\theta_{0}$, and let $\alpha
_{\theta_{0}}(\varphi)$ denote the corresponding angle $\alpha$ as a
function of $\varphi$. We shall compute the values of $\theta_{\theta_{0}}$, 
$\frac {d}{d\varphi}\theta_{\theta_{0}}$ and $\alpha_{\theta_{0}}$ for $%
0\leq \varphi\leq\pi/2$ and $0\leq\theta_{0}\leq\pi/3$.

Elimination of the arc-length parameter $s$ in the system (\ref{ODE*}) is
achieved by using the third equation to rewrite the first two equations with 
$\varphi$ as the independent variable. In standard (explicit) form the new
system reads : 
\begin{align}
\alpha^{\prime} & =-\frac{1}{4}\sqrt{1+(\sin^{2}\varphi)\theta^{\prime2}}+%
\frac{\cot\alpha}{2U^{\ast}}(\ \frac{\partial U^{\ast}}{\partial\varphi }+%
\frac{\partial U^{\ast}}{\partial\theta}\theta^{\prime})  \notag \\
\theta^{\prime\prime} & =\frac{1}{2U^{\ast}}[\frac{\partial U^{\ast}}{%
\partial\theta}\csc^{2}\alpha\csc^{2}\varphi-(4U^{\ast}\cot\varphi +\frac{%
\partial U^{\ast}}{\partial\varphi}\csc^{2}\alpha)\theta^{\prime }
\label{ODE3} \\
& +(\frac{\partial U^{\ast}}{\partial\theta}\csc^{2}\alpha)\theta^{\prime
2}-(U^{\ast}\sin2\varphi+\frac{\partial U^{\ast}}{\partial\varphi}\csc
^{2}\alpha\sin^{2}\varphi)\theta^{\prime3}]  \notag
\end{align}
where $\alpha^{\prime},\theta^{\prime},\theta^{\prime\prime}$ means
differentiation with respect to $\varphi$.

We consider the power series expansions at $\varphi=0:$ 
\begin{align}
\theta_{\theta_{0}}(\varphi) & =\theta_{0}+\varphi(f_{0}+f_{1}\varphi
+f_{2}\varphi^{2}+...)  \label{series2} \\
\alpha_{\theta_{0}}(\varphi) & =\varphi(g_{0}+g_{1}\varphi+g_{2}\varphi
^{2}+...)  \notag
\end{align}
whose coefficients can be calculated recursively as functions of $\theta_{0}$
by the method of undetermined coefficients. This is similar to the
calculation of the expansions in (\ref{A9}) using the system $ODE^{\ast}$.
Let 
\begin{equation}
\theta_{\theta_{0}}^{[n]}(\varphi)\text{, \ }\alpha_{\theta_{0}}^{[n]}(%
\varphi)\text{, }\frac{d}{d\varphi}\theta_{\theta_{0}}^{[n]}(\varphi )\text{%
\ }   \label{polynom}
\end{equation}
be the polynomials in $\varphi$ of degree $n+1$ (resp. $n$ for the third
polynomial), where the first two are obtained by substituting the calculated
expressions for\ $f_{i},g_{i}$ (as functions of $\theta_{0}$), $0\leq i\leq
n $, into (\ref{series2}) and truncating higher order terms. The last
polynomial is the derivative of $\theta_{\theta_{0}}^{[n]}(\varphi)$. Then
for sufficiently small $\varphi$ the \textquotedblright
true\textquotedblright\ functions $\theta_{\theta_{0}}(\varphi)$, $%
\alpha_{\theta_{0}}(\varphi)$ and $\frac{d}{d\varphi}\theta_{\theta_{0}}(%
\varphi)=\theta_{\theta_{0}}^{\prime }(\varphi)$ will be closely
approximated by the polynomials (\ref{polynom}), and the approximations can
be made arbitrarily accurate by taking $n$ sufficiently large, i.e. by
computing enough coefficients $f_{i},g_{i}$.

Given a value of $\theta_{0}$ with $0<\theta_{0}<\pi/3$, we fix a small $%
\varphi_{0}$ and compute the polynomials (\ref{polynom}), to be regarded as
approximations of $\theta_{\theta_{0}}(\varphi)$, $\alpha_{\theta_{0}}(%
\varphi)$ and $\theta_{\theta_{0}}^{\prime}(\varphi)$ on the interval $%
0\leq\varphi\leq\varphi_{0}$. In particular, the values of the three
polynomials at $\varphi=\varphi_{0}$ will serve as (approximate) initial
values for the functions $\alpha,\theta$ and $\theta^{\prime}$, whose
further development on the interval $\varphi_{0}\leq\varphi\leq\pi/2$ is
governed by the system (\ref{ODE3}). This allows us to obtain numerical
solutions for $\theta$ and $\alpha$ on this interval, using any of the
standard iterative methods. Pieced together, these data furnish us with the $%
C^{1}$-data of the triple collision shape curves $\Gamma_{\theta_{0}}^{\ast}$
within the interval $0\leq\varphi\leq\pi/2$.

\subsubsection{C$^{1}$-data for a selection of triple collision motions}

As in Section 8.5.3 we perform symbolic computations to calculate
successively the coefficients $f_{i}$ and $g_{i}$ of the expansions (\ref%
{series2}). As before, these are trigonometric polynomials of $3\theta_{0}$,
namely polynomials of $\sin3\theta_{0}$ and $\cos3\theta_{0}$, and we have
calculated the exact expressions for $i\leq9$. Beyond that they tend to be
rather untractable in their exact form. The exact expressions for the first
few coefficients are listed below for the sake of reference : 
\begin{align}
f_{0} & =\frac{5(10+\sqrt{13})}{232}\sin3\theta_{0},\text{ \ }f_{1}=\frac{%
25(113+20\sqrt{13})}{53824}\sin6\theta_{0}\text{ }  \notag \\
\text{\ }f_{2} & =\frac{15(544702206+58374421\sqrt{13})}{201243199488}%
\sin3\theta_{0}  \notag \\
& +\frac{5875625(1390+313\sqrt{13})}{201243199488}\sin3\theta_{0}\cos
6\theta_{0}  \label{table1f} \\
f_{3} & =\frac{2(1909168577687-51730231240\sqrt{13})}{1318947929444352}%
\sin6\theta_{0}  \notag \\
& +\frac{1422740625(17969+4520\sqrt{13})}{1318947929444352}\sin6\theta
_{0}\cos6\theta_{0}  \notag
\end{align}
\begin{align}
\text{ \ }g_{1} & =\frac{-15(1+3\sqrt{13})}{1856}\cos3\theta_{0}\text{, \ } 
\notag \\
g_{2} & =\frac{543509+352091\sqrt{13}}{29280256}+\frac{-8925(49+31\sqrt {13})%
}{29280256}\cos6\theta_{0}  \label{table1g} \\
g_{3} & =\frac{10(253093537+56946315\sqrt{13})}{134162132992}\cos3\theta _{0}
\notag \\
& +\frac{-3525375(893+359\sqrt{13})}{134162132992}\cos3\theta_{0}\cos
6\theta_{0}  \notag
\end{align}
The coefficient $g_{0}$ equals the constant $a_{0}$ in (\ref{roots}) and
hence is omitted here. To illustrate the (decreasing) magnitude of the
coefficients of the trigonometric polynomials $f_{i},g_{i}$ we list a few
approximate expressions 
\begin{align*}
f_{1} & \approx(8.6)10^{-2}\sin6\theta_{0}\text{, \ }f_{3}\approx
(2.6)10^{-3}\sin6\theta_{0}+(3.7)10^{-2}\sin6\theta_{0}\cos6\theta_{0}\text{%
, \ } \\
g_{1} & \approx-(9.6)10^{-2}\cos3\theta_{0}\text{, \ }g_{3}\approx
(3.1)10^{-2}\cos3\theta_{0}-(5.7)10^{-2}\cos3\theta_{0}\cos6\theta_{0}
\end{align*}
In this way one obtains the approximating polynomials (\ref{polynom}) for $%
n=9$, say. Hence, to obtain approximate numerical data for the solutions $%
\Gamma_{\theta_{0}}^{\ast}$ of the system (\ref{ODE3}), we have chosen $%
\varphi_{0}=0.05$ and $\theta_{0}=k\pi/300,k=0,1,...,100$, and we have
computed the numerical solutions using the Runge-Kutta method. These 101
solution curves outline the general behavior of the shape curve $\Gamma
_{\theta_{0}}^{\ast}$ of triple collision motions parametrized by the
initial longitude angle $\theta_{0}$. We refer to Table 1, Table 2 and Table
3 which list the calculated values of $\alpha_{\theta_{0}}(\varphi)$, $%
\theta _{\theta_{0}}(\varphi)$ and $\frac{d}{d\varphi}\theta_{\theta_{0}}(%
\varphi)$ for $\theta_{0}=k\pi/30,k=0,1,..,10$, and for 6 different values
of $\varphi.$ All angles are measured in radians.

\ \ \ \ \ \ \ \ \ \ \ \ \ \ \ \ \ \ 

\qquad\qquad Table 1 : Inclination angle $\alpha_{\theta_{0}}(\varphi)$, for 
$\theta_{0}=k\pi/30$\qquad\allowbreak

\qquad\qquad\qquad\allowbreak\allowbreak

$ 
\begin{tabular}{|l|l|l|l|l|l|l|}
\hline
$\ k\ \setminus\varphi\ \ $ & $\frac{\pi}{4}$ & $\frac{3\pi}{8}$ & $\frac{%
15\pi}{32}$ & $\frac{63\pi}{128}$ & $\frac{255\pi}{512}$ & $\frac{1023\pi}{%
2048}$ \\ \hline
0 & .1947 & .2292 & .2072 & .1926 & .1883 & .1871 \\ \hline
1 & .2132 & .3470 & .5938 & .7243 & .7705 & .7835 \\ \hline
2 & .2464 & .4554 & .7917 & .9694 & 1.0379 & 1.0586 \\ \hline
3 & .2742 & .5170 & .8895 & 1.0914 & 1.1747 & 1.2014 \\ \hline
4 & .2943 & .5542 & .9465 & 1.1643 & 1.2597 & 1.2922 \\ \hline
5 & .3083 & .5780 & .9827 & 1.2121 & 1.3179 & 1.3566 \\ \hline
6 & .3181 & .5938 & 1.0067 & 1.2247 & 1.3599 & 1.4054 \\ \hline
7 & .3248 & .6043 & 1.0226 & 1.267? & 1.3904 & 1.4433 \\ \hline
8 & .3292 & .6110 & 1.0328 & 1.2816 & 1.4116 & 1.4723 \\ \hline
9 & .3317 & .6147 & 1.0385 & 1.2898 & 1.4243 & 1.4917 \\ \hline
10 & .3325 & .6159 & 1.0403 & 1.2925 & 1.4285 & 1.4989 \\ \hline
\end{tabular}
$

\bigskip\newpage

\qquad\qquad Table 2 : Longitude angle $\theta_{\theta_{0}}(\varphi)$, for $%
\theta_{0}=k\pi/30\qquad$

\qquad\qquad\qquad

$ 
\begin{tabular}{|l|l|l|l|l|l|l|}
\hline
$\ k\ \setminus\varphi\ \ $ & $\frac{\pi}{4}$ & $\frac{3\pi}{8}$ & $\frac{%
15\pi}{32}$ & $\frac{63\pi}{128}$ & $\frac{255\pi}{512}$ & $\frac{1023\pi}{%
2048}$ \\ \hline
0 & 0 & 0 & 0 & 0 & 0 & 0 \\ \hline
1 & .23558 & .40978 & .66916 & .77595 & .80954 & .81866 \\ \hline
2 & .42158 & .61693 & .83930 & .92403 & .95180 & .95968 \\ \hline
3 & .55924 & .73215 & .90859 & .97447 & .99697 & 1.0037 \\ \hline
4 & .66448 & .80830 & .94747 & .99907 & 1.0173 & 1.0231 \\ \hline
5 & .74914 & .86481 & .97349 & 1.0137 & 1.0283 & 1.0333 \\ \hline
6 & .82059 & .91025 & .99302 & 1.0236 & 1.0350 & 1.0391 \\ \hline
7 & .88344 & .94901 & 1.0089 & 1.0310 & 1.0394 & 1.0426 \\ \hline
8 & .94075 & .98370 & 1.0226 & 1.0370 & 1.0425 & 1.0448 \\ \hline
9 & .99475 & 1.0160 & 1.0352 & 1.0423 & 1.0450 & 1.0462 \\ \hline
10 & 1.0472 & 1.0472 & 1.0472 & 1.0472 & 1.0472 & 1.0472 \\ \hline
\end{tabular}
$

\qquad\ \ \ \ \ \ \ \ \ \ \ \ \ \newline
\ \ \ \ 

\qquad\qquad Table 3 : Values of $\frac{d}{d\varphi}\theta_{\theta_{0}}(%
\varphi)$, for $\theta_{0}=k\pi/30\qquad$

\qquad\qquad\qquad

$ 
\begin{tabular}{|l|l|l|l|l|l|l|}
\hline
$\ k\ \setminus\varphi\ \ $ & $\frac{\pi}{4}$ & $\frac{3\pi}{8}$ & $\frac{%
15\pi}{32}$ & $\frac{63\pi}{128}$ & $\frac{255\pi}{512}$ & $\frac{1023\pi}{%
2048}$ \\ \hline
0 & 0 & 0 & 0 & 0 & 0 & 0 \\ \hline
1 & .29933 & .63143 & 1.2517 & 1.7182 & 1.9466 & 2.0210 \\ \hline
2 & .40322 & .60768 & .99427 & 1.3887 & 1.6602 & 1.7708 \\ \hline
3 & .39183 & .50135 & .76959 & 1.0998 & 1.3930 & 1.5427 \\ \hline
4 & .34216 & .40251 & .60029 & .87390 & 1.1710 & 1.3623 \\ \hline
5 & .28289 & .31737 & .46599 & .68801 & .97142 & 1.2023 \\ \hline
6 & .22295 & .24306 & .35361 & .52754 & .78004 & 1.0406 \\ \hline
7 & .16476 & .17645 & .25526 & .38365 & .58941 & .85626 \\ \hline
8 & .10857 & .11502 & .16580 & .25043 & .39595 & .62716 \\ \hline
9 & .05389 & .05675 & .08166 & .12369 & .19908 & .33781 \\ \hline
10 & 0 & 0 & 0 & 0 & 0 & 0 \\ \hline
\end{tabular}
$

\subsection{An outlook on the general case}

Finally, let us\ briefly consider the more general case of non-equal masses
and/or non-vanishing total energy $h$. Namely, by Theorem \ref{1994-Th6},
the system (\ref{ODE*}) must be replaced by 
\begin{equation}
\left\{ 
\begin{array}{c}
\frac{d\alpha}{ds}=-\frac{1}{2}+\frac{1}{4}\frac{u(s)}{u(s)+h\rho}\left(
1+2\cot\alpha\frac{d}{d\mathbf{\tau}^{\ast}}\ln(U^{\ast})\right) \medskip \\ 
\mathcal{K}_{g}^{\ast}\sin^{2}\alpha=\ \frac{1}{2}\frac{u(s)}{u(s)+h\rho }\ 
\frac{d}{d\mathbf{\nu}^{\ast}}\ln(U^{\ast})\medskip\mathstrut \\ 
\vspace{0in}\smallskip\strut\nolinebreak1=(\frac{d\varphi}{ds})^{2}+(\sin
^{2}\varphi)(\frac{d\theta}{ds})^{2}\medskip\medskip%
\end{array}
\right.   \label{ODE4}
\end{equation}
where $u(s)=U^{\ast}(\Gamma^{\ast}(s))$ and $\Gamma^{\ast}(s)$ is the
arc-length parametrized shape curve (\ref{A0}) on the unit sphere $S^{2}(1)$%
. Now the size function $\rho$ in the moduli space $\bar{M}$ appears
explicitly, so the system involves the two \textquotedblright
auxiliary\textquotedblright\ functions $\alpha(s),\rho(s)$ which are still
related by (\ref{frame5}) and (\ref{rho}). Their initial value at triple
collision is $\alpha(0)=\rho(0)=0$, and the shape curve $\Gamma^{\ast}$
emanates from the physical center $\mathbf{\hat{p}}_{0}$ $=\Gamma^{\ast}(0)$%
, namely the minimumspoint of $U^{\ast}$ on the northern hemisphere. Due to
the space-time scaling symmetries of the Newtonian equation (cf. Chapter 7),
for $h\neq0$ there are essentially only two cases, $h>0$ and $h<0$, and our
case $h=0$ may be viewed as the limiting case between negative and positive
energies. However, for $h\neq0$ we may scale and assume $h=\pm1$.

We point out the open problem of finding the appropriate version of Theorem
G (or G$_{1})$ in the two cases $h=\pm1$ or when the masses $m_{i}$ are
unequal. For this purpose, it is natural to try first the following two
special cases.

\begin{itemize}
\item $h=\pm1$ and equal masses (i.e., $m_{i}=1/3$). Many results in Section
8.5 still apply and there are symmetries as before, e.g. it suffices to
consider shape curves whose angular direction at the north pole is in the
range $0\leq\theta_{0}\leq\pi/3$. However, the initial value problem is
"essentially" singular, in the sense that the solutions of (\ref{ODE4}) are
singular at $s=0$. For example, if we assume a series expansion 
\begin{equation*}
\alpha=a_{0}s^{q}(1+a_{1}s+..),\text{ \ }\rho=\rho_{0}s^{p}(1+r_{1}s+..) 
\end{equation*}
then we would have $q=1$ and $\rho$ would have the leading exponent $%
p=1/2a_{0}$. Moreover, from the first equation of (\ref{ODE4}) it follows
that $a_{0}$ has the value from (\ref{roots}), but this equation\ also tells
us that $\Gamma^{\ast}(s)$ is singular at $s=0$.

\item $h=0$ and the masses are not equal. The system (\ref{ODE4}) is the
same as (\ref{ODE*}). Note that the mass distribution $\left\{ m_{i}\right\} 
$ affects the system (\ref{ODE4}) solely via the potential function $%
U^{\ast} $, but the trigonometric series development of $U^{\ast}$, similar
to that in Section 8.2, remains to be done for non-equal masses. We also
seek a convenient coordinate system on the sphere near the point $\mathbf{%
\hat{p}}_{0}$. In Section 6.6.1 we actually worked out a series expansion of 
$U^{\ast}$ centered at $\mathbf{\hat{p}}_{0}$, involving coefficients which
are symmetric functions of the masses, but here we rather need its spherical
polar coordinate version.
\end{itemize}

On the other hand, in Chapter 6 there are expressions for $U^{\ast}$ in
terms of spherical polar coordinates $(\varphi,\theta)$ centered at the
north pole $\mathcal{N}$. Therefore, one approach is to find the coordinates 
$(\hat{\varphi},\hat{\theta})$ of $\mathbf{\hat{p}}_{0}$ in this coordinate
system and then use spherical trigonometry to determine the transformation
from $(\varphi,\theta)$ to spherical polar coordinates\ $(\bar{\varphi},\bar{%
\theta})$ centered at\ $\mathbf{\hat{p}}_{0}$. For this purpose, we recall 
\begin{equation}
\cos\hat{\varphi}=\frac{\sqrt{3}\sqrt{\bar{m}}}{\hat{m}}\text{ \ \ \ cf. (%
\ref{z0})}   \label{phys1}
\end{equation}
and the value of $\hat{\theta}$ (which depends on the choice of zero
meridian, $\theta=0$) can be determined from the longitude differences $\hat{%
\omega}_{i}=$ $\pm(\hat{\theta}-\theta_{i})$, with $0\leq\hat{\omega}%
_{i}\leq\pi$, $i=1,2,3$, where $\theta_{i}$ denotes the longitude angle of
the binary collision points $\mathbf{\hat{b}}_{i}$. To this end, consider
the spherical triangle with vertices $\mathcal{N}$, $\mathbf{\hat{p}}_{0}$
and $\mathbf{\hat{b}}_{i}$ and note that $\hat{\omega}_{i}$ is the angle at
the vertex $\mathcal{N}$. The arc opposite to $\mathcal{N}$, connecting $%
\mathbf{\hat{p}}_{0}$ and $\mathbf{\hat{b}}_{i}$, has length 
\begin{equation*}
2\sigma_{i}=2\arccos\sqrt{\frac{I_{i}}{1-m_{i}}}=2\arccos\sqrt{1-\frac{%
\hat
{m}_{i}}{(1-m_{i})\hat{m}}}\text{ \ \ cf. (\ref{intrins6}), (\ref{phys}%
)}
\end{equation*}
and the two other sides have length $\pi/2$ and $\hat{\varphi}$. Hence, by
the spherical cosine law we deduce the formula 
\begin{equation}
\cos\hat{\omega}_{i}=\frac{\cos(2\sigma_{i})}{\sin\hat{\varphi}}=\frac{1-%
\vspace{0in}\frac{2\hat{m}_{i}}{(1-m_{i})\hat{m}}}{\sqrt {1-\frac{3\bar{m}}{%
\hat{m}^{2}}}}   \label{phys2}
\end{equation}

Knowing the expansion of $U^{\ast}$ as a trigonometric series in terms of
the angles $\bar{\varphi},\bar{\theta}$, we believe the initial value
problem at $s=0$ for the system (\ref{ODE*}) can be investigated in the same
way as we did for equal masses. But this time we expect regularity issues to
depend crucially on the mass distribution. \ \ \ \ 

\ \ \ 

\ \ \textbf{NOTE} : The figures belonging to this memoir are available at
the website http://www.math.ntnu.no/\symbol{126}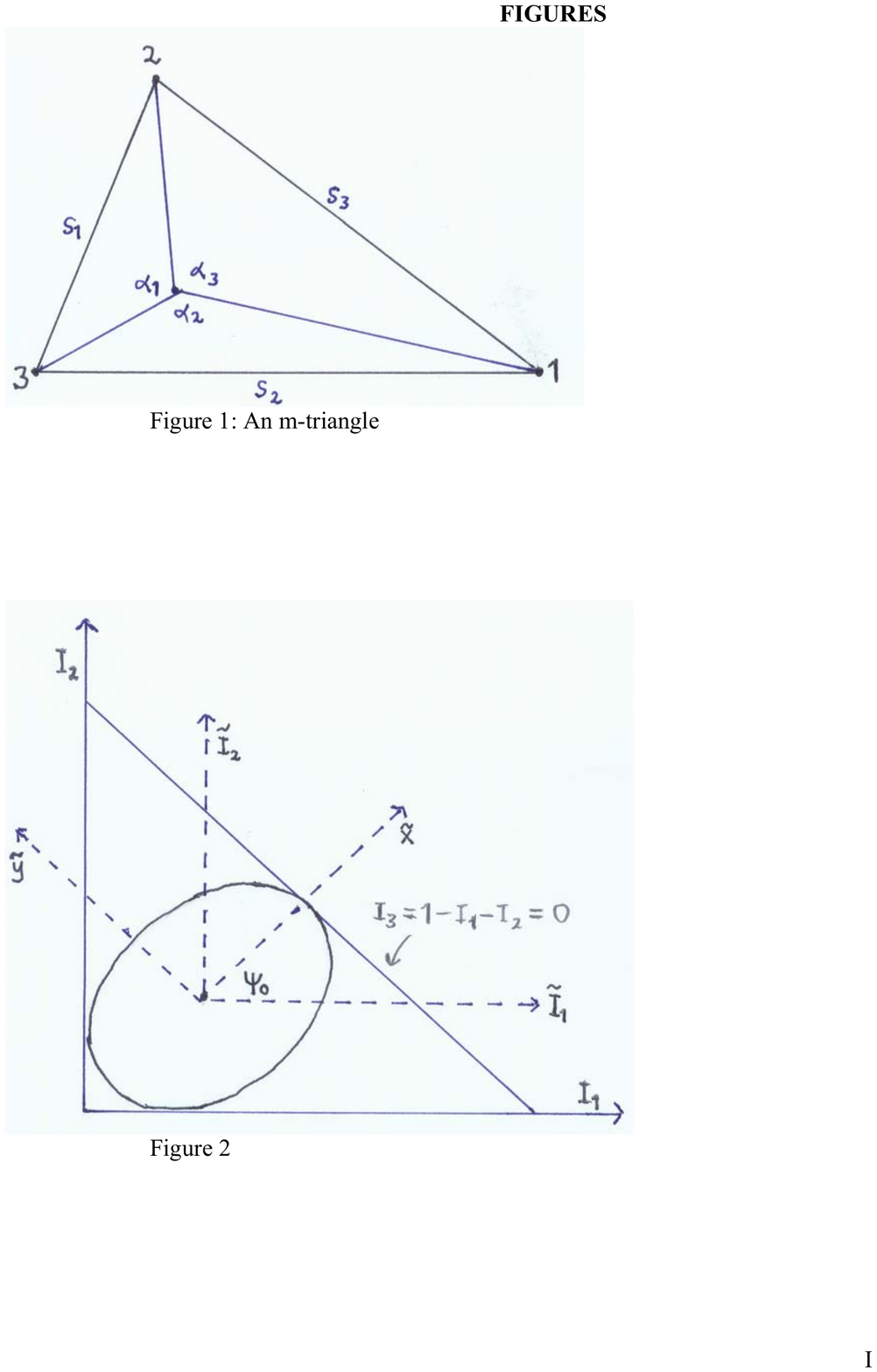

\bigskip

\end{document}